%% file: paper.tex
\documentclass[10pt,letterpaper]{article}
\pdfoutput=1
\usepackage[margin=1in]{geometry}
\usepackage[utf8]{inputenc}
\usepackage[T1]{fontenc}
\usepackage{microtype}
\usepackage{amsfonts}
\usepackage{amssymb}
\usepackage{amsmath}
\usepackage{amsthm}
\usepackage{fullpage}
\usepackage{booktabs}
\usepackage[inline]{enumitem}
\usepackage[hidelinks,bookmarks,bookmarksopen,bookmarksnumbered]{hyperref}
\usepackage[capitalize]{cleveref}
\usepackage{mathtools}
\usepackage{tikz}
\usepackage[font=small,labelfont=bf]{caption}
\usepackage{titlesec}
\usepackage{thmtools}
\usepackage{wrapfig}
\usepackage{tablefootnote}
\usepackage{thm-restate}
\usepackage{soul}
\usepackage[draft]{fixme}
\usepackage[norefs,nocites,nomsgs]{refcheck}  %

\input{settings}

\input{macros}

\begin{document}

\title{Lempel--Ziv (LZ77) Factorization in Sublinear Time}

\author{
  \large Dominik Kempa\thanks{Supported by the
  NSF CAREER Award 2337891 and the
  Simons Foundation Junior Faculty Fellowship.}\\[-0.3ex]
  \normalsize Department of Computer Science,\\[-0.3ex]
  \normalsize Stony Brook University,\\[-0.3ex]
  \normalsize Stony Brook, NY, USA\\[-0.3ex]
  \normalsize \texttt{kempa@cs.stonybrook.edu}
  \and
  \large Tomasz Kociumaka\\[-0.3ex]
  \normalsize Max Planck Institute for Informatics,\\[-0.3ex]
  \normalsize Saarland Informatics Campus,\\[-0.3ex]
  \normalsize Saarbrücken, Germany\\[-0.3ex]
  \normalsize \texttt{tomasz.kociumaka@mpi-inf.mpg.de}
}

\date{\vspace{-0.5cm}}
\maketitle
\input{abstract}

\thispagestyle{empty}
\setcounter{page}{0}
\clearpage

\input{intro}

\input{prelim}

\input{overview}
\input{tools}

\input{index}

\input{lpf}

\input{lz}

\bibliographystyle{alphaurl}
\bibliography{paper}

\end{document}

%% file: macros.tex
\newcommand{\bigO}{\mathcal{O}}
\newcommand{\Oh}{\bigO}
\DeclareMathOperator{\polylog}{polylog}

\DeclareMathOperator*{\argmin}{arg\,min}
\newcommand{\ceil}[1]{\left\lceil #1 \right\rceil}
\newcommand{\floor}[1]{\left\lfloor #1 \right\rfloor}

\newcommand{\dd}{\mathinner{.\,.}}

\newcommand{\Pat}{P}
\newcommand{\Text}{T}
\newcommand{\Textinf}{\Text^{\infty}}
\newcommand{\Textlen}{n}
\newcommand{\AlphabetSize}{\sigma}
\newcommand{\IntegerAlphabet}{[0 \dd \AlphabetSize)}
\newcommand{\emptystring}{\varepsilon}
\newcommand{\BitvectorMin}{B_{\rm min}}
\newcommand{\deltatext}{\delta_{\rm text}}
\newcommand{\SSS}{\mathsf{S}}
\newcommand{\SSSSize}{n'}

\newcommand{\Z}{\mathbb{Z}}
\newcommand{\Zz}{\Z_{\ge 0}}
\newcommand{\Zn}{\Zz}
\newcommand{\Zp}{\Z_{>0}}

\newcommand{\SA}[1]{\mathrm{SA}_{#1}}
\newcommand{\ISA}[1]{\mathrm{ISA}_{#1}}

\newcommand{\LPF}[1]{\mathrm{LPF}_{#1}}
\newcommand{\LPnF}[1]{\mathrm{LPnF}_{#1}}
\newcommand{\LPFMinOcc}[1]{\mathrm{LPFMinOcc}_{#1}}
\newcommand{\LPnFMinOcc}[1]{\mathrm{LPnFMinOcc}_{#1}}

\newcommand{\SuffixTreeSym}{\mathcal{T}_{\rm st}}
\newcommand{\SuffixTree}[1]{\SuffixTreeSym(#1)}
\newcommand{\SDepth}[1]{\mathrm{sdepth}(#1)}
\newcommand{\Index}[1]{\mathrm{index}(#1)}
\newcommand{\Str}[1]{\mathrm{str}(#1)}
\newcommand{\Repr}[1]{\mathrm{repr}(#1)}
\newcommand{\MinOcc}[1]{\mathrm{minocc}(#1)}

\newcommand{\LCE}[3]{\mathrm{LCE}_{#1}(#2,#3)}
\newcommand{\lcp}[2]{\mathrm{lcp}(#1,#2)}
\newcommand{\per}{\mathrm{per}}
\newcommand{\revstr}[1]{\overline{#1}}
\newcommand{\Successor}[2]{\mathrm{succ}_{#1}(#2)}
\newcommand{\PowStr}[2]{\mathrm{Pow}(#1, #2)}
\newcommand{\Int}[3]{\mathrm{int}(#1,#2,#3)}
\newcommand{\EncodeSeq}{\mathrm{enc}}

\newcommand{\LZSizeSym}{z}
\newcommand{\LZNonOvSizeSym}{z_{\rm no}}

\newcommand{\LZSize}[1]{\LZSizeSym(#1)}
\newcommand{\LZNonOvSize}[1]{\LZNonOvSizeSym(#1)}

\newcommand{\OccTwo}[2]{\mathrm{Occ}(#1, #2)}
\newcommand{\RangeBegTwo}[2]{\mathrm{RangeBeg}(#1, #2)}
\newcommand{\RangeEndTwo}[2]{\mathrm{RangeEnd}(#1, #2)}

\newcommand{\DistPrefixPat}[4]{\mathrm{DistPrefix}(#1,#2,#3,#4)}
\newcommand{\DistPrefixPos}[4]{\mathrm{DistPrefix}(#1,#2,#3,#4)}
\newcommand{\DistPrefixes}[3]{\mathcal{D}(#1, #2, #3)}

\newcommand{\RootPos}[3]{\mathrm{root}(#1,#2,#3)}
\newcommand{\RootPat}[2]{\mathrm{root}(#1,#2)}
\newcommand{\HeadPos}[3]{\mathrm{head}(#1,#2,#3)}
\newcommand{\HeadPat}[2]{\mathrm{head}(#1,#2)}
\newcommand{\TailPos}[3]{\mathrm{tail}(#1,#2,#3)}
\newcommand{\TailPat}[2]{\mathrm{tail}(#1,#2)}
\newcommand{\ExpPos}[3]{\mathrm{exp}(#1,#2,#3)}
\newcommand{\ExpPat}[2]{\mathrm{exp}(#1,#2)}
\newcommand{\TypePos}[3]{\mathrm{type}(#1,#2,#3)}
\newcommand{\TypePat}[2]{\mathrm{type}(#1,#2)}
\newcommand{\RunEndFullPos}[3]{e^{\rm full}(#1,#2,#3)}
\newcommand{\RunEndFullPat}[2]{e^{\rm full}(#1,#2)}
\newcommand{\RunEndPos}[3]{e(#1,#2,#3)}
\newcommand{\RunEndPat}[2]{e(#1,#2)}
\newcommand{\RunMinEndPos}[3]{e_{\rm min}(#1,#2,#3)}

\newcommand{\RName}{\mathsf{R}}
\newcommand{\RMinusName}{\RName^{-}}
\newcommand{\RPlusName}{\RName^{+}}
\newcommand{\RTwo}[2]{\RName(#1, #2)}
\newcommand{\RThree}[3]{\RName_{#1}(#2, #3)}
\newcommand{\RFour}[4]{\RName_{#1,#2}(#3, #4)}
\newcommand{\RFive}[5]{\RName_{#1,#2,#3}(#4, #5)}
\newcommand{\RMinusTwo}[2]{\RMinusName(#1, #2)}
\newcommand{\RMinusThree}[3]{\RMinusName_{#1}(#2, #3)}
\newcommand{\RMinusFour}[4]{\RMinusName_{#1,#2}(#3, #4)}
\newcommand{\RMinusFive}[5]{\RMinusName_{#1,#2,#3}(#4, #5)}
\newcommand{\RPlusTwo}[2]{\RPlusName(#1, #2)}
\newcommand{\RPlusThree}[3]{\RPlusName_{#1}(#2, #3)}
\newcommand{\RPlusFour}[4]{\RPlusName_{#1,#2}(#3, #4)}
\newcommand{\RPlusFive}[5]{\RPlusName_{#1,#2,#3}(#4, #5)}

\newcommand{\RPrimName}{\RName'}
\newcommand{\RPrimMinusName}{\RPrimName^{-}}
\newcommand{\RPrimPlusName}{\RPrimName^{+}}
\newcommand{\RPrimTwo}[2]{\RPrimName(#1, #2)}
\newcommand{\RPrimMinusTwo}[2]{\RPrimMinusName(#1, #2)}
\newcommand{\RPrimPlusTwo}[2]{\RPrimPlusName(#1, #2)}
\newcommand{\RPrimMinusThree}[3]{\RPrimMinusName_{#1}(#2, #3)}
\newcommand{\RPrimPlusThree}[3]{\RPrimPlusName_{#1}(#2, #3)}
\newcommand{\RMinName}{\RName_{\rm min}}
\newcommand{\RMinMinusTwo}[2]{\RMinName^{-}(#1, #2)}

\newcommand{\Rank}[3]{\mathsf{rank}_{#1,#2}(#3)}
\newcommand{\Select}[3]{\mathsf{select}_{#1,#2}(#3)}
\newcommand{\RMQ}[3]{\mathsf{rmq}_{#1}(#2, #3)}
\newcommand{\PrefixRank}[3]{\mathsf{prefix\mbox{-}rank}_{#1}(#2,#3)}
\newcommand{\PrefixSelect}[3]{\mathsf{prefix\mbox{-}select}_{#1}(#2,#3)}
\newcommand{\PrefixRMQ}[5]{\mathsf{prefix\mbox{-}rmq}_{#1, #2}(#3, #4, #5)}
\newcommand{\TwoSidedRangeCount}[3]{\mathsf{two\mbox{-}sided\mbox{-}rcount}_{#1}(#2, #3)}
\newcommand{\ThreeSidedRangeCount}[4]{\mathsf{three\mbox{-}sided\mbox{-}rcount}_{#1}(#2, #3, #4)}
\newcommand{\RangeSelect}[3]{\mathsf{rselect}_{#1}(#2, #3)}
\newcommand{\ThreeSidedRMQ}[5]{\mathsf{three\mbox{-}sided\mbox{-}rmq}_{#1,#2}(#3, #4, #5)}

\newcommand{\InsertSubseq}[2]{\mathsf{insert}(#1,#2)}
\newcommand{\DeletePos}[2]{\mathsf{delete}(#1,#2)}
\newcommand{\DeleteSubseq}[2]{\mathsf{delete}(#1,#2)}

\newcommand{\NavCore}[1]{\mathrm{NavCore}(#1)}
\newcommand{\NavNonperiodic}[2]{\mathrm{NavNonperiodic}(#1, #2)}
\newcommand{\NavPeriodic}[1]{\mathrm{NavPeriodic}(#1)}
\newcommand{\MinOccIndexCore}[1]{\mathrm{MinOccIndexCore}(#1)}
\newcommand{\MinOccIndexNonperiodic}[1]{\mathrm{MinOccIndexNonperiodic}(#1)}
\newcommand{\MinOccIndexPeriodic}[1]{\mathrm{MinOccIndexPeriodic}(#1)}
\newcommand{\MinOccIndex}[1]{\mathrm{MinOccIndex}(#1)}

\newcommand{\LexSorted}[2]{\mathrm{LexSorted}(#1,#2)}
\newcommand{\RunsMinusLexSortedTwo}[2]{\mathrm{RunsLexSorted}^{-}(#1,#2)}
\newcommand{\RunsMinusLexSortedThree}[3]{\mathrm{RunsLexSorted}^{-}_{#1}(#2,#3)}
\newcommand{\RunsMinusTextSortedTwo}[2]{\mathrm{RunsTextSorted}^{-}(#1,#2)}
\newcommand{\RunsMinusTextSortedThree}[3]{\mathrm{RunsTextSorted}^{-}_{#1}(#2,#3)}

\newcommand{\MinPosBitvectorMinusTwo}[2]{\mathrm{MinPosBitvector}^{-}(#1,#2)}
\newcommand{\MinPosBitvectorMinusFour}[4]{\mathrm{MinPosBitvector}^{-}_{#1,#2}(#3,#4)}
\newcommand{\MinPosBitvectorMinusFive}[5]{\mathrm{MinPosBitvector}^{-}_{#1,#2,#3}(#4,#5)}

%% file: abstract.tex
\begin{abstract}
  Lempel--Ziv (LZ77) factorization is a fundamental problem in string
  processing: Greedily partition a given string $\Text$ from left to right
  into blocks (called \emph{phrases}) so that each phrase is either the leftmost occurrence
  of a single letter or the longest prefix of the unprocessed suffix
  that has another occurrence earlier in the text.
  This simple routine has numerous applications.
  Most importantly, the LZ77 factorization is the central component and the computational
  bottleneck of most existing compression algorithms (utilized in
  formats like ${\tt zip}$, ${\tt pdf}$, and ${\tt png}$). 
  LZ77 is also  a widely used algorithmic tool for the detection of
  repetitions and periodicities in strings, and the centerpiece of many
  powerful compressed indexes that enable computation directly over compressed data.
  LZ77 factorization is one of the
  most studied problems in string processing. In the 47 years since
  its inception, numerous efficient algorithms were
  developed for different models of computation, including parallel, GPU,
  external-memory, and quantum. Remarkably, however, the
  complexity of the most basic problem is still not settled: All
  existing algorithms in the RAM model run in $\Omega(\Textlen)$ time, which is a
  $\Theta(\log n)$ factor away from the lower bound of
  $\Omega(\Textlen / \log \Textlen)$ (following simply from the necessity to read the entire input, which takes $\Theta(\Textlen / \log \Textlen)$ space for any $\Text \in \{{\tt 0}, {\tt 1}\}^{\Textlen}$).
  Sublinear-time algorithms are known for nearly all other fundamental problems on strings, but LZ77 seems resistant to all currently known
  techniques.
  
  We present the first $o(\Textlen)$-time algorithm for constructing the LZ77
  factorization, breaking the linear-time barrier present
  for nearly 50 years.
  More precisely, we show that, in the standard RAM model, it is
  possible to compute the LZ77 factorization of a given length-$n$
  string $\Text \in \{{\tt 0}, {\tt 1}\}^{\Textlen}$ in
  $\bigO(\Textlen / \sqrt{\log \Textlen}) = o(\Textlen)$ time and using the
  optimal $\bigO(\Textlen / \log \Textlen)$ working space. Our
  algorithm generalizes to larger alphabets $\Sigma = \IntegerAlphabet$, where
  $\AlphabetSize = \Textlen^{\bigO(1)}$.
  The runtime and working space then become $\bigO((\Textlen \log
  \AlphabetSize) / \sqrt{\log \Textlen})$ and $\bigO(\Textlen /
  \log_{\AlphabetSize} \Textlen)$, respectively.  To achieve this sublinear-time LZ77
  algorithm, we prove a more general result: We show that, for any
  constant $\epsilon \in (0, 1)$ and string $\Text \in
  \IntegerAlphabet^{\Textlen}$, in $\bigO((\Textlen \log
  \AlphabetSize) / \sqrt{\log \Textlen})$ time and using
  $\bigO(\Textlen / \log_{\AlphabetSize} \Textlen)$ working space,
  we can  construct an index of optimal size $\bigO(\Textlen /
  \log_{\AlphabetSize} \Textlen)$ that, given any substring $\Pat =
  \Text[j \dd j + \ell)$ specified with a pair $(j,\ell)$, computes
  the leftmost occurrence of $\Pat$ in $\Text$ in
  $\bigO(\log^{\epsilon} \Textlen)$ time. In other words, we solve
  the indexing/online variant of the LZ77 problem, where we can
  efficiently query the phrase length starting at any position.
  Our solution is based on a new type
  of queries that we call \emph{prefix range minimum queries} or \emph{prefix RMQ}. 
  After developing  an efficient solution for these queries, we provide a general
  reduction showing that any new tradeoff for the prefix RMQ implies
  a new tradeoff for an index finding leftmost occurrences
  (and hence a new LZ77 factorization algorithm).
\end{abstract}

%% file: intro.tex
\section{Introduction}\label{sec:intro}

The Lempel--Ziv (LZ77) factorization~\cite{LZ76,LZ77} is one of the most fundamental concepts in data
compression.
In this method, we partition the input string $\Text$ into a
sequence of blocks $\Text = f_1 f_2 \cdots f_z$. Each block, called a
``\emph{phrase}'', is either \begin{enumerate*}[label=(\alph*)]
\item\label{it:firstoccletter}the first occurrence of a letter or 
\item\label{it:prevocc}a substring that has an earlier occurrence in $\Text$. 
\end{enumerate*}
We then encode each phrase
$f_j$ either explicitly (in case~\ref{it:firstoccletter}) or as a pair $(\ell, i)$,
where $\ell = |f_j|$ and $i$ is the position of an earlier
occurrence of $f_j$ (in case~\ref{it:prevocc}).
Such representation needs $\bigO(z)$ space and the greedy approach,
where $\Text$ is decomposed from left to right into the longest possible phrases,
has been shown to minimize the number of phrases $z$~\cite[Theorem~1]{LZ76}.

Due to its excellent practical performance, strong theoretical
guarantees, and numerous applications, the above algorithm went on to
become one of the most widely used compression methods.
In 2004, LZ77 was named the IEEE Milestone~\cite{milestone},
and in 2021 Jacob Ziv was awarded the IEEE Medal of Honor~\cite{medal} (the highest
IEEE recognition) for his work on LZ77 and its variant
LZ78~\cite{LZ78}. 
Below, we list some of the applications of LZ77.

\begin{itemize}
\item LZ77 is the most common compression method: 57 out of 207
  compressors in the Large Text Compression Benchmark~\cite{ltcb} use it as their main algorithm.
  It is used in
  ${\tt png}$, ${\tt pdf}$, ${\tt zip}$, ${\tt xz}$, ${\tt 7z}$,
  ${\tt gz}$, and ${\tt arj}$ formats (to name a few), and in virtually all modern web
  browsers and web servers~\cite{brotli}.

\item LZ77 underlies
  compressed text indexes supporting
  random access~\cite{Rytter03,BLRSRW15,balancing,blocktree,KempaS22},
  longest common extension (LCE) queries~\cite{NishimotoMFCS,tomohiro-lce,dynstr,KempaS22},
  rank and select queries~\cite{PereiraNB17,Prezza19,blocktree}, pattern
  matching~\cite{GagieGKNP12,KreftN13,GagieGKNP14,hybrid,Valenzuela16,BilleEGV18,NishimotoIIBT20,ChristiansenEKN21,%
  KociumakaNO22}, and suffix array functionality~\cite{collapsing};\footnote{Some
    of the indexes use the closely related notion of grammars~\cite{charikar} or
    string attractors~\cite{attractors}, instead of building directly on LZ77.
    However, since computing the smallest grammar and the smaller string attractor is NP-complete~\cite{charikar,attractors},
    the LZ77-based approximations, such as~\cite{charikar,Rytter03,Jez16}, are used in most cases.}
  see~\cite{NavarroMeasures,NavarroIndexes} for a recent survey.

\item LZ77, together with the closely related grammar
  compression~\cite{charikar}, is the framework
  of algorithms operating directly on compressed data to solve many central problems, including
  longest common subsequence and edit distance~\cite{HermelinLLW13,Tiskin15,Gawrychowski12},
  Hamming distance~\cite{AbboudBBK17,GaneshKLS22}, exact~\cite{Gaw11,Jez2015,GanardiG22} and
  approximate pattern matching~\cite{GagieGP15,BringmannWK19,CKW20,Charalampopoulos22},
  and matrix-vector multiplication~\cite{FerraginaMGKNST22}.

\item LZ77 is one of the most widely used measures of
  repetitiveness~\cite{NavarroIndexes,NavarroMeasures}, and it comes with solid
  mathematical foundations: As shown
  in~\cite{charikar,Rytter03,GNPlatin18,attractors,resolution,KempaS22,delta}, LZ77 is up to logarithmic factors
  equivalent to grammar compression~\cite{charikar}, LZ-End~\cite{kreft2010navarro},
  run-length-encoded Burrows--Wheeler Transform~\cite{bwt}, macro schemes~\cite{macro}, collage
  systems~\cite{collage}, string attractors~\cite{attractors}, and
  substring complexity~\cite{delta}. However, whilst many of
  the those measures
  are NP-hard to optimize~\cite{charikar,macro,collage,attractors}, LZ77 can be constructed in linear time~\cite{RodehPE81}.
  Moreover, it is one of the smallest measures in practice~\cite{kreft2010navarro,NavarroIndexes}.

\item LZ77 factorization is the central tool used for
  efficient detection of regularities in strings: repetitions~\cite{Crochemore86},
  runs (maximal repeats)~\cite{main1989detecting,KolpakovK99,ChenPS07,CrochemoreI08},
  repeats with a fixed gap~\cite{KolpakovK00}, approximate repetitions~\cite{KolpakovK03},
  tandem repeats~\cite{GusfieldS04}, sequence alignments~\cite{CrochemoreLZ02},
  local periods~\cite{DuvalKKLL04}, and seeds~\cite{KociumakaKRRW20}.
  These regularities, in turn, have applications in bioinformatics,
  data mining, and combinatorics; see~\cite{gusfield,CrochemoreIR09,Al-HafeedhCIKSTY12}.
\end{itemize}

In nearly all applications above, finding the LZ77 factorization is the \ul{computational bottleneck}. This applies in
data compression~\cite{brotli,ZuH14}, detection of repetitions~\cite{ChenPS07,Al-HafeedhCIKSTY12}, as well as
index construction: given the LZ77 factorization of the input text, recent
algorithms~\cite{resolution,collapsing} can construct text indexes in \emph{compressed time}
(i.e., $\bigO(z \polylog \Textlen)$, where $\Textlen = |\Text|$), which for highly repetitive texts is orders of magnitude
smaller than the original (uncompressed) text~\cite{NavarroIndexes}. Thus,
LZ77 factorization is the dominant step.

Algorithms for efficient LZ77 factorization are known in nearly all models of computation, 
including parallel~\cite{Naor91,CrochemoreR91,FarachM95,KleinW05,ShunZ13,HanLN22},
GPU~\cite{OzsoyS11,OzsoySC14,ZuH14},
external memory~\cite{KarkkainenKP14,KosolobovVNP20}, and
quantum~\cite{quantumlz} models.
LZ77 factorization has also been studied in the dynamic setting~\cite{NishimotoIIBT20} and for general (non-integer) alphabets,
where the known bounds for factorizing the length-$\Textlen$ string with $\AlphabetSize$ distinct characters are
$\Theta(\Textlen \log \AlphabetSize)$ symbol comparisons (for ordered alphabets)~\cite{Kosolobov15b} or
$\Theta(\Textlen \AlphabetSize)$ symbol equality tests (for unordered alphabets)~\cite{EllertGG23}.

In this paper, we focus on the most fundamental and most studied variant, i.e., LZ77 factorization in the static setting in the standard RAM model~\cite{Hagerup98} with the word size $w \ge \log \Textlen$.
In this model, the input text $\Text \in \IntegerAlphabet^{\Textlen}$ of length $\Textlen$ over integer alphabet
$\Sigma = \IntegerAlphabet$ is represented
using $\Textlen \log \AlphabetSize$ bits, or $\Theta(\Textlen / \log_{\AlphabetSize} \Textlen)$ machine
words.\footnote{Unless indicated otherwise, we measure the space in machine words.}
The trivial lower bound for the runtime in this model, following
from the necessity to read the input, is $\Omega(\Textlen / \log_{\AlphabetSize} \Textlen)$.
Since the number of LZ77 phrases satisfies $z = \bigO(\Textlen / \log_{\AlphabetSize} \Textlen)$ for every
text $\Text \in \IntegerAlphabet^{\Textlen}$~\cite[Theorem~2]{LZ76}, an algorithm running in
$\bigO(\Textlen / \log_{\AlphabetSize} \Textlen)$ is hypothetically plausible.

The first efficient algorithm for constructing the LZ77 factorization was proposed in 1981~\cite{RodehPE81}.
The algorithm is based on suffix trees~\cite{Weiner73} and achieves
$\bigO(\Textlen \log \AlphabetSize)$ time and $\bigO(\Textlen)$ space.
The first $\bigO(\Textlen)$-time algorithm (independent of the alphabet size) was given in~\cite{CrochemoreI08}.
Numerous other linear or near linear-time algorithms using $\Theta(\Textlen)$ space in the worst case
followed~\cite{OhlebuschG11,KempaP13,KarkkainenKP13b,GotoB13,GotoB14,FischerIK15,LiuNCW16,Hong0B23},
aiming to reduce the runtime or space usage in practice.
The first algorithm to reduce the space complexity achieved
$\bigO(\Textlen \log^3 \Textlen)$ time in the optimal
$\bigO(\Textlen/\log_{\AlphabetSize} \Textlen)$ space~\cite{OkanoharaS08}. 
Subsequent works lowered the time (while keeping the $\bigO(\Textlen / \log_{\AlphabetSize} \Textlen)$ space) to
$\bigO(\Textlen \log^2 \Textlen)$~\cite{Starikovskaya12},
$\bigO(\Textlen \log \Textlen \cdot (1 + \tfrac{\log \AlphabetSize}{(\log \log \Textlen)^2}))$~\cite{KreftN13},
$\bigO(\Textlen \log \Textlen)$ \cite{KarkkainenKP13,YamamotoIBIT14,PolicritiP15},
$\bigO(\Textlen (\log \AlphabetSize + \log \log \Textlen))$~\cite{Kosolobov15},
$\bigO(\Textlen \log \AlphabetSize)$~\cite{OhlebuschG11},
$\bigO(\Textlen \log \log \AlphabetSize)$~\cite{BelazzouguiP16},
randomized $\bigO(\Textlen)$~\cite{BelazzouguiP16},
and finally, by combining~\cite{KopplS16} and~\cite{MunroNN17}, to deterministic $\bigO(\Textlen)$ time.
There also exist algorithms whose runtime depends on $z$:
The procedures in~\cite{Kempa19,Ellert23}
achieve the time complexity of $\bigO(\Textlen / \log_{\AlphabetSize} \Textlen + z \polylog \Textlen)$.\footnote{The complexity
of the algorithm in~\cite{Kempa19} is originally stated as
$\bigO(\Textlen / \log_{\AlphabetSize} \Textlen + z \polylog \Textlen + r \polylog \Textlen)$, where $r$ is the
number of equal-letter runs in the Burrows--Wheeler transform (BWT)~\cite{bwt} of $\Text$, but this can be
simplified to $\bigO(\Textlen / \log_{\AlphabetSize} \Textlen + z \polylog \Textlen)$ due to the more recent upper bound $r = \bigO(z \log^{2} \Textlen)$~\cite{resolution}.}
For sufficiently small $z$,
this is $\bigO(\Textlen / \log_{\AlphabetSize} \Textlen)$, but
these algorithms still require $\Omega(\Textlen)$ time in the worst case.
Summing up, all prior algorithms to compute the LZ77 factorization
need $\Omega(\Textlen)$ time in the worst case.
Given the fundamental role of LZ77, we thus ask:
\[
   \text{\emph{Can we compute the LZ77 factorization of a string
   $\Text \in \IntegerAlphabet^{\Textlen}$ faster than in $\bigO(\Textlen)$ time?}}
\]

\paragraph{Our Results}
After nearly 50 years since the invention of LZ77, we present the first algorithm to compute the LZ77 factorization in $o(\Textlen)$ time.
For a binary alphabet ($\AlphabetSize = 2$),
our algorithm runs in $\bigO(\Textlen / \sqrt{\log \Textlen})$ time and uses the optimal $\bigO(\Textlen / \log \Textlen)$
space. For an integer alphabet $\Sigma = \IntegerAlphabet$, it runs in
$\bigO((\Textlen \log \AlphabetSize) / \sqrt{\log \Textlen})$ time and uses
$\bigO(\Textlen / \log_{\AlphabetSize} \Textlen)$ space.
We obtain the same complexities for a variant of LZ77 that
prohibits overlaps between phrases and their previous occurrences.\footnote{The number of phrases $z_{\rm no}$ in this variant
satisfies $z_{\rm no} = \bigO(\Textlen / \log_{\AlphabetSize} \Textlen)$; see \cref{th:lz-nonov-size}.} (This variant is sometimes
preferred in practice since it simplifies the usage of the factorization.)
All our
algorithms are deterministic.

\begin{theorem}[LZ77 factorization, \cref{sec:lz}]\label{th:main-lz-theorem}
  Given the $\bigO(\Textlen / \log_{\AlphabetSize} \Textlen)$-space representation of a text
  $\Text \in \IntegerAlphabet^{\Textlen}$, %
  the overlapping and non-overlapping LZ77 factorization of $\Text$ can be constructed
  in $\bigO((\Textlen \log \AlphabetSize) / \sqrt{\log \Textlen})$ time and
  $\bigO(\Textlen / \log_{\AlphabetSize} \Textlen)$ working space.
\end{theorem}

We achieve this result as a simple corollary of a much more general tool that we develop.
Namely, we propose the first index with sublinear construction
that can quickly locate the leftmost occurrences of substrings of $\Text$. More precisely, we show (in \cref{sec:minocc-index})
that, given any constant $\epsilon \in (0, 1)$
and the $\bigO(\Textlen / \log_{\AlphabetSize} \Textlen)$-space representation of $\Text \in \IntegerAlphabet^{\Textlen}$,
where $2 \leq \AlphabetSize < \Textlen^{1/7}$, in $\bigO((\Textlen \log \AlphabetSize) / \sqrt{\log \Textlen})$ time\footnote{We actually achieve
a slightly better time of $\bigO(\Textlen \min(1, \log \AlphabetSize / \sqrt{\log \Textlen}))$, but for simplicity we use the basic bound. If $\AlphabetSize \ge \Textlen^{1/7}$, \cref{th:main-lz-theorem} follows from standard linear-time solutions~\cite{CrochemoreI08,CrochemoreT11} because $\log \AlphabetSize = \Theta(\log \Textlen)$.} and
$\bigO(\Textlen / \log_{\AlphabetSize} \Textlen)$ working space, we can construct an index that, for any position
$j \in [1 \dd \Textlen]$ and any length $\ell \in [1 \dd \Textlen+1-j]$,\footnote{For $i,j\in \mathbb{Z}$, denote $[i\dd j] =
\{k \in \mathbb{Z} : i \le k \le j\}$, $[i
\dd j)=\{k \in \mathbb{Z} : i \le k <
j\}$, and $(i\dd j]={\{k \in \mathbb{Z}: i
< k \le j\}}$.} in
$\bigO(\log^{\epsilon} \Textlen)$ time returns the position
$\min \OccTwo{\Pat}{\Text}$ for $\Pat = \Text[j \dd j + \ell)$, where
$\OccTwo{\Pat}{\Text} = \{i \in [1 \dd \Textlen] : i + |\Pat| \leq \Textlen + 1\text{ and }\Text[i \dd i + |\Pat|) = \Pat\}$
consists of the starting positions of the occurrences of $\Pat$ in $\Text$.
Our index also works for \emph{explicit} patterns: given the $\bigO(m / \log_{\AlphabetSize} \Textlen)$-space representation of
any pattern $\Pat \in \IntegerAlphabet^{m}$ satisfying $\OccTwo{\Pat}{\Text} \neq \emptyset$, we can in
$\bigO(\log^{\epsilon} \Textlen + m / \log_{\AlphabetSize} \Textlen)$ time compute $\min \OccTwo{\Pat}{\Text}$.
Observe that, given such data structure, it is easy to compute the overlapping and non-overlapping versions of LZ77 simply by binary searching the length of each phrase. 
More precisely, computing the length $\ell$ and the position $i$ of the previous occurrence of a phrase starting at any position $j$ in $\Text$
takes $\bigO(\log \ell \cdot \log^{\epsilon} \Textlen)$ time.
Across all $z$ phrases of total length $\Textlen$, this sums up to $\bigO(z \log(\Textlen/z)\cdot \log^{\epsilon}\Textlen)$ since logarithm is a concave function.
Due to $z = \bigO(\Textlen / \log_{\AlphabetSize} \Textlen)$, it thus suffices to set $\epsilon < \tfrac{1}{2}$ to achieve the worst-case running time of $\bigO(\Textlen / \log_{\AlphabetSize} \Textlen \cdot \log\log_{\AlphabetSize} \Textlen\cdot \log^{\epsilon}\Textlen)=
\bigO((\Textlen \log \AlphabetSize) / \sqrt{\log \Textlen})$. This method works for both the overlapping
and non-overlapping variants of LZ77.
However, rather than applying the above strategy directly, we go one step further %
and generalize the above idea so that, after additional sublinear preprocessing,
the computation of the pair $(\ell,i)$ in the above scenario takes $\bigO(\log^{\epsilon} \Textlen)$ time (rather than
$\bigO(\log \ell\cdot \log^{\epsilon} \Textlen)$). In other words, we obtain a data structure %
that provides $\bigO(\log^{\epsilon} \Textlen)$-time
access to the so-called \emph{Longest Previous Factor (LPF)}~\cite{CrochemoreIS08} and the \emph{Longest Previous
non-overlapping Factor (LPnF)}~\cite{CrochemoreT11} arrays (\cref{def:LPF,def:LPnF}).
The following result applied with $\epsilon \leq \tfrac{1}{2}$, and combined with \cref{th:lz-size,th:lz-nonov-size},
thus
yields \cref{th:main-lz-theorem}.

\begin{theorem}[LPF and LPnF Index, \cref{sec:lpf-indexes}]
  Given any constant $\epsilon \in (0, 1)$ and the $\bigO(\Textlen / \log_{\AlphabetSize} \Textlen)$-space
  representation of a text $\Text \in \IntegerAlphabet^{\Textlen}$, where $2 \leq \AlphabetSize < \Textlen^{1/7}$, we
  can in $\bigO((\Textlen \log \AlphabetSize) / \sqrt{\log \Textlen})$ time and
  $\bigO(\Textlen / \log_{\AlphabetSize} \Textlen)$ working space construct a data structure of size
  $\bigO(\Textlen / \log_{\AlphabetSize} \Textlen)$ that, given any
  $j \in [1 \dd \Textlen]$, in $\bigO(\log^{\epsilon} \Textlen)$ time
  returns $\LPF{\Text}[j]$, $\LPnF{\Text}[j]$, $\LPFMinOcc{\Text}[j]$, and $\LPnFMinOcc{\Text}[j]$,
  which are defined as the length $\ell$ and the leftmost occurrence of the longest previous (non-overlapping) factor $T[j\dd j+\ell)$.
\end{theorem}

As discussed above, the central technical result of our paper %
is a space-efficient index that quickly locates leftmost occurrences of substrings in the text
and admits a sublinear-time construction algorithm. The main obstacle
to obtaining such an index using prior techniques is that locating leftmost occurrences is typically
achieved using \emph{Range Minimum Queries (RMQ)} (see \cref{sec:rmq}) on top of the suffix array.
Although RMQ queries only add $\bigO(m)$ bits on top of the length-$m$ array they augment~\cite{FischerH11},
their construction needs $\Omega(m)$ time, which prevents achieving $o(\Textlen)$-time construction for a
length-$\Textlen$ text. We instead exploit a sampling-based approach~\cite{sss}, where the idea is to first
carefully compute a sample $\SSS \subseteq [1 \dd \Textlen]$ of representative text positions within nonperiodic
regions of the text (periodic regions are handled separately) such that
$|\SSS| = \bigO(\Textlen / \log_{\AlphabetSize} \Textlen)$~\cite{sss}, and then reduce the queries on the text
to orthogonal range queries on a set of points defined by the set $\SSS$
(\cref{lm:nonperiodic-pat-occ}). The natural query corresponding to finding
leftmost occurrences is then a 4-sided orthogonal RMQ query. Indeed, a reduction to orthogonal range queries
underlies some of the fastest indexes of size $\bigO(z \polylog \Textlen)$~\cite{ChristiansenEKN21,KociumakaNO22,collapsing}, but
it does not lead to an efficient solution in our scenario because no fast construction is known for efficient orthogonal RMQ data structures (such as~\cite{Nekrich21}).
We instead
propose to replace the general RMQ queries on a plane with a new type of query we
call \emph{prefix RMQ}. Given an array of integers $A[1 \dd m]$ and sequence $S$ of $m$ strings
over alphabet $\Sigma$, the prefix RMQ query with arguments $b, e \in [0 \dd m]$ and
$X \in \Sigma^{*}$ asks to compute the position that minimizes the value $A[i]$ among all indices $i \in (b \dd e]$
for which $X$ is a prefix of $S[i]$; see \cref{def:prefix-rmq}.
This variant of a 4-sided RMQ query is precisely the specialization
that we need to support on $\SSS$ (see \cref{lm:nonperiodic-pat-occ-min}). We
propose a space-efficient data structure for prefix RMQ queries and describe its fast construction
(Sections~\ref{sec:rmq} and~\ref{sec:prefix-rmq}).
Furthermore, by carefully handling periodic regions of the text, where
we again prove that the orthogonal RMQ queries have a special structure that supports faster
queries (Sections~\ref{sec:range-queries},~\ref{sec:three-sided-rmq}, and~\ref{sec:dyn-rmq}),
we achieve the following very general reduction from prefix RMQ queries. %
Plugging our specific tradeoff from \cref{th:prefix-rmq} to this
reduction yields our main index (\cref{th:minocc-index}). %
Observe that this reduction is very efficient: aside from an extra $\bigO(\log \log \Textlen)$ term
in the query time, prefix RMQ dominate all the complexities.

\begin{theorem}[Index for Leftmost Occurrences, \cref{sec:minocc-index}]
  Consider a data structure answering prefix RMQ queries (\cref{def:prefix-rmq})
  that, for any sequence of $k$ length-$\ell$ strings over alphabet $\IntegerAlphabet$,
  achieves the following complexities:
  \begin{enumerate}
  \item Space usage $S(k,\ell,\AlphabetSize)$,
  \item Preprocessing time $P_t(k,\ell,\AlphabetSize)$,
  \item Preprocessing space $P_s(k,\ell,\AlphabetSize)$, and
  \item Query time $Q(k,\ell,\AlphabetSize)$.
  \end{enumerate}
  For every $\Text \in \IntegerAlphabet^{n}$ with $2 \leq \AlphabetSize < \Textlen^{1/7}$,
  there exists $k = \bigO(\Textlen / \log_{\AlphabetSize} \Textlen)$ and $\ell = \bigO(\log_{\AlphabetSize} \Textlen)$
  such that, given the $\bigO(\Textlen / \log_{\AlphabetSize} \Textlen)$-space representation of $\Text$,
  we can in $\bigO(\Textlen / \log_{\AlphabetSize} \Textlen + P_t(k,\ell,\AlphabetSize))$ time
  and $\bigO(\Textlen / \log_{\AlphabetSize} \Textlen + P_s(k,\ell,\AlphabetSize))$ working space build
  a data structure of size $\bigO(\Textlen / \log_{\AlphabetSize} \Textlen + S(k,\ell,\AlphabetSize))$ that
  supports the following queries:
  \begin{itemize}
  \item Given any position $j \in [1 \dd \Textlen]$ and any length $\ell \in [1\dd \Textlen + 1-j]$,
    in $\bigO(\log \log \Textlen + Q(k,\ell,\AlphabetSize))$ time compute the
    position $\min \OccTwo{\Pat}{\Text}$, where $\Pat = \Text[j \dd j + \ell)$.
  \item Given the packed representation\footnote{By a ``packed'' representation of a string $S \in \IntegerAlphabet^{m}$,
    we mean its $\bigO(m / \log_{\AlphabetSize} \Textlen)$-space encoding in memory; see \cref{sec:prelim}.}
    of any pattern $\Pat \in \IntegerAlphabet^{m}$ that satisfies
    $\OccTwo{\Pat}{\Text} \neq \emptyset$,
    in $\bigO(\log \log \Textlen + Q(k,\ell,\AlphabetSize) + m / \log_{\AlphabetSize} \Textlen)$
    time compute the position $\min \OccTwo{\Pat}{\Text}$.
  \end{itemize}
\end{theorem}

\paragraph{Related Work}

Ellert~\cite{Ellert23} described an $\bigO(\Textlen / \log_{\AlphabetSize} \Textlen)$-time
algorithm that computes a 3-approximation of LZ77. Two $\bigO(z)$-working-space algorithms constructing
a 2-approximation and a $(1+\epsilon)$-approximation running in $\bigO(\Textlen \log \Textlen)$
and $\bigO(\Textlen \log^2 \Textlen)$ time, respectively, were proposed in~\cite{FischerGGK15}.
A practical external-memory approximation was described in~\cite{KosolobovVNP20}.

\emph{Rightmost} LZ77 is a variant of LZ77 where the encoding of every phrase refers to its rightmost previous occurrence.
An $\bigO(\Textlen \log \Textlen)$-time and $\bigO(\Textlen)$-space algorithm for this
problem was given in~\cite{AmirLU02,Larsson14}.
This has been improved to $\bigO(\Textlen + (\Textlen \log\AlphabetSize) / \log \log \Textlen)$ time and
$\bigO(\Textlen)$ space in~\cite{bitlz}, and further to
$\bigO(\Textlen \log \log \AlphabetSize  + (\Textlen \log \AlphabetSize) / \sqrt{\log \Textlen})$ (deterministic) or
$\bigO(\Textlen + (\Textlen \log \AlphabetSize) / \sqrt{\log \Textlen})$ (randomized) time, while using the optimal
$\bigO(\Textlen / \log_{\AlphabetSize} \Textlen)$ space, in~\cite{BelazzouguiP16}.
An $(1+\epsilon)$-approximation of the rightmost LZ77 
can be constructed in $\bigO(\Textlen (\log z + \log \log \Textlen))$ time and $\bigO(\Textlen)$ space; see~\cite{BilleCFG17}.

A variant of LZ77 that requires phrases to have earlier occurrences ending at phrase boundaries is called
\emph{LZ-End}~\cite{kreft2010navarro}.
It was proved to achieve an approximation ratio of $\bigO(\log^{2}(\Textlen/z))$ in~\cite{KempaS22}.
This was recently improved by a factor $\Theta(\log \log (\Textlen/z))$~\cite{GawrychowskiKM23}.
Algorithms computing LZ-End in $\bigO(\Textlen \ell_{\max} (\log \AlphabetSize + \log \log \Textlen))$ time
and $\bigO(\Textlen)$ space, or in $\bigO(\Textlen \ell_{\max} \log^{1+\epsilon} \Textlen)$
time and $\bigO(\Textlen / \log_{\AlphabetSize} \Textlen)$ space (where
$\ell_{\max}$ is the length of the longest phrase and $\epsilon > 0$ is any positive constant)
were given in~\cite{KreftN13}.
A construction running in $\bigO(\Textlen)$ time and space was then given in~\cite{KempaK17}.
The same time and space were achieved for the rightmost variant of LZ-End in~\cite{EllertFP23}.
Lastly, an LZ-End factorization algorithm running in $\bigO(\Textlen \log \ell_{\max})$
expected time and $\bigO(z_{\rm end} + \ell_{\max})$ working space (where $z_{\rm end}$ is the size of the
LZ-End factorization) was given in~\cite{KempaK17b}.

\paragraph{Organization of the Paper}

After introducing the basic notation and tools in \cref{sec:prelim}, we
give a technical overview of the paper in \cref{sec:overview}. In \cref{sec:tools},
we then introduce the tools to answer prefix RMQ and various other abstract queries.
\cref{sec:minocc-index} describes our index for leftmost occurrences. In \cref{sec:lpf-indexes}, we show how to use this
index to construct a data structure that can query the LPF and LPnF arrays. Finally, in \cref{sec:lz}, we
obtain sublinear LZ77 factorization algorithms.

%% file: prelim.tex
\section{Preliminaries}\label{sec:prelim}

\begin{wrapfigure}{R}{0.31\textwidth}
  \vspace{-.1cm}
  \begin{tikzpicture}[yscale=0.35]
    \foreach \x [count=\i] in {a, aababa, aababababaababa,
      aba, abaababa, abaababababaababa, ababa, ababaababa,
      abababaababa, ababababaababa, ba, baababa,
      baababababaababa, baba, babaababa, babaababababaababa,
      bababaababa, babababaababa, bbabaababababaababa}
        \draw (1.9, -\i) node[right]
          {$\texttt{\x}$};
    \draw(1.9,0) node[right] {\scriptsize $\Text[\SA{\Text}[i]\dd \Textlen]$};
    \foreach \x [count=\i] in {b, b, b, b, b, b, a, b, b,
                               a, a, a, a, a, a, b, a, a,\$}
      \draw (0.7, -\i) node {\footnotesize $\i$};
    \draw(0.7,0) node{\scriptsize $i$};
    \foreach \x [count=\i] in {19,14,5,17,12,3,15,10,
                               8,6,18,13,4,16,11,2,9,7,1}
      \draw (1.4, -\i) node {$\x\vphantom{\textbf{\underline{7}}}$};
    \draw(1.4,0) node{\scriptsize $\SA{\Text}[i]$};
  \end{tikzpicture}

  \vspace{-1.5ex}
  \caption{A list of all sorted suffixes of $\Text =
    \texttt{bbabaababababaababa}$ along with
    the suffix array.}\label{fig:example}
  \vspace{-1.3cm}
\end{wrapfigure}

\paragraph{Basic Definitions}

A \emph{string} is a finite sequence of characters from a given
\emph{alphabet} $\Sigma$.  The length of a string $S$ is denoted $|S|$. For $i
\in [1\dd |S|]$, the $i$th character of $S$ is denoted $S[i]$.
A~\emph{substring} of $S$ is a string of the form $S[i \dd j) =
S[i]S[{i+1}]\cdots S[{j-1}]$ for some $1\le i \le j \le |S|+1$. 
Substrings of the form $S[1\dd j)$ and $S[i\dd |S|{+}1)$ are called
\emph{prefixes} and \emph{suffixes}, respectively. We use $\revstr{S}$
to denote the \emph{reverse} of $S$, i.e., $S[|S|]\cdots S[2]S[1]$.
We denote the \emph{concatenation} of two strings $U$ and $V$, that
is, $U[1]\cdots U[|U|]V[1]\cdots V[|V|]$, by $UV$ or $U\cdot V$.
Furthermore, $S^k = \bigodot_{i=1}^k S$ is the concatenation of $k \in
\Zz$ copies of $S$; note that $S^0 = \emptystring$ is the \emph{empty
string}. A nonempty string $S$ is said to be \emph{primitive} if it
cannot be written as $S = U^k$, where $k \geq 2$.  An integer $p\in
[1\dd |S|]$ is a \emph{period} of $S$ if $S[i] = S[i + p]$ holds for
every $i \in [1 \dd |S|-p]$. We denote the shortest period of $S$ as
$\per(S)$.  For every $S \in \Sigma^{+}$, we define the
infinite power $S^{\infty}$ so that $S^{\infty}[i] = S[1 + (i-1) \bmod |S|]$
for $i \in \Z$.  In particular, $S = S^{\infty}[1 \dd |S|]$.  
By $\lcp{U}{V}$
we denote the length of
the longest common prefix
of $U$ and $V$. For any
string $S \in \Sigma^{*}$ and any $j_1, j_2 \in [1 \dd |S|+1]$, we
denote $\LCE{S}{j_1}{j_2} = \lcp{S[j_1 \dd |S|]}{S[j_2 \dd |S|]}$.
We use $\preceq$ to denote the order on $\Sigma$, extended to the
\emph{lexicographic} order on $\Sigma^*$ so that $U,V\in \Sigma^*$
satisfy $U \preceq V$ if and only if either \begin{enumerate*}[label=(\alph*)] \item $U$ is a prefix of $V$, or \item
$U[1 \dd i) = V[1 \dd i)$ and $U[i]\prec V[i]$ holds for some $i\in
[1\dd \min(|U|,|V|)]$. \end{enumerate*}

\paragraph{Suffix Array}

For any string $\Text \in \Sigma^{\Textlen}$ (of length $\Textlen \geq 1$), the \emph{suffix
array} $\SA{\Text}[1 \dd \Textlen]$ of $\Text$ is a permutation of $[1 \dd \Textlen]$
such that $\Text[\SA{\Text}[1] \dd \Textlen] \prec \Text[\SA{\Text}[2] \dd \Textlen] \prec \cdots
\prec \Text[\SA{\Text}[\Textlen] \dd \Textlen]$, i.e., $\SA{\Text}[i]$ is the starting position
of the lexicographically $i$th suffix of $\Text$; see \cref{fig:example}
for an example.  The \emph{inverse suffix array} $\ISA{\Text}[1 \dd \Textlen]$ (also
denoted $\SA{\Text}^{-1}[1 \dd \Textlen]$)
is the inverse permutation of $\SA{\Text}$, i.e., $\ISA{\Text}[j] = i$ holds if
and only if $\SA{\Text}[i] = j$. Intuitively, $\ISA{\Text}[j]$ stores the
lexicographic \emph{rank} of $\Text[j \dd \Textlen]$ among the suffixes of~$\Text$.

\begin{definition}\label{def:occ}
  For any $\Text\in \Sigma^{\Textlen}$ and $\Pat \in \Sigma^*$,
  we define
  \vspace{-2ex}
  \begin{align*}
    \OccTwo{\Pat}{\Text}
      &\,{=}\, \{j \in [1 \dd \Textlen] : j + |\Pat| \leq \Textlen + 1\text{ and }\Text[j \dd j + |\Pat|) = \Pat\},\\
    \RangeBegTwo{\Pat}{\Text}
      &\,{=}\, |\{j \in [1 \dd \Textlen] : \Text[j \dd \Textlen] \prec \Pat\}|,\\
    \RangeEndTwo{\Pat}{\Text}
      &\,{=}\, \RangeBegTwo{\Pat}{\Text} + |\OccTwo{\Pat}{\Text}|.
  \end{align*}
\end{definition}

In other words, $\OccTwo{\Pat}{\Text}$ consists of the starting positions of the (exact) occurrences of $\Pat$ in $\Text$, with the convention that $\OccTwo{\emptystring}{\Text}=[1\dd \Textlen]$ holds if \mbox{$\Textlen = |\Text| > 0$}.
The two values $\RangeBegTwo{\Pat}{\Text}$ and $\RangeEndTwo{\Pat}{\Text}$ are defined to be endpoints of the so-called \emph{SA-interval} representing the occurrences of $\Pat$ in $\Text$. 
Formally, 
$\OccTwo{\Pat}{\Text} =
\{\SA{\Text}[i] : i \in (\RangeBegTwo{\Pat}{\Text} \dd
\RangeEndTwo{\Pat}{\Text}]\}$ holds for every pattern $\Pat \in \Sigma^{*}$, including when $\Pat=\emptystring$ and when $\OccTwo{\Pat}{\Text}=\emptyset$.

\paragraph{Lempel--Ziv Compression}

A fragment $\Text[j \dd j + \ell)$ of $\Text$ is a \emph{previous factor} if
it has an earlier occurrence in $\Text$, i.e., $\LCE{\Text}{i}{j}\geq \ell$
holds for some $i\in [1 \dd j)$.  An \emph{LZ77-like factorization}
of $\Text$ is a decomposition
$\Text = f_1 \cdots f_z$ into non-empty
\emph{phrases} such that each phrase $f_k$ with $|f_k| > 1$ is a
previous factor. In the underlying \emph{LZ77-like representation},
every phrase $f_k = \Text[j \dd j + \ell)$ that is a previous factor is
encoded as $(i, \ell)$, where $i \in [1 \dd j)$ satisfies $\LCE{\Text}{i}{j}
\geq \ell$ (and is chosen arbitrarily in case of multiple
options); if $f_k = \Text[j]$ is not a previous factor, we encode it
as $(\Text[j], 0)$.

The LZ77 factorization~\cite{LZ77} of a string
$\Text$ is then just an LZ77-like factorization constructed by greedily
factorizing $\Text$ from left to right into the longest possible phrases.  More
precisely, the $k$th phrase $f_k$ is the longest previous factor
starting at position $1 + |f_1 \cdots f_{k - 1}|$; if no previous
factor starts there, then $f_k$ consists of a single character. 
This greedy construction yields the smallest LZ77-like factorization of~$\Text$ \cite[Theorem~1]{LZ76}.
We denote the number of phrases in the LZ77 factorization of $\Text$ by $\LZSize{\Text}$.
For example, the text of
\cref{fig:example} has LZ77 factorization $\Text=\texttt{b}\cdot
\texttt{b}\cdot \texttt{a}\cdot \texttt{ba} \cdot \texttt{aba}\cdot
\texttt{bababa} \cdot \texttt{ababa}$ with $\LZSize{\Text} = 7$ phrases, and its
LZ77 representation is $ (\texttt{b},0), (1,1), (\texttt{a},0), (2,2),
(3,3), (7,6), (10,5)$.

A variant of LZ77 factorization in which we additionally require
that the earlier occurrence of every phrase does not overlap the phrase
itself is called the \emph{non-overlapping LZ77}. We denote the number
of phrases in this variant by $\LZNonOvSize{\Text}$.
The non-overlapping LZ77 factorization of the text of \cref{fig:example} is $\Text=\texttt{b}\cdot
\texttt{b}\cdot \texttt{a}\cdot \texttt{ba} \cdot \texttt{aba}\cdot
\texttt{baba} \cdot \texttt{baababa}$ with $\LZNonOvSize{\Text} = 7$ phrases.

\paragraph{String Synchronizing Sets}
String synchronizing sets~\cite{sss} allow for a locally-consistent selection of positions in a given text $\Text$.
The underlying parameter $\tau$ governs the context size (with respect to which the selection is consistent) and the achievable size of the synchronizing set. 

\begin{definition}[$\tau$-synchronizing set~\cite{sss}]\label{def:sss}
  Let $\Text\in \Sigma^{\Textlen}$ be a string and let $\tau \in
  [1 \dd \lfloor\frac{\Textlen}{2}\rfloor]$ be a parameter. A set $\SSS
  \subseteq [1 \dd \Textlen - 2\tau + 1]$ is called a
  \emph{$\tau$-synchronizing set} of $\Text$ if it satisfies the
  following \emph{consistency} and \emph{density} conditions:
  \begin{enumerate}
  \item\label{def:sss-consistency}
    If $\Text[i \dd i + 2\tau) = \Text[j\dd j + 2\tau)$, then $i \in \SSS$
    holds if and only if $j \in \SSS$
    (for $i, j \in [1 \dd \Textlen - 2\tau + 1]$),
  \item\label{def:sss-density}
    $\SSS\cap[i \dd i + \tau) = \emptyset$ if and only if
    $i \in \RTwo{\tau}{\Text}$ (for $i \in [1 \dd \Textlen - 3\tau + 2]$),
    where
    \[
      \RTwo{\tau}{\Text} := \{i \in [1 \dd \Textlen - 3\tau + 2] :
      \per(\Text[i \dd i + 3\tau - 2]) \leq \tfrac{1}{3}\tau\}.
    \]
  \end{enumerate}
\end{definition}

\begin{remark}\label{rm:sss-size}
  In most applications, we want to minimize $|\SSS|$. Note, however, that
  the density condition imposes a lower bound
  $|\SSS| = \Omega(\frac{\Textlen}{\tau})$ for strings of length
  $\Textlen \ge 3\tau-1$ that do not contain substrings of length $3\tau - 1$
  with period at most $\frac{1}{3}\tau$.  Thus, we cannot hope to achieve an
  upper bound improving in the worst case upon the following ones.
\end{remark}

\begin{theorem}[{\cite[Proposition~8.10]
      {sss}}]\label{th:sss-existence-and-construction}
  For every string $\Text$ of length $\Textlen$ and parameter $\tau \in
  [1 \dd \lfloor\frac{\Textlen}{2}\rfloor]$, there exists a $\tau$-synchronizing
  set $\SSS$ of size $|\SSS| =
  \bigO\left(\frac{\Textlen}{\tau}\right)$.
  Moreover, if $\Text \in \IntegerAlphabet^{\Textlen}$, where
  $\AlphabetSize = \Textlen^{\bigO(1)}$, such $\SSS$ can be deterministically
  constructed in $\bigO(\Textlen)$ time.
\end{theorem}

\begin{theorem}[{\cite[Theorem~8.11]{sss}}]\label{th:sss-packed-construction}
  For every constant $\mu < \tfrac{1}{5}$, given the packed
  representation of a text $\Text \in \IntegerAlphabet^{\Textlen}$
  and a positive integer $\tau \leq \mu\log_\AlphabetSize \Textlen$,
  one can deterministically construct in $\bigO(\frac{\Textlen}{\tau})$
  time a $\tau$-synchronizing set of size $\bigO(\frac{\Textlen}{\tau})$.
\end{theorem}

\paragraph{Rank and Selection Queries}\label{sec:rank-select}

\begin{definition}\label{def:rank-select}
  For a string $S\in \Sigma^n$, we define:
  \begin{description}[style=sameline,itemsep=1ex]
  \item[Rank query $\Rank{S}{a}{j}$:] Given $a\in \Sigma$ and $j\in
    [0\dd n]$, compute $|\{i\in [1\dd j]: S[i]=a\}|$.
  \item[Selection query $\Select{S}{a}{r}$:] Given $a\in \Sigma$ and
    $r\in [1\dd \Rank{S}{a}{n}]$, find the $r$th smallest element of
    $\{i\in [1\dd n] : S[i]=a\}$.
  \end{description}
\end{definition}

\begin{theorem}[Rank and selection queries in
    bitvectors~\cite{WaveletSuffixTree,Clark98,Jac89,MunroNV16}]\label{th:bin-rank-select}
  For every string $S\in \{{\tt 0},{\tt 1}\}^*$, there exists a data structure of
  $\bigO(|S|)$ bits answering rank and selection queries in $\bigO(1)$
  time. Moreover, given the packed representations of $m$ binary
  strings of total length $n$, the data structures for all these
  strings can be constructed in $\bigO(m + n/\log n)$ time.
\end{theorem}

\paragraph{Model of Computation}

We use the standard word RAM model of computation~\cite{Hagerup98}
with $w$-bit \emph{machine words}, where $w \ge \log \Textlen$, and all
standard bit-wise and arithmetic operations taking $\bigO(1)$ time.
Unless explicitly stated otherwise, we measure the space complexity in machine words.

In the RAM model, strings are usually represented as arrays, with each
character occupying one memory cell (or a constant number of memory cells if $\AlphabetSize>n$).
 A single character, however, only needs
$\lceil \log \AlphabetSize \rceil$ bits, which might be much less
than $w$. We can therefore store (the \emph{packed representation} of)
a string $S \in \IntegerAlphabet^{m}$ using
$\bigO(\lceil \tfrac{m \log \AlphabetSize}{w} \rceil)$ words.

%% file: overview.tex
\section{Technical Overview}\label{sec:overview}

Consider a text $\Text \in \IntegerAlphabet^{\Textlen}$.
We assume that $2 \leq \AlphabetSize < \Textlen^{1/7}$; larger alphabet sizes $\AlphabetSize$ satisfy $\log \AlphabetSize = \Theta(\log \Textlen)$, and, in that case, most of the problems considered in this paper can be solved using standard large-alphabet techniques.
For example, whenever $\AlphabetSize = \Textlen^{\Theta(1)}$, the LZ77 factorization algorithm from~\cite{CrochemoreI08} runs in $\bigO(\Textlen) = \bigO(\Textlen / \log_{\AlphabetSize} \Textlen)$ time and $\bigO(n) = \bigO(\Textlen / \log_{\AlphabetSize} \Textlen)$ space.

\subsection{Index for Leftmost Occurrences}\label{sec:minocc-overview}

We now outline how, given any constant $\epsilon \in (0, 1)$ along with the $\bigO(\Textlen / \log_{\AlphabetSize} \Textlen)$-space representation of $\Text \in \IntegerAlphabet^{\Textlen}$, in $\bigO(\Textlen \log \AlphabetSize / \sqrt{\log \Textlen})$ time 
and $\bigO(\Textlen / \log_{\AlphabetSize} \Textlen)$ working space, we can construct an index of
size $\bigO(\Textlen / \log_{\AlphabetSize} \Textlen)$ that, given any position
$j \in [1 \dd \Textlen]$ and any length $\ell \in [1\dd \Textlen-j+1]$, in
$\bigO(\log^{\epsilon} \Textlen)$ time returns the position $\min \OccTwo{\Pat}{\Text}$ (see \cref{def:occ}) for
$\Pat = \Text[j \dd j + \ell)$. Our index can also compute $\min \OccTwo{\Pat}{\Text}$ in
$\bigO(\log^{\epsilon} \Textlen + m / \log_{\AlphabetSize} \Textlen)$ time given the $\bigO(m / \log_{\AlphabetSize} \Textlen)$-space
representation of any $\Pat \in \IntegerAlphabet^{m}$ satisfying $\OccTwo{\Pat}{\Text} \neq \emptyset$,
which is of independent interest.

\paragraph{The Index Core}

Let $\mu \in (0, \tfrac{1}{6})$ be a positive constant such that $\tau := \mu \log_{\AlphabetSize} \Textlen$ is a positive integer.
Such $\mu$ exists because $\log_{\AlphabetSize}\Textlen >7$ follows from $\AlphabetSize < \Textlen^{1/7}$.
Consider integers $j \in [1 \dd \Textlen]$ and $\ell \in [1\dd \Textlen+1-j]$, and a pattern $\Pat \in \IntegerAlphabet^{m}$
of length $m > 0$ given in $\bigO(1 + m / \log_{\AlphabetSize} \Textlen)$ space.

We begin by observing that the number of strings
$X \in \IntegerAlphabet^{*}$ satisfying $|X| < 3\tau - 1$
is bounded by $\AlphabetSize^{3\tau} \leq %
\Textlen^{1/2}$. Thus,
we can precompute and store $\min \OccTwo{X}{\Text}$ for all of them.
This computation is easily done in
$\bigO(\Textlen / \log_{\AlphabetSize} \Textlen)$ exploiting the packed representation of $\Text$;
see \cref{pr:minocc-index-core-construction}. Using this lookup table, we can answer queries
when $\ell < 3\tau - 1$ and $m < 3\tau - 1$.

Let us now assume that $\ell \geq 3\tau - 1$ and $m \geq 3\tau - 1$. The computation
of $\min \OccTwo{\Text[j \dd j + \ell)}{\Text}$ works differently, depending on whether
$\per(\Text[j \dd j + \ell)) \leq \tfrac{1}{3}\tau$ (or equivalently $j \in \RTwo{\tau}{\Text}$; see \cref{def:sss}).
For explicit patterns $\Pat$, we also proceed depending on whether $\per(\Pat) \leq \tfrac{1}{3}\tau$, in which
case we call $\Pat$ \emph{$\tau$-periodic}; see \cref{def:periodic-pattern}. To distinguish these cases,
we store the packed representation of $\Text$ (to retrieve $\Text[j \dd j + \ell)$) and a lookup table keeping $\per(X)$ for every
$X \in \IntegerAlphabet^{3\tau-1}$. 
Such table takes $\bigO(\AlphabetSize^{3\tau}) = \bigO(\Textlen^{1/2})$
space and is easily computed in $\bigO(\Textlen / \log_{\AlphabetSize} \Textlen)$~time.

We collect the above structures in a component of our index called the index \emph{core} (\cref{sec:minocc-index-core}).

\paragraph{The Nonperiodic Patterns and Positions}

Assume that $j \not\in \RTwo{\tau}{\Text}$ and $\Pat$ is $\tau$-nonperiodic. Assume that we computed
(using \cref{th:sss-packed-construction} in $\bigO(\Textlen / \log_{\AlphabetSize} \Textlen)$ time)
a $\tau$-synchronizing set $\SSS$ of $\Text$ satisfying $|\SSS| = \bigO(\tfrac{\Textlen}{\tau}) = 
\bigO(\Textlen / \log_{\AlphabetSize} \Textlen)$. By the density condition of $\SSS$
(\cref{def:sss}\eqref{def:sss-density}), the successor $\Successor{\SSS}{j}$ in $\SSS$ of every position
$j \in [1 \dd \Textlen - 3\tau + 2] \setminus \RTwo{\tau}{\Text}$ satisfies $\Successor{\SSS}{j} - j < \tau$.
This implies that the substring $D := \Text[j \dd \Successor{\SSS}{j} + 2\tau)$
(called the \emph{distinguishing prefix} of
$\Text[j \dd j + \ell)$; see \cref{def:dist-prefixes})
satisfies $|D| \leq 3\tau - 1$ (\cref{lm:dist-prefixes}). Moreover, by the consistency condition of $\SSS$
(\cref{def:sss}\eqref{def:sss-consistency}), every occurrence $j' \in \OccTwo{D}{\Text}$ satisfies $j' + \deltatext \in \SSS$, where $\deltatext = |D| - 2\tau$.

\begin{description}[style=sameline,itemsep=1ex,font={\normalfont\itshape}]
\item[Observation: Computation of $\min \OccTwo{\Text[j \dd j + \ell)}{\Text}$ and $\min \OccTwo{\Pat}{\Text}$
  can be efficiently reduced to a 4-sided RMQ query.]
By the discussion above,
we can characterize $\OccTwo{\Text[j \dd j + \ell)}{\Text}$ as a set of all positions of the form $s - \deltatext$, where
$s \in \SSS$ satisfies
\begin{enumerate*}[label=(\arabic*)]
  \item\label{it:prefix} $\Textinf[s - \deltatext \dd s + 2\tau) = D$ (or equivalently, $s - \deltatext \in \OccTwo{D}{\Text}$),
and \item\label{it:suffix} $s \in \OccTwo{\Text[j + \deltatext \dd j + \ell)}{\Text}$.
\end{enumerate*}
Analogous %
characterization holds for $\Pat$ %
(\cref{lm:nonperiodic-pat-occ}).
Consider now a sequence
$(s_i)_{i \in [1 \dd |\SSS|]}$ containing positions in $\SSS$ sorted according
to the lexicographical order of the corresponding suffixes.
This sequence can be
constructed in $\bigO(\Textlen / \log_{\AlphabetSize} \Textlen)$ time
using \cref{th:sss-lex-sort}.
Given a pair $(j,\ell)$ (resp.\ $\bigO(m / \log_{\AlphabetSize} \Textlen)$-space representation of $\Pat$), we can in
$\bigO(\log \log \Textlen)$ (resp.\ $\bigO(\log \log \Textlen + m / \log_{\AlphabetSize} \Textlen)$) time compute the
boundaries of a range $(b\dd e]$ such that $(s_i)_{i \in (b\dd e]}$ consists of all positions
in the set $\SSS \cap \OccTwo{\Text[j + \deltatext \dd j + \ell)}{\Text}$ (resp.\ $\SSS \cap \OccTwo{\Pat[1+\deltatext \dd m]}{\Text}$).
This is easily achieved using weighted
ancestor queries on a compact trie of suffixes starting in~$\SSS$; see \cref{pr:nav-index-nonperiodic}.
Then, $\{s_i\}_{i \in (b \dd e]}$ is the set of all $s \in \SSS$ satisfying condition \ref{it:suffix}.
To satisfy condition \ref{it:prefix}, we need to select all $s \in \SSS$ for which
$s - \deltatext \in \OccTwo{D}{\Text}$. 
If each position $s_i \in \SSS$ is represented as a point $(i,D_i)$, where $\revstr{D_i} =
\Textinf[s - \tau \dd s + 2\tau)$, then the sought subset consists of all $s_i \in \SSS$ such that $\revstr{D}$ is a prefix of
$D_i$, which is equivalent to
$\revstr{D} \preceq D_i \prec \revstr{D}c^{\infty}$, where
$c = \AlphabetSize - 1$. Thus, finding
$\min \OccTwo{\Text[j \dd j + \ell)}{\Text}$ (resp.\ $\min \OccTwo{\Pat}{\Text}$)
reduces to the 4-sided range RMQ query in a rectangle obtained by
intersecting ranges $(b\dd e]$ and $[\revstr{D}, \revstr{D}c^{\infty})$.

\end{description}

The bottleneck of the above reduction is answering the orthogonal RMQ queries because
it would require a data structure supporting 4-sided RMQ in $\bigO(\log^{\epsilon} \Textlen)$ time.
The only such structure~\cite{Nekrich21} is not known to admit a sufficiently fast construction algorithm.
Thus, we take a different approach by
directly answering the query characterizing the set $\OccTwo{\Text[j \dd j + \ell)}{\Text}$
and $\OccTwo{\Pat}{\Text}$. As noted above, $\Textinf[s_i - \deltatext \dd s_i + 2\tau) = D$ holds if and only if
$\revstr{D}$ is a prefix of $D_i$. Taking into account the constraint $i \in (b \dd e]$ on the other axis,
we obtain a \emph{prefix
range minimum query} (\cref{def:prefix-rmq}). Thus, with $A_{\SSS}[1 \dd n']$ and $A_{\rm str}[1 \dd n']$ defined as arrays of length $n' = |\SSS|$ such that $A_{\SSS}[i] = s_i$ and $A_{\rm str}[i] = \revstr{D_i}$, it holds
$\min \OccTwo{\Pat}{\Text} = A_{\SSS}[\PrefixRMQ{A_{\SSS}}{A_{\rm str}}{b}{e}{\revstr{D}}] - \deltatext$; see
\cref{lm:nonperiodic-pat-occ-min}.

To our knowledge, prefix range minimum queries have not been studied before.
We develop a solution that not only answers these queries in $\bigO(\log^{\epsilon} \Textlen)$ time but also admits efficient construction.
Our data structure
is described in \cref{sec:rmq,sec:prefix-rmq}. In addition to solving a novel type of query, this requires
improvements of existing structures for standard RMQ (\cref{sec:rmq}). Our tradeoff used to obtain
the main results is \cref{th:prefix-rmq}. Applied to arrays $A_{\SSS}[1 \dd n']$ and $A_{\rm str}[1 \dd n']$, it yields a data structure of size $\bigO(\Textlen / \log_{\AlphabetSize} \Textlen)$ that answers prefix RMQ in
$\bigO(\log^{\epsilon} \Textlen)$ time, and can be constructed in $\bigO((\Textlen \log \AlphabetSize) / \sqrt{\log \Textlen})$ time
and $\bigO(\Textlen / \log_{\AlphabetSize} \Textlen)$ working space. We remark, however, that other tradeoffs for prefix RMQ
will automatically yield new tradeoffs for indexes for leftmost occurrences (\cref{th:minocc-index-general}), and hence also
new LPF/LPnF indexes and LZ77 factorization algorithms (see \cref{sec:lpf-indexes,sec:lz}).

\paragraph{The Periodic Patterns and Positions}

Let us now assume that $j \in \RTwo{\tau}{\Text}$ and $\Pat$ is $\tau$-periodic.
First, observe that $\Text[j \dd j + \ell)$ and $\Pat$ are both prefixed with a string $X\in \IntegerAlphabet^{3\tau-1}$ satisfying $\per(X) \leq \tfrac{1}{3}\tau$.
Thus, $\OccTwo{\Text[j \dd j + \ell)}{\Text}, \OccTwo{\Pat}{\Text} \subseteq \RTwo{\tau}{\Text}$ holds by \cref{def:sss}.
The central property of $\RTwo{\tau}{\Text}$ is that every maximal block of positions in $\RTwo{\tau}{\Text}$ corresponds 
to a \emph{$\tau$-run}, i.e., a maximal fragment $Y$ of $\Text$ satisfying $|Y| \geq 3\tau - 1$ and $\per(Y) \leq \tfrac{1}{3}\tau$. The gap between $|Y|$ and $\per(Y)$ ensures that $\tau$-runs overlap by fewer than
$\tfrac{2}{3}\tau$ symbols, so the number of $\tau$-runs is $\bigO(\tfrac{\Textlen}{\tau}) =
\bigO(\Textlen / \log_{\AlphabetSize} \Textlen)$ (see \cref{lm:runs}). We
represent $\tau$-runs by their starting positions, denoted~$\RPrimTwo{\tau}{\Text}$~(see~\cref{sec:minocc-index-periodic-prelim}).

To efficiently process $\tau$-runs, we introduce the following definitions.
Let $x \in \RTwo{\tau}{\Text}$, and let $\Text[y]$ be the position immediately following of the $\tau$-run containing $\Text[x \dd x + 3\tau - 1)$.
By $y-x \geq 3\tau - 1$ and $p := \per(\Text[x \dd y)) \leq \tfrac{1}{3}\tau$, we can uniquely write $\Text[x \dd y) = H'H^{k}H''$, where $H = \min\{\Text[x + \delta \dd x + \delta + p) : \delta \in [0 \dd p)\}$ is the so-called \emph{Lyndon root} of the run
and $H'$ (resp.\ $H''$) is a proper suffix (resp.\ prefix) of $H$. 
We denote
$\RunEndPos{x}{\tau}{\Text} = y$,
$\RootPos{x}{\tau}{\Text} = H$, and
$\RunEndFullPos{x}{\tau}{\Text} = y - |H''|$. We also
let $\TypePos{x}{\tau}{\Text} = -1$ if $\Text[y] \prec \Text[y-p]$ and
$\TypePos{x}{\tau}{\Text} = +1$~otherwise. The above definitions naturally generalize to $\tau$-periodic patterns
(see \cref{sec:minocc-index-periodic-prelim}). Let us focus on the computation of
$\min \OccTwo{\Pat}{\Text}$ (the computation of $\min \OccTwo{\Text[j \dd j + \ell)}{\Text}$ proceeds similarly).
The query algorithm differs depending on whether the periodic prefix of $\Pat$ ends before $|\Pat|$, i.e.,
$\RunEndPat{\Pat}{\tau} \leq |\Pat|$. In that case, we call $\Pat$ \emph{partially periodic}. Otherwise, i.e.,
when $\RunEndPat{\Pat}{\tau} = |\Pat| + 1$, we call $\Pat$ \emph{fully periodic}.

\begin{description}[style=sameline,itemsep=1ex,font={\normalfont\itshape}]
\item[Observation 1: Computation of $\min \OccTwo{\Pat}{\Text}$ in the partially periodic case
  can be efficiently reduced to a 3-sided RMQ query.] Observe that if $\RunEndPat{\Pat}{\tau} \leq |\Pat|$, then
  the end of the periodic prefix of $\Pat$ has to align with the end of a periodic $\tau$-run in $\Text$ with the same root,
  i.e., if $j \in \OccTwo{\Pat}{\Text}$, then $j \in \RTwo{\tau}{\Text}$, $\RootPos{j}{\tau}{\Text} = \RootPat{\Pat}{\tau}$,
  and $\RunEndPos{j}{\tau}{\Text}-j = \RunEndPat{\Pat}{\tau}-1$. 
  At the same time, the remaining suffix of $\Pat$ must follow
  the run in the text, i.e., 
  we need to have $j + \deltatext \in \OccTwo{\Pat'}{\Text}$, where $\deltatext = \RunEndPat{\Pat}{\tau}-1$ and $\Pat' = \Pat(\deltatext \dd |\Pat|]$.
  If we sorted every $\tau$-run $\Text[x \dd y)$ first according to its root and then according to $\Text[y \dd \Textlen]$,
  then runs satisfying the second criterion would form a range. 
  This, however, would make it difficult to make sure the
  end of the periodic substring in the text is properly aligned with the periodic prefix of $\Pat$ solely based on the length
  of the $\tau$-run. Sorting every $\tau$-run $\Text[x \dd y)$ first according to $\RootPos{x}{\tau}{\Text}$ and
  then according to the suffix $\Text[\RunEndFullPos{x}{\tau}{\Text} \dd \Textlen]$
  solves the alignment problem. Identifying the range in the resulting sorted sequence of
  $\tau$-runs (\cref{def:runs-minus-lex-sorted}) then again reduces to a weighted ancestor query
  (\cref{pr:nav-index-periodic}\eqref{pr:nav-index-periodic-it-2c}). To
  simultaneously also align according to the root, we slightly back away in both the pattern and the text by a multiple of
  $|\RootPat{\Pat}{\tau}|$. After identifying the range containing $\tau$-runs with the right-context matching $\Pat$,
  it remains to only select runs $\Text[x \dd y)$ whose length is at least the length of the periodic prefix of $\Pat$.
  Then, $\min\OccTwo{\Pat}{\Text}$ is located in the leftmost selected $\tau$-run.
  This corresponds to a 3-sided RMQ query (\cref{lm:partially-periodic-pat-occ-min}).
  To answer it efficiently, we exploit the fact that both the maximum and the sum
  of the $y$-coordinates of the corresponding point-set are bounded by $\Textlen$; see \cref{sec:three-sided-rmq}.

\item[Observation 2: Computation of $\min \OccTwo{\Pat}{\Text}$ in the fully periodic case can
  be efficiently reduced to a rank query on a bitvector.] Let us denote
  $\RMinusTwo{\tau}{\Text} := \{j \in \RTwo{\tau}{\Text} : \TypePos{j}{\tau}{\Text} = -1\}$
  and focus on computing $\min \OccTwo{\Pat}{\Text} \cap \RMinusTwo{\tau}{\Text}$ (the other
  minimum is computed analogously). Let $\RMinMinusTwo{\tau}{\Text}$ be the set of all
  $j \in \RMinusTwo{\tau}{\Text}$ satisfying $j = \min \OccTwo{\Text[j \dd \RunEndPos{j}{\tau}{\Text})}{\Text} \cap \RMinusTwo{\tau}{\Text}$
  (\cref{def:rmin}). The key idea is to store a bitvector marking all $i \in [1 \dd \Textlen]$ such that
  $\SA{\Text}[i] \in \RMinMinusTwo{\tau}{\Text}$ (\cref{def:min-pos-bitvector-minus}). Because
  the values $\RunEndPos{j}{\tau}{\Text} - j$ increase within every block of $\SA{\Text}$ entries containing all
  $j \in \RMinusTwo{\tau}{\Text} \cap \OccTwo{\Pat[1 \dd 3\tau {-} 1]}{\Text}$ (\cref{lm:R-lex-block-pos}),
  it follows that, given the $\SA{\Text}$-range corresponding to $\Pat$, the
  computation of $\min\OccTwo{\Pat}{\Text}$ reduces to a rank/select query on
  the above bitvector (see \cref{lm:bmin-bit,lm:fully-periodic-pat-min-occ}). The main challenge is thus
  computing the bitvector. This step is one of the most technically challenging parts of our data structure.
  The construction is a complex algorithm that first prepares the set of ``events''
  (\cref{pr:emin-text-order,pr:emin-any-order}) and then computes partial bitvectors
  using the sweeping technique (\cref{pr:min-bv-first,pr:min-bv-rest}).
  This requires developing numerous new combinatorial results
  (Lemmas~\ref{lm:rmin-text-block}, \ref{lm:rmin-block-size}, \ref{lm:rmin-equivalence}, \ref{lm:emin}, \ref{lm:sweep-step}, and
  \ref{lm:sweep-init}), and efficient solutions for offline range counting
  (\cref{sec:range-queries}) and dynamic one-sided RMQ (\cref{sec:dyn-rmq}).
  Combining all these ingredients, we achieve the optimal $\bigO(\Textlen / \log_{\AlphabetSize} \Textlen)$-time construction
  (\cref{pr:min-bv}).
\end{description}

\paragraph{Summary of New Techniques} Our key technical contributions can be summarized as follows:
\begin{itemize}
\item We define a new query called \emph{prefix RMQ} and develop an efficient solution (\cref{sec:prefix-rmq}), improving
  the construction of small-alphabet RMQ on the way to this result (\cref{sec:rmq}).
\item We show how to use the above to find leftmost occurrences of nonperiodic patterns~(\cref{sec:minocc-index-nonperiodic}).
\item To handle periodic patterns, we prove numerous new combinatorial results characterizing leftmost occurrences of
  substrings, and we develop efficient solutions for
  offline range counting (\cref{sec:range-queries}),
  three-sided RMQ (\cref{sec:three-sided-rmq}), and
  dynamic one-sided RMQ (\cref{sec:dyn-rmq}).
\item Using the above queries, we show how to compute leftmost occurrences of periodic patterns (\cref{sec:minocc-index-periodic});
  combined with the above result, this gives an optimal-size index constructible in
  $\bigO((\Textlen \log \AlphabetSize) / \sqrt{\log \Textlen})$ time and $\bigO(\Textlen / \log_{\AlphabetSize} \Textlen)$ working space.
  This reduction is very efficient and depends almost entirely on the tradeoff for prefix RMQ queries (\cref{th:minocc-index-general}).
\item Using the above index, we design a structure that provides random access to the LPF and LPnF arrays (see \cref{sec:lpf-overview}).
\end{itemize}
Putting everything together, we get the first $o(\Textlen)$-time LZ77 factorization after nearly 50 years.

\subsection{Index for Longest Previous Factors}\label{sec:lpf-overview}

We outline how, given any constant $\epsilon \in (0, 1)$ along with
the $\bigO(\Textlen / \log_{\AlphabetSize} \Textlen)$-space representation of $\Text \in \IntegerAlphabet^{\Textlen}$,
in $\bigO(\Textlen \log \AlphabetSize / \sqrt{\log \Textlen})$ time 
and $\bigO(\Textlen / \log_{\AlphabetSize} \Textlen)$ working space, we can construct an index of
size $\bigO(\Textlen / \log_{\AlphabetSize} \Textlen)$ that, given $i \in [1 \dd \Textlen]$, returns
$\LPF{\Text}[i]$ and $\LPFMinOcc{\Text}[i]$ (\cref{def:LPF}).
For simplicity, we focus on the LPF array allowing self-overlaps;
the computations for the non-overlapping variant are similar.

We begin by observing that $\LPF{\Text}[i] \geq \LPF{\Text}[i-1] - 1$ holds for  every $i \in [2 \dd \Textlen]$.
This implies that, given any indexes $i, k \in [1 \dd \Textlen]$ satisfying $i \leq k$, we can 
use the values $\LPF{\Text}[i]$ and $\LPF{\Text}[k]$ to compute a common lower and upper bound for $\LPF{\Text}[j]$ with $j\in [i\dd k]$.
More precisely, we can then compute
$\ell_{\min}$ and $\ell_{\max}$ such that $\LPF{\Text}[j] \in [\ell_{\min} \dd \ell_{\max}]$
and $\ell_{\max} - \ell_{\min} = \bigO((\LPF{\Text}[k] - \LPF{\Text}[i]) + (k-i))$.%

Let $b = \Theta(\log^{3} \Textlen)$ and $b' = \Theta(\log^{6} \Textlen)$ be integers.
We partition $\LPF{\Text}$ into blocks of size~$b$ and store the $\LPF{\Text}[\cdot]$ values at all block boundaries.
Additionally, every block $\LPF{\Text}((i-1)b \dd ib]$ satisfying
$\LPF{\Text}[ib] - \LPF{\Text}[(i-1)b] \geq b'-b$ has all the $\LPF{\Text}[\cdot]$ values explicitly stored in our data structure, and the index $i$ of the block is marked in a bitvector.
At query time, we first check if the queried position $x$ is in one of the stored blocks. If so, we have the answer.
Otherwise, utilizing the above observation, we compute $\ell_{\min}$ and $\ell_{\max}$, and then determine $\LPF{\Text}[j]$
using binary search and the index for leftmost occurrences
(constructed for $\epsilon' = \epsilon/2$).
Due to our choice of $b$ and $b'$,
this takes $\bigO(\log^{\epsilon'} \Textlen \cdot \log (\ell_{\max} - \ell_{\min}+1)) =
\bigO(\log^{\epsilon/2} \Textlen \cdot \log b') = \bigO(\log^{\epsilon/2} \Textlen \cdot \log (\log^{6} \Textlen)) =
\bigO(\log^{\epsilon} \Textlen)$ time.
Once the value $\LPF{\Text}[x]$ is computed, we obtain $\LPFMinOcc{\Text}[x]$ in $\bigO(\log^{\epsilon'} \Textlen)$ time.

Once the index for leftmost occurrences is constructed, computing the $\LPF{\Text}[\cdot]$ values at all block boundaries costs $\bigO(\Textlen/b\cdot \log^{1+\epsilon'}\Textlen)=o(\Textlen/\log_{\AlphabetSize} \Textlen)$ time.
It remains to bound the number of blocks $\LPF{\Text}((i-1)b \dd ib]$ with $\LPF{\Text}[ib] - \LPF{\Text}[(i-1)b] \ge b'-b$
and the total time to process them. For this, observe that the function $f(i) = i + \LPF{\Text}[i]$
is nondecreasing and does not exceed $\Textlen + 1$. This way, we
can bound the number of above blocks by $\bigO(\Textlen / b') = \bigO(\Textlen / \log^{6} \Textlen)$.
Computing all the $\LPF{\Text}[\cdot]$ values in these blocks takes $\bigO((\Textlen / b') \cdot b\cdot \log^{1+\epsilon'} \Textlen) =
\bigO(\Textlen / \log^{3} \Textlen\cdot \log^{1+\epsilon'} \Textlen) = o(\Textlen / \log_{\AlphabetSize} \Textlen)$ time.
The index construction is thus dominated by the time to build the index for leftmost occurrences.

%% file: tools.tex
\section{Auxiliary Tools}\label{sec:tools}
In this section, we develop efficient data structures for a few abstract problems.
\input{tools/rmq.tex}

\input{tools/prefix-rmq.tex}

\input{tools/range-queries.tex}

\input{tools/three-sided-rmq.tex}

\input{tools/dynamic-rmq.tex}

%% file: tools/rmq.tex
\subsection{Range Minimum Queries}\label{sec:rmq}

We start with the textbook Range Minimum Queries (RMQ). 

\begin{definition}[Range Minimum Queries, RMQ]\label{def:rmq}
  Let $A \in \Zn^m$ be a sequence of $m$ non-negative integers.
  For every $b, e \in [0 \dd m]$
  we define\footnote{We assume that $\argmin\{ f(x) : x\in S\}$ returns the
  \emph{smallest} $y \in S$ such that $f(y) = \min\{ f(x) : x\in S\}$.}
  \[
    \RMQ{A}{b}{e} :=      \arg\min \{A[i] : i \in (b \dd e]\}.
  \]
\end{definition}
Classic RMQ data structures take $\Oh(m)$ space and answer queries in constant time~\cite{HT84,GBT84}.
More recently, the structure size has been improved from $\Oh(m)$ machine words to $\Oh(m)$ bits.

\begin{lemma}[{Fischer and Heun~\cite{FischerH11}}]\label{lm:rmq:fh}
  For every array $A[1\dd m]$ of $m$ integers, there is a data structure of $\Oh(m)$ bits that answers range minimum queries over $A$ in $\Oh(1)$ time and can be constructed in $\Oh(m)$ time.
\end{lemma}

Unfortunately, the construction algorithm of~\cite{FischerH11} is unable to achieve sublinear running time if $A$ consists of small integers.
Gao, He, and Nekrich addressed this issue in~\cite[Lemma 7]{Gao0N20}, but their data structure takes $\Oh(m \log \log n)$ bits.
In the following lemma, we show how to simultaneously achieve optimal size, construction time, and query time in the \emph{systematic} setting, where the query algorithm is assumed to have random access to the array $A$.

\begin{lemma}\label{lm:rmq:packed}
  For every array $A[1\dd m]$ of $m$ integers in $[0\dd\sigma)$, there is a data structure of $\Oh(m)$ bits that answers range minimum queries over $A$ in $\Oh(1)$ time using $\Oh(1)$ comparisons between elements of $A$.
  It can be constructed in $\Oh(m\log \sigma / \log n)$ time assuming that $A$ is given in the packed representation of $A$ and with access to a universal table of $o(n)$ bits precomputed in $o(n)$ time.
\end{lemma}
\begin{proof}
  We henceforth assume that $\sigma$ is a power of two (otherwise, it can be increased to the nearest power of two) and that $\sigma \le \sqrt{n}$ (otherwise, the result immediately follows from \cref{lm:rmq:fh}).
  
  Let us partition $A$ into blocks of $\tau=\lfloor\frac{\log n}{2\log \sigma}\rfloor$ elements (the last block can be shorter).
  Formally, let $m'=\lceil \frac{m}{\tau}\rceil$ be the number of blocks and, for $j\in [1\dd m']$, let $B[j]$ denote the $j$th block of $A$, corresponding to entries $A[i]$ with $i\in ((j-1)\tau\dd \min(j\tau,m)]$.
  Moreover, let $A'[1\dd m']$ be an array whose $j$th entry is the minimum value in the $j$th block $B[j]$.

  Our data structure consists of two components:
  \begin{enumerate}
    \item A packed sequence $R[1\dd m']$ whose $j$th element $R[j]$ is the RMQ data structure of \cref{lm:rmq:fh} for $B[j]$.
    \item The RMQ data structure of \cref{lm:rmq:fh} for $A'$.
  \end{enumerate}
  Each entry $R[j]$ occupies $\Oh(\tau)$ bits, so the sequence $R$ uses $\Oh(m'\tau)=\Oh(m)$ bits in total.
  Moreover, the RMQ data structure for $A'$ takes $\Oh(m')\le \Oh(m)$ bits.

  At query time, given a range $(b\dd e]$, we first compute $b' = \ceil{\frac{b}{\tau}}$ and $e'=\floor{\frac{e}{\tau}}$. 
  If $b'>e'$, then the entire query falls within the $b'$th block, and it can be answered using $R[b']$.
  Otherwise, we decompose $(b\dd e]$ into three subranges: $(b\dd b'\tau]$, $(b'\tau\dd e'\tau]$, and $(e'\tau\dd e]$,
  compute the answer for each nonempty subrange independently, and finally compare at most three elements of the original sequence $A$ to determine which subrange minimum is also the minimum within the entire range $(b\dd e]$.
  As for $(b\dd b'\tau]$ and $(e'\tau\dd e]$, we simply use $R[b']$ and $R[e'+1]$, respectively.
  As for $(b'\tau\dd e'\tau]$, we use the RMQ data structure for $A'$ to determine $a':=\RMQ{A'}{b'}{e'}$.
  This means $\RMQ{A}{b'\tau}{e'\tau}\in ((a'-1)\tau\dd a']$ can be retrieved using $R[a']$.
  Overall, the query algorithm takes $\Oh(1)$ time and compares at most three elements of $A$.

  At construction time, we use a universal table that stores the data structure of \cref{lm:rmq:fh} for every sequence of at most $\tau$ elements.
  It takes $\Oh(\sigma^\tau \cdot \tau)=o(\sigma^{2\tau})=o(n)$ bits and can be constructed in $\Oh(\sigma^\tau \cdot \tau)=o(\sigma^{2\tau})=o(n)$ time.
  Using this table, each element of the sequence $R$ can be constructed in $\Oh(1)$ time, for a total of $\Oh(m')$.
  Similarly, the RMQ data structure for $A'$ can be constructed in $\Oh(m')$.
  Overall, the construction algorithm takes $\Oh(m')=\Oh(m/\tau)=\Oh(m\log \sigma / \log n)$ time.
\end{proof}

%% file: tools/prefix-rmq.tex
\subsection{Prefix Range Minimum Queries}\label{sec:prefix-rmq}
As indicated in \cref{sec:overview}, our index for leftmost occurrences relies on a new query type: \emph{prefix range minimum queries}.
In this section, we formally define these queries and provide a series of efficient data structures for answering them.

\begin{definition}[Prefix RMQ]\label{def:prefix-rmq}
  Let $A \in \Zn^m$ be a sequence of $m$ nonnegative integers and $S \in (\Sigma^{*})^m$ be
  a sequence of $m$ strings over alphabet $\Sigma$.
  For every $b, e \in [0 \dd m]$ and $X \in \Sigma^{*}$
  we define
  \[
    \PrefixRMQ{A}{S}{b}{e}{X} :=
      \arg\min \{A[i] : i \in (b \dd e]\text{ and }
      X\text{ is a prefix of }S[i]\}.
  \]
  We assume that $\PrefixRMQ{A}{S}{b}{e}{X}=\arg\min\emptyset = \infty$ if there is no $i \in (b \dd e]$ for which $X$ is a prefix of $S[i]$.
\end{definition}

All our solutions reduce prefix RMQ to the following prefix rank and selection queries, introduced in~\cite{breaking} also in the context of compact text indexes.

\begin{definition}[Prefix Rank and Selection Queries,~\cite{breaking}]
  Let $S \in (\Sigma^{*})^m$ be a sequence of strings over alphabet~$\Sigma$.
\begin{description}[style=sameline,itemsep=1ex]
\item[$\PrefixRank{S}{j}{X}$:]
 Given $X \in \Sigma^{*}$ and $j \in [0 \dd m]$, compute
 $|\{i \in [1 \dd j]: X\text{ is a prefix of }S[i]\}|$.
\item[$\PrefixSelect{S}{r}{X}$:]
 Given $X \in \Sigma^{*}$ and $r \in [1\dd \PrefixRank{S}{m}{X}]$,
 find the $r$th smallest element of
 $\{i \in [1 \dd m] : X\text{ is a prefix of }S[i]\}$.
\end{description}
\end{definition}

The following proposition presents the state-of-the-art trade-off for prefix rank and selection queries. 
Our goal will be to achieve the same bounds for prefix RMQ.

\begin{proposition}[{\cite[Proposition 4.3]{breaking}}]\label{pr:prefix-rs}
  For all integers $h,m,\ell,\sigma \in\Zp$ satisfying $h\ge 2$ and $m\ge \sigma^\ell \ge 2$, and for every sequence $S\in ([0\dd \sigma)^\ell)^{\le m}$, there exists a data structure of size $\Oh(m\log_h (h\ell))$ that answers prefix rank queries in $\Oh(h\log\log m \log_h(h\ell))$ time and prefix selection queries in $\Oh(h\log_h(h\ell))$ time.
  Moreover, it can be constructed in $\Oh(m\min(\ell,\sqrt{\log m})\log_{h}(h\ell))$ time using $\Oh(m\log_h (h\ell))$ space given the packed representation of $S$ and the parameter $h$.
\end{proposition}

In the context of a fixed instance of the RMQ problem, for every $X\in \Sigma^*$, let $S_X$ denote the subsequence of $S_X$ to denote the subsequence of $S$ consisting of elements prefixed with $X$ and $m_X$ to denote the number of these elements.
Moreover, let $A_X$ denote the subsequence of $A$ consisting of the entries $A[i]$ such that $X$ is a prefix of $S[i]$.
We repeatedly use prefix rank and selection queries to convert indices in $S_X$ or $A_X$ into indices in $S$ or $A$, respectively, and vice versa. 
In particular, $S[i] = S_{S[i]}[\PrefixRank{S}{i}{S[i]}]$ and $A[i] = A_{S[i]}[\PrefixRank{S}{i}{S[i]}]$ holds for every $i\in [1\dd m_{\varepsilon}]$. 
Conversely, $S_X[r]=S[\PrefixSelect{S}{r}{X}]$ and $A_X[r]=A[\PrefixSelect{S}{r}{X}]$ holds for every $X\in \Sigma^*$ and $r\in [1\dd m_X]$. 
Consequently, the answer to every prefix RMQ query can be expressed as $\PrefixRMQ{A}{S}{b}{e}{X}=\RMQ{A_X}{\PrefixRank{S}{b}{X}}{\PrefixRank{S}{e}{X}}$.

The most challenging task on the way to achieving the trade-off of \cref{pr:prefix-rs} is to match the construction-time bound.
We address this in a series of auxiliary results that build on top of each other to achieve faster and faster construction.
We start with a simple version with $\Oh(m\ell)$-time construction algorithm; here, we do not need to assume that $S$ is given in the packed representation.

\begin{lemma}\label{lm:prefix-rmq:simple}
  For all integers $m,\ell,\sigma \in\Zp$ satisfying $m\ge \sigma^\ell \ge 2$, and for all equal-length sequences $A\in \Zn^{\le m}$ and $S\in ([0\dd \sigma)^\ell)^{\le m}$, there exists a data structure of $\Oh(m\ell+\sigma^\ell\log m)$ bits that answers prefix RMQ queries in $\Oh(1)$ time using $\Oh(1)$ prefix rank and selection queries on $S$.
  It can be constructed in $\Oh(m\ell)$ time using $\Oh(m)$ space assuming that $S$ is given in the packed representation.
\end{lemma}
\begin{proof}
  For every $X\in [0\dd \sigma)^{\le \ell}$, we store an instance $R_X$ of the RMQ data structure of \cref{lm:rmq:fh} constructed for $A_X$. This data structure takes $\Oh(m_X)$ bits.
  The total size of these components is $\Oh(m\ell)$ bits since each string $S[i]$ has $\ell+1$ different prefixes.
  Additionally, we use $\Oh(\sigma^{\ell}\log m)$ bits for pointers to these components.
  
  Given a query $\PrefixRMQ{A}{S}{b}{e}{X}$, we first compute $b'=\PrefixRank{S}{b}{X}$ and $e'=\PrefixRank{S}{e}{X}$;
  observe that $\{A[i] : i\in (b\dd e] \text{ and }X\text{ is a prefix of }S[i]\}=\{A_X[i] : i\in (b'\dd e']\}$.
  If the range $(b'\dd e]$ is empty, then we output $\infty$.
  Otherwise, we retrieve $a'=\RMQ{A_X}{b'}{e'}$ using $R_X$ and output $a:=\PrefixSelect{S}{a'}{X}$.
  Since the query time of \cref{lm:rmq:fh} is constant, the overall query time is also constant except for the prefix rank and selection queries.

  At construction time, we iterate over prefix lengths $k\in [0\dd \ell]$.
  At each iteration, we explicitly build the arrays $A_X$ for $X\in [0\dd \sigma)^k$; for this, we process each pair $(A[i],S[i])$ in the left-to-right order, retrieve the prefix $X= S[i][1\dd k]$, and append $A[i]$ to the array $A_X$ (implemented as an extendible vector, initially empty).
  Once we process the entire sequences $A$ and $S$, we use the construction algorithm of \cref{lm:rmq:fh} to build $R_X$ for every $X\in [0\dd \sigma)^k$.
  Each iteration takes $\Oh(m)$ time and space, for a total of $\Oh(m\ell)$ time and $\Oh(m)$ working space.
\end{proof}

Next, we develop a variant whose construction algorithm uses a universal table shared by many instances of the data structure. This variant is useful when $\ell$ is large compared to $\log m$.
\begin{lemma}\label{lm:prefix-rmq:packed}
  For all integers $n,m,\ell,\sigma \in\Zp$ satisfying $n\ge m\ge \sigma^\ell \ge 2$ and all equal-length sequences $A\in [0\dd m)^{\le m}$ and $S\in ([0\dd \sigma)^\ell)^{\le m}$, there exists a data structure of $\Oh(m\ell + \sigma^\ell\log m)$ bits that answers prefix RMQ queries in $\Oh(1)$ time using $\Oh(1)$ prefix rank and selection queries on $S$ and comparisons between elements of~$A$.  
  It can be constructed in $\Oh(m+m \log^2 m/ \log n)$ time using $\Oh(m)$ space assuming that $S$ is given in the packed representation and with access to universal tables of size $o(n)$ precomputed in $o(n)$ time.
\end{lemma}
\begin{proof}
  We henceforth assume that $m \le \sqrt[3]{n}$; otherwise, $\Oh(m\ell)\le \Oh(m\log m)\le \Oh(m\log^ m / \log n)$ and the result follows immediately from \cref{lm:prefix-rmq:simple}.
  We proceed as in the proof of \cref{lm:prefix-rmq:simple}, but each component $R_X$ is constructed using \cref{lm:rmq:packed}
  instead of \cref{lm:rmq:fh}. The total size of these components is $\Oh(m\ell)$ bits, and pointers to them occupy $\Oh(\sigma^\ell \log m)$ bits on top of that.

  The query algorithm works as in the proof of \cref{lm:prefix-rmq:simple}, but now $\RMQ{A_X}{b'}{e'}$ needs to make $\Oh(1)$ access queries to the array $A_X$.
  To retrieve a value $A_X[i']$, we compute $i:=\PrefixSelect{S}{i'}{X}$ and retrieve $A[i]$.
  Overall, the query algorithm still takes $\Oh(1)$ time, it also needs $\Oh(1)$ access queries to the array $A$ on top of $\Oh(1)$ prefix rank and select queries.

  In the construction algorithm, we still iterate over $k\in [0\dd \ell]$ and build $A_X$ for each $X\in [0\dd \sigma)^k$.
  However, the sequence $A_X$ is now constructed in the packed representation, and we build it along with (the packed representation of) the subsequence $S_X$ consisting of the elements of $S$ prefixed with $X$.
  In the algorithm, we interpret each character in $[0\dd \sigma)$ as a binary string of length $\ceil{\log \sigma}$, and each string $X\in [0\dd \sigma)^k$ as a string in $\{{\tt 0},{\tt 1}\}^{k\ceil{\log \sigma}}$.
  Thus, for each $Y\in \{{\tt 0},{\tt 1}\}^{\le \ell\ceil{\log \sigma}}$, we can define $A_Y$ and $S_Y$ analogously to $A_X$ and $S_Y$, respectively.
  Our construction algorithm, in fact, iterates over $k\in [0\dd \ell\ceil{\log \sigma}]$ and, at each iteration, constructs the packed representations of $A_Y$ and $S_Y$ for all $Y\in \{{\tt 0},{\tt 1}\}^k$. 
  Whenever $k$ is an integer multiple of $\ceil{\log \sigma}$ and  $Y$ can be interpreted as $X\in [0\dd \sigma)^{k/\log \sigma}$, we use the construction algorithm of \cref{lm:rmq:packed} to $R_X$, which can be equivalently interpreted as the RMQ data structure for $A_Y$.
  At every iteration other than the last one, for every pair of non-empty sequences $(A_Y,S_Y)$, we build the analogous pairs $(A_{Y0},S_{Y0})$ and $(A_{Y1},S_{Y1})$. 

  To implement this operation efficiently, we use a precomputed table that, given a pair of sequences $(A',S')$ of equal length at most $\tau:=\floor{\frac{\log n}{3\log m}}$ and a position $k\in [1\dd \ell\ceil{\log \sigma}]$,
  splits the input sequences into $(A'_0,S'_0)$ and $(A'_1,S'_1)$ depending on whether the $k$-bit of $S'[i]$ is equal to $0$ or $1$.
  Each entry occupies $\tau(\log m + \ell \log \sigma)=\Oh(\tau\log m)$ bits and can be computed in $\Oh(\tau\cdot \ell) \le \Oh(\tau\log m)$ time.
  The total number of entries is $\Oh((m\cdot \sigma^\ell)^\tau \cdot \ell)\le \Oh(m^{2\tau}\log m) \le \Oh(n^{2/3}\log n)$.
  Overall, the table uses $\Oh(n^{2/3}\log n \cdot \tau \log m)=\Oh(n^{2/3}\log^2 n)=o(n)$ bits and takes $\Oh(n^{2/3}\log^2 n)=o(n)$ construction time.

  With this precompute table, the sequences $(A_{Y0},S_{Y0})$ and $(A_{Y1},S_{Y1})$ can be constructed in $\Oh(1+m_Y/\tau)$ time for each $Y\in \{{\tt 0},{\tt 1}\}^{\ell\ceil{\log \sigma}}$, which is $\Oh(m+m\ell \log \sigma/\tau)=\Oh(m+m\log^2 m / \log n)$ in total.
  The construction algorithm of \cref{lm:rmq:packed} takes $\Oh(1+m_X\log m/\log n)$ time for each $X\in [0\dd \sigma)^\ell$, which is $\Oh(m+m\ell \log m / \log n)\le \Oh(m+m\log^2 m / \log n)$ in total.

  The working space of the construction algorithm remains at $\Oh(m)$ machine words because the sequences $(A_Y,S_Y)$ can be discarded as soon as $(A_{Y0},S_{Y0})$ and $(A_{Y1},S_{Y1})$ are constructed.
\end{proof}

The following variant builds upon the previous one to achieve competitive construction time also when $\ell$ is small compared to $\log m$ and, in particular, $\ell \log \sigma \le \log m$.

\begin{lemma}\label{lm:prefix-rmq:shallow}
  For all integers $n,m,\ell,\sigma \in\Zp$ satisfying $n\ge m\ge \sigma^\ell \ge 2$ and all equal-length sequences $A\in [0\dd m)^{\le m}$ and $S\in ([0\dd \sigma)^\ell)^{\le m}$, there exists a data structure of $\Oh(m\ell + \sigma^\ell\log m)$ bits that answers prefix RMQ queries in $\Oh(1)$ time using $\Oh(1)$ prefix rank and selection queries on $S$ and comparisons between the elements of~$A$.  
  It can be constructed in $\Oh(m+m (\ell \log \sigma)^2/ \log n)$ time using $\Oh(m)$ space assuming that $S$ is given in the packed representation and with access to universal tables of size $o(n)$ precomputed in $o(n)$ time.
\end{lemma}
\begin{proof}
  We henceforth assume that $m \ge \sigma^{\ell}\cdot \ceil{\log m}$; otherwise, $\log m = \Oh(\ell\log \sigma)$ and the result immediately follows from \cref{lm:prefix-rmq:packed}.
  We partition $A$ and $S$ into blocks of $\tau:=\sigma^{\ell}\cdot \ceil{\log m}$ elements each (the last blocks can be shorter); the number of blocks is $m':= \ceil{\frac{m}{\tau}}$.
  For each $j\in [1\dd m']$, let $A^{(j)}$ and $S^{(j)}$ denote the $j$th block of $A$ and $S$, respectively, with the values in $A^{(j)}$ replaced by ranks within $A^{(j)}$, which are integers in $[0\dd \tau)$.
  
  Additionally, for each $X\in [0\dd \sigma)^{\ell}$, define an array $A'_X[1\dd m']$ such that, for every $j\in [1\dd m']$,
   \[A'_X[j]=\min\{A[i]:i\in ((j-1)\tau\dd j\tau] \text{ and }X\text{ is a prefix of }S[i]\},\]
  under the standard assumption that $\min \emptyset = \infty$ is represented with an integer than any entry in $A$.
  Moreover, for each $X\in [0\dd \sigma)^{\ell}$, define an array $P_X[1\dd m']$ with $P_X[j]=\PrefixRank{S}{(j-1)\tau}{X}$ for every $j\in [1\dd m']$, assuming that $\PrefixRank{S}{0}{X}=0$.

  Our data structure consists of the following components:
  \begin{enumerate}
    \item the prefix RMQ data structure of \cref{lm:prefix-rmq:packed} for $A^{(j)}$ and $S^{(j)}$ for each $j\in [1\dd m']$;
    \item the RMQ data structure of \cref{lm:rmq:fh} for $A'_X$ for each $X\in [0\dd \sigma)^{\ell}$;
    \item the $P_X$ table for each $X\in [0\dd \sigma)^{\ell}$.
  \end{enumerate}

  Each prefix RMQ data structure takes $\Oh(\tau \ell + \sigma^{\ell}\log \tau)=\Oh(\tau\ell)$ bits, for a total of $\Oh(m\ell)$ bits.
  Each RMQ data structure and $P_X$ table takes $\Oh(m')$ and $\Oh(m'\log m)$ bits, respectively, for a total $\Oh(m'\sigma^{\ell} \log m)\le \Oh(m'\tau)=\Oh(m)$ bits.
  Overall, the size of our data structure is $\Oh(m\ell)$ bits.

  To apply the query algorithm of \cref{lm:prefix-rmq:packed}, we need to implement prefix rank and selection queries on $S^{(j)}$ as well as comparisons between elements of $A^{(j)}$.
  For this, observe that $\PrefixRank{S^{(j)}}{i}{X}=\PrefixRank{S}{(j-1)\tau+i}{X}-P_X[j]$, $\PrefixSelect{S^{(j)}}{r}{X}=\PrefixSelect{S}{r+P_X[j]}{X}-(j-1)\tau$, and $A^{(j)}[i]<A^{(j)}[i] \Longleftrightarrow A[i+(j-1)\tau] < A[i'+(j-1)\tau]$.
  Consequently, prefix RMQ queries on $A^{(j)}$ and $S^{(j)}$ can be implemented in $\Oh(1)$ time using $\Oh(1)$ prefix rank and selection queries on $S$ as well comparisons between elements of $A$.

  Given a query $\PrefixRMQ{A}{S}{b}{e}{X}$, we first compute $b'=\ceil{\frac{b}{\tau}}$ and $e'=\floor{\frac{e}{\tau}}$.
  If $b'>e'$, then the entire query falls within the $b'$th block, and it can be answered using a prefix RMQ query on $A^{(b')}$ and $S^{(b')}$; formally, $\PrefixRMQ{A}{S}{b}{e}{X} = (b'-1)\tau + \PrefixRMQ{A^{(b')}}{S^{(b')}}{b-(b'-1)\tau}{e-(b'-1)\tau}{X}$.
  Otherwise, we decompose $(b\dd e]$ into three subranges: $(b\dd b'\tau]$, $(b'\tau\dd e'\tau]$, and $(e'\tau\dd e]$, compute the answer for each nonempty subrange, and finally compare at most three elements of $A$ to determine which subrange prefix minimum is also the prefix minimum within the entire range $(b\dd e]$. 
  As for $(b\dd b'\tau]$ and $(e'\tau\dd e]$, we simply use the components of \cref{lm:prefix-rmq:packed} for $\PrefixRMQ{A^{(b')}}{S^{(b')}}{\cdot}{\cdot}{\cdot}$ and $\PrefixRMQ{A^{(e'+1)}}{S^{(e'+1)}}{\cdot}{\cdot}{\cdot}$ queries, respectively.
  As for $(b'\tau\dd e'\tau]$, we use the RMQ data structure for $A'_X$ to determine $a':= \RMQ{A'_X}{b'}{e'}$.
  This means that $\PrefixRMQ{A}{S}{b'\tau}{e'\tau}{X}\in \{\infty\}\cup ((a'-1)\tau\dd a'\tau]$ can be retrieved using a $\PrefixRMQ{A^{(a')}}{S^{(a')}}{\cdot}{\cdot}{\cdot}$ query.
  Overall, the query algorithm takes $\Oh(1)$ time, performs $\Oh(1)$ prefix rank and selection queries on $S$, and compares $\Oh(1)$ elements of $A$.

  It remains to describe the construction algorithm. 
  The sequences $S^{(j)}$ are built in $\Oh(m)$ time simply by splitting the sequence $S$. 
  The arrays $A^{(j)}$ can also be built in $\Oh(m)$ time by inspecting the entries $A[i]$ ordered by increasing values and keeping track of the number of already processed entries in each block.
  On top of these ingredients, each data structure of \cref{lm:prefix-rmq:packed} is constructed in $\Oh(\tau+\tau\log^2\tau/\log n)\le \Oh(\tau+\tau (\ell \log \sigma)^2/\log n + \tau \log^2 \log m/\log n)=\Oh(\tau+\tau(\ell \log \sigma)^2/\log n)$ time,
  for a total of $\Oh(m+m(\ell\log\sigma)^2/\log n)$ time across all blocks.
  The space complexity of this phase is $\Oh(m)$.
  
  The arrays $A'_X$ are initialized $\infty$ values and constructed first for $|X|=\ell$.
  For this, we scan the entries of $A$ and $S$ from left to right.
  For each index $i$ contained within the block $j=\ceil{\frac{i}{\tau}}$, we set $A'_{S[i]}[j]:=\min(A[i],A'_{S[i]}[j])$.
  Then, we process the strings $X\in [0\dd \sigma)^{<\ell}$ in the order of decreasing lengths, setting $A'_X[j]=\min_{c\in [0\dd \sigma)} A'_{Xc}[j]$ for each $j\in [1\dd m']$.
  Overall, constructing the arrays $A'_X$ takes $\Oh(m+m'\sigma^\ell)=\Oh(m)$ time and space.

  To construct the arrays $P_X$, we scan the sequence $S$ from left to right keeping track, for each $X\in [0\dd \sigma)^\ell$, the number $q_X$ of entries equal to $X$ encountered so far.
  These values are initialized with $0$s and $q_{S[i]}$ is increment when we process $S[i]$.
  Before we start processing the $j$th block, that is, before processing $S[(j-1)\tau+1]$, we set $P_X[j]:=q_X$ for $X\in [0\dd \sigma)^\ell$; for  $X\in [0\dd \sigma)^{<\ell}$ in the decreasing order of lengths, we set $P_X[j]=\sum_{c\in [0\dd \sigma)}P_{Xc}[j]$.
  Overall, constructing the arrays $P_X$ takes $\Oh(m+m'\sigma^\ell)=\Oh(m)$ time and space.
  The entire construction algorithm takes $\Oh(m+m(\ell\log\sigma)^2/\log n)$ time and $\Oh(m)$ space.
\end{proof}

Our final implementation combines multiple instances of the data structure of \cref{lm:prefix-rmq:shallow} using a high-level scheme inspired by \cref{lm:prefix-rmq:simple}. 
This is similar to how the state-of-the-art wavelet tree construction algorithms~\cite{WaveletSuffixTree,MunroNV16} are designed.

\begin{proposition}\label{pr:prefix-rmq}
  For all integers $m,\ell,\sigma \in\Zp$ satisfying $m\ge \sigma^\ell \ge 2$ and all equal-length sequences $A\in [0\dd m)^{\le m}$ and $S\in ([0\dd \sigma)^\ell)^{\le m}$, there exists a data structure of $\Oh(m\ell + \sigma^\ell\log m)$ bits that answers prefix RMQ queries in $\Oh(1)$ time using $\Oh(1)$ prefix rank and selection queries on $S$ and comparisons between the elements of~$A$.
  It can be constructed in $\Oh(m+m\ell\log\sigma/\sqrt{\log m})$ time using $\Oh(m)$ space assuming that $S$ is given in the packed representation.
\end{proposition}
\begin{proof}
  We henceforth assume that $\sqrt{\log m} \le \ell\log\sigma$; otherwise, the result follows immediately from \cref{lm:prefix-rmq:shallow} with $n=m$ because then $m(\ell\log\sigma)^2/\log n \le m\ell\log\sigma/\sqrt{\log m}$.
  Moreover, we assume that $\log \sigma \le\sqrt{\log m}$; otherwise, the result follows immediately from \cref{lm:prefix-rmq:simple}
  because then $m\ell \le m\ell\log\sigma/\sqrt{\log m}$.

  Let $\alpha = \floor{\sqrt{\log m}/\log \sigma}$.
  For every $X\in [0\dd \sigma)^{\ell}$, define a sequence $S'_X[1\dd m_X]$ such that $S'_X[i]=S_X[i](|X|\dd \min(\ell,|X|+\alpha)]$;
  in other words, $S'_X[i]$ consists of the first $\alpha$ characters following the prefix $X$ of $S_X[i]$ (or fewer characters if $|X|>\ell-\alpha)$. 
  Moreover, we define $A'_X\in [0\dd m_X)^{m_X}$ to be a sequence obtained from $A_X\in \Zn^{m_X}$ by replacing each value $A_X[i]$ with its rank among all the values present in $A_X$.
  For every prefix $Y\in [0\dd \sigma)^{<\ell}$ whose length is divisible by $\alpha$, we store the prefix RMQ data structure of \cref{lm:prefix-rmq:shallow} constructed for $A'_Y$ and $S'_Y$ with parameter $n:=m$.

  We analyze the size of these components separately for each length $k\in [0\dd \ell)$ divisible by $\alpha$.
  If $k\le \ell-\alpha$, then each component takes $\Oh((m_Y+\sigma^{\alpha})\alpha+\sigma^{\alpha}\log (m_Y+\sigma^\alpha)) \le \Oh(m_Y\alpha+\sigma^{\alpha}\log m)$ bits.
  Across all $Y\in [0\dd \sigma)^k$, this sums up to $\Oh(m\alpha+\sigma^{k+\alpha}\log m)$ bits.
  If $k > \ell-\alpha$, on the other hand, each component takes $\Oh((m_Y+\sigma^{\ell-k})(\ell-k)+\sigma^{\ell-k}\log (m_Y+\sigma^{\ell-k})) \le \Oh(m_Y\alpha+\sigma^{\ell-k}\log m)$ bits.
  Across all $Y\in [0\dd \sigma)^k$, this sums up to $\Oh(m\alpha+\sigma^{\ell}\log m)$ bits.
  Taking into account the contribution of each length $k\in [0\dd \ell)$ divisible by $\alpha$, the total size of our data structure is $\Oh(m\alpha \ceil{\ell/\alpha}+\sigma^{\ell}\log m)=\Oh(m\ell+\sigma^{\ell}\log m)$.
  Additionally, pointers to the individual components take $\Oh(\sigma^{\ell}\log m)$ bits in total.

  Before we can use the query algorithm of \cref{lm:prefix-rmq:shallow}, we need to explain how to implement prefix rank and selection queries on $S'_Y$ and how to compare elements of $A'_Y$.
  As for the latter, observe that comparing elements of $A'_Y$ is equivalent to comparing elements of $A_Y$ and that $A_Y[r]=A[\PrefixSelect{S}{r}{Y}]$ holds for every $r\in [1\dd m_Y]$.
  Consequently, a comparison between two elements of $A'_Y$ can be implemented in $\Oh(1)$ time using two prefix selection queries on $S$ and one comparison between two elements of $A$.
  As for the prefix rank and selection queries on $S'_Y$, consider a string $Z\in [0\dd \sigma)^{\le \alpha}$ and observe that $\PrefixRank{S'_Y}{i}{Z}=\PrefixRank{S}{\PrefixSelect{S}{i}{Y}}{YZ}$ holds for every $i\in [1\dd m_Y]$,
  and $\PrefixSelect{S'_Y}{r}{Z}=\PrefixRank{S}{\PrefixSelect{S}{r}{YZ}}{Y}$ holds for every $r\in [1\dd m_{YZ}]$.
  Consequently, each prefix rank and selection query on $S'_Y$ can be implemented in $\Oh(1)$ time using one prefix rank query and one prefix rank selection query on $S$.
  Overall, we conclude that the query algorithm of \cref{lm:prefix-rmq:shallow} lets us answer prefix RMQ queries on $A_Y$ and $S'_Y$ in $\Oh(1)$ time using $\Oh(1)$ prefix rank and selection queries on $S$ and comparisons between the elements of~$A$.

  Given a query $\PrefixRMQ{A}{S}{b}{e}{X}$, we express $X$ as $X=YZ$, where $|Z|=|X|\bmod \alpha$,
  and compute $b'=\PrefixRank{S}{b}{Y}$ and $e'=\PrefixRank{S}{e}{Y}$;
  observe that $\{A[i]: i\in (b\dd e]\text{ and }X\text{ is a prefix of }S[i]\}=\{A_Y[i] : i\in (b'\dd e']\text{ and }Z\text{ is a prefix of }S'_Y[i]\}$. 
  Consequently, we use the component of \cref{lm:prefix-rmq:shallow} to determine $a':=\PrefixRMQ{A'_Y}{S'_Y}{b'}{e'}{Z}$.
  If $a'=\infty$, we report $\infty$. Otherwise, we report $a:=\PrefixSelect{S}{a'}{Y}$.
  Overall, the query algorithm is implemented $\Oh(1)$ time using $\Oh(1)$ prefix rank and selection queries on $S$ and comparisons between the elements of~$A$.

  The construction algorithm starts with building the universal tables necessary for \cref{lm:prefix-rmq:shallow}; this takes $o(m)$ time and space.
  Next, we iterate over integers $k\in [0\dd\ell)$ that are multiples of~$\alpha$.
  To build $A_Y$ and $S_Y$ for all $Y\in [0\dd \sigma)^k$, for each index $i\in [1\dd m_{\varepsilon}]$ in the left-to-right order,
  we extract $Y:= S[i][1\dd k]$ and append $A[i]$ and $S[i]$ to the arrays $A_X$ and $S_X$, respectively (both implemented as extendible vectors, initially empty).
  In order to transform $A_Y$ to $A'_Y$, we sort tuples $(Y,A_Y[i],i)$ across $Y\in [0\dd \sigma)^k$ and $i\in [1\dd m_Y]$ using a linear-time sorting algorithm. 
  Finally, we extract $S'_Y$ from $S_Y$ by taking $S'_Y[i]=S_Y[i](k\dd \min(k+\alpha,\ell)]$ for each  $Y\in [0\dd \sigma)^k$ and $i\in [1\dd m_Y]$.
  Overall, we take $\Oh(m)$ time and space to build $A'_Y$ and $S'_Y$ for all $Y\in [0\dd \sigma)^k$.
  As a final step, we build the data structure of \cref{lm:prefix-rmq:shallow} on top each pair $(A'_Y,S'_Y)$.
  If $k\le\ell-\alpha$, this takes $\Oh((m_Y+\sigma^{\alpha})\cdot (1+(\alpha \log \sigma)^2/\log m))=\Oh(m_Y+\sigma^{\alpha})$ time and space per instance, for a total of $\Oh(m+\sigma^{k+\alpha})$.
  Otherwise, the construction algorithm takes $\Oh((m_Y+\sigma^{\ell-k})\cdot (1+((\ell-k) \log \sigma)^2/\log m))=\Oh(m_Y+\sigma^{\ell-k})$ time and space per instance, for a total of $\Oh(m+\sigma^{\ell})$.
  Across all integers $k\in [0\dd\ell)$ that are multiples of~$\alpha$, the construction time is $\Oh(m\cdot \ceil{\ell/\alpha}+\sigma^{\ell})=\Oh(m+m\ell\log\sigma/\sqrt{\log m})$.  
\end{proof}

The data structure of \cref{pr:prefix-rs} is formulated as a reduction to prefix rank and selection queries, and it does account for the space occupied by the array $A$.
The following theorem describes a stand-alone solution incorporating these components and the overheads necessary to use them.   

\begin{theorem}\label{th:prefix-rmq}
  For all integers $h,m,\ell,\sigma \in\Zp$ satisfying $h\ge 2$ and $m\ge \sigma^\ell \ge 2$, and for all equal-length sequences $A\in [0\dd m)^{\le m}$ and $S\in ([0\dd \sigma)^\ell)^{\le m}$, there exists a data structure of size $\Oh(m\log_h (h\ell))$ that answers prefix RMQ queries in $\Oh(h\log\log m \log_h(h\ell))$ time.
  Moreover, it can be constructed in $\Oh(m\min(\ell,\sqrt{\log m})\log_{h}(h\ell))$ time using $\Oh(m\log_h (h\ell))$ space assuming that $S$ is given in the packed representation.
\end{theorem}
\begin{proof}
  Our solution consists of three components:
  \begin{enumerate}
    \item the array $A$,
    \item the data structure of \cref{pr:prefix-rs} for prefix rank and selection queries on $S$, and
    \item the data structure of \cref{pr:prefix-rmq} (if $\sqrt{\log m} < \ell$) or \cref{lm:prefix-rmq:simple} (otherwise) for prefix RMQ queries on $A$ and $S$.
  \end{enumerate}
  The array $A$ takes $\Oh(m)$ space, the component for prefix rank and selection queries takes $\Oh(m\log_h(h\ell))$ space,
  and the component for prefix RMQ queries takes $\Oh(m\ell+\sigma^{\ell}\log m)\le \Oh(m\log m)$ bits, which is $\Oh(m)$ machine words.
  The overall size of our solution is $\Oh(m\log_h(h\ell))$.

  Each query is answered in $\Oh(1)$ time using the component of \cref{pr:prefix-rmq} or \cref{lm:prefix-rmq:simple}, but this query algorithm also issues $\Oh(1)$ prefix rank and selection queries on $S$ and comparisons between elements of $A$.
  These auxiliary queries are implemented in $\Oh(h\log\log m \log_h(h\ell))$, $\Oh(h \log_h(h\ell))$, and $\Oh(1)$ time, respectively, using the other two components.
  The overall query time is $\Oh(h\log\log m \log_h(h\ell))$.

  Constructing the array $A$ trivially takes $\Oh(m)$ time, and building the component of \cref{pr:prefix-rs} takes $\Oh(m\min(\ell,\sqrt{\log m})\log_{h}(h\ell))$ and $\Oh(m\log_h (h\ell))$ space.
  If $\ell \le \sqrt{\log m}$, then we use \cref{lm:prefix-rmq:simple} for prefix RMQ queries, and this component is built using $\Oh(m\ell)=\Oh(m\min(\ell,\sqrt{\log m}))$ time and $\Oh(m)$ space.
  Otherwise, we use \cref{pr:prefix-rmq}, which takes $\Oh(m+m\ell\log \sigma / \sqrt{\log m})=\Oh(m\sqrt{\log m})=\Oh(m\min(\ell,\sqrt{\log m}))$ time and $\Oh(m)$ space.
  Overall, the construction algorithm uses $\Oh(m\min(\ell,\sqrt{\log m})\log_{h}(h\ell))$ time and $\Oh(m\log_h (h\ell))$ space.
\end{proof}

Instantiated with $h=\ceil{\ell^{\epsilon/2}}$, \cref{th:prefix-rmq} immediately implies the following result:
\begin{corollary}\label{cr:prefix-rmq}
  For all integers $m,\ell,\sigma \in\Zp$ satisfying $m\ge \sigma^\ell \ge 2$, every constant $\epsilon > 0$, and for all equal-length sequences $A\in [0\dd m)^{\le m}$ and $S\in ([0\dd \sigma)^\ell)^{\le m}$, there exists a data structure of size $\Oh(m)$ that answers prefix RMQ queries in $\Oh(\ell^{\epsilon/2}\log \log m)=\Oh(\log^{\epsilon}m)$ time.
  Moreover, it can be constructed in $\Oh(m\min(\ell,\sqrt{\log m}))$ time using $\Oh(m)$ space assuming that $S$ is given in the packed representation.
\end{corollary}

%% file: tools/range-queries.tex
\subsection{Offline Range Counting}\label{sec:range-queries}

Let $A[1 \dd m]$ be an array of $m \geq 0$ nonnegative integers. We define
the following queries on $A$:

\begin{description}[style=sameline,itemsep=1ex]
\item[Range counting:] Let $b,e \in [0 \dd m]$ and $v \geq 0$. We define
  \begin{itemize}
  \item $\TwoSidedRangeCount{A}{e}{v} := |\{j \in (0 \dd e] : A[j] \geq v\}|$,
  \item $\ThreeSidedRangeCount{A}{b}{e}{v} := |\{j \in (b \dd e] : A[j] \geq v\}|$.
  \end{itemize}
\item[Range selection:] Let $v \geq 0$ and $r \in [1 \dd \TwoSidedRangeCount{A}{m}{v}]$.
  We define $\RangeSelect{A}{r}{v}$ as the $r$th smallest element of
  $\{i \in (0 \dd m] : A[i] \geq v\}$.
\end{description}

\begin{definition}\label{def:int}
  Let $m > 0$, $\AlphabetSize > 1$. For every $X \in \IntegerAlphabet^{\leq m}$,
  by $\Int{m}{\AlphabetSize}{X}$ we denote an integer
  constructed by appending $2m - 2|X|$ zeros to $X$ and $|X|$ $c$s (where $c = \AlphabetSize - 1$)
  to $X$, and then interpreting the resulting string as a base-$\AlphabetSize$
  representation of a number in $[0 \dd \AlphabetSize^{2m})$.
\end{definition}

\begin{lemma}\label{lm:int}
  Let $m > 0$ and $\AlphabetSize > 1$.
  For all $X, X' \in \IntegerAlphabet^{\leq m}$,
  $X \prec X'$ implies $\Int{m}{\AlphabetSize}{X} <
  \Int{m}{\AlphabetSize}{X'}$.
\end{lemma}
\begin{proof}
  For any $x \in [0 \dd \AlphabetSize^{2m})$, let $S(x) \in \IntegerAlphabet^{2m}$
  denote the string obtained by interpreting $x$ written in base $\AlphabetSize$ as a
  string, with zeros appended on the left to pad the string to length $2m$.
  Observe that for any $x, x' \in [0 \dd \AlphabetSize^{2m})$, $x < x'$ holds
  if and only if $S(x) \prec S(x')$.

  Denote $x = \Int{m}{\AlphabetSize}{X}$, $x' = \Int{m}{\AlphabetSize}{X'}$, $s = S(x)$, and $s' = S(x')$.
  By the above, to prove the claim it suffices to show that $s \prec s'$. Consider two cases:
  \begin{itemize}
  \item First, assume that there exists $\ell < \min(|X|, |X'|)$ such that $X[1 \dd \ell] = X'[1 \dd \ell]$ and
    $X[\ell + 1] \prec X'[\ell + 1]$. Observe that by \cref{def:int}, $X$ (resp.\ $X'$) is a prefix of $s$ (resp.\ $s'$).
    This immediately implies $s \prec s'$.
  \item Let us now assume that $X$ is a proper prefix of $X'$. Let $Y$ be such that $XY = X'$, and let $Z = {\tt 0}^{|Y|}$.
    By \cref{def:int}, $XZ$ (resp.\ $XY$) is a prefix of $s$ (resp.\ $s'$). If $Y \neq Z$, then we must have
    $Y \prec Z$, and hence we obtain $s \prec s'$. Let us consider the remaining case when $Y = Z$.
    Note that by \cref{def:int}, we then have $s = X' \cdot {\tt 0}^{2m - |X| - |X'|} \cdot {\tt c}^{|X|}$ and $s' =
    X' \cdot {\tt 0}^{2m - 2|X'|} \cdot {\tt c}^{|X'|}$, where ${\tt c} = \AlphabetSize - 1$. By $|X| < |X'|$, we thus
    obtain $s \prec s'$.
    \qedhere
  \end{itemize}
\end{proof}

\begin{proposition}\label{pr:packed-succ}
  Let $\alpha \in (0,1)$ be a constant. Let $u \geq 1$. In $\bigO(u^{\alpha})$ time we
  can construct a data structure, such that given the pointer to a packed
  representation of a bitvector $B \in \{{\tt 0}, {\tt 1}\}^{m}$ and any pair $(i,j)$
  satisfying $0 \leq i \leq j \leq m$, we can  compute
  $i' = \min\{t \in (i \dd j] : B[t] = {\tt 1}\} \cup \{j+1\}$
  in $\bigO(1 + (i'-i)/\log u)$ time.
\end{proposition}
\begin{proof}

  Let $\alpha' \in (0, \alpha)$ be a constant such that $b := \alpha' \log u$ is a positive integer.
  It is easy to see that such $\alpha'$ exists for all $u \geq u_{\min}$, where $u_{\min}$ is some constant.
  Let $L_{\rm msb}$ be a mapping such that for any $Y \in \{{\tt 0}, {\tt 1}\}^{b}$, $L_{\rm msb}$ maps
  $Y$ into the value $\min\{i \in [1 \dd b] : Y[i] = 1\} \cup \{b+1\}$.

  The data structure consists of a single component: the mapping $L_{\rm msb}$ stored in plain form. When
  accessing $L_{\rm msb}$, we convert all $Y \in \{{\tt 0}, {\tt 1}\}^{b}$ into a number in $[0 \dd 2^b)$.
  Thus, $L_{\rm msb}$ needs $\bigO(2^b) = \bigO(u^{\alpha'}) = \bigO(u^{\alpha})$ space.

  Denote $b' = \lfloor \log u \rfloor$. At query time, first in $\bigO(1+k)$ time we compute the
  largest integer $k \geq 0$ such that $i + kb' \leq j$ and $B(i \dd i+kb']$ contains only zeros. Then,
  $i' \in (i+kb' \dd i+(k+1)b']$. Using $L_{\rm msb}$, we then determine
  $i'$ in $\bigO(1/\alpha') = \bigO(1)$ time. In total, this takes
  $\bigO(1 + k) = \bigO(1 + (i'-i)/\log u)$ time.

  The construction of $L_{\rm msb}$ (and hence the whole structure) takes
  $\bigO(2^b \cdot b) = \bigO(u^{\alpha'} \log u) = \bigO(u^{\alpha})$ time.
\end{proof}

\begin{proposition}\label{pr:packed-copy}
  Let $\alpha \in (0,1)$ be a constant. Let $u \geq 1$. In $\bigO(u^{\alpha})$ time we can
  construct a data structure such that, given the packed representation
  of any bitvector $B \in \{{\tt 0},{\tt 1}\}^{m}$ and an integer $k > 0$, we can compute
  the packed representation of a bitvector $B^{k} \in \{{\tt 0},{\tt 1}\}^{mk}$ in
  $\bigO(1 + mk / \log u)$ time.
\end{proposition}
\begin{proof}

  Let $\alpha' \in (0, \alpha/2)$ be a constant such that $b := \alpha' \log u$ is a positive integer.
  Such $\alpha'$ exists for all $u \geq u_{\min}$, where $u_{\min}$ is some constant.
  Let $L_{\rm pow}$ be a mapping such that for every string $X \in \{{\tt 0}, {\tt 1}\}^{\leq b}$,
  $L_{\rm pow}$ maps a string $X$ to $X^{p}$, where $p = \lfloor \tfrac{\log u}{|X|} \rfloor$.
  Note that $b = \alpha' \log u < \tfrac{\alpha}{2} \log u < \tfrac{1}{2} \log u$ implies that $p \geq 2$.

  The data structure consists of a single component: the lookup table $L_{\rm pow}$. When accessing
  $L_{\rm pow}$, the input string $X \in \{{\tt 0}, {\tt 1}\}^{\leq b}$ is mapped into an integer $\Int{b}{2}{X} \in
  [0 \dd 2^{2b})$ (\cref{def:int}). Similarly, each of the values $Y$ is a string of length
  not exceeding $\log u$, and hence we encode it as an integer $\Int{\lceil \log u \rceil}{2}{Y}$ using
  $2\lceil \log u \rceil = \bigO(\log u)$ bits. The lookup table thus needs $\bigO(2^{2b}) = \bigO(2^{2\alpha'\log u})
  = \bigO(u^{2\alpha'}) = \bigO(u^{\alpha})$ space.

  At query time, we consider two cases:
  \begin{itemize}
  \item First, assume $b < m$. Then,
    $\alpha' < m / \log u$, i.e., $m / \log u = \Omega(1)$. Thus,
    copying/appending a packed representation of $B$ takes
    $\bigO(1 + m / \log u) = \bigO(m / \log u)$ time. The query thus takes
    $\bigO(km / \log u)$ time.
  \item Let us now assume that $m \leq b$.
    First, using $L_{\rm pow}$, in $\bigO(1)$ time we obtain the packed
    representation of $X' := X^{k'}$, where $k' = \lfloor \tfrac{\log u}{m} \rfloor$.
    Then, letting $k'' = \lceil \tfrac{mk}{|X'|} \rceil$, in $\bigO(1 + k'')$ time
    we compute the packed representation of the string $X'' := X'^{k''}$. Finally, in $\bigO(1 + mk/\log u)$ time we
    extract its prefix of length $mk$. In total, this takes $\bigO(1 + mk/\log u + k'')$ time.
    To bound $k''$, note that by $m \leq b = \alpha' \log u < \tfrac{1}{2} \log u$, it follows
    that $|X'| = m\lfloor \tfrac{\log u}{m} \rfloor \geq m(\tfrac{\log u}{m} - 1) = \log u - m \geq \tfrac{1}{2}\log u$,
    and hence $k'' = \lceil \tfrac{mk}{|X'|} \rceil \leq 1 + \tfrac{mk}{|X'|} \leq 1 + \tfrac{2mk}{\log u} = \bigO(1 + mk/\log u)$.
  \end{itemize}

  Each of the entries of the lookup table $L_{\rm pow}$ can be computed in $\bigO(\log u)$ time.
  Thus, its construction (including the initialization), takes $\bigO(2^{2b} + 2^b \log u) =
  \bigO(u^{2\alpha'} + u^{\alpha'} \log u) = \bigO(u^{\alpha})$ time.
\end{proof}

\begin{proposition}\label{pr:packed-rank}
  Let $\alpha \in (0,1)$ be a constant. Let $u \geq 1$.
  Given the packed representation of a bitvector $B \in \{{\tt 0}, {\tt 1}\}^{m}$,
  we can in $\bigO(m / \log u)$ time augment it with a support for $\bigO(1)$-time
  rank queries (\cref{def:rank-select}) occupying additional $\bigO(m / \log u)$ space,
  assuming $\bigO(u^{\alpha})$ time and space preprocessing shared by
  all instances of the structure.
\end{proposition}
\begin{proof}

  Let $\alpha' \in (0, \alpha)$ be a constant such that $b := \alpha' \log u$ is a positive integer.
  Such $\alpha'$ exists for all $u \geq u_{\min}$, where $u_{\min}$ is some constant.
  Let $L_{\rm cnt}$ be a mapping such that for every string $X \in \{{\tt 0}, {\tt 1}\}^{b}$,
  $L_{\rm cnt}$ maps $X$ to $\Rank{X}{1}{b}$. Let also $t := \lceil \log u \rceil$.

  The result of the preprocessing is the lookup table $L_{\rm cnt}$. When accessing
  $L_{\rm cnt}$, any $X \in \{{\tt 0}, {\tt 1}\}^{b}$ is represented as an
  integer in $[0 \dd 2^b)$. Thus, $L_{\rm cnt}$ needs $\bigO(2^b) = \bigO(2^{\alpha' \log u}) = \bigO(u^{\alpha'})
  = \bigO(u^{\alpha})$ space.

  To augment a given $B \in \{{\tt 0}, {\tt 1}\}^{m}$ with support for rank queries,
  we precompute an array $R[1 \dd m']$ (with $m' = \lfloor \tfrac{m}{t} \rfloor$) defined by
  $R[i] = \Rank{B}{1}{it}$. Given $R$ and $L_{\rm cnt}$,
  we can then compute $\Rank{B}{1}{j}$ for any $j \in [0 \dd m]$ in $\bigO(1)$ time as
  $\Rank{B}{1}{j} = R[i] + \Rank{B'}{1}{\delta}$, where $i = \lfloor \tfrac{j}{t} \rfloor$,
  $\delta = j - it$, and $B' = B(it \dd j]$.
  Computing the second term using $L_{\rm cnt}$ takes $\bigO(t/b) = \bigO(\alpha') = \bigO(1)$ time.

  Lastly, we note that the preprocessing can be done in
  $\bigO(b2^b) = \bigO(u^{\alpha'} \log u) = \bigO(u^{\alpha})$ time.
\end{proof}

\begin{proposition}\label{pr:offline-range-queries-two-sided-sorted}
  Let $\alpha \in (0,1)$ be a constant. Let $u \geq 1$. In $\bigO(u^{\alpha})$ time we
  can construct a data structure that answers the following query: given
  any array $A[1 \dd m]$ of $m$ nonnegative integers satisfying
  $\sum_{i=1}^{m} A[i] = s$, and any arrays
  $Q_{\rm val}[1 \dd q]$ and $Q_{\rm pos}[1 \dd q]$ satisfying $Q_{\rm val}[i] \in \Zn$ and
  $Q_{\rm pos}[i] \in [0 \dd m]$ for $i \in [1 \dd q]$, and
  $Q_{\rm pos}[i] \leq Q_{\rm pos}[i+1]$ for $i \in [1 \dd m)$, in
  $\bigO(m + q + s/\log u)$ time compute the array $A_{\rm ans}[1 \dd q]$ defined by
  $A_{\rm ans}[i] = \TwoSidedRangeCount{A}{Q_{\rm pos}[i]}{Q_{\rm val}[i]}$.
\end{proposition}
\begin{proof}

  The data structure consists of the following components:
  \begin{enumerate}
  \item The data structure for bounded successor queries from \cref{pr:packed-succ} using $\bigO(u^{\alpha})$ space.
  \item The data structure for copying packed bitvectors from \cref{pr:packed-copy} using $\bigO(u^{\alpha})$ space.
  \item The result of preprocessing for rank queries from \cref{pr:packed-rank} using $\bigO(u^{\alpha})$ space.
  \end{enumerate}
  In total, the structure needs $\bigO(u^{\alpha})$ space.

  \paragraph{Implementation of queries}

  Let $A[1 \dd m]$ be an array of $m$ nonnegative integers satisfying
  $\sum_{i=1}^{m} A[i] = s$. Let also $Q_{\rm val}[1 \dd q]$ and $Q_{\rm pos}[1 \dd q]$ be such
  that $Q_{\rm val}[i] \in \Zn$ and
  $Q_{\rm pos}[i] \in [0 \dd m]$ holds for $i \in [1 \dd q]$, and
  $Q_{\rm pos}[i] \leq Q_{\rm pos}[i+1]$ holds for $i \in [1 \dd m)$.
  Denote $y = \lfloor \log u \rfloor$ and $s' = \lfloor \tfrac{s}{y} \rfloor$.
  To compute the array $A_{\rm ans}[1 \dd m]$ (defined as in the claim),
  we proceed in five steps:
  \begin{enumerate}

  \item For every $k \geq 0$, denote $m_k = \TwoSidedRangeCount{A}{m}{ky}$ and let $P_k[1 \dd m_k]$ be such for
    for every $i \in [1 \dd m_k]$, $P_k[i] = \RangeSelect{A}{i}{ky}$. Let $k_{\max} = \max\{k \geq 0 : m_k > 0\}$.
    Observe that since elements in $A$ are nonnegative, it follows that $\max_{i \in [1 \dd m]} A[i] \leq s$,
    and hence $k_{\max} = \lfloor \tfrac{1}{y} \max_{i \in [1 \dd m]} A[i] \rfloor \leq \lfloor \tfrac{s}{y} \rfloor = s'$.
    We compute the arrays $P_k$ for $k \in [0 \dd k_{\max}]$ as follows. First, we set $m_0 = m$ and $P_0[i] = i$ for
    all $i \in [1 \dd m_0]$. For $k > 0$, we iterate over $P_{k-1}$ and add to $P_k$ only values $P_{k-1}[i]$ satisfying
    $A[P_{k-1}[i]] \geq ky$. Note that during this procedure we can also determine $k_{\max}$.
    To bound the total time, first observe that since a position $i \in [1 \dd m]$ occurs in
    $\lceil \tfrac{A[i]+1}{y} \rceil$ arrays, it follows that 
    that $\sum_{k \geq 0} m_k = \sum_{i \in [1 \dd m]}\lceil \tfrac{A[i]+1}{y} \rceil
    \leq 2m + \sum_{i \in [1 \dd m]} \lfloor \tfrac{A[i]}{y} \rfloor \leq
    2m + \lfloor \sum_{i \in [1 \dd m]}\tfrac{A[i]}{y} \rfloor = 2m + \lfloor \tfrac{1}{y} \sum_{i \in [1 \dd m]}A[i] \rfloor
    = 2m + s'$. Thus, this step takes $\bigO(m + s')$ time.

  \item For any $v \geq 0$, we define a bitvector $B_v[1 \dd m_k]$, where $k = \lfloor \tfrac{v}{y} \rfloor$ such that for
    every $i \in [1 \dd m_k]$, $B_v[i] = 1$ holds if and only if $A[P_k[i]] \geq v$. For any $k \geq 0$, we then let
    $B'_k = B_{ky} B_{ky+1} \cdots B_{ky+y-1}$. We compute the packed representation of bitvectors $B'_k$ for
    $k \in [0 \dd k_{\max}]$ as follows. Let $k \in [0 \dd k_{\max}]$.
    \begin{enumerate}
    \item For any $\delta \in [1 \dd y)$, let $C_{\delta}[1 \dd m_k]$ denote a bitvector such that for any
      $i \in [1 \dd m_k]$, $C_{\delta}[i] = 1$ holds if and only if $A[P_k[i]] = ky + \delta - 1$.
      Let $C' = C_1 C_2 \cdots C_{y-1}$. For every $\delta \in [1 \dd y)$, let $r_{\delta} = \Rank{C_{\delta}}{1}{m_k}$.
      Note that $r_1 + \ldots + r_{y-1} \leq m_k$. We compute the packed representation of $C'$.
      To this end, we first initialize $C' = {\tt 0}^{(y-1)m_k}$ in $\bigO(1 + (y-1)m_k / \log u) = \bigO(m_k)$ time.
      For every $i \in [1 \dd m_k]$, if $\delta = A[P_k[i]] \bmod y$ satisfies $\delta + 1 < y$, then we set
      $C'[\delta m_k + i] = 1$. In total, the computation of a packed representation of $C'$ takes $\bigO(m_k)$ time.
    \item We then repeatedly execute the following sequence of steps,
      maintaining the invariant at the beginning of each iteration:
      \begin{itemize}
      \item $\delta \in [0 \dd y]$,
      \item $B_{\rm out}$ stores the packed representation of $B_{ky}B_{ky+1}\cdots B_{ky+\delta-1}$,
      \item If $\delta < y$, then $B_{\rm cur}$ stores the packed representation of $B_{ky+\delta}$.
      \end{itemize}
      To ensure the invariant holds at the beginning of the first iteration, we set
      $\delta := 0$, $B_{\rm out} := \emptystring$, and $B_{\rm cur} := B_{ky} = {\tt 1}^{m_k}$ in $\bigO(1+m_k/\log u)$.
      As long as $\delta < y$, we then execute the following steps:
      \begin{enumerate}
      \item We compute $\delta' = \min\{t \in (\delta \dd y) : \Rank{C_t}{1}{m_k} > 0\} \cup \{y\}$.
        To this end, we first let $b = \delta \cdot m_k$, then
        using \cref{pr:packed-succ}, we compute $e = \min\{t \in (b \dd |C'|] : C'[t] = 1\} \cup \{|C'| + 1\}$,
        and then set $\delta' = \lceil \tfrac{e}{m_k} \rceil$.
        This takes $\bigO(1 + (e-b) / \log u) = \bigO(1 + (\delta' - \delta) \cdot m_k / \log u)$ time.
        Observe that for every $t \in [\delta \dd \delta')$, we then have $B_{ky+t} = B_{ky+\delta}$.
      \item Using \cref{pr:packed-copy}, in $\bigO(1 + (\delta' - \delta) \cdot m_k / \log u)$ time, we append the packed
        representation of $B_{\rm cur}^{\delta'-\delta} = B_{ky+\delta}^{\delta'-\delta}$ to $B_{\rm out}$.
        After this update, we have $B_{\rm out} = B_{ky} B_{ky+1} \cdots B_{ky+\delta'-1}$.
      \item We now consider two cases. If $\delta' = y$, we set $\delta := \delta'$ and go back to the beginning of the loop.
        Let us thus assume that $\delta' < y$. Then, in preparation for the next iteration, we update $B_{\rm cur}$ so that
        it is equal to $B_{ky+\delta'}$. To this end, by repeatedly using \cref{pr:packed-succ} on $C'$,
        in $\bigO(r_{\delta'} + m_k / \log u)$
        total time we compute the set $Q = \{j \in [1 \dd m_k] : C_{\delta'}[j] = 1\}$, and then for every
        $j \in Q$, we set $B_{\rm cur}[j] := 0$. After these updates, it holds $B_{\rm cur} = B_{ky+\delta'}$.
        Finally, we set $\delta := \delta'$.
      \end{enumerate}
      When the algorithm stops, we have $\delta = y$, and hence $B_{\rm out}$ stores the packed representation of
      $B_{ky} B_{ky+1} \cdots B_{ky + y - 1} = B'_k$. To bound the total time, let $(\delta_i)_{i \in [0 \dd p]}$ denote
      a sequence satisfying $\delta_0 < \dots < \delta_p$ and $\{\delta_0, \ldots, \delta_p\} =
      \{0\} \cup \{t \in (0 \dd y) : \Rank{C_t}{1}{m_k} > 0\} \cup \{y\}$. Note that $p \leq \min(y, m_k)$
      and $\sum_{i=1}^{p}\delta_i - \delta_{i-1} = \delta_p - \delta_0 = y$. Let also $r_y = 0$ and note that
      $\sum_{i=1}^{p} r_{\delta_i} \leq m_k$. The total time spent in the above algorithm is:
      \begin{align*}
        \textstyle\sum_{i=1}^{p} \bigO(1 + r_{\delta_i} + \tfrac{m_k}{\log u} + \tfrac{(\delta_{i}-\delta_{i-1}) \cdot m_k}{\log u})
          &= \bigO(p + \textstyle\sum_{i=1}^{p} r_{\delta_i} + p\tfrac{m_k}{\log u} +
              \textstyle\sum_{i=1}^{p} \tfrac{(\delta_{i}-\delta_{i-1})m_k}{\log u})\\
          &= \bigO(p + m_k + \tfrac{ym_k}{\log u}) = \bigO(m_k).
      \end{align*}
    \end{enumerate}
    In summary, the computation of the packed representation of $B'_k$ takes $\bigO(m_k)$ time.
    Summing over all $k \in [0 \dd k_{\max}]$, we spend
    $\sum_{k \in [0 \dd k_{\max}]} \bigO(m_k) = \bigO(\sum_{k \geq 0} m_k)
    = \bigO(m + s')$ time.

  \item Using \cref{pr:packed-rank}, for every $k \in [0 \dd k_{\max}]$, we augment
    the bitvector $B'_k$ with support for $\bigO(1)$-time rank queries. Recall
    that for any $k \in [0 \dd k_{\max}]$, $|B'_k| = ym_k$. Thus, in total,
    we spend $\sum_{k \in [0 \dd k_{\max}]} \bigO(1 + \tfrac{ym_k}{\log u})
    = \sum_{k \in [0 \dd k_{\max}]} \bigO(1 + m_k) = \bigO((1 + k_{\max}) + \sum_{k \geq 0} m_k)
    = \bigO(m + s')$ time.

  \item For any $k \in [0 \dd k_{\max}]$, let $q_k = |\{i \in [1 \dd q] : \lfloor \tfrac{Q_{\rm val}[i]}{y} \rfloor = k\}|$
    and $Q_k[1 \dd q_k]$ be an array containing all elements of the set
    $\{i \in [1 \dd q] : \lfloor \tfrac{Q_{\rm val}[i]}{y} \rfloor = k\}$ in increasing order. We
    compute the arrays $Q_k$ for all $k \in [0 \dd k_{\max}]$ with a single scan of $Q_{\rm val}[1 \dd q]$.
    This takes $\bigO(1 + k_{\max} + q) = \bigO(s' + q)$ time.

  \item We are now ready to compute the array $A_{\rm ans}[1 \dd q]$. First, in $\bigO(q)$ time we scan $Q_{\rm val}[1 \dd q]$,
    and for every $i \in [1 \dd q]$ satisfying $Q_{\rm val}[i] \geq yk_{\max}$, we set $A_{\rm ans}[i] = 0$.
    To answer the remaining queries,
    observe that for any $j \in [0 \dd m]$ and $v \in [0 \dd yk_{\max})$,
    letting $k = \lfloor \tfrac{v}{y} \rfloor$, $\delta = v \bmod y$, and
    $j' = |\{t \in [1 \dd m_k] : P_k[t] \leq j\}|$, it holds
    \[
      \TwoSidedRangeCount{A}{j}{v} = \Rank{B_{v}}{1}{j'} =
      \Rank{B'_{k}}{1}{\delta m_k + j'} - \Rank{B'_{k}}{1}{\delta m_k}.
    \]
    We thus proceed as follows. Let $k \in [0 \dd k_{\max}]$. By performing
    a synchronous scan of arrays $P_k[1 \dd m_k]$ and the sequence
    $Q_{\rm pos}[Q_k[1]], Q_{\rm pos}[Q_k[2]], \dots, Q_{\rm pos}[Q_k[q_k]]$,
    in $\bigO(m_k + q_k)$ time we determine, for every $i \in [1 \dd q_k]$,
    the value $c_i = |\{t \in [1 \dd m_k] : P_k[t] \leq Q_{\rm pos}[Q_k[i]]\}|$. Note
    that here we utilize the fact that elements in the array $Q_{\rm pos}[1 \dd q]$ and in the sequence
    $Q_{\rm pos}[Q_k[1]], Q_{\rm pos}[Q_k[2]], \ldots, Q_{\rm pos}[Q_k[q_k]]$,
    are in non-decreasing order.
    By the above observation, letting $\delta = Q_{\rm val}[Q_k[i]] \bmod y$, we can then
    in $\bigO(1)$ time compute
    \begin{align*}
      A_{\rm ans}[Q_k[i]]
        &= \TwoSidedRangeCount{A}{Q_{\rm pos}[Q_k[i]]}{Q_{\rm val}[Q_k[i]]}\\
        &= \Rank{B'_{k}}{1}{\delta m_k + c_i} - \Rank{B'_{k}}{1}{\delta m_k}.
    \end{align*}
    Over all $k \in [0 \dd k_{\max}]$, we spend
    $\sum_{k \in [0 \dd k_{\max}]} \bigO(m_k + q_k) =
    \bigO(\sum_{k \in [0 \dd k_{\max}]} m_k + \sum_{k \in [0 \dd k_{\max}]} q_k)
    = \bigO(m + s' + q)$ time. Adding the initial scan of $Q_{\rm val}[1 \dd q]$
    results in total time $\bigO(m + s' + q)$.
  \end{enumerate}
  In total, computing the array $A_{\rm ans}[1 \dd q]$ takes
  $\bigO(m + q + s') = \bigO(m + q + s/\log u)$ time.

  \paragraph{Construction algorithm}

  The components of the data structure are constructed as follows:
  \begin{enumerate}
  \item First, in $\bigO(u^{\alpha})$ time we construct the data structure for successor queries from \cref{pr:packed-succ}.
  \item Next, also in $\bigO(u^{\alpha})$ time we construct the structure for copying packed bitvectors from \cref{pr:packed-copy}.
  \item Finally, in $\bigO(u^{\alpha})$ time we construct and store the result of preprocessing from \cref{pr:packed-rank}.
  \end{enumerate}
  In total, the construction takes $\bigO(u^{\alpha})$ time.
\end{proof}

\begin{theorem}\label{th:offline-range-queries-two-sided}
  Let $\alpha \in (0,1)$ be a constant. Let $u \geq 1$. In $\bigO(u^{\alpha})$ time we
  can construct a data structure that answers the following query: given
  any array $A[1 \dd m]$ of $m$ nonnegative integers satisfying
  $\sum_{i=1}^{m} A[i] = s$, and any arrays
  $Q_{\rm val}[1 \dd q]$ and $Q_{\rm pos}[1 \dd q]$ satisfying $Q_{\rm val}[i] \in \Zn$ and
  $Q_{\rm pos}[i] \in [0 \dd m]$ for $i \in [1 \dd q]$, in
  $\bigO(m + q + s/\log u)$ time compute the array $A_{\rm ans}[1 \dd q]$ defined by
  $A_{\rm ans}[i] = \TwoSidedRangeCount{A}{Q_{\rm pos}[i]}{Q_{\rm val}[i]}$.
\end{theorem}
\begin{proof}

  The data structure consists of a single component:
  the structure from \cref{pr:offline-range-queries-two-sided-sorted}.

  The queries are implemented as follows.
  Let $A[1 \dd m]$ be an array of $m$ nonnegative integers satisfying
  $\sum_{i=1}^{m} A[i] = s$. Let also $Q_{\rm val}[1 \dd q]$ and $Q_{\rm pos}[1 \dd q]$ be such
  that $Q_{\rm val}[i] \in \Zn$ and
  $Q_{\rm pos}[i] \in [0 \dd m]$ holds for $i \in [1 \dd q]$.
  To compute the array $A_{\rm ans}[1 \dd m]$, we proceed in four steps:
  \begin{enumerate}
  \item We compute the array $Q_{\rm perm}[1 \dd q]$ containing the permutation of $\{1, \ldots, q\}$ such
    that for every $i, j \in [1 \dd q]$, $i < j$ implies $Q_{\rm pos}[Q_{\rm perm}[i]] < Q_{\rm pos}[Q_{\rm perm}[j]]$,
    or $Q_{\rm pos}[Q_{\rm perm}[i]] = Q_{\rm pos}[Q_{\rm perm}[j]]$ and $Q_{\rm perm}[i] < Q_{\rm perm}[j]$.
    To this end, in $\bigO(q)$ time we compute the array $Q_{\rm sort}[1 \dd q]$ defined by
    $Q_{\rm sort}[i] = (Q_{\rm pos}[i], i)$. We then sort it lexicographically.
    The first coordinate is an integer in $[0 \dd m]$, and the second coordinate is an integer in $[1 \dd q]$.
    Thus, using a radix sort, we can sort it in $\bigO(m + q)$ time. The resulting array contains $Q_{\rm perm}$ on the second
    coordinate.
  \item In $\bigO(q)$ time we compute the arrays $Q'_{\rm pos}[1 \dd q]$ and $Q'_{\rm val}[1 \dd q]$ defined by
    $Q'_{\rm pos}[i] = Q_{\rm pos}[Q_{\rm perm}[i]]$ and $Q'_{\rm val}[i] = Q_{\rm val}[Q_{\rm perm}[i]]$.
  \item Using \cref{pr:offline-range-queries-two-sided-sorted}, in $\bigO(m + q + s/\log u)$ time we compute the array
    $A'_{\rm ans}[1 \dd q]$ defined by $A'_{\rm ans}[i] = \TwoSidedRangeCount{A}{Q'_{\rm pos}[i]}{Q'_{\rm val}[i]}
    = \TwoSidedRangeCount{A}{Q_{\rm pos}[Q_{\rm perm}[i]]}{Q_{\rm val}[Q_{\rm perm}[i]]}$.
  \item For every $i \in [1 \dd q]$, we set $A_{\rm ans}[Q_{\rm perm}[i]] = A'_{\rm ans}[i]$.
    This takes $\bigO(q)$ time.
  \end{enumerate}
  In total, the computation of $A_{\rm ans}$ takes $\bigO(m + q + s/\log u)$ time.

  Lastly, we note that the structure of
  \cref{pr:offline-range-queries-two-sided-sorted} takes $\bigO(u^{\alpha})$ time to build.
\end{proof}

\begin{theorem}\label{th:offline-range-queries-three-sided}
  Let $\alpha \in (0,1)$ be a constant. Let $u \geq 1$. In $\bigO(u^{\alpha})$ time we
  can construct a data structure that answers the following query: given
  any array $A[1 \dd m]$ of $m$ nonnegative integers satisfying
  $\sum_{i=1}^{m} A[i] = s$, and any arrays
  $Q_{\rm val}[1 \dd q]$, $Q_{\rm beg}[1 \dd q]$, and $Q_{\rm end}[1 \dd q]$
  satisfying $Q_{\rm val}[i] \in \Zn$, $Q_{\rm beg}[i] \in [0 \dd m]$, and
  $Q_{\rm end}[i] \in [0 \dd m]$ for $i \in [1 \dd q]$, in
  $\bigO(m + q + s/\log u)$ time compute the array $A_{\rm ans}[1 \dd q]$ defined by
  $A_{\rm ans}[i] = \ThreeSidedRangeCount{A}{Q_{\rm beg}[i]}{Q_{\rm end}[i]}{Q_{\rm val}[i]}$.
\end{theorem}
\begin{proof}

  The data structure consists of a single component:
  the structure from \cref{th:offline-range-queries-two-sided}.

  The queries are implemented as follows.
  Let $A[1 \dd m]$, $Q_{\rm val}[1 \dd q]$, $Q_{\rm beg}[1 \dd q]$, and $Q_{\rm end}[1 \dd q]$
  be as in the claim. To compute $A_{\rm ans}[1 \dd m]$, we proceed in three steps:
  \begin{enumerate}
  \item Using \cref{th:offline-range-queries-two-sided},
    in $\bigO(m + q + s/\log u)$ time
    compute an array $A'_{\rm ans}[1 \dd q]$ defined by
    $A'_{\rm ans}[i] = \TwoSidedRangeCount{A}{Q_{\rm end}[i]}{Q_{\rm val}[i]}$.
  \item Similarly as above, compute an array $A''_{\rm ans}[1 \dd q]$ defined by
    $A''_{\rm ans}[i] = \TwoSidedRangeCount{A}{Q_{\rm beg}[i]}{Q_{\rm val}[i]}$.
  \item For every $i \in [1 \dd q]$, set
    $A_{\rm ans}[i] = 0$ (if $Q_{\rm beg}[i] \geq Q_{\rm end}[i]$)
    or $A_{\rm ans}[i] = A'_{\rm ans}[i] - A''_{\rm ans}[i]$ (otherwise).
  \end{enumerate}
  In total, the computation of $A_{\rm ans}$ takes $\bigO(m + q + s/\log u)$ time.

  Lastly, we note that the structure of
  \cref{th:offline-range-queries-two-sided} takes $\bigO(u^{\alpha})$ time to build.
\end{proof}

%% file: tools/three-sided-rmq.tex
\subsection{Three-Sided RMQ}\label{sec:three-sided-rmq}

\begin{definition}[Three-sided RMQ]\label{def:three-sided-rmq}
  Let $A[1 \dd m]$ and $B[1 \dd m]$ be two arrays of $m \geq 0$ nonnegative
  integers.
  For every $b, e \in [0 \dd m]$ and $v \geq 0$ such that
  $\ThreeSidedRangeCount{B}{b}{e}{v} > 0$, we
  define
  \[
    \ThreeSidedRMQ{A}{B}{b}{e}{v} :=
      \argmin_{i \in (b \dd e]:\,B[i] \geq v}A[i].
  \]
\end{definition}

\begin{theorem}\label{th:three-sided-rmq}
  Arrays $A[1 {\dd} m']$ and $B[1 {\dd} m']$ of $m' \in [0 {\dd} m]$ nonnegative
  integers such that
  $\max_{i=1}^{m'}A[i] = \bigO(m \log m)$ and
  $\sum_{i=1}^{m'}B[i] = \bigO(m \log m)$ can be preprocessed in
  $\bigO(m)$ time so that three-sided RMQ queries on $A$ and $B$
  can be answered in $\bigO(\log \log m)$ time.
\end{theorem}
\begin{proof}

  We use the following definitions.
  Let $y = \alpha \log m$ be a positive integer, where $\alpha \in (0, 1]$ is a constant.
  Let also $x = \beta (\log m)/(\log \log m)$ be a positive integer, where $\beta \in (0, \tfrac{1}{6}]$ is a constant.
  We also assume that $4 \leq x < y$. It is easy to see that such $\alpha$ and $\beta$ exist for all $m \geq m_{\min}$,
  where $m_{\min}$ is some constant.

  For any $k \geq 0$, by $P_k[1 \dd m_k]$, where $m_k := \TwoSidedRangeCount{B}{m}{ky}$ (see \cref{sec:range-queries}),
  we denote the array
  defined by $P_k[i] = \RangeSelect{B}{i}{ky}$. We then define $A_k[1 \dd m_k]$ and $B_k[1 \dd m_k]$ to be such that
  for every $i \in [1 \dd m_k]$, it holds $A_k[i] = A[P_k[i]]$ and $B_k[i] = B[P_k[i]]$. We also let
  $k_{\max} = \max\{k \geq 0 : m_k > 0\}$.
  Note that because all elements of $B$
  are nonnegative, and $\sum_{i=1}^{m'}B[i] \in \bigO(m \log m)$, it follows that
  $\max_{i \in [1 \dd m']} B[i] \in \bigO(m \log m)$, and
  hence,
  \begin{align*}
    k_{\max}
      &= \lfloor \tfrac{1}{y} \max_{i \in [1 \dd m']} B[i] \rfloor = \bigO(m).
  \end{align*}
  Note also that each $i \in [1 \dd m']$ occurs in
  $\lceil \tfrac{B[i] + 1}{y} \rceil$ arrays. Therefore,
  \begin{align*}
    \textstyle\sum_{k \geq 0}m_k
      &= \textstyle\sum_{i=1}^{m'} \lceil \tfrac{B[i] + 1}{y} \rceil\\
      &\leq 2m' + \textstyle\sum_{i=1}^{m'} \lfloor \tfrac{B[i]}{y} \rfloor\\
      &\leq 2m' + \tfrac{1}{y}\textstyle\sum_{i=1}^{m'}B[i] = \bigO(m).
  \end{align*}

  Let $k \in [0 \dd k_{\max}]$. Denote $m^{x}_k := \lceil \tfrac{m_k}{x} \rceil$,
  and consider any $i \in [1 \dd m^{x}_k]$. We then define:
  \begin{itemize}
  \item $b_{k,i} := (i-1)x$, $e_{k,i} := \min(m_k, ix)$, and $m_{k,i} := e_{k,i} - b_{k,i}$.
    Note that by $x < y$, it holds $m_{k,i} < y$.
  \item We let $A_{k,i}[1 \dd m_{k,i}]$ be an array defined so that for every $j \in (b_{k,i} \dd e_{k,i}]$,
    it holds $A_{k,i}[j - b_{k,i}] = |\{a \in \mathcal{A} : a < A_k[j]\}|$, where $\mathcal{A} =
    \{A_k[t] : t \in (b_{k,i} \dd e_{k,i}]\}$.
    In other words, $A_{k,i}$ stores the elements of $A_k(b_{k,i} \dd e_{k,i}]$ in the rank space. Then, for every
    $j, j' \in (b_{k,i} \dd e_{k,i}]$, $A_k[j] < A_k[j']$ holds if and only if
    $A_{k,i}[j-b_{k,i}] < A_{k,i}[j'-b_{k,i}]$.
    Simultaneously, for every $j \in [1 \dd m_{k,i}]$, we have $A_{k,i}[j] \in [0 \dd x) \subseteq [0 \dd y)$.
  \item We let $B_{k,i}[1 \dd m_{k,i}]$ be an array defined so
    that for every $j \in (b_{k,i} \dd e_{k,i}]$,
    it holds $B_{k,i}[j - b_{k,i}] = \min(y - 1, B_k[j] - ky)$. Note that since for every $j \in [1 \dd m_k]$,
    it holds $B_k[j] \geq ky$,
    we obtain that for every $j \in [1 \dd m_{k,i}]$, we have $B_{k,i}[j] \in [0 \dd y)$.
  \end{itemize}

  Denote $a_{\rm inf} = 1 + \max_{i \in [1 \dd m']}A[i]$ and $k'_{\max} = \max\{k \geq 0 : m_k \geq y\}$.
  Let $v \in [0 \dd y \cdot k'_{\max}]$, and let $k = \lfloor \tfrac{v}{y} \rfloor$ and
  $m^{y}_k := \lfloor \tfrac{m_k}{y} \rfloor$. We let
  $M^{\rm pos}_{v}[1 \dd m^{y}_k]$ and $M^{\rm val}_{v}[1 \dd m^{y}_k]$ to be such that for every $i \in [1 \dd m^{y}_k]$:
  \begin{itemize}
  \item
    If $\ThreeSidedRangeCount{B_{k}}{(i-1)y}{iy}{v} = 0$, then we
    define $M^{\rm val}_{v}[i] = a_{\rm inf}$, and leave
    $M^{\rm pos}_{v}[i]$ undefined.
  \item Otherwise,
    we let $M^{\rm pos}_{v}[i] = \ThreeSidedRMQ{A_{k}}{B_{k}}{(i-1)y}{iy}{v}$,
    and $M^{\rm val}_{v}[i] = A_k[M^{\rm pos}_{v}[i]]$.
  \end{itemize}

  For every sequence $(a_i)_{i \in [1 \dd t]}$ satisfying $t \leq 3x$ and $a_i \in [0 \dd y)$
  (for all $i \in [1 \dd t]$),
  we define $\EncodeSeq(a_1, \ldots, a_t)$
  as an integer in $[0 \dd y^{3x})$ obtained by appending $3x - t$ zeros at the end of the sequence
  $(a_i)_{i \in [1 \dd t]}$ and interpreting the resulting sequence as digits (with $a_1$ being the most
  significant digit) of a number in base $y$.
  Note, that if $t < 3x$, then $\EncodeSeq(a_1, \ldots, a_t) = \EncodeSeq(a_1, \ldots, a_t, 0)$, i.e.,
  the encoding does not include the sequence length, but in our applications,
  we will always be able to identify the sequence length.

  Let $m_{\rm short} \in [1 \dd x]$. Consider any arrays
  $A_{\rm short}[1 \dd m_{\rm short}]$ and $B_{\rm short}[1 \dd m_{\rm short}]$ such
  that for every $i \in [1 \dd m_{\rm short}]$, it holds $A_{\rm short}[i] \in [0 \dd x)$
  (note that by $x < y$, this implies
  $A_{\rm short}[i] \in [0 \dd y)$) and $B_{\rm short}[i] \in [0 \dd y)$.
  Let also $b, e \in [0 \dd x]$ and $\delta \in [0 \dd y)$.
  We define $L_{\rm rmq}$ to be a mapping such that for every
  $m_{\rm short}$, $A_{\rm short}$, $B_{\rm short}$, $b$, $e$, and
  $\delta$ as above, the mapping $L_{\rm rmq}$ maps the integer
  \[
    \EncodeSeq(b, e, \delta, m_{\rm short}, A_{\rm short}[1], \ldots, A_{\rm short}[m_{\rm short}],
    B_{\rm short}[1], \ldots, B_{\rm short}[m_{\rm short}]) \in [0 \dd y^{3x})
  \]
  to either
  $\ThreeSidedRMQ{A_{\rm short}}{B_{\rm short}}{b}{e}{\delta}
  \in [1 \dd m_{\rm short}] \subseteq [1 \dd x]$ (if
  $\ThreeSidedRangeCount{B_{\rm short}}{b}{e}{\delta} > 0$), or to $0$ (otherwise).
  Note that we can apply the above encoding since by $4 \leq x$ and $m_{\rm short} \leq x$,
  the sequence has length $4 + 2m_{\rm short} \leq 4 + 2x \leq 3x$. Note also
  that all elements of the encoded sequence are from $[0 \dd y)$.

  For every $k \in [0 \dd k_{\max}]$, let $E_{k}[1 \dd m^{x}_k]$ be an array defined by
  \[
    E_{k}[i] = \EncodeSeq(m_{k,i}, A_{k,i}[1], \ldots, A_{k,i}[m_{k,i}], B_{k,i}[1], \ldots, B_{k,i}[m_{k,i}]).
  \]
  Note that the above use of sequence encoding is well-defined, since all sequences are of length not exceeding
  $3x$, and consist of integers in $[0 \dd y)$.

  Lastly, by $A_{\rm pow}[0 \dd 3x)$ we denote the array defined by $A_{\rm pow}[i] = y^i$.

  \paragraph{Components}

  The structure consists of eight components:
  \begin{enumerate}
  \item A plain representation of $A[1 \dd m']$ using $\bigO(m') = \bigO(m)$ space.
  \item A plain representation of $B[1 \dd m']$ also using $\bigO(m') = \bigO(m)$ space.
  \item For every $k \in [0 \dd k_{\max}]$, we store a plain representation of the array
    $P_k[1 \dd m_k]$ using $\bigO(m_k)$ space. Each array is augmented
    with a static predecessor data structure from~\cite[Proposition~2]{wexp},
    and hence achieves linear space and
    $\bigO(\log \log m)$ query time. To space to store
    all the arrays $P_k$ (including the associated predecessor data
    structures) is $\bigO((k_{\max} + 1) + \sum_{k \geq 0}m_k)
    = \bigO(m)$ space.
  \item For every $v \in [0 \dd y \cdot k'_{\max})$, we store a plain representation of array $A^{\rm val}_{v}$,
    augmented with a data structure for RMQ queries. We assume that the structure returns the position of the leftmost
    minimum in the query range. Using for example~\cite{FischerH11} achieves linear preprocessing time (and hence also
    linear space) and $\bigO(1)$ query time. The array $A^{\rm val}_{v}$ has size $m^{y}_k = \lfloor \tfrac{m_k}{y} \rfloor$,
    where $k = \lfloor \tfrac{v}{y} \rfloor$. Thus, in total, the space for all arrays and the associated RMQ structures is
    \begin{align*}
      y \cdot \textstyle\sum_{k\in [0 \dd k'_{\max}]} m^{y}_k
          &= y \cdot \textstyle\sum_{k \in [0 \dd k'_{\max}]} \lfloor \tfrac{m_k}{y} \rfloor\\
          &\leq y \cdot \tfrac{1}{y}\textstyle\sum_{k \in [0 \dd k'_{\max}]} m_k\\
          &= \textstyle\sum_{k \in [0 \dd k'_{\max}]} m_k
          = \bigO(m).
     \end{align*}
  \item For every $v \in [0 \dd y \cdot k'_{\max})$, we store a plain representation of array $A^{\rm pos}_{v}$.
    The array $A^{\rm pos}_{v}$ has size $m^{y}_k$, where $k = \lfloor \tfrac{v}{y} \rfloor$.
    Thus, in total, all arrays need $\bigO(y \cdot \sum_{k \in [0 \dd k'_{\max}]} m^{y}_k) = \bigO(m)$ space.
  \item We store a plain representation of the array $A_{\rm pow}[0 \dd 3x)$ using
    $\bigO(x) = \bigO(\tfrac{\log m}{\log \log m}) = \bigO(m)$ space.
  \item For every $k \in [0 \dd k_{\max}]$, we store a plain representation of the array
    $E_k[1 \dd m^{x}_k]$ using $\bigO(m^{x}_k) = \bigO(m_k)$ space. By
    the above analysis, all arrays in total need $\bigO((k_{\max} + 1) +
    \sum_{k \geq 0}m_k) = \bigO(m)$ space.
  \item The plain representation of lookup table $L_{\rm rmq}[0 \dd y^{3x})$.
    By definitions of $x$ and $y$, the space for $L_{\rm rmq}$ is
    \begin{align*}
    \bigO(y^{3x})
      = \bigO\left(2^{(\log (\alpha\log m)) \cdot \tfrac{3\beta \log m}{\log \log m}}\right)
      = \bigO\left(2^{(\log \log m) \cdot \tfrac{\log m}{2 \log \log m}}\right)
      = \bigO(m^{1/2}) = \bigO(m).
    \end{align*}
  \end{enumerate}

  In total, the data structure takes $\bigO(m)$ space.

  \paragraph{Implementation of queries}

  We develop the final query algorithm gradually in five steps:
  \begin{itemize}

  \item Let $k \in [0 \dd k_{\max}]$, $i \in [1 \dd m^{x}_k]$, $\delta \in [0 \dd y)$, and
    $b, e \in [b_{k,i} \dd e_{k,i}]$. First, we show how in
    $\bigO(1)$ time check if $\ThreeSidedRangeCount{B_{k}}{b}{e}{ky+\delta} > 0$,
    and if so, return $\ThreeSidedRMQ{A_{k}}{B_{k}}{b}{e}{ky+\delta}$.
    The query algorithm is based on the observation (following from the definition of arrays
    $A_{k,i}$ and $B_{k,i}$), that, letting $b' = b - b_{k,i}$ and $e' = e - b_{k,i}$, it holds:
    \begin{align*}
      \ThreeSidedRangeCount{B_{k}}{b}{e}{ky+\delta} = \ThreeSidedRangeCount{B_{k,i}}{b'}{e'}{\delta}.
    \end{align*}
    Moreover, if $\ThreeSidedRangeCount{B_{k}}{b}{e}{ky+\delta} > 0$, then:
    \begin{align*}
      \ThreeSidedRMQ{A_{k}}{B_{k}}{b}{e}{ky+\delta} =
        b_{k,i} + \ThreeSidedRMQ{A_{k,i}}{B_{k,i}}{b'}{e'}{\delta}.
    \end{align*}
    The query thus proceeds in three steps as follows:
    \begin{enumerate}
    \item In $\bigO(1)$ time, compute $t := b_{k,i} = (i-1)x$,
      $b' := b - t$, and $e' := e - t$.
    \item In $\bigO(1)$ time, compute 
    \begin{align*}
      v &:= \EncodeSeq(b', e', \delta,
            m_{k,i}, A_{k,i}[1], \ldots, A_{k,i}[m_{k,i}], B_{k,i}[1], \ldots, B_{k,i}[m_{k,i}])\\
        &=  b' \cdot A_{\rm pow}[3x-1] + e' \cdot A_{\rm pow}[3x-2] +
            \delta \cdot A_{\rm pow}[3x-3] + E_{k}[i] / A_{\rm pow}[3].
    \end{align*}
    \item In $\bigO(1)$ time, we compute $v' = L_{\rm rmq}[v]$. If $v' = 0$ then by definition of $L_{\rm rmq}$ and the above
      observation, it follows
      that $\ThreeSidedRMQ{A_{k}}{B_{k}}{b}{e}{ky+\delta} =
      \ThreeSidedRangeCount{B_{k,i}}{b'}{e'}{\delta} = 0$.
      Otherwise, the definition of $L_{\rm rmq}$ and the above observation yields
      $\ThreeSidedRMQ{A_{k}}{B_{k}}{b}{e}{ky+\delta} = b_{k,i} + \ThreeSidedRMQ{A_{k,i}}{B_{k,i}}{b'}{e'}{\delta} = t + v'$.
    \end{enumerate}
    In total, the query takes $\bigO(1)$ time.

  \item Let $k \in [0 \dd k_{\max}]$, $\delta \in [0 \dd y)$, $b \in [0 \dd m_k]$, and $\ell > 0$
    be such that $b + \ell \leq m_k$. Denote $e = b + \ell$.
    We now generalize the above query algorithm so that
    in $\bigO(1 + \ell / x)$ time we can check if $\ThreeSidedRangeCount{B_{k}}{b}{e}{ky+\delta} > 0$,
    and if so, return $\ThreeSidedRMQ{A_{k}}{B_{k}}{b}{e}{ky+\delta}$.
    If $b \geq e$, then we immediately return that $\ThreeSidedRangeCount{B_{k}}{b}{e}{ky+\delta} = 0$. Let us
    thus assume that $b < e$.
    Observe the range $(b \dd e]$ overlaps at most $2 + \lfloor \ell/x \rfloor$ subarrays $A_{k,i}$ of $A_{k}$. Thus,
    we can partition $(b \dd e]$ into $\bigO(1 + \ell/x)$ subranges,
    such that for each such subrange $(b' \dd e']$, using the algorithm described above,
    we can in $\bigO(1)$ time check if
    $\ThreeSidedRangeCount{B_{k}}{b'}{e'}{ky+\delta} > 0$, and if so, return
    $\ThreeSidedRMQ{A_{k}}{B_{k}}{b'}{e'}{ky+\delta}$. During this algorithm, we
    keep track of whether there exists a subrange $(b' \dd e']$ such that
    $\ThreeSidedRangeCount{B_{k}}{b'}{e'}{ky+\delta} > 0$.
    \begin{itemize}
    \item If there is no such subrange, then we return that
      $\ThreeSidedRangeCount{B_{k}}{b}{e}{ky+\delta} = 0$.
    \item If there exists only one subrange $(b' \dd e']$
      satisfying $\ThreeSidedRangeCount{B_{k}}{b'}{e'}{ky+\delta} > 0$, then we return
      $\ThreeSidedRMQ{A_{k}}{B_{k}}{b'}{e'}{ky+\delta} = \ThreeSidedRMQ{A_{k}}{B_{k}}{b}{e}{ky+\delta}$ as the answer.
    \item If there are at least two such subranges $(b' \dd e']$ and $(b'' \dd e'']$, then we need to identify block with
      the smallest value. Note that for any two such subranges $(b' \dd e']$ and $(b'' \dd e'']$, we
      can identify the leftmost position of the minimum by comparing
      $A_k[i_{\min}]$ and $A_k[i'_{\min}]$, where $i_{\min} = \ThreeSidedRMQ{A_{k}}{B_{k}}{b'}{e'}{ky+\delta}$ and
      $i'_{\min} = \ThreeSidedRMQ{A_{k}}{B_{k}}{b''}{e''}{ky+\delta}$. Since for every $j \in [1 \dd m_{k}]$, we
      defined $A_k[j] = A[P_k[j]]$, we can access $A_k$ in $\bigO(1)$ time.
    \end{itemize}
    In total, the query takes $\bigO(1 + \ell/x)$ time.

  \item Let $k \in [0 \dd k'_{\max}]$, $\delta \in [0 \dd y)$, and $i, j \in [0 \dd m^{y}_k]$.
    Denote $b = iy$, $e = jy$.
    We now show how in $\bigO(1)$ time check if
    $\ThreeSidedRangeCount{B_{k}}{b}{e}{ky+\delta} > 0$, and if so, return
    $\ThreeSidedRMQ{A_{k}}{B_{k}}{b}{e}{ky+\delta}$. The algorithm proceeds as follows.
    If $i \geq j$, then we return that $\ThreeSidedRangeCount{B_{k}}{b}{e}{ky+\delta} = 0$. Let us thus
    assume that $i < j$.
    Denote $v = ky+\delta \in [0 \dd y \cdot k'_{\max})$
    The algorithm proceeds in two steps:
    \begin{enumerate}
    \item Using the RMQ structure over for
      $A^{\rm val}_{v}[1 \dd m^{y}_{k}]$, in $\bigO(1)$ time
      we compute $p = \argmin_{t \in (i \dd j]} A^{\rm val}_{v}[t]$.
    \item In $\bigO(1)$ time we then lookup the value $a := A^{\rm val}_{v}[p]$. If $a = a_{\rm inf}$, then
      by definition of $A^{\rm val}_{v}$, it follows that $\ThreeSidedRangeCount{B_{k}}{b}{e}{ky+\delta} = 0$,
      and hence the query algorithm is complete. Let us thus assume that $a \neq a_{\rm inf}$. Observe, that
      then $\ThreeSidedRangeCount{B_{k}}{b}{e}{ky+\delta} > 0$ and:
      \begin{align*}
        \ThreeSidedRMQ{A_{k}}{B_{k}}{b}{e}{ky+\delta} = \ThreeSidedRMQ{A_{k}}{B_{k}}{(p-1)y}{py}{ky+\delta}.
      \end{align*}
      Recall that by definition of $A^{\rm pos}_{v}$, it holds
      $A^{\rm pos}_{v}[p] = \ThreeSidedRMQ{A_{k}}{B_{k}}{(p-1)y}{py}{v} =
      \ThreeSidedRMQ{A_{k}}{B_{k}}{(p-1)y}{py}{ky+\delta}$. Thus, we return
      the answer in $\bigO(1)$ time.
    \end{enumerate}
    In total, the query takes $\bigO(1)$ time.

  \item Let $k \geq 0$, $\delta \in [0 \dd y)$, and $b, e \in [0 \dd m_k]$.
    We now show how to combine the previous two query subprocedures so that in $\bigO(\log \log m)$ time
    we can check if $\ThreeSidedRangeCount{B_{k}}{b}{e}{ky+\delta} > 0$, and if so, return
    $\ThreeSidedRMQ{A_{k}}{B_{k}}{b}{e}{ky+\delta}$. The algorithm proceeds as follows.
    If $b \geq e$, then we return that $\ThreeSidedRangeCount{B_{k}}{b}{e}{ky+\delta} = 0$. Let us
    thus assume that $b < e$. Denote $\ell = e - b$. If $\ell < y$, then we check if $\ThreeSidedRangeCount{B_{k}}{b}{e}{ky+\delta} > 0$,
    and if so, return $\ThreeSidedRMQ{A_{k}}{B_{k}}{b}{y}{ky+\delta}$ using the algorithm described above in
    $\bigO(1 + \ell/x) = \bigO(1 + y/x) = \bigO(\log \log m)$ time. Let us thus assume that $\ell \geq y$. Observe,
    that then we have $m_k \geq \ell \geq y$, and hence $k \in [0 \dd k'_{\max}]$. The query then proceeds as follows:
    \begin{enumerate}
    \item In $\bigO(1)$ time we compute
      $i := \lceil \tfrac{b}{y} \rceil \in [0 \dd m^{y}_k]$ and
      $j := \lfloor \tfrac{e}{y} \rfloor \in [0 \dd m^{y}_k]$.
      Observe that $b \leq iy \leq jy \leq e$. Let also
      $\ell_{\rm left} = iy - b$ and $\ell_{\rm right} = e - jy$,
      and note that $\ell_{\rm left}, \ell_{\rm right} < y$. We thus obtain
      a decomposition of $(b \dd e]$ into a disjoint union of three intervals
      $(b \dd iy]$, $(iy \dd jy]$, and $(jy \dd e]$.
      Observe, that at least one of these three intervals is nonempty.
    \item We initialize $\mathcal{I}$ to $\emptyset$.
      Using the query algorithm described above, in $\bigO(1 + \ell_{\rm left}/x) =
      \bigO(1 + y/x) = \bigO(\log \log m)$ time we check if
      $\ThreeSidedRangeCount{B_{k}}{b}{iy}{ky+\delta} > 0$, and if so, we
      compute and add to $\mathcal{I}$ the position $\ThreeSidedRMQ{A_{k}}{B_{k}}{b}{iy}{ky+\delta}$.
    \item Using the query described above,
      in $\bigO(1)$ time we check if
      $\ThreeSidedRangeCount{B_{k}}{iy}{jy}{ky+\delta} > 0$, and if so,
      we compute and add to $\mathcal{I}$ the position
      $\ThreeSidedRMQ{A_{k}}{B_{k}}{iy}{jy}{ky+\delta}$.
    \item Similarly as above, in $\bigO(1 + \ell_{\rm right}/x) = \bigO(\log \log m)$ time we check if
      $\ThreeSidedRangeCount{B_{k}}{jy}{e}{ky+\delta} > 0$, and if so, we
      compute and add to $\mathcal{I}$
      the position $\ThreeSidedRMQ{A_{k}}{B_{k}}{jy}{e}{ky+\delta}$.
    \item If $|\mathcal{I}| = 1$, then we return the only element of $\mathcal{I}$ as the answer.
      Otherwise, we locate the leftmost minimum
      by comparing the values at the corresponding indexes in $A_k$.
      For example, if $q = 2$ and $\mathcal{I} = \{i_1, i_2\}$, where $i_1 < i_2$,
      then we first compute
      $a_1 := A_k[i_1] = A[P_k[i_1]]$ and
      $a_2 = A_k[i_2] = A[P_k[i_2]]$. If $a_1 \leq a_2$,
      then we return $i_1$ as the answer. Otherwise, we return $i_2$.
      This final step takes $\bigO(1)$ time.
    \end{enumerate}
    In total, the query takes $\bigO(\log \log m)$ time.

  \item Let $b, e \in [0 \dd m']$ and $v \geq 0$. We now finalize the query algorithm by showing how
    in $\bigO(\log \log m)$ time to check if $\ThreeSidedRangeCount{B}{b}{e}{v} > 0$, and if
    so, return $\ThreeSidedRMQ{A}{B}{b}{e}{v}$. First, in $\bigO(1)$ time we compute
    $k := \lfloor \tfrac{v}{y} \rfloor$ and $\delta = v \bmod y$. Observe that letting
    $b' := |\{j \in [1 \dd m_k] : P_k[j] \leq b\}|$ and $e' := |\{j \in [1 \dd m_k] : P_k[j] \leq e\}|$, it holds:
    \begin{align*}
      \ThreeSidedRangeCount{B}{b}{e}{v} = \ThreeSidedRangeCount{B_{k}}{b'}{e'}{ky+\delta}.
    \end{align*}
    Moreover, if $\ThreeSidedRangeCount{B}{b}{e}{v} > 0$, then
    \begin{align*}
      \ThreeSidedRMQ{A}{B}{b}{e}{v} = P_k[\ThreeSidedRMQ{A_{k}}{B_{k}}{b'}{e'}{ky+\delta}].
    \end{align*}
    The query algorithm thus proceeds as follows:
    \begin{enumerate}
    \item In $\bigO(1)$ time we compute $k := \lfloor \tfrac{v}{y} \rfloor$ and $\delta = v \bmod y$.
      If $k > k_{\max}$, then $m_k = 0$. This implies that $b' = e' = 0$, and hence
      also $\ThreeSidedRangeCount{B_{k}}{b'}{e'}{ky+\delta} = 0$. Thus, we return
      that $\ThreeSidedRangeCount{B}{b}{e}{v} = 0$. Let us now assume that $k \leq k_{\max}$.
    \item Using the predecessor structure over $P_k[1 \dd m_k]$, in $\bigO(\log \log m_k) = \bigO(\log \log m)$ time
      we compute $b'$ and $e'$ as defined above.
      If $b' \geq e'$, then we return that $\ThreeSidedRangeCount{B}{b}{e}{v} = 0$. Let us thus
      assume that $b' < e'$.
    \item Using the above query, in $\bigO(\log \log m)$ time we check if
      $\ThreeSidedRangeCount{B_{k}}{b'}{e'}{ky+\delta} > 0$, and if so we compute
      $i_{\min} := \ThreeSidedRMQ{A_{k}}{B_{k}}{b'}{e'}{ky+\delta}$.
      If
      $\ThreeSidedRangeCount{B_{k}}{b'}{e'}{ky+\delta} = 0$, then by the above we return that
      $\ThreeSidedRangeCount{B}{b}{e}{v} = 0$. Otherwise we return $P_k[i_{\min}] = 
      \ThreeSidedRMQ{A}{B}{b}{e}{v}$.
    \end{enumerate}
    In total, the query takes $\bigO(\log \log m)$ time.
  \end{itemize}

  \paragraph{Construction algorithm}

  We construct each of the components of the data structure as follows:
  \begin{enumerate}

  \item The array $A[1 \dd m']$ is stored in $\bigO(m') = \bigO(m)$ time.

  \item Similarly, we save $B[1 \dd m']$ in $\bigO(m') = \bigO(m)$ time.

  \item To compute the arrays $P_k[1 \dd m_k]$ for $k \in [0 \dd k_{\max}]$ and augment them with predecessor support,
    we proceed as follows.
    We first set $P_0[i] = i$ for all $i \in [1 \dd m']$. For $k \in [1 \dd k_{\max}]$, $P_k$ is then
    computed by iterating over $P_{k-1}$ and including only elements $P_{k-1}[i]$
    satisfying $A[P_{k-1}[i]] \geq ky$. By $\sum_{k \geq 0} m_k = \bigO(m)$, in total we spend $\bigO(m)$ time.
    We then augment all arrays $P_k$ with the predecessor structures. Since the arrays are sorted,
    using~\cite[Proposition~2]{wexp}, we spend $\bigO(m)$ total time.

  \item\label{th:three-sided-rmq-construction-it-4}
    Next, we construct the arrays $A^{\rm val}_{v}$ for $v \in [0 \dd y \cdot k'_{\max}]$ and augment then with RMQ
    data structures. Consider $k \in [0 \dd k'_{\max}]$.
    Note that we then have $m_k \geq y$.
    We first present how to compute the arrays
    $A^{\rm val}_{ky}[1 \dd m^{y}_k], A^{\rm val}_{ky+1}[1 \dd m^{y}_k], \ldots, A^{\rm val}_{ky+y-1}[1 \dd m^{y}_k]$
    in $\bigO(m_k)$ time.
    \begin{enumerate}
    \item In $\bigO(m_k)$ time we compute $A_k[1 \dd m_k]$ and $B_k[1 \dd m_k]$ as $A_k[i]=A[P_k[i]]$ and $B_k[i]=B[P_k[i]]$.
    \item In $\bigO(m_k + y) = \bigO(m_k)$ time we compute lists $L_0, L_1, \ldots, L_{y-1}$, where
      $L_{y-1} = \{(j,A_k[j]) : j \in [1 \dd m_k]\text{ and }B_k[j] \geq (k+1)y - 1\}$, and
      for every $\delta \in [0 \dd y-1)$,
      \[
        L_{\delta} = \{(j,A_k[j]) : j \in [1 \dd m_k]\text{ and }B_k[j] = ky+\delta\}.
      \]
    \item\label{th:three-sided-rmq-construction-it-4c}
      We then compute the arrays $A^{\rm val}_{ky+\delta}[1 \dd m^{y}_k]$ for $\delta=y-1,y-2,\ldots,0$ (in this order) as follows.
      We first initialize $A^{\rm val}_{ky+\delta}[1 \dd m^{y}_k]$ as follows. If $\delta = y-1$,
      we simply set $A^{\rm val}_{ky+\delta}[i]$ for all $i \in [1 \dd m^{y}_k]$.
      If $\delta < y-1$, we instead copy $A^{\rm val}_{ky+\delta+1}$ to $A^{\rm val}_{ky+\delta}$.
      We then go through the list $L_{\delta}$ and for every $(j,a) \in L_{\delta}$, we
      first compute $i := \lceil \tfrac{j}{y} \rceil$, and then replace
      $A^{\rm val}_{ky+\delta}[i]$ with $\min(A^{\rm val}_{ky+\delta}[i], a)$.
      The computation of $A^{\rm val}_{ky+\delta}[1 \dd m^{y}_k]$ takes $\bigO(m^{y}_k + |L_{\delta}|)$ time.
      Since $|L_0| + \ldots + |L_{y-1}| = m_k$, over all $\delta \in [0 \dd y)$, we spend
      $\bigO(y \cdot m^{y}_k + m_k) = \bigO(m_k)$ time.
    \end{enumerate}
    In total, the computation of
    $A^{\rm val}_{ky}[1 \dd m^{y}_k], A^{\rm val}_{ky+1}[1 \dd m^{y}_k], \ldots, A^{\rm val}_{ky+y-1}[1 \dd m^{y}_k]$
    takes $\bigO(m_k)$ time. Augmenting all arrays with the RMQ data structures
    from~\cite{FischerH11} takes $\bigO(m_k)$ extra time.
    Summing over all $k \in [0 \dd k'_{\max}]$, we spend
    $\bigO(\sum_{k \in [0 \dd k'_{\max}]} m_k) = \bigO(m)$ time.

  \item Assume that the arrays $A^{\rm val}_{v}$ for $v \in [0 \dd y \cdot k'_{\max}]$ have been computed.
    The computation of arrays $A^{\rm pos}_{v}$ for $v \in [0 \dd y \cdot k'_{\max}]$ proceeds similarly as in
    Step~\ref{th:three-sided-rmq-construction-it-4}, except in Step~\ref{th:three-sided-rmq-construction-it-4c},
    we proceed slightly differently. First, the array $A^{\rm pos}_{ky+\delta}$ is initialized to $A^{\rm pos}_{ky+\delta+1}$
    when $\delta < y-1$, and the array $A^{\rm pos}_{(k+1)y-1}$ is left uninitialized.
    Second, when we iterate over
    $L_{\delta}$, for every $(j,a) \in L_{\delta}$, we first compute $i := \lceil \tfrac{j}{y} \rceil$, and then
    compare $a$ with $A^{\rm val}_{ky+\delta}[i]$. We replace $A^{\rm pos}_{ky+\delta}[i]$ with $j$ if $a < A^{\rm val}_{ky+\delta}[i]$,
    or $a = A^{\rm val}_{ky+\delta}[i]$ and $j < A^{\rm pos}_{ky+\delta}[i]$. Otherwise, we leave $A^{\rm pos}_{ky+\delta}[i]$ unchanged.
    Similarly as above, we spend $\bigO(m)$ total time.

  \item Next, we construct the array $A_{\rm pow}[0 \dd 3x)$ in
    $\bigO(x) = \bigO(\tfrac{\log m}{\log \log m})  = \bigO(m)$ time.

  \item Next, we construct the arrays $E_k[1 \dd m^{x}_k]$ for all $k \in [0 \dd k_{\max}]$.
    We proceed in four steps:
    \begin{enumerate}

    \item For every $k \in [0 \dd k_{\max}]$, let $S_k[1 \dd m_k]$ be an array
      containing the permutation of $[1 \dd m_k]$ such that
      for every $j, j' \in [1 \dd m_k]$, $j < j'$ implies that
      $A_k[S_k[j]] < A_k[S_k[j']]$, or $A_k[S_k[j]] = A_k[S_k[j']]$ and
      $S_k[j] < S_k[j']$. Note that this implies
      $A_k[S_k[1]] \leq A_k[S_k[2]] \leq \cdots \leq A_k[S_k[m_k]]$.
      We compute the arrays $S_k$ for $k \in [0 \dd k_{\max}]$ as follows.
      To compute $S_0$, we create a sequence of length $m'$ containing at the $i$th
      position the pair $(A[i], i)$.
      Recall that $\max_{i \in [1 \dd m']} A[i] = \bigO(m \log m)$.
      Thus, using a 4-round radix sort, we
      can sort it in $\bigO(m)$ time. The resulting sequence contains
      $S_0$ on the second coordinate.
      Let now $k > 0$ and assume that we already computed $S_{k-1}$.
      To compute $S_k$:
      \begin{enumerate}
      \item First, we compute an array
        $A_{\rm next}[1 \dd m_{k-1}]$
        defined by $A_{\rm next}[i] = \TwoSidedRangeCount{A_{k-1}}{i}{ky}$.
        Note that given access to $A_{k-1}$ (which we can achieve via $A$ and
        $P_{k-1}$), computation of $A_{\rm next}$ takes $\bigO(m_{k-1})$ time.
        Observe that $A_{\rm next}$ gives the position of every element of
        $A_{k-1}$ that occurs in $A_{k}$, i.e.,
        for every $i \in [1 \dd m_{k-1}]$, $A_{k-1}[i] \geq ky$ implies
        $A_{k-1}[i] = A_{k}[A_{\rm next}[i]]$.
      \item We then construct $S_k$. First, set $p := 0$.
        For every $i=1, \ldots, m_{k-1}$, we check if
        $A_{k-1}[S_{k-1}[i]] \geq yk$. If so, we increment $p$ and set
        $S_{k}[p] := A_{\rm next}[S_{k-1}[i]]$. This takes
        $\bigO(m_{k-1})$ time.
      \end{enumerate}
      In total, the computation of arrays $S_k$ for $k \in [0 \dd k_{\max}]$
      takes $\bigO(m) + \sum_{k \in [0 \dd k_{\max})} \bigO(m_{k}) =
      \bigO((k_{\max} + 1) + m + \sum_{k \geq 0} m_k) = \bigO(m)$ time.

    \item In the second step, we construct the arrays $A_{k,i}$ for every
      $k \in [0 \dd k_{\max}]$ and $i \in [1 \dd m^{x}_{k}]$.
      Let us fix $k \in [0 \dd k_{\max}]$. We compute the arrays
      $A_{k,i}$ for all $i \in [1 \dd m^{x}_{k}]$,
      where $m^{x}_k = \lceil \tfrac{m_k}{x} \rceil$ as follows:
      \begin{enumerate}
      \item For every $i \in [1 \dd m^{x}_{k}]$, we initialize the arrays
        $\mathcal{A}_{\max}[i] = -\infty$ and $\mathcal{A}_{\rm size}[i] = 0$.
      \item We iterate over all values $t \in [1 \dd m_k]$ in increasing order,
        and in each iteration we proceed as follows. Let $j = S_k[t]$,
        $i = \lceil \tfrac{j}{x} \rceil \in [1 \dd m^{x}_k]$, and
        $b = (i-1)x$. We set $A_{k,i}[j - b] := \mathcal{A}_{\rm size}[i]$.
        Then, if $a := A_k[j] = A[P_k[j]]$ satisfies
        $a > \mathcal{A}_{\rm max}[i]$, we set $\mathcal{A}_{\rm max}[i] := a$
        and increment $\mathcal{A}_{\rm size}[i]$.
      \end{enumerate}
      The computation of all arrays $A_{k,i}$, where $i \in [1 \dd m^{x}_{k}]$
      takes $\bigO(m_k)$ time. Thus, over all $k \in [0 \dd k_{\max}]$,
      we spend $\sum_{k \in [0 \dd k_{\max}]} \bigO(m_k) =
      \bigO((k_{\max} + 1) + \sum_{k \geq 0} m_k) = \bigO(m)$ time.

    \item Next, we compute the arrays $B_{k,i}$ for every
      $k \in [0 \dd k_{\max}]$ and $i \in [1 \dd m^{x}_{k}]$.
      Let us fix $k \in [0 \dd k_{\max}]$. For every
      $j \in [1 \dd m_k]$, we first compute $i = \lceil \tfrac{j}{x} \rceil$
      and $b = (i-1)x$. We then set $B_{k,i}[j - b] := \min(y-1, B_k[j] - ky)
      = \min(y-1, B[P_k[j]] - ky)$.
      This takes $\bigO(m_k)$ time, and hence over all $k \in [0 \dd k_{\max}]$,
      we spend $\bigO((k_{\max} + 1) + \sum_{k \geq 0} m_k) = \bigO(m)$ time.

    \item Finally, we compute the arrays $E_k[1 \dd m^{x}_{k}]$ for all
      $k \in [0 \dd k_{\max}]$. Let us fix $k \in [0 \dd k_{\max}]$ and
      $i \in [1 \dd m^{x}_{k}]$. First, we compute $b = (i-1)x$,
      $e = \min(ix, m_k)$, and $\ell = e - b$. Recall that above we computed
      $A_{k,i}$ and $B_{k,i}$. Given these arrays and $A_{\rm pow}[0 \dd 3x)$,
      we can compute
      $\EncodeSeq(\ell, A_{k,i}[1], \ldots, A_{k,i}[\ell], B_{k,i}[1], \ldots, B_{k,i}[\ell])$
      in $\bigO(\ell)$ time. Over all $i \in [1 \dd m^{x}_{k}]$, this takes $\bigO(m_k)$ time.
      Thus, over $k \in [0 \dd k_{\max}]$, we spend
      $\bigO((k_{\max} + 1) + \sum_{k \geq 0} m_k) = \bigO(m)$ time.
    \end{enumerate}
    In total, we spend $\bigO(m)$ time.

  \item To construct the lookup table $L_{\rm rmq}[0 \dd y^{3x})$, we enumerate
    every combination of integers $b, e \in [0 \dd x]$, $\delta \in [0 \dd y)$,
    and $m_{\rm short} \in [1 \dd x]$. For a fixed combination of these integers, we then enumerate all arrays
    $A_{\rm short}[1 \dd m_{\rm short}]$ and $B_{\rm short}[1 \dd m_{\rm short}]$ such that $A_{\rm short}$ (resp.\ $B_{\rm short}$)
    contains integers in $[0 \dd x)$ (resp.\ $[0 \dd y)$).
    For each such combination of $b, e, \delta, m_{\rm short}, A_{\rm short}$, and $B_{\rm short}$:
    \begin{enumerate}
    \item In $\bigO(x)$ time
      we check if $\ThreeSidedRangeCount{B_{\rm short}}{b}{e}{\delta} = 0$. If so,
      we let $r := 0$.  Otherwise, we compute
      $r := \ThreeSidedRMQ{A_{\rm short}}{B_{\rm short}}{b}{e}{\delta}$.
    \item We compute $t = \EncodeSeq(b,e,\delta,m_{\rm short},A_{\rm short}[1], \ldots, A_{\rm short}[m_{\rm short}],
      B_{\rm short}[1], \ldots, B_{\rm short}[m_{\rm short}])$. Utilizing the array $A_{\rm pow}$, this
      takes $\bigO(x)$ time. We then write $L_{\rm rmq}[t] := r$.
    \end{enumerate}
    For every combination of parameters, we spend $\bigO(x)$ time. Since every combination
    results in a different value $t$, the number of combinations is bounded by $\bigO(y^{3x})$.
    In total, we thus spend $\bigO(y^{3x} x) = \bigO(m^{1/2} \tfrac{\log m}{\log \log m}) = \bigO(m)$ time.
  \end{enumerate}
  In total, the construction takes $\bigO(m)$ time.
\end{proof}

%% file: tools/dynamic-rmq.tex
\subsection{Dynamic One-Sided RMQ}\label{sec:dyn-rmq}

\begin{proposition}\label{pr:dynamic-predecessor}
  Let $\alpha \in (0, 1)$ be a constant. Let $u \geq 1$ and
  $h > 0$ be such that $h = \bigO(\log u)$.
  Consider a set $\mathcal{S} \subseteq [0 \dd h) \times \Zp$,
  and denote ${\rm keys}(\mathcal{S}) := \{k : (k,v) \in \mathcal{S}\}$.
  After $\bigO(u^{\alpha})$-time preprocessing,
  we can in $\bigO(1)$ time support the following operations on the initially empty $\mathcal{S}$:
  \begin{enumerate}
  \item (Insert)
    Given any $(k, v) \in [0 \dd h) \times \Zp$ such that
    $x \not\in {\rm keys}(\mathcal{S})$,
    insert $(k,v)$ into $\mathcal{S}$.
  \item (Delete)
    Given any $k \in {\rm keys}(\mathcal{S})$ remove from $\mathcal{S}$
    the (unique) pair $(k',v') \in \mathcal{S}$ satisfying $k' = k$.
  \item (Predecessor)
    Given any $q \in [0 \dd h)$, return the pair $(k,v) \in \mathcal{S} \cup \{(-1,0)\}$
    satisfying \[k = \max\{k' : (k',v') \in \mathcal{S} \cup \{(-1,0)\}\text{ and }k' \leq q\}.\]
  \item (Successor)
    Given any $q \in [0 \dd h)$, return the pair $(k,v) \in \mathcal{S} \cup \{(h,0)\}$
    satisfying \[k = \min\{k' : (k',v') \in \mathcal{S} \cup \{(h,0)\}\text{ and }k' \geq q\}.\]
  \end{enumerate}
\end{proposition}
\begin{proof}

  Let $\alpha' \in (0, \alpha)$ be a constant such that $b := \alpha' \log u$ is a positive integer.
  It is easy to see that such $\alpha'$ exists for all $u \geq u_{\min}$, where $u_{\min}$ is some constant.
  Let $L_{\rm pred}$ (resp.\ $L_{\rm succ}$) be a mapping such that for any
  $X \in \{{\tt 0}, {\tt 1}\}^{b}$ and any $q \in [1 \dd b]$,
  $L_{\rm pred}$ (resp.\ $L_{\rm succ}$) maps the pair $(X,q)$ into the value
  $\max\{i \in [1 \dd q] : X[i] = 1\} \cup \{0\}$
  (resp.\ $\min\{i \in [q \dd b] : X[i] = 1\} \cup \{b+1\}$).

  The result of the preprocessing consists of three components:
  \begin{enumerate}
  \item First, we store the lookup table $L_{\rm pred}$. When accessing
    $L_{\rm pred}$, any pair $(X,q) \in \{{\tt 0}, {\tt 1}\}^{b} \times [1 \dd b]$ is represented as an
    integer in $[0 \dd 2^{b+b'})$, where $b' = \lceil \log b \rceil$. Thus,
    $L_{\rm pred}$ needs $\bigO(2^{b+b'}) = \bigO(2^{\alpha' \log u} \cdot \log b)
    = \bigO(u^{\alpha'} \log u) = \bigO(u^{\alpha})$ space.
  \item Encoded similarly as above, we also store the lookup table $L_{\rm succ}$.
  \item Lastly, we store an array $A_{\rm val}[0 \dd h)$. During the execution of the sequence
    of operations, we will maintain the invariant that if $k \in {\rm keys}(\mathcal{S})$, then $A_{\rm val}[k] = v$,
    where $v \in \Zp$ is such that $(k,v) \in \mathcal{S}$.
    The array $A_{\rm val}$ needs $\bigO(\log u) = \bigO(u^{\alpha})$ space.
  \end{enumerate}
  In total, the result of preprocessing needs $\bigO(u^{\alpha})$ space.

  The queries are implemented as follows. During the sequence of operations, we maintain
  a bitvector $B_{\rm keys} \in \{{\tt 0}, {\tt 1}\}^{h}$ such that for every $k \in [0 \dd h)$,
  $k \in {\rm keys}(\mathcal{S})$ holds if and only if $B_{\rm keys}[k+1] = {\tt 1}$. The bitvector
  $B_{\rm keys}$ in stored in packed form as an integer $x_{\rm keys} \in [0 \dd 2^h)$ encoded using
  $\bigO(1)$ machine words. We also maintain the invariant for the array
  $A_{\rm val}$ (as described above). Then:
  \begin{itemize}
  \item To insert a pair $(k,v)$ into $\mathcal{S}$, in $\bigO(1)$ time we update $x_{\rm keys}$ using
    bit operations, and then set $A_{\rm val}[k] := v$.
  \item To delete a pair with a given key $k$, we set the $(k+1)$st bit of $x_{\rm keys}$
    to $0$ in $\bigO(1)$ time.
  \item Given an integer $q \in [0 \dd h)$, we can easily compute its predecessor in $\mathcal{S}$ in
    $\bigO(1 + h/b) = \bigO(1 + 1/\alpha') = \bigO(1)$ time using bit operations and the lookup table
    $L_{\rm pred}$.
  \item The successor operation is implemented similarly as above in $\bigO(1)$ time.
  \end{itemize}

  The lookup tables $L_{\rm pred}$ and $L_{\rm succ}$ are easy to construct in
  $\bigO(2^b \cdot b^2) = \bigO(u^{\alpha'} \log^2 u) =
  \bigO(u^{\alpha})$ time.
\end{proof}

\begin{proposition}\label{pr:narrow-range-max}
  Let $\alpha \in (0, 1)$ be a constant. Let $u \geq 1$ and
  $h > 0$ be such that $h = \bigO(\log u)$.
  Consider a set $\mathcal{S} \subseteq [0 \dd h) \times \Zp$.
  After $\bigO(u^{\alpha})$-time preprocessing,
  we can in $\bigO(m)$ total time execute any sequence of $m$ operations
  (on the initially empty $\mathcal{S}$)
  of the following type:
  \begin{itemize}
  \item Insert a given pair $(x,y) \in [0 \dd h) \times \Zp$ into $\mathcal{S}$.
  \item Given any $q \in [0 \dd h)$, return
    $\max \{y : (x,y) \in \mathcal{S} \cup \{(h,0)\}\text{ and }x \geq q\}$.
  \end{itemize}
\end{proposition}
\begin{proof}

  Observe that if there exist $(x_1, y_1), (x_2, y_2) \in \mathcal{S}$
  satisfying $(x_1, y_1) \neq (x_2, y_2)$, $x_1 \leq x_2$, and $y_1 \leq y_2$,
  then for every $\mathcal{S}' \subseteq [0 \dd h) \times \Zp$
  satisfying $\mathcal{S} \subseteq \mathcal{S}'$, and every $q \in [0 \dd h)$, it holds
  \[
    \max \{y : (x,y) \in \mathcal{S}' \cup \{(h,0)\}\text{ and }x \geq q\} =
    \max\{y : (x,y) \in \mathcal{S'} \cup \{(h,0)\} \setminus \{(x_1,y_1)\}\text{ and }x \geq q\}.
  \]
  In other words, in such case, we can delete $(x_1,y_1)$ from $\mathcal{S}$, since regardless of what elements are added to
  $\mathcal{S}$, $(x_1,y_1)$ will never affect the result of any query. Therefore, in such cases, we say that
  $(x_1, y_1)$ is \emph{redundant}. We call any set $\mathcal{S}$ \emph{non-redundant} if it does not contain any redundant
  elements. Observe that every set can be made non-redundant by repeatedly removing redundant pairs, and moreover,
  the resulting non-redundant
  set does not depend on the order in which we remove redundant pairs. For any $\mathcal{S}$, let thus ${\rm nonred}(\mathcal{S})$ denote
  the result of repeatedly removing redundant pairs until the set is non-redundant.
  In our structure we will maintain ${\rm nonred}(\mathcal{S})$
  rather than $\mathcal{S}$. Since for every $\mathcal{S}_1$ and $\mathcal{S}_2$, it holds
  ${\rm nonred}(\mathcal{S}_1 \cup \mathcal{S}_2) = {\rm nonred}({\rm nonred}(\mathcal{S}_1) \cup \mathcal{S}_2)$, we can
  remove the redundant elements as soon as they appear. The advantage of maintaining $\mathcal{S}' := {\rm nonred}(\mathcal{S})$ instead
  of $\mathcal{S}$, is that the query from the claim on such a set reduces to a successor query, i.e.,
  for every $q \in [0 \dd h)$, it holds
  \[
    \max\{y : (x,y) \in \mathcal{S}' \cup \{(h,0)\}\text{ and }x \geq q\}
    = y_{\rm succ},
  \]
  where $(x_{\rm succ},y_{\rm succ}) \in \mathcal{S}' \cup \{(h,0)\}$ is such that
  $x_{\rm succ} = \min\{x : (x,y) \in \mathcal{S}' \cup \{(h,0)\}\text{ and }x \geq q\}$.

  With the above in mind, our data structure consists of a single component: the result of
  preprocessing from \cref{pr:dynamic-predecessor}. It needs $\bigO(u^{\alpha})$ space.

  The operations are implemented as follows:
  \begin{itemize}
  \item Let $(x,y) \in [0 \dd h) \times \Zp$ and $\mathcal{S}'$ denote
    the current non-redundant subset of $\mathcal{S}$. Note that this implies $|\mathcal{S}'| \leq h$.
    The insertion operation is implemented in three steps:
    \begin{enumerate}
    \item First, we check if $(x,y)$ would be redundant after inserting into $\mathcal{S}'$.
      To this end, using successor query in $\bigO(1)$ time
      we obtain a pair $(x', y') \in \mathcal{S}'$ such that
      $x' = \min \{a : (a,b) \in \mathcal{S}'\cup\{(h,0)\}\text{ and }a \geq x\}$.
      If $y \leq y'$, then $(x,y)$ would be redundant,
      and we conclude the insertion algorithm. Let us thus assume that $y > y'$.
    \item Next, we perform the removal stage, i.e., we remove from
      $\mathcal{S'}$ all pairs
      that are redundant in $\mathcal{S}' \cup \{(x,y)\}$.
      To this end, we repeatedly execute the following step. First, using the predecessor query in
      $\bigO(1)$ time we check
      if $\mathcal{S}'$ contains a pair with the first coordinate not exceeding $x$. If not, we finish the
      removal stage. Otherwise, we obtain a pair $(x'', y'') \in \mathcal{S}'$ satisfying
      $x'' = \max\{a : (a,b) \in \mathcal{S}' \cup \{(h,0)\}\text{ and }a \leq x\}$.
      If $y'' \leq y$, then in $\bigO(1)$ time we remove the pair $(x'',y'')$ from $\mathcal{S}'$.
      Otherwise, we finish the removal stage.
    \item Finally, in $\bigO(1)$ time
      we insert $(x,y)$ into the current set.
    \end{enumerate}
    It is easy to check that the resulting set is non-redundant.
    In total, we spend $\bigO(1 + d)$ time, where $d$
    is the number of deleted pairs.
  \item The query is implemented as described above, i.e., given $q \in [0 \dd h)$, using
    successor query in $\bigO(1)$ time we compute a pair
    $(x_{\rm succ},y_{\rm succ}) \in \mathcal{S}' \cup \{(h,0)\}$ such that
    $x_{\rm succ} = \min\{x : (x,y) \in \mathcal{S}' \cup \{(h,0)\}\text{ and }x \geq q\}$.
    We then have $\max\{y : (x,y) \in \mathcal{S}' \cup \{(h,0)\}\text{ and }x \geq q\}
    = y_{\rm succ}$.
  \end{itemize}
  Assume now that we performed $m$ operations (each of which is either a
  query or an insertion). The total time spent during queries is clearly
  $\bigO(m)$. The total time for insertions is
  $\bigO(m + d_{\rm all})$, where $d_{\rm all}$ is the total
  number of pairs removed during the insertion operations. Since we insert
  at most $m$ elements, it follows that $d_{\rm all} \leq m$. Consequently,
  the total time for $m$ operations in $\bigO(m)$.

  By \cref{pr:dynamic-predecessor}, the preprocessing for the above
  component takes $\bigO(u^{\alpha})$ time.
\end{proof}

%% file: index.tex
\section{Index for Leftmost Occurrences}\label{sec:minocc-index}

Let $\epsilon \in (0, 1)$ be a constant and let
$\Text \in \Sigma^{\Textlen}$, where $\Sigma = \IntegerAlphabet$ and
$2 \leq \AlphabetSize < \Textlen^{1/7}$.
We assume that $\Text[\Textlen]$ does not occur in
$\Text[1 \dd \Textlen)$.\footnote{If the text does not end with a unique symbol, we first modify the
text by increasing the alphabet by one and appending the unique
symbol. This change has no effect on leftmost occurrences of the original text, and the space usage,
and construction time and working space of the index remains asymptotically the same.
Note that appending a unique symbol at the end of $\Text$ may
not be possible if $\AlphabetSize = 2^k$ for some $k \geq 1$, since
the number of bits per character does not accommodate a new symbol. In that case, we need to compute the new packed
representation of text, where each character uses one more bit. Such representation is easy to compute in
$\bigO(\Textlen / \log_{\AlphabetSize} \Textlen)$ time using lookup tables.}
In this section, we show how,
given a packed representation of $\Text$, in
$\bigO(\Textlen \min(1, \log \AlphabetSize / \sqrt{\log \Textlen}))$ time
and using $\bigO(\Textlen / \log_{\AlphabetSize} \Textlen)$ working space construct
a data structure of size $\bigO(\Textlen / \log_{\AlphabetSize} \Textlen)$ that:
(1) given a substring of $\Text$ represented by its starting position and length,
finds its leftmost occurrence in
$\Text$ in $\bigO(\log^{\epsilon} \Textlen)$ time; and
(2) given the packed representation of any pattern $\Pat \in \IntegerAlphabet^{m}$
that satisfies $\OccTwo{\Pat}{\Text} \neq \emptyset$, in
$\bigO(\log^{\epsilon} \Textlen + m / \log_{\AlphabetSize} \Textlen)$ time find
its leftmost occurrence in $\Text$.
We also derive a general reduction depending on prefix RMQ queries.

We assume that $\epsilon$, $\AlphabetSize$, and $\Text$ (and hence also $\Textlen$) are fixed for the duration of this section.
Additionally, we fix a constant $\mu \in (0, \tfrac{1}{6})$
such that $\mu \log_{\AlphabetSize} \Textlen$ is a positive integer. Such constant
exists by $2 \leq \AlphabetSize < \Textlen^{1/7}$. Observe also that we then
have $3\mu \log_{\AlphabetSize} \Textlen - 1 \leq \Textlen$.

\input{index/prelim.tex}

\input{index/core.tex}

\input{index/nonperiodic.tex}

\input{index/periodic.tex}

\input{index/final.tex}

\input{index/summary.tex}

\input{index/applications.tex}

%% file: index/prelim.tex
\subsection{Preliminaries}\label{sec:minocc-index-prelim}

\begin{definition}[$\tau$-periodic and $\tau$-nonperiodic patterns]\label{def:periodic-pattern}
  Let $\Pat \in \Sigma^{m}$ and $\tau \geq 1$. We say that $\Pat$ is
  \emph{$\tau$-periodic} if it holds $m \geq 3\tau - 1$ and
  $\per(\Pat[1 \dd 3\tau - 1]) \leq \tfrac{1}{3}\tau$. Otherwise, it
  is called \emph{$\tau$-nonperiodic}.
\end{definition}

%% file: index/core.tex
\subsection{The Index Core}\label{sec:minocc-index-core}

\input{index/core/nav.tex}

\input{index/core/structure.tex}

\input{index/core/queries.tex}

\input{index/core/construction.tex}

%% file: index/core/nav.tex
\subsubsection{Basic Navigation Primitives}\label{sec:minocc-index-core-nav}

\begin{proposition}\label{pr:nav-index-core}
  Let $\tau = \mu\log_{\AlphabetSize} \Textlen$.
  Given the packed representation of the text
  $\Text$, we can in $\bigO(\Textlen / \log_{\AlphabetSize} \Textlen)$ time
  construct a data structure, denoted $\NavCore{\Text}$, that supports
  the following queries:
  \begin{enumerate}
  \item\label{pr:nav-index-core-it-1}
    Given any $j \in [1 \dd \Textlen - 3\tau + 2]$, in $\bigO(1)$ time check if $j \in \RTwo{\tau}{\Text}$.
  \item\label{pr:nav-index-core-it-2}
    Given a packed representation of any $\Pat \in \Sigma^{m}$ satisfying $m \geq 3\tau - 1$, in
    $\bigO(1)$ time check if $\Pat$ is $\tau$-periodic (\cref{def:periodic-pattern}).
  \end{enumerate}
\end{proposition}
\begin{proof}
  $\NavCore{\Text}$ contains a single component: the structure from~\cite[Section~6.1.1]{breaking}. It needs
  $\bigO(\Textlen / \log_{\AlphabetSize} \Textlen)$ space, answers both queries in $\bigO(1)$ time, and
  its construction takes $\bigO(\Textlen / \log_{\AlphabetSize} \Textlen)$ time.
\end{proof}

%% file: index/core/structure.tex
\subsubsection{The Data Structure}\label{sec:minocc-index-core-structure}

\paragraph{Definitions}

Let $\tau = \mu\log_{\AlphabetSize} \Textlen$, where $\mu$ is as defined at
the beginning of \cref{sec:minocc-index}.
Let $L_{\rm minocc}$ be a mapping such that for every nonempty pattern $\Pat \in \IntegerAlphabet^{<3\tau-1}$
satisfying $\OccTwo{\Pat}{\Text} \neq \emptyset$, $L_{\rm minocc}$ maps the string $\Pat$ to
the position $\min \OccTwo{\Pat}{\Text}$.

\paragraph{Components}

The data structure, denoted $\MinOccIndexCore{\Text}$, consists of three components:
\begin{enumerate}
\item The structure $\NavCore{\Text}$ from \cref{pr:nav-index-core}. It needs
  $\bigO(\Textlen / \log_{\AlphabetSize} \Textlen)$ space.
\item The packed representation of $\Text$ using
  $\bigO(\Textlen / \log_{\AlphabetSize} \Textlen)$ space.
\item The lookup table $L_{\rm minocc}$. When accessing $L_{\rm minocc}$, strings
  $\Pat \in \IntegerAlphabet^{<3\tau-1}$ are converted to small integers
  using the mapping $\Int{3\tau}{\AlphabetSize}{\Pat}$ (\cref{def:int}).
  By $\Int{3\tau}{\AlphabetSize}{\Pat} \in [0 \dd \AlphabetSize^{6\tau})$, $L_{\rm minocc}$
  needs $\bigO(\AlphabetSize^{6\tau}) = \bigO(\Textlen^{6\mu}) = \bigO(\Textlen / \log_{\AlphabetSize} \Textlen)$ space.
\end{enumerate}

In total, $\MinOccIndexCore{\Text}$ needs
$\bigO(\Textlen / \log_{\AlphabetSize} \Textlen)$ space.

%% file: index/core/queries.tex
\subsubsection{Implementation of Queries}\label{sec:minocc-index-core-queries}

\begin{proposition}\label{pr:minocc-index-core-query-pat}
  Let $\tau = \mu\log_{\AlphabetSize} \Textlen$. Let $\Pat \in \Sigma^{m}$ be a nonempty pattern
  satisfying
  $m < 3\tau - 1$ and $\OccTwo{\Pat}{\Text} \neq \emptyset$.
  Given $\MinOccIndexCore{\Text}$ (\cref{sec:minocc-index-core-structure}) and
  the packed representation of $\Pat$, we can in $\bigO(1)$
  time compute $\min \OccTwo{\Pat}{\Text}$.
\end{proposition}
\begin{proof}
  The query is answered in $\bigO(1)$ time using the lookup table $L_{\rm minocc}$
  (see \cref{sec:minocc-index-core-structure}).
\end{proof}

\begin{proposition}\label{pr:minocc-index-core-query-pos}
  Let $\tau = \mu\log_{\AlphabetSize} \Textlen$, $j \in [1 \dd \Textlen]$, and
  $0 < \ell < 3\tau - 1$ be such that $j + \ell \leq \Textlen + 1$. Given
  $\MinOccIndexCore{\Text}$ (\cref{sec:minocc-index-core-structure}) and
  the pair $(j,\ell)$, we can in $\bigO(1)$ time
  compute $\min \OccTwo{\Pat}{\Text}$, where $\Pat = \Text[j \dd j + \ell)$.
\end{proposition}
\begin{proof}
  The algorithm proceeds in two steps:
  \begin{enumerate}
  \item Using the packed representation of $\Text$ (stored as part of
    $\MinOccIndexCore{\Text}$; see \cref{sec:minocc-index-core-structure}),
    in $\bigO(1)$ time we obtain the packed representation of $\Pat$.
  \item Using \cref{pr:minocc-index-core-query-pat}, we compute
    $\min \OccTwo{\Pat}{\Text}$ in $\bigO(1)$ time.
  \end{enumerate}
  In total, we spend $\bigO(1)$ time.
\end{proof}

%% file: index/core/construction.tex
\subsubsection{Construction Algorithm}\label{sec:minocc-index-core-construction}

\begin{proposition}\label{pr:minocc-index-core-construction}
  Given the packed representation of $\Text$, we can construct
  $\MinOccIndexCore{\Text}$ (\cref{sec:minocc-index-core-structure}) in
  $\bigO(\Textlen / \log_{\AlphabetSize} \Textlen)$ time.
\end{proposition}
\begin{proof}
  Let $\tau = \mu\log_{\AlphabetSize} \Textlen$ be as defined at the beginning of
  \cref{sec:minocc-index-core-structure}.
  We construct the components of $\MinOccIndexCore{\Text}$ (\cref{sec:minocc-index-core-structure}) as follows:
  \begin{enumerate}
  \item Using \cref{pr:nav-index-core}, we construct $\NavCore{\Text}$
    in $\bigO(\Textlen / \log_{\AlphabetSize} \Textlen)$ time.
  \item In $\bigO(\Textlen / \log_{\AlphabetSize} \Textlen)$ time we save
    the packed representation of $\Text$.
  \item The lookup table $L_{\rm minocc}$ is constructed as follows.
    Let $\ell = 3\tau - 2$. For any $i \in [1 \dd \Textlen]$, denote
    $B(i) := \Text[i \dd \min(n + 1, i + 2\ell - 1))$.
    The construction proceeds in three steps:
    \begin{enumerate}
    \item We set the initial value in $L_{\rm minocc}$ for every nonempty pattern
      $\Pat \in \IntegerAlphabet^{\leq \ell}$ to $\Textlen + 1$.
      Including the allocation of the array (see \cref{sec:minocc-index-core-structure}),
      this takes $\bigO(\AlphabetSize^{6\tau}) =
      \bigO(\Textlen^{6\mu}) = \bigO(\Textlen / \log_{\AlphabetSize} \Textlen)$ time.
    \item
      Denote $m = \lceil \tfrac{\Textlen}{\ell} \rceil$. Let $A_{\rm blk}[1 \dd m]$ be an array
      containing the permutation of $\{1 + i\ell : i \in [0 \dd m)\}$
      such that for every $j, j' \in [1 \dd m]$, $j < j'$
      implies that, letting $b = A_{\rm blk}[j]$ and $b' = A_{\rm blk}[j]$, it holds $B(b) \prec B(b')$, or
      $B(b) = B(b')$ and $b < b'$. We compute the array $A_{\rm blk}[1 \dd m]$ as follows.
      First, in $\bigO(m) = \bigO(\Textlen / \ell) = \bigO(\Textlen / \log_{\AlphabetSize} \Textlen)$ time
      we initialize the array $A_{\rm sort}[1 \dd m]$ such that for every $i \in [0 \dd m)$,
      $A_{\rm sort}[i+1] = (\Int{2\ell}{\AlphabetSize}{B(1 + i\ell)}, 1 + i\ell)$.
      We then sort $A_{\rm sort}$ lexicographically.
      By \cref{def:int}, the first coordinate is in range $[0 \dd \AlphabetSize^{4\ell}) \subseteq
      [0 \dd \Textlen^{12\mu}) \subseteq [0 \dd \Textlen^2)$, and the second coordinate is in $[1 \dd \Textlen]$.
      Thus, using a 6-round
      radix sort, the sorting takes $\bigO(m + \sqrt{n}) = \bigO(\Textlen / \log_{\AlphabetSize} \Textlen)$
      time. By \cref{lm:int}, the resulting array contains
      $A_{\rm blk}$ on the second coordinate.
    \item For every $i \in [1 \dd m]$ such that either $i = 1$, or $i > 1$ and $B(A_{\rm blk}[i-1]) \neq B(A_{\rm blk}[i])$,
      we enumerate all $\Theta(\ell^2)$ nonempty substrings of $B(A_{\rm blk}[i])$ of length not exceeding $\ell$, and update
      $L_{\rm minocc}$. In other words, letting $b = A_{\rm blk}[i]$ and $e = \min(n + 1, b + 2\ell - 1)$, we enumerate
      all pairs $(s, t)$ satisfying $t \leq \ell$ and $b \leq s < s + t \leq e$, and for each such pair, we set
      $L_{\rm minocc}[q] := \min(L_{\rm minocc}[q], b)$, where
      $\Pat = \Text[s \dd s+t)$ and $q = \Int{3\tau}{\AlphabetSize}{\Pat}$. The number of inspected positions $i \in [1 \dd m - 2]$
      is bounded by the number of distinct substrings of $\Text$ of length $2\ell-1$, which in turn is bounded by
      $\AlphabetSize^{2\ell-1} \leq \AlphabetSize^{6\tau} = \Textlen^{6\mu}$. For each inspected position, we spend $\bigO(\ell^2)
      = \bigO(\log^2 \Textlen)$ time, and hence in total, we spend $\bigO(\Textlen^{6\mu} \log^2 \Textlen) =
      \bigO(\Textlen / \log_{\AlphabetSize} \Textlen)$ time.
    \end{enumerate}
    In total, we spend $\bigO(\Textlen / \log_{\AlphabetSize} \Textlen)$ time.
    To prove that the above procedure correctly computes
    $L_{\rm minocc}$, first observe that for every $(s, t)$ such that $1 \leq s < s + t \leq \Textlen + 1$ and $t \leq \ell$,
    $\Text[s \dd s + t)$ occurs in $B(1 + i\ell)$, where $i = \lfloor \tfrac{s-1}{\ell} \rfloor$. On the other hand,
    we skip inspecting a block if and only if it has another occurrence to the left. Thus, the algorithm never misses
    the leftmost occurrence of any substring.
  \end{enumerate}
  In total, the construction takes
  $\bigO(\Textlen / \log_{\AlphabetSize} \Textlen)$ time.
\end{proof}

%% file: index/nonperiodic.tex
\subsection{The Nonperiodic Patterns and Positions}\label{sec:minocc-index-nonperiodic}

\input{index/nonperiodic/prelim.tex}

\input{index/nonperiodic/nav.tex}

\input{index/nonperiodic/structure.tex}

\input{index/nonperiodic/queries.tex}

\input{index/nonperiodic/construction.tex}

%% file: index/nonperiodic/prelim.tex
\subsubsection{Preliminaries}\label{sec:minocc-index-nonperiodic-prelim}

\begin{definition}\label{def:lex-sorted}
  Let $P \subseteq [1 \dd \Textlen]$ and $q = |P|$. By
  $\LexSorted{P}{\Text}$, we denote a sequence $(a_i)_{1 \in [1 \dd q]}$
  containing all positions from $P$ such that for every
  $i, j \in [1 \dd q]$, $i < j$ implies
  $\Text[a_i \dd \Textlen] \prec \Text[a_j \dd \Textlen]$.
\end{definition}

\begin{lemma}\label{lm:sss-max}
  Let $\tau \geq 1$ be such that $3\tau - 1 \leq \Textlen$ and $\SSS$
  be a $\tau$-synchronizing set of $\Text$. Then, it holds
  $\SSS \neq \emptyset$ and $\max \SSS \geq \Textlen - 3\tau + 2$.
\end{lemma}
\begin{proof}
  By $3\tau - 1 \leq \Textlen$ and the uniqueness of
  $\Text[\Textlen]$ in $\Text$ (see \cref{sec:minocc-index}), we 
  have $\Textlen - 3\tau + 2 \in [1 \dd \Textlen - 3\tau + 2] \setminus \RTwo{\tau}{\Text}$.
  Thus, by the density condition (\cref{def:sss}\eqref{def:sss-density}), it follows
  that $[\Textlen - 3\tau + 2 \dd \Textlen - 2\tau + 2) \cap \SSS \neq \emptyset$.
  Hence, $\SSS \neq \emptyset$ and $\max \SSS \geq \Textlen - 3\tau + 2$.
\end{proof}

\begin{definition}\label{def:dist-prefixes}
  Let $\tau \geq 1$ be such that $3\tau - 1 \leq \Textlen$, and
  $\SSS$ be a $\tau$-synchronizing set of $\Text$.
  For every $j \in [1 \dd \Textlen - 3\tau + 2] \setminus \RTwo{\tau}{\Text}$,
  we denote $\DistPrefixPos{j}{\tau}{\Text}{\SSS} :=
  \Text[j \dd \Successor{\SSS}{j} + 2\tau)$, where
  $\Successor{\SSS}{j} := \min\{j' \in \SSS : j' \geq j\}$.
  We then let
  \[
    \DistPrefixes{\tau}{\Text}{\SSS}
      := \{\DistPrefixPos{j}{\tau}{\Text}{\SSS} : j \in [1 \dd \Textlen -
          3\tau + 2] \setminus \RTwo{\tau}{\Text}\}.
  \]
\end{definition}

\begin{remark}\label{rm:dist-prefixes}
  Note that $\Successor{\SSS}{j}$ in \cref{def:dist-prefixes}
  is well-defined for every $j \in [1 \dd \Textlen - 3\tau + 2]
  \setminus \RTwo{\tau}{\Text}$, since by
  \cref{lm:sss-max}, it holds $\SSS \neq \emptyset$ and
  $\max \SSS \geq \Textlen - 3\tau + 2$.
\end{remark}

\begin{lemma}\label{lm:dist-prefixes}
  Let $\tau \geq 1$ be such that $3\tau - 1 \leq \Textlen$ and $\SSS$
  be a $\tau$-synchronizing set of $\Text$. 
  \begin{enumerate}
  \item\label{lm:dist-prefixes-it-1}
    It holds $\DistPrefixes{\tau}{\Text}{\SSS} \subseteq \IntegerAlphabet^{\leq 3\tau - 1}$.
  \item\label{lm:dist-prefixes-it-2}
    $\DistPrefixes{\tau}{\Text}{\SSS}$ is prefix-free, i.e.,
    for $D, D' \in \DistPrefixes{\tau}{\Text}{\SSS}$,
    $D \neq D'$ implies that $D$ is not a prefix~of~$D'$.
  \end{enumerate}
\end{lemma}
\begin{proof} 
  1. Let $D \in \DistPrefixes{\tau}{\Text}{\SSS}$. By \cref{def:dist-prefixes}, there
  exists $j \in [1 \dd \Textlen - 3\tau + 2] \setminus \RTwo{\tau}{\Text}$ such
  that $D = \DistPrefixPos{j}{\tau}{\Text}{\SSS}$, i.e.,  $D = \Text[j \dd s + 2\tau)$,
  where $s = \Successor{\SSS}{j}$. By \cref{def:sss}\eqref{def:sss-density}, we then
  have $[j \dd j + \tau) \cap \SSS \neq \emptyset$. This implies
  $s - j = \Successor{\SSS}{j} - j < \tau$, and hence $|D| = (s - j) + 2\tau
  \leq 3\tau - 1$.

  2. Suppose that there exist $D, D' \in \DistPrefixes{\tau}{\Text}{\SSS}$ such
  that $D \neq D'$ and $D$ is a prefix of $D'$. By \cref{def:dist-prefixes}, there
  exist $j, j' \in [1 \dd \Textlen - 3\tau + 2] \setminus \RTwo{\tau}{\Text}$
  such that $D = \DistPrefixPos{j}{\tau}{\Text}{\SSS}$ and $D' = \DistPrefixPos{j'}{\tau}{\Text}{\SSS}$, i.e.,
  $D = \Text[j \dd s + 2\tau)$ and $D' = \Text[j' \dd s' + 2\tau)$, where
  $s = \Successor{\SSS}{j}$ and $s' = \Successor{\SSS}{j'}$. Since $D$ is a prefix of $D'$ and $D \neq D'$,
  we have $s - j = |D| - 2\tau < |D'| - 2\tau = s' - j'$. Denote $\delta = s - j$.
  Observe that $D$ being a prefix of $D'$ implies that
  $\Text[j + \delta \dd j + \delta + 2\tau) = \Text[j' + \delta \dd j' + \delta + 2\tau)$.
  Since $j + \delta = s \in \SSS$, it thus follows by \cref{def:sss}\eqref{def:sss-consistency} that
  $j' + \delta \in \SSS$. By $s - j < s' - j'$, it follows that $j' + \delta = j' + (s - j) < j' + (s' - j') = s'$.
  Note, however, that by $s' = \Successor{\SSS}{j'}$, it holds $[j' \dd s') \cap \SSS = \emptyset$. This
  contradicts $j' + \delta \in \SSS$.
\end{proof}

\begin{lemma}\label{lm:dist-prefix-existence}
  Let $\tau \geq 1$ be such that $3\tau - 1 \leq \Textlen$ and $\SSS$ be
  a $\tau$-synchronizing set of $\Text$. Let $\Pat \in \Sigma^{m}$ be
  a $\tau$-nonperiodic pattern (\cref{def:periodic-pattern}) such
  that $m \geq 3\tau - 1$ and $\OccTwo{\Pat}{\Text} \neq \emptyset$.
  Then, there exists a unique $D \in \DistPrefixes{\tau}{\Text}{\SSS}$ (\cref{def:dist-prefixes})
  that is a prefix of $\Pat$.
\end{lemma}
\begin{proof}
  Consider any $j \in \OccTwo{\Pat}{\Text}$ (such position exists by
  the assumption $\OccTwo{\Pat}{\Text} \neq \emptyset$). Since $\Pat$
  is $\tau$-nonperiodic, it follows by $m \geq 3\tau - 1$ and \cref{def:periodic-pattern}
  that $j \in [1 \dd \Textlen - 3\tau + 2]$ and
  $\per(\Text[j \dd j + 3\tau - 1)) = \per(\Pat[1 \dd 3\tau - 1]) > \tfrac{1}{3}\tau$.
  By \cref{def:sss}, we thus have $j \in [1 \dd \Textlen - 3\tau + 2] \setminus \RTwo{\tau}{\Text}$
  and $[j \dd j + \tau) \cap \SSS \neq \emptyset$.
  Consequently, letting $j' = \Successor{\SSS}{j}$ (see \cref{def:dist-prefixes}), it holds
  $j' - j < \tau$. Therefore, letting $D = \Text[j \dd j' + 2\tau)$, it
  holds $|D| = j' - j + 2\tau \leq 3\tau - 1 \leq m$, and hence $D$ is a prefix of $\Pat$.
  On the other hand, by \cref{def:dist-prefixes}, we have $D \in \DistPrefixes{\tau}{\Text}{\SSS}$.
  To finish the proof of the first claim, it remains to observe that 
  since $\DistPrefixes{\tau}{\Text}{\SSS}$ is prefix-free (\cref{lm:dist-prefixes}\eqref{lm:dist-prefixes-it-2}),
  it follows that no other string from $\DistPrefixes{\tau}{\Text}{\SSS}$ can be a prefix of $\Pat$.
\end{proof}

\begin{definition}\label{def:dist-prefix-pat}
  Let $\tau \geq 1$ be such that $3\tau - 1 \leq \Textlen$ and $\SSS$ be
  a $\tau$-synchronizing set of $\Text$. For every $\tau$-nonperiodic
  pattern $\Pat \in \Sigma^{m}$ satisfying $m \geq 3\tau - 1$
  and $\OccTwo{\Pat}{\Text} \neq \emptyset$, by $\DistPrefixPat{\Pat}{\tau}{\Text}{\SSS}$
  we denote the unique $D \in \DistPrefixes{\tau}{\Text}{\SSS}$ (\cref{def:dist-prefixes})
  that is a prefix of $\Pat$ (such $D$ exists by \cref{lm:dist-prefix-existence}).
\end{definition}

%% file: index/nonperiodic/nav.tex
\subsubsection{Basic Navigation Primitives}\label{sec:minocc-index-nonperiodic-nav}

\begin{proposition}\label{pr:nav-index-nonperiodic}
  Let $\tau = \mu\log_{\AlphabetSize} \Textlen$ and $\SSS$ be a $\tau$-synchronizing
  set of $\Text$ satisfying $|\SSS| = \bigO(\tfrac{\Textlen}{\tau})$.
  Denote $(s_i)_{i \in [1 \dd n']} = \LexSorted{\SSS}{\Text}$ (\cref{def:lex-sorted}).
  Given the set $\SSS$, and the packed representation of the text
  $\Text$, we can in $\bigO(\Textlen / \log_{\AlphabetSize} \Textlen)$ time
  construct a data structure, denoted $\NavNonperiodic{\Text}{\SSS}$,
  that supports the following queries:
  \begin{enumerate}
  \item\label{pr:nav-index-nonperiodic-it-1}
    Given the packed representation of any string $X \in \IntegerAlphabet^{\leq 3\tau - 1}$,
    in $\bigO(1)$ time return the packed representation of string $\revstr{X}$ (see \cref{sec:prelim}).
  \item\label{pr:nav-index-nonperiodic-it-2}
    Let $j \in [1 \dd \Textlen - 3\tau + 2] \setminus \RTwo{\tau}{\Text}$ and
    $\ell \geq 3\tau - 1$ be such that $j + \ell \leq \Textlen + 1$. Denote
    $D = \DistPrefixPos{j}{\tau}{\Text}{\SSS}$ (\cref{def:dist-prefixes})
    and $\deltatext = |D| - 2\tau$.
    \begin{enumerate}
    \item\label{pr:nav-index-nonperiodic-it-2a}
      Given $j$, in $\bigO(1)$ time compute
      the packed representation of $D$.
    \item\label{pr:nav-index-nonperiodic-it-2b}
      Given $j$ and $\ell$,
      in $\bigO(\log \log \Textlen)$ time compute $(b,e)$ such
      that $b = |\{i \in [1 \dd n'] : \Text[s_i \dd \Textlen] \prec \Text[j + \deltatext \dd j + \ell)\}|$
      and $e - b = |\{i \in [1 \dd n'] :
      \Text[j + \deltatext \dd j + \ell)\text{ is a prefix of }\Text[s_i \dd \Textlen]\}|$.
    \end{enumerate}
  \item\label{pr:nav-index-nonperiodic-it-3}
    Let $\Pat \in \Sigma^{m}$ be a $\tau$-nonperiodic pattern satisfying
    $m \geq 3\tau - 1$ and $\OccTwo{\Pat}{\Text} \neq \emptyset$. Denote
    $D = \DistPrefixPat{\Pat}{\tau}{\Text}{\SSS}$ (\cref{def:dist-prefix-pat}), $\deltatext = |D| - 2\tau$, and
    $\Pat' = \Pat(\deltatext \dd m]$.
    \begin{enumerate}
    \item\label{pr:nav-index-nonperiodic-it-3a}
      Given the packed representation of $\Pat$, in $\bigO(1)$ time
      compute the packed representation~of~$D$.
    \item\label{pr:nav-index-nonperiodic-it-3b}
      Given the packed representation of $\Pat$, in $\bigO(m / \log_{\AlphabetSize} \Textlen + \log \log \Textlen)$ time
      compute $(b,e)$ such that $b = |\{i \in [1 \dd n'] : \Text[s_i \dd \Textlen] \prec \Pat'\}|$
      and $e-b = |\{i \in [1 \dd n'] : \Pat'\text{ is a prefix of }\Text[s_i \dd \Textlen]\}|$.
    \end{enumerate}
  \end{enumerate}
\end{proposition}
\begin{proof}

  $\NavNonperiodic{\Text}{\SSS}$ consists of two components:
  \begin{enumerate}
  \item The component of the pattern matching index to handle nonperiodic patterns from~\cite[Section~6.3.1]{breaking} with
    $\SSS$ as the underlying $\tau$-synchronizing set, except we do not store the structure for prefix rank and select
    queries. Note that this index requires that $|\SSS| = \bigO(\tfrac{\Textlen}{\tau})$,
    which holds here. The index needs $\bigO(\Textlen / \log_{\AlphabetSize} \Textlen)$ space.
  \item The component of the compressed suffix tree (CST) to handle nonperiodic nodes from~\cite[Section~7.2.1]{breaking} with
    $\SSS$ as the underlying $\tau$-synchronizing set, except, similarly as above, we do not store the structure for
    prefix rank and select queries. Similarly as above, $\SSS$ satisfies the required space bound. This
    component needs $\bigO(\Textlen / \log_{\AlphabetSize} \Textlen)$ space.
  \end{enumerate}
  In total, $\NavNonperiodic{\Text}{\SSS}$ needs $\bigO(\Textlen / \log_{\AlphabetSize} \Textlen)$ space.

  \paragraph{Implementation of queries}

  All queries except for~\ref{pr:nav-index-nonperiodic-it-2b} and~\ref{pr:nav-index-nonperiodic-it-3b} are
  standard navigation queries of the above indexes.
  Query~\ref{pr:nav-index-nonperiodic-it-2b} is performed as follows:
  \begin{enumerate}
  \item First, using Query~\ref{pr:nav-index-nonperiodic-it-2a}, we compute the packed representation of
    $D = \DistPrefixPos{j}{\tau}{\Text}{\SSS}$. We then let $\deltatext = |D| - 2\tau$. Note that by
    $j \in [1 \dd \Textlen - 3\tau + 2] \setminus \RTwo{\tau}{\Text}$, it follows that $j + \delta \in \SSS$.
  \item Denote $(s^{\rm text}_i)_{i \in [1 \dd n']}$ be a sequence containing $\SSS$ in sorted order.
    Using the rank query on a bitvector marking the positions in $\SSS$, in $\bigO(1)$ time we compute $i \in [1 \dd n']$
    such that $s^{\rm text}_i = j + \delta$.
  \item Using the mapping between sequences $(s^{\rm text}_i)_{i \in [1 \dd n']}$ and $(s_i)_{i \in [1 \dd n']}$, in
    $\bigO(1)$ time we compute $i' \in [1 \dd n']$ such that $s_{i'} = s^{\rm text}_{i}$.
  \item In $\bigO(1)$ time we locate the $i'$th leftmost leaf $v$ in the compact trie $\mathcal{T}_{\SSS}$ containing the substrings
    $\{\Text[s_i \dd \Textlen]\}_{i \in [1 \dd n']}$ using~\cite[Proposition~4.1]{breaking}.
  \item Using the weighted ancestor query on $\mathcal{T}_{\SSS}$,
    in $\bigO(\log \log \Textlen)$ time we then compute the most shallow ancestor
    of $v$ whose string depth is at least $\ell$. The range of leaves stored in this
    node contains the answer.
  \end{enumerate}
  In total, the query takes $\bigO(\log \log \Textlen)$ time.

  Query~\ref{pr:nav-index-nonperiodic-it-3b} is performed by executing the
  first step of~\cite[Proposition~6.4]{breaking}, and then returning the answer (without doing a prefix rank query).

  \paragraph{Construction algorithm}

  Excluding the component for prefix rank and selection queries, the construction of both indexes takes
  $\bigO(\Textlen / \log_{\AlphabetSize} \Textlen)$ time (see~\cite[Proposition~6.5 and Proposition~7.14]{breaking}).
\end{proof}

%% file: index/nonperiodic/structure.tex
\subsubsection{The Data Structure}\label{sec:minocc-index-nonperiodic-structure}

\paragraph{Definitions}

Let $\tau = \mu\log_{\AlphabetSize} \Textlen$, where $\mu$ is as defined at
the beginning of \cref{sec:minocc-index}.
Let $\SSS$ be a $\tau$-synchronizing set of $\Text$ with $|\SSS| = \bigO(\tfrac{\Textlen}{\tau})$
computed using \cref{th:sss-packed-construction}. Denote $\SSSSize = |\SSS|$.
Note that it holds $\tau \geq 1$ and $3\tau - 1 \leq \Textlen$ (see the beginning of \cref{sec:minocc-index}).
Thus, by \cref{lm:sss-max}, we have $\SSS \neq \emptyset$, i.e., $\SSSSize \geq 1$.
Let $A_{\SSS}[1 \dd n']$ denote the array containing the sequence
$\LexSorted{\SSS}{\Text}$ (\cref{def:lex-sorted}). Let
$A_{\rm str}[1 \dd \SSSSize]$ be an array defined by $A_{\rm str}[i] =
\revstr{D_i}$, where $D_i = \Textinf[A_{\SSS}[i] - \tau \dd
A_{\SSS}[i] + 2\tau)$.

\paragraph{Components}

The data structure, denoted $\MinOccIndexNonperiodic{\Text}$, to handle
$\tau$-nonperiodic patterns and positions in
$[1 \dd \Textlen - 3\tau + 2] \setminus \RTwo{\tau}{\Text}$,
consists of two components:
\begin{enumerate}
\item The data structure $\NavNonperiodic{\Text}{\SSS}$ from \cref{pr:nav-index-nonperiodic}.
  It needs $\bigO(\Textlen / \log_{\AlphabetSize} \Textlen)$ space.
\item The plain representations of arrays $A_{\SSS}[1 \dd n']$ and $A_{\rm str}[1 \dd n']$ augmented
  with the data structure from \cref{cr:prefix-rmq}. By $n' =
  \bigO(\tfrac{\Textlen}{\tau}) = \bigO(\Textlen / \log_{\AlphabetSize} \Textlen)$
  and $\AlphabetSize^{3\tau} = \bigO(\sqrt{\Textlen}) = \bigO(\Textlen / \log \Textlen)$,
  both the arrays, and the augmentation of \cref{cr:prefix-rmq} need
  $\bigO(\Textlen / \log_{\AlphabetSize} \Textlen)$ space.
\end{enumerate}

In total, $\MinOccIndexNonperiodic{\Text}$ needs
$\bigO(\Textlen / \log_{\AlphabetSize} \Textlen)$ space.

%% file: index/nonperiodic/queries.tex
\subsubsection{Implementation of Queries}\label{sec:minocc-index-nonperiodic-queries}

\paragraph{Combinatorial Properties}

\begin{lemma}\label{lm:nonperiodic-pat-occ}
  Let $\tau \geq 1$ be such that $3\tau - 1 \leq \Textlen$, and let
  $\SSS$ be a $\tau$-synchronizing set of $\Text$. Denote
  $(s_i)_{i \in [1 \dd n']} = \LexSorted{\SSS}{\Text}$ (\cref{def:lex-sorted}).
  Let $D \in \DistPrefixes{\tau}{\Text}{\SSS}$ (\cref{def:dist-prefixes})
  and let $\Pat \in \Sigma^{m}$ be a $\tau$-nonperiodic pattern having $D$
  as a prefix. Denote $\deltatext = |D| - 2\tau$,
  $\Pat' = \Pat(\deltatext \dd m]$, and let
  $(b, e)$ be such that $b = |\{i \in
  [1 \dd \SSSSize] : \Text[s_i \dd \Textlen] \prec \Pat'\}|$
  and $e - b = |\{i \in [1 \dd \SSSSize] :
  \Pat'\text{ is a prefix of }\Text[s_i \dd \Textlen]\}|$. Then,
  \begin{align*}
    \OccTwo{\Pat}{\Text} = \{s_i - \deltatext :
      i \in (b \dd e]\text{ and }
      s_i - \deltatext \in \OccTwo{D}{\Text}\}.
  \end{align*}
\end{lemma}
\begin{proof}

  Denote $A = \{s_i - \deltatext : i \in (b \dd e]\text{ and }s_i - \deltatext \in \OccTwo{D}{\Text}\}$.
  Observe that for every $X, Y \in \Sigma^{*}$, letting $X' = Xc^{\infty}$ (where $c = \max \Sigma$), it holds that
  $X$ is a prefix of $Y$ if and only if $X \preceq Y \prec X'$. Consequently, letting $\Pat'' = \Pat'c^{\infty}$, it holds
  $e - b = |\{i \in [1 \dd n'] : \Pat' \preceq \Text[s_i \dd \Textlen] \prec \Pat''\}|$. Combining with
  $b = |\{i \in [1 \dd n'] : \Text[s_i \dd \Textlen] \prec \Pat'\}|$, it follows that $n' - e = |\{i \in [1 \dd n'] :
  \Pat'' \preceq \Text[s_i \dd \Textlen]\}|$. Thus, by $(s_i)_{i \in [1 \dd n']} =
  \LexSorted{\SSS}{\Text}$ (\cref{def:lex-sorted}), we obtain that
  for every $i \in [1 \dd n']$, $s_i \in \OccTwo{\Pat'}{\Text}$ holds if and only if $i \in (b \dd e]$.

  First, we prove that $\OccTwo{\Pat}{\Text} \subseteq A$.
  Let $j \in \OccTwo{\Pat}{\Text}$.
  Denote $j' = j + \deltatext$.
  \begin{itemize}
  \item First, observe that since $D$ is a prefix of $\Pat$, we have $j \in \OccTwo{D}{\Text}$.
  \item Second, note that since $\Pat'$ is a suffix of $\Pat$ and $\deltatext + |\Pat'| = m$,
    it follows by $j \in \OccTwo{\Pat}{\Text}$, that $j' = j + \deltatext \in \OccTwo{\Pat'}{\Text}$.
  \item Finally, we prove that $j' \in \SSS$.
    By \cref{def:dist-prefixes}, there exists
    $i \in [1 \dd \Textlen-3\tau+2] \setminus \RTwo{\tau}{\Text}$,
    such that $i \in \OccTwo{D}{\Text}$ and $\Successor{\SSS}{i} = i + |D| - 2\tau = i + \deltatext$.
    Thus, $i + \deltatext \in \SSS$.
    By $j \in \OccTwo{\Pat}{\Text}$ and $D$ being a prefix of $\Pat$, we have $j \in \OccTwo{D}{\Text}$.
    Consequently, $\Text[j + \deltatext \dd j + |D|) = \Text[i + \deltatext \dd i + |D|)$, and
    hence by the consistency of $\SSS$ (\cref{def:sss}\eqref{def:sss-consistency}), $j' = j + \deltatext \in \SSS$.
  \end{itemize}
  By $j' \in \SSS$, there exists $i \in [1 \dd n']$ such that $s_i = j'$.
  Above we proved that $j' \in \OccTwo{\Pat'}{\Text}$. Thus, by $s_i = j'$, and the earlier characterization of the
  set $\{s_i\}_{i \in (b \dd e]}$, we obtain $i \in (b \dd e]$.
  Lastly, note that by above we also have $j = j' - \deltatext = s_i - \deltatext \in \OccTwo{D}{\Text}$. Putting everything
  together, we have thus proved that there exists $i \in (b \dd e]$ such that $j = s_i - \deltatext$ and
  $s_i - \deltatext \in \OccTwo{D}{\Text}$, i.e., $j \in A$.

  We now prove that $A \subseteq \OccTwo{\Pat}{\Text}$. Let $j \in A$.
  To prove $j \in \OccTwo{\Pat}{\Text}$, by \cref{def:occ},
  we need to show that $j \in [1 \dd \Textlen]$, $j + m \leq \Textlen + 1$,
  and $\Text[j \dd j + m) = \Pat$. We proceed as follows:
  \begin{itemize}
  \item First, note that by $j \in A$, there exists $i \in (b \dd e]$ such
    that $j = s_i - \deltatext$ and $s_i - \deltatext \in \OccTwo{D}{\Text}$.
    In particular, by \cref{def:occ}, this implies $j = s_i - \deltatext \in [1 \dd \Textlen]$.
  \item Next, observe that by $i \in (b \dd e]$ and the earlier
    characterization of the set $\{s_{b+1}, \ldots, s_{e}\}$,
    we have $j + \deltatext = s_i \in \OccTwo{\Pat'}{\Text}$. By \cref{def:occ},
    this implies $j + m = j + \deltatext + |\Pat'| = s_i + |\Pat'| \leq \Textlen + 1$.
  \item Finally, note that above we proved that $D$ (resp.\ $\Pat'$) is a prefix (resp.\ suffix)
    of $\Text[j \dd j + m)$. Since the same is true for $\Pat$, and it holds $|D| + |\Pat'| = (\deltatext + 2\tau) + (m - \deltatext)
    = m + 2\tau > m$, we thus have $\Text[j \dd j + m) = \Pat$.
  \end{itemize}
  This includes the proof of $j \in \OccTwo{\Pat}{\Text}$.
\end{proof}

\begin{lemma}\label{lm:nonperiodic-pat-occ-table}
  Let $\tau \geq 1$ be such that $3\tau - 1 \leq \Textlen$, and let
  $\SSS$ be a $\tau$-synchronizing set of $\Text$. Denote
  $(s_i)_{i \in [1 \dd n']} = \LexSorted{\SSS}{\Text}$ (\cref{def:lex-sorted}).
  Let $A_{\SSS}[1 \dd n']$ and $A_{\rm str}[1 \dd n']$ be defined by
  \begin{itemize}
  \item $A_{\SSS}[i] = s_i$,
  \item $A_{\rm str}[i] = \revstr{D_i}$,
    where $D_i = \Textinf[s_i - \tau \dd s_i + 2\tau)$.
  \end{itemize}
  Let $D \in \DistPrefixes{\tau}{\Text}{\SSS}$ (\cref{def:dist-prefixes})
  and let $\Pat \in \Sigma^{m}$ be a $\tau$-nonperiodic pattern having $D$
  as a prefix. Denote $\deltatext = |D| - 2\tau$,
  $\Pat' = \Pat(\deltatext \dd m]$, and let
  $(b, e)$ be such that $b = |\{i \in
  [1 \dd \SSSSize] : \Text[s_i \dd \Textlen] \prec \Pat'\}|$
  and $e - b = |\{i \in [1 \dd \SSSSize] :
  \Pat'\text{ is a prefix of }\Text[s_i \dd \Textlen]\}|$. Then,
  \begin{align*}
    \OccTwo{\Pat}{\Text} = \{A_{\SSS}[i] - \deltatext :
      i \in (b \dd e]\text{ and }
      \revstr{D}\text{ is a prefix of }A_{\rm str}[i]\}.
  \end{align*}
\end{lemma}
\begin{proof}

  We begin by proving two properties of the string $D$.
  \begin{itemize}
  \item First, we show that $|D| \leq 3\tau - 1$.
    By $D \in \DistPrefixes{\tau}{\Text}{\SSS}$ and \cref{def:dist-prefixes}, it follows
    that there exists $j \in [1 \dd \Textlen - 3\tau + 2] \setminus \RTwo{\tau}{\Text}$ such that, letting
    $s = \Successor{\SSS}{j}$, it holds $D = \Text[j \dd s + 2\tau)$. By
    $j \in [1 \dd \Textlen - 3\tau + 2] \setminus \RTwo{\tau}{\Text}$ and
    the density condition (\cref{def:sss}\eqref{def:sss-density}), it follows that $[j \dd j + \tau) \cap \SSS \neq \emptyset$.
    Consequently, $\Successor{\SSS}{j} - j = s - j < \tau$. Thus, it holds $|D| = (s - j) + 2\tau \leq 3\tau - 1$.
  \item Second, we prove that no nonempty suffix of $\Text$ is a proper prefix of $D$. Suppose that this is not the case.
    Observe that then $\Text[\Textlen]$ occurs in $D[1 \dd |D|)$. On the other hand, by \cref{def:dist-prefixes},
    $D$ is a substring of $\Text$. Thus, we obtain that $\Text[\Textlen]$ occurs in $\Text[1 \dd \Textlen)$. This contradicts
    the assumption about $\Text[\Textlen]$ (see the beginning of \cref{sec:minocc-index}).
  \end{itemize}

  Next, we prove that for every $i \in [1 \dd \SSSSize]$, $s_i - \deltatext \in \OccTwo{D}{\Text}$
  holds if and only if $\revstr{D}$ is a prefix of $A_{\rm str}[i]$.
  \begin{itemize}
  \item Let $i \in [1 \dd \SSSSize]$ be such that $s_i - \deltatext \in \OccTwo{D}{\Text}$.
    By \cref{def:occ}, this implies that $s_i - \deltatext \in [1 \dd \Textlen]$, $s_i - \deltatext + |D| \leq \Textlen + 1$,
    and $\Text[s_i - \deltatext \dd s_i - \deltatext + |D|) = D$. Equivalently, by $|D| = 2\tau + \deltatext$, we obtain
    that $s_i - \deltatext \geq 1$, $s_i - \deltatext + |D| = s_i + 2\tau \leq \Textlen + 1$, and
    $\Text[s_i - \deltatext \dd s_i - \deltatext + |D|) = \Text[s_i - \deltatext \dd s_i + 2\tau) = D$.
    By $|D| \leq 3\tau$, we thus obtain that $D$ is a suffix of $\Text[s_i - \tau \dd s_i + 2\tau) =
    \Textinf[s_i - \tau \dd s_i + 2\tau)$.
    By definition of $A_{\rm str}[i]$, this implies that $\revstr{D}$ is a prefix of $A_{\rm str}[i]$.
  \item Let us now consider $i \in [1 \dd \SSSSize]$ such that $\revstr{D}$ is a prefix of $A_{\rm str}[i]$.
    By definition of $A_{\rm str}[i]$, this implies that $D$ is a suffix of $\Textinf[s_i - \tau \dd s_i + 2\tau)$.
    Equivalently, by $|D| = 2\tau + \deltatext$, we obtain that $D = \Textinf[s_i - \deltatext \dd s_i + 2\tau)$.
    To prove $s_i - \deltatext \in \OccTwo{D}{\Text}$, it thus remains (see \cref{def:occ}) to show that
    $s_i - \deltatext \geq 1$ and $s_i + 2\tau \leq \Textlen + 1$. Suppose that $s_i - \deltatext < 1$.
    This implies
    that the substring $Y := \Textinf[s_i - \deltatext \dd \Textlen]$ is a prefix of $D$ (and hence also a prefix
    of $\Pat$). However, since $s_i \geq 1$, it
    follows that $|Y| \leq \deltatext$, which implies
    that $|Y| \leq \deltatext < \tau < 3\tau - 1 \leq \Textlen$ and $|Y| \leq \deltatext < \tau < |D|$.
    Thus, $Y$ is a nonempty suffix of text that
    is a proper prefix of $D$. This contradicts the property of string $D$ proved above, and hence
    we obtain $s_i - \deltatext \geq 1$. Suppose now that $s_i + 2\tau > \Textlen + 1$.
    Note that by $s_i - \deltatext \in [1 \dd \Textlen]$, this implies that
    $Y' := \Text[s_i - \deltatext \dd \Textlen]$ is a nonempty suffix of $\Text$ that is a prefix of $D$.
    Moreover, by $s_i + 2\tau > \Textlen + 1$, it follows that $|Y'| < |D|$, i.e., $Y'$ is a proper prefix of $D$.
    This again contradicts the property of $D$ proved above. Hence, $s_i + 2\tau \leq \Textlen + 1$.
    Putting everything together, we thus obtain $s_i - \deltatext \in \OccTwo{D}{\Text}$.
  \end{itemize}

  By putting together the above equivalence and \cref{lm:nonperiodic-pat-occ}, we obtain that
  \begin{align*}
    \OccTwo{\Pat}{\Text}
      &= \{s_i - \deltatext : i \in (b \dd e]\text{ and }s_i - \deltatext \in \OccTwo{D}{\Text}\}\\
      &= \{A_{\SSS}[i] - \deltatext : i \in (b \dd e]\text{ and }\revstr{D}\text{ is a prefix of }A_{\rm str}[i]\}.
      \qedhere
  \end{align*}
\end{proof}

\begin{lemma}\label{lm:nonperiodic-pat-occ-min}
  Let $\tau \geq 1$ be such that $3\tau - 1 \leq \Textlen$, and let
  $\SSS$ be a $\tau$-synchronizing set of $\Text$. Denote
  $(s_i)_{i \in [1 \dd n']} = \LexSorted{\SSS}{\Text}$ (\cref{def:lex-sorted}), and let
  $A_{\SSS}[1 \dd n']$ and $A_{\rm str}[1 \dd n']$ be defined by
  \begin{itemize}
  \item $A_{\SSS}[i] = s_i$,
  \item $A_{\rm str}[i] = \revstr{D_i}$,
    where $D_i = \Textinf[s_i - \tau \dd s_i + 2\tau)$.
  \end{itemize}
  Let $D \in \DistPrefixes{\tau}{\Text}{\SSS}$ (\cref{def:dist-prefixes})
  and let $\Pat \in \Sigma^{m}$ be a $\tau$-nonperiodic pattern having $D$
  as a prefix. Let us also assume $\OccTwo{\Pat}{\Text} \neq \emptyset$.
  Denote $\deltatext = |D| - 2\tau$,
  $\Pat' = \Pat(\deltatext \dd m]$, and let
  $(b, e)$ be such that $b = |\{i \in
  [1 \dd \SSSSize] : \Text[s_i \dd \Textlen] \prec \Pat'\}|$
  and $e - b = |\{i \in [1 \dd \SSSSize] :
  \Pat'\text{ is a prefix of }\Text[s_i \dd \Textlen]\}|$.
  Then, there exists $i \in (b \dd e]$
  such that $\revstr{D}$ is a prefix of $A_{\rm str}[i]$, and it holds:
  \begin{align*}
    \min \OccTwo{\Pat}{\Text} =
      A_{\SSS}[\PrefixRMQ{A_{\SSS}}{A_{\rm str}}{b}{e}{\revstr{D}}] - \deltatext.
  \end{align*}
\end{lemma}
\begin{proof}

  To show the first claim, observe now that by the assumption $\OccTwo{\Pat}{\Text} \neq \emptyset$
  and \cref{lm:nonperiodic-pat-occ-table}, it follows that there exists
  $i \in (b \dd e]$ such that $\revstr{D}$ is a prefix of $A_{\rm str}[i]$.
  In particular, this proves that
  $\PrefixRMQ{A_{\SSS}}{A_{\rm str}}{b}{e}{\revstr{D}}$ is well-defined (see \cref{def:prefix-rmq}).

  To prove the second claim, we combine \cref{lm:nonperiodic-pat-occ-table} and \cref{def:prefix-rmq} to obtain:
  \begin{align*}
    \min \OccTwo{\Pat}{\Text}
      &= \min \{A_{\SSS}[i] - \deltatext : i \in (b \dd e]\text{ and }\revstr{D}\text{ is a prefix of }A_{\rm str}[i]\}\\
      &= \min \{A_{\SSS}[i] : i \in (b \dd e]\text{ and }\revstr{D}\text{ is a prefix of }A_{\rm str}[i]\} - \deltatext\\
      &= A_{\SSS}[\PrefixRMQ{A_{\SSS}}{A_{\rm str}}{b}{e}{\revstr{D}}] - \deltatext.
      \qedhere
  \end{align*}
\end{proof}

\paragraph{Query Algorithms}

\begin{proposition}\label{pr:minocc-index-nonperiodic-query-pos}
  Let $\tau = \mu\log_{\AlphabetSize} \Textlen$, $j \in [1 \dd \Textlen - 3\tau + 2] \setminus \RTwo{\tau}{\Text}$,
  and $\ell \geq 3\tau - 1$ be such that $j + \ell \leq \Textlen + 1$. Given
  $\MinOccIndexNonperiodic{\Text}$ (\cref{sec:minocc-index-nonperiodic-structure}) and
  the pair $(j,\ell)$, we can in $\bigO(\log^{\epsilon} \Textlen)$ time compute $\min \OccTwo{\Pat}{\Text}$, where
  $\Pat = \Text[j \dd j + \ell)$.
\end{proposition}
\begin{proof}
  Observe that $j \in [1 \dd \Textlen - 3\tau + 2] \setminus \RTwo{\tau}{\Text}$ implies
  that $\Pat$ is $\tau$-nonperiodic (see \cref{def:periodic-pattern}). Clearly, we also
  have $\OccTwo{\Pat}{\Text} \neq \emptyset$. Note that we also have
  $3\tau - 1 \leq \ell \leq \Textlen + 1 - j \leq \Textlen$.
  Let $\SSS$ be the $\tau$-synchronizing set of $\Text$ as defined
  in \cref{sec:minocc-index-nonperiodic-structure}, and let
  $(s_i)_{i \in [1 \dd n']} = \LexSorted{\SSS}{\Text}$ (\cref{def:lex-sorted}).
  The algorithm proceeds in four steps:
  \begin{enumerate}
  \item Using \cref{pr:nav-index-nonperiodic}\eqref{pr:nav-index-nonperiodic-it-2a},
    in $\bigO(1)$ time we compute
    the packed representation of the string
    $D := \DistPrefixPos{j}{\tau}{\Text}{\SSS}$ (\cref{def:dist-prefixes}).
    In $\bigO(1)$ time we then calculate $\deltatext = |D| - 2\tau$.
    Note that by \cref{lm:dist-prefixes}\eqref{lm:dist-prefixes-it-1}, it holds $|D| \leq 3\tau - 1$.
  \item Using \cref{pr:nav-index-nonperiodic}\eqref{pr:nav-index-nonperiodic-it-1},
    in $\bigO(1)$ time we compute the packed representation of $\revstr{D}$.
  \item Denote $\Pat' = \Pat(\deltatext \dd \ell]$.
    Using \cref{pr:nav-index-nonperiodic}\eqref{pr:nav-index-nonperiodic-it-2b},
    in $\bigO(\log \log \Textlen)$ time, we compute the pair $(b, e)$ defined by
    $b = |\{i \in [1 \dd \SSSSize] : \Text[s_i \dd \Textlen] \prec
    \Pat'\}|$ and $e - b = |\{i \in [1 \dd \SSSSize] :
    \Pat'\text{ is a prefix of }\Text[s_i \dd \Textlen]\}|$.
  \item Using \cref{cr:prefix-rmq}, in $\bigO(\log^{\epsilon} \Textlen)$ time we compute
    and return as the answer the position
    $j_{\min} := A_{\SSS}[\PrefixRMQ{A_{\SSS}}{A_{\rm str}}{b}{e}{\revstr{D}}] - \deltatext$,
    where $A_{\SSS}[1 \dd n']$ and $A_{\rm str}[1 \dd n']$ are as in \cref{sec:minocc-index-nonperiodic-structure}.
    By \cref{lm:nonperiodic-pat-occ-min}, it holds $j_{\min} = \min \OccTwo{\Pat}{\Text}$.
  \end{enumerate}
  In total, we spend $\bigO(\log^{\epsilon} \Textlen)$ time.
\end{proof}

\begin{proposition}\label{pr:minocc-index-nonperiodic-query-pat}
  Let $\tau = \mu\log_{\AlphabetSize} \Textlen$. Let $\Pat \in \Sigma^{m}$ be a $\tau$-nonperiodic
  pattern satisfying $m \geq 3\tau - 1$ and $\OccTwo{\Pat}{\Text} \neq \emptyset$.
  Given $\MinOccIndexNonperiodic{\Text}$ (\cref{sec:minocc-index-nonperiodic-structure}) and
  the packed representation of $\Pat$, we can in $\bigO(m / \log_{\AlphabetSize} \Textlen + \log^{\epsilon} \Textlen)$
  time compute $\min \OccTwo{\Pat}{\Text}$.
\end{proposition}
\begin{proof}
  Observe that $m \geq 3\tau - 1$ and $\OccTwo{\Pat}{\Text} \neq \emptyset$.
  implies that $m \leq \Textlen$. Thus, $3\tau - 1 \leq \Textlen$.
  Let $\SSS$ be the $\tau$-synchronizing set of $\Text$ as defined
  in \cref{sec:minocc-index-nonperiodic-structure}, and let
  $(s_i)_{i \in [1 \dd n']} = \LexSorted{\SSS}{\Text}$ (\cref{def:lex-sorted}).
  The algorithm proceeds in four steps:
  \begin{enumerate}
  \item Using \cref{pr:nav-index-nonperiodic}\eqref{pr:nav-index-nonperiodic-it-3a},
    in $\bigO(1)$ time we compute
    the packed representation of the string
    $D := \DistPrefixPat{\Pat}{\tau}{\Text}{\SSS}$ (\cref{def:dist-prefix-pat}).
    In $\bigO(1)$ time we then calculate $\deltatext = |D| - 2\tau$.
    Note that by \cref{lm:dist-prefixes}\eqref{lm:dist-prefixes-it-1}, it holds $|D| \leq 3\tau - 1$.
  \item Using \cref{pr:nav-index-nonperiodic}\eqref{pr:nav-index-nonperiodic-it-1},
    in $\bigO(1)$ time we compute the packed representation of $\revstr{D}$.
  \item Denote $\Pat' = \Pat(\deltatext \dd m]$.
    Using \cref{pr:nav-index-nonperiodic}\eqref{pr:nav-index-nonperiodic-it-3b},
    in $\bigO(\log \log \Textlen)$ time, we compute the pair $(b, e)$ defined by
    $b = |\{i \in [1 \dd \SSSSize] : \Text[s_i \dd \Textlen] \prec
    \Pat'\}|$ and $e - b = |\{i \in [1 \dd \SSSSize] :
    \Pat'\text{ is a prefix of }\Text[s_i \dd \Textlen]\}|$.
  \item Using \cref{cr:prefix-rmq}, in $\bigO(\log^{\epsilon} \Textlen)$ time we compute
    and return as the answer the position
    $j_{\min} := A_{\SSS}[\PrefixRMQ{A_{\SSS}}{A_{\rm str}}{b}{e}{\revstr{D}}] - \deltatext$,
    where $A_{\SSS}[1 \dd n']$ and $A_{\rm str}[1 \dd n']$ are as in \cref{sec:minocc-index-nonperiodic-structure}.
    By \cref{lm:nonperiodic-pat-occ-min}, it holds $j_{\min} = \min \OccTwo{\Pat}{\Text}$.
  \end{enumerate}
  In total, we spend $\bigO(\log^{\epsilon} \Textlen)$ time.
\end{proof}

%% file: index/nonperiodic/construction.tex
\subsubsection{Construction Algorithm}\label{sec:minocc-index-nonperiodic-construction}

\begin{theorem}[{\cite[Theorem~4.3]{sss}}]\label{th:sss-lex-sort}
  Given the packed representation of text $\Text \in \IntegerAlphabet^{\Textlen}$
  and its $\tau$-synchronizing set $\SSS$ of
  size $|\SSS| = \bigO(\tfrac{\Textlen}{\tau})$ for $\tau = \bigO(\log_{\AlphabetSize} \Textlen)$,
  we can compute the sequence $\LexSorted{\SSS}{\Text}$
  (\cref{def:lex-sorted}) in $\bigO(\tfrac{\Textlen}{\tau})$ time.
\end{theorem}

\begin{proposition}\label{pr:minocc-index-nonperiodic-construction}
  Given the packed representation of $\Text$, we can construct
  $\MinOccIndexNonperiodic{\Text}$ (\cref{sec:minocc-index-nonperiodic-structure}) in
  $\bigO(\Textlen \min(1, \log \AlphabetSize / \sqrt{\log \Textlen}))$ time and
  using $\bigO(\Textlen / \log_{\AlphabetSize} \Textlen)$ working space.
\end{proposition}
\begin{proof}
  Let $\tau = \mu\log_{\AlphabetSize} \Textlen$ be as defined at the beginning of
  \cref{sec:minocc-index-nonperiodic-structure}.
  We construct the components of $\MinOccIndexNonperiodic{\Text}$ (\cref{sec:minocc-index-nonperiodic-structure}) as follows:
  \begin{enumerate}
  \item Using \cref{th:sss-packed-construction}, we construct the $\tau$-synchronizing set
    $\SSS$ satisfying $|\SSS| = \bigO(\tfrac{\Textlen}{\tau}) = \bigO(\Textlen / \log_{\AlphabetSize} \Textlen)$.
    This takes $\bigO(\Textlen / \log_{\AlphabetSize} \Textlen)$ time. Then,
    using $\SSS$ and the packed representation of $\Text$ as input,
    we construct $\NavNonperiodic{\Text}{\SSS}$ in
    $\bigO(\Textlen / \log_{\AlphabetSize} \Textlen)$ time
    using \cref{pr:nav-index-nonperiodic}.
  \item Denote $n' = |\SSS| \geq 1$ and $(s_i)_{1 \in [1 \dd n']} = \LexSorted{\SSS}{\Text}$ (\cref{def:lex-sorted}).
    Using \cref{th:sss-lex-sort}, in $\bigO(\tfrac{\Textlen}{\tau}) =
    \bigO(\Textlen / \log_{\AlphabetSize} \Textlen)$ time, we compute
    the array $A_{\SSS}[1 \dd n']$ defined by $A_{\SSS}[i] = s_i$.
    Using \cref{pr:nav-index-nonperiodic}\eqref{pr:nav-index-nonperiodic-it-1}, in $\bigO(n')
    = \bigO(\Textlen / \log_{\AlphabetSize} \Textlen)$ time, we then initialize
    the array $A_{\rm str}[1 \dd n']$ defined by $A_{\rm str}[i] = \revstr{D_i}$, where
    $D_i = \Textinf[A_{\SSS}[i] - \tau \dd A_{\SSS}[i] + 2\tau)$. Using \cref{cr:prefix-rmq}, in
    $\bigO(\Textlen \min(1, \log \AlphabetSize / \sqrt{\log \Textlen}))$ time and using
    $\bigO(\Textlen / \log_{\AlphabetSize} \Textlen)$ working space
    we then construct the structure for prefix RMQ queries on arrays $A_{\SSS}[1 \dd n']$ and
    $A_{\rm str}[1 \dd n']$.
  \end{enumerate}
  In total, the construction takes
  $\bigO(\Textlen \min(1, \log \AlphabetSize / \sqrt{\log \Textlen}))$ time
  and uses $\bigO(\Textlen / \log_{\AlphabetSize} \Textlen)$ working space.
\end{proof}

%% file: index/periodic.tex
\subsection{The Periodic Patterns and Positions}\label{sec:minocc-index-periodic}

\input{index/periodic/prelim.tex}

\input{index/periodic/nav.tex}

\input{index/periodic/structure.tex}

\input{index/periodic/queries.tex}

\input{index/periodic/construction.tex}

%% file: index/periodic/prelim.tex
\subsubsection{Preliminaries}\label{sec:minocc-index-periodic-prelim}

In this section, we review the basic properties of $\tau$-periodic patterns
and positions in $\RTwo{\tau}{\Text}$.
Following the basic combinatorial toolbox, we present the new
definitions and notation specific to the index for leftmost
occurrences.
We prove only the new results; the proofs of the remaining
claims can be found, e.g., in~\cite{collapsing}.

\paragraph{Notation and Definitions for Patterns}

Let $\tau \geq 1$ and $\Pat \in \Sigma^{m}$ be a $\tau$-periodic
pattern. Denote $p =
\per(\Pat[1 \dd 3\tau - 1])$.  We define $\RootPat{\Pat}{\tau}
:= \min\{\Pat[1 + t \dd 1 + t + p) : t \in [0 \dd p)\}$ and
$\RunEndPat{\Pat}{\tau} := 1 + p +
\lcp{\Pat[1 \dd m]}{\Pat[1 + p \dd m]}$.  Observe that then we can
write $\Pat[1 \dd \RunEndPat{\Pat}{\tau}) = H' H^{k} H''$, where $H =
\RootPat{\Pat}{\tau}$, and $H'$ (resp.\ $H''$) is a proper
suffix (resp.\ prefix) of $H$. This factorization is unique, since the
opposite would contradict the synchronization property of primitive
strings~\cite[Lemma~1.11]{AlgorithmsOnStrings}. We denote
$\HeadPat{\Pat}{\tau} := |H'|$, $\ExpPat{\Pat}{\tau} := k$, and
$\TailPat{\Pat}{\tau} := |H''|$.  We then let
$\RunEndFullPat{\Pat}{\tau} := 1 + s + kp = \RunEndPat{\Pat}{\tau}
- \TailPat{\Pat}{\tau}$, where $s = \HeadPat{\Pat}{\tau}$.  Finally,
we define $\TypePat{\Pat}{\tau} = +1$ if $\RunEndPat{\Pat}{\tau} \leq
|\Pat|$ and
$\Pat[\RunEndPat{\Pat}{\tau}] \succ \Pat[\RunEndPat{\Pat}{\tau} - p]$,
and $\TypePat{\Pat}{\tau} = -1$ otherwise.

\begin{lemma}\label{lm:periodic-pat-lce}
  Let $\tau \geq 1$ and $\Pat \in \Sigma^{+}$ be a $\tau$-periodic
  pattern. For every $\Pat'
  \in \Sigma^{+}$, $\lcp{\Pat}{\Pat'} \geq 3\tau - 1$ holds if and
  only if $\Pat'$ is $\tau$-periodic, $\RootPat{\Pat'}{\tau} =
  \RootPat{\Pat}{\tau}$, and $\HeadPat{\Pat'}{\tau} =
  \HeadPat{\Pat}{\tau}$.  Moreover, if $\RunEndPat{\Pat}{\tau}
  \leq |\Pat|$ and $\lcp{\Pat}{\Pat'} \geq \RunEndPat{\Pat}{\tau}$
  (which holds, in particular, when $\Pat$ is a prefix of $\Pat'$),
  then:
  \begin{itemize}
  \item $\RunEndPat{\Pat'}{\tau} = \RunEndPat{\Pat}{\tau}$,
  \item $\TailPat{\Pat'}{\tau} = \TailPat{\Pat}{\tau}$,
  \item $\RunEndFullPat{\Pat'}{\tau} =
    \RunEndFullPat{\Pat}{\tau}$,
  \item $\ExpPat{\Pat'}{\tau} = \ExpPat{\Pat}{\tau}$,
  \item $\TypePat{\Pat'}{\tau} = \TypePat{\Pat}{\tau}$.
  \end{itemize}
\end{lemma}

\begin{lemma}\label{lm:pat-lex}
  Let $\tau \geq 1$ and
  $\Pat_1, \Pat_2 \in \Sigma^{+}$ be $\tau$-periodic patterns such
  that $\RootPat{\Pat_1}{\tau} = \RootPat{\Pat_2}{\tau}$
  and $\HeadPat{\Pat_1}{\tau} =
  \HeadPat{\Pat_2}{\tau}$. Denote $t_1 =
  \RunEndPat{\Pat_1}{\tau} - 1$ and $t_2 = \RunEndPat{\Pat_2}{\tau}
  - 1$. Then, it holds $\lcp{\Pat_1}{\Pat_2} \geq \min(t_1,
  t_2)$. Moreover:
  \begin{enumerate}
  \item\label{lm:pat-lex-it-1} If $\TypePat{\Pat_1}{\tau} \neq
    \TypePat{\Pat_2}{\tau}$ or $t_1 \neq t_2$, then $\Pat_1 \neq
    \Pat_2$ and $\lcp{\Pat_1}{\Pat_2} = \min(t_1, t_2)$,
  \item\label{lm:pat-lex-it-2} If $\TypePat{\Pat_1}{\tau} \neq
    \TypePat{\Pat_2}{\tau}$, then $\Pat_1 \prec \Pat_2$ if and only if
    $\TypePat{\Pat_1}{\tau} < \TypePat{\Pat_2}{\tau}$,
  \item\label{lm:pat-lex-it-3} If $\TypePat{\Pat_1}{\tau} = -1$,
    then $t_1 < t_2$ implies $\Pat_1 \prec \Pat_2$,
  \item\label{lm:pat-lex-it-4} If $\TypePat{\Pat_1}{\tau} =
    +1$, then $t_1 < t_2$ implies $\Pat_1 \succ \Pat_2$,
  \item\label{lm:pat-lex-it-5} If $\TypePat{\Pat_1}{\tau} =
    \TypePat{\Pat_2}{\tau}  = -1$ and $t_1 \neq
    t_2$, then $t_1 < t_2$ if and only if $\Pat_1 \prec \Pat_2$,
  \item\label{lm:pat-lex-it-6} If $\TypePat{\Pat_1}{\tau} =
    \TypePat{\Pat_2}{\tau} = +1$ and $t_1 \neq t_2$, then
    $t_1 < t_2$ if and only if $\Pat_1 \succ \Pat_2$.
  \end{enumerate}
\end{lemma}

\paragraph{Notation and Definitions for Positions}

Let $\tau \in [1 \dd \floor{\frac{\Textlen}{2}}]$.
Observe that if $j \in \RTwo{\tau}{\Text}$,
then $\Text[j \dd \Textlen]$ is $\tau$-periodic
(\cref{def:periodic-pattern}). Letting $\Pat = \Text[j \dd \Textlen]$, we
denote:
\begin{itemize}
\item $\RootPos{j}{\tau}{\Text} := \RootPat{\Pat}{\tau}$,
\item $\HeadPos{j}{\tau}{\Text} := \HeadPat{\Pat}{\tau}$,
\item $\ExpPos{j}{\tau}{\Text} := \ExpPat{\Pat}{\tau}$,
\item $\TailPos{j}{\tau}{\Text} := \TailPat{\Pat}{\tau}$,
\item $\RunEndPos{j}{\tau}{\Text} := j + \RunEndPat{\Pat}{\tau} - 1$,
\item $\RunEndFullPos{j}{\tau}{\Text} := j + \RunEndFullPat{\Pat}{\tau} - 1$,
\item $\TypePos{j}{\tau}{\Text} := \TypePat{\Pat}{\tau}$.
\end{itemize}
Observe, that, letting
$s = \HeadPos{j}{\tau}{\Text}$,
$H = \RootPos{j}{\tau}{\Text}$,
$p = |H|$, and
$k = \ExpPos{j}{\tau}{\Text}$, it holds:
$\RunEndPos{j}{\tau}{\Text} = j + p + \LCE{\Text}{j}{j+p}$ and
$\RunEndFullPos{j}{\tau}{\Text} = j + s + kp = \RunEndPos{j}{\tau}{\Text} - \TailPos{j}{\tau}{\Text}$.

Let $H \in \Sigma^{+}$ and $s \in \Zz$.
We will repeatedly refer to the following subsets of $\RTwo{\tau}{\Text}$:
\begin{itemize}
\item $\RMinusTwo{\tau}{\Text} := \{j \in \RTwo{\tau}{\Text} : \TypePos{j}{\tau}{\Text} = -1\}$,
\item $\RPlusTwo{\tau}{\Text} := \RTwo{\tau}{\Text} \setminus \RMinusTwo{\tau}{\Text}$,
\item $\RThree{H}{\tau}{\Text} := \{j \in \RTwo{\tau}{\Text} : \RootPos{j}{\tau}{\Text} = H\}$,
\item $\RMinusThree{H}{\tau}{\Text} := \RMinusTwo{\tau}{\Text} \cap \RThree{H}{\tau}{\Text}$,
\item $\RPlusThree{H}{\tau}{\Text} := \RPlusTwo{\tau}{\Text} \cap \RThree{H}{\tau}{\Text}$,
\item $\RFour{s}{H}{\tau}{\Text} := \{j \in \RThree{H}{\tau}{\Text} : \HeadPos{j}{\tau}{\Text} = s\}$,
\item $\RMinusFour{s}{H}{\tau}{\Text} := \RMinusTwo{\tau}{\Text} \cap \RFour{s}{H}{\tau}{\Text}$,
\item $\RPlusFour{s}{H}{\tau}{\Text} := \RPlusTwo{\tau}{\Text} \cap \RFour{s}{H}{\tau}{\Text}$,
\item $\RFive{s}{k}{H}{\tau}{\Text} := \{j \in \RFour{s}{H}{\tau}{\Text} : \ExpPos{j}{\tau}{\Text} = k\}$,
\item $\RMinusFive{s}{k}{H}{\tau}{\Text} := \RMinusTwo{\tau}{\Text} \cap \RFive{s}{k}{H}{\tau}{\Text}$,
\item $\RPlusFive{s}{k}{H}{\tau}{\Text} := \RPlusTwo{\tau}{\Text} \cap \RFive{s}{k}{H}{\tau}{\Text}$.
\end{itemize}

Maximal blocks of positions from $\RTwo{\tau}{\Text}$ play an important role
in our data structure. The starting positions of these blocks are
defined as
\begin{align*}
  \RPrimTwo{\tau}{\Text} := \{j \in \RTwo{\tau}{\Text} : j-1 \not\in \RTwo{\tau}{\Text}\}.
\end{align*}
We then denote:
\begin{itemize}
\item $\RPrimMinusTwo{\tau}{\Text} := \RPrimTwo{\tau}{\Text} \cap \RMinusTwo{\tau}{\Text}$,
\item $\RPrimPlusTwo{\tau}{\Text} := \RPrimTwo{\tau}{\Text} \cap \RPlusTwo{\tau}{\Text}$,
\item $\RPrimMinusThree{H}{\tau}{\Text} := \RPrimTwo{\tau}{\Text} \cap \RMinusThree{H}{\tau}{\Text}$,
\item $\RPrimPlusThree{H}{\tau}{\Text} := \RPrimTwo{\tau}{\Text} \cap \RPlusThree{H}{\tau}{\Text}$.
\end{itemize}

\begin{lemma}\label{lm:R-text-block}
  Let $\tau \in [1 \dd \floor{\frac{\Textlen}{2}}]$. For every $j \in
  \RTwo{\tau}{\Text}$ such that $j-1 \in \RTwo{\tau}{\Text}$, it holds
  \begin{itemize}
  \item $\RootPos{j-1}{\tau}{\Text} = \RootPos{j}{\tau}{\Text}$,
  \item $\RunEndPos{j-1}{\tau}{\Text} = \RunEndPos{j}{\tau}{\Text}$,
  \item $\TailPos{j-1}{\tau}{\Text} = \TailPos{j}{\tau}{\Text}$,
  \item $\RunEndFullPos{j-1}{\tau}{\Text} = \RunEndFullPos{j}{\tau}{\Text}$,
  \item $\TypePos{j-1}{\tau}{\Text} = \TypePos{j}{\tau}{\Text}$.
  \end{itemize}
\end{lemma}

\begin{lemma}\label{lm:end}
  Let $\tau \in [1 \dd \floor{\frac{\Textlen}{2}}]$. For every $j \in
  \RTwo{\tau}{\Text}$, it holds
  \[
    \RunEndPos{j}{\tau}{\Text} =
      \max\{j' \in [j \dd \Textlen] : [j \dd j'] \subseteq \RTwo{\tau}{\Text}\} +
      3\tau - 1.
  \]
\end{lemma}

\begin{lemma}\label{lm:periodic-pos-lce}
  Let $\tau \in [1 \dd \lfloor \tfrac{\Textlen}{2} \rfloor]$ and
  $j \in [1 \dd \Textlen]$.
  \begin{enumerate}
  \item\label{lm:periodic-pos-lce-it-1}
    Let $\Pat \in \Sigma^{+}$ be a $\tau$-periodic pattern. Then,
    the following conditions are equivalent:
    \begin{itemize}
    \item $\lcp{\Pat}{\Text[j \dd \Textlen]} \geq 3\tau - 1$,
    \item $j \in \RTwo{\tau}{\Text}$, $\RootPos{j}{\tau}{\Text} =
      \RootPat{\Pat}{\tau}$, and $\HeadPos{j}{\tau}{\Text} =
      \HeadPat{\Pat}{\tau}$.
    \end{itemize}
    Moreover, if, letting $t = \RunEndPat{\Pat}{\tau} - 1$, it holds
    $\lcp{\Pat}{\Text[j \dd \Textlen]} > t$, then:
    \begin{itemize}
    \item $\RunEndPat{\Pat}{\tau} - 1 = \RunEndPos{j}{\tau}{\Text} - j$,
    \item $\TailPat{\Pat}{\tau} = \TailPos{j}{\tau}{\Text}$,
    \item $\RunEndFullPat{\Pat}{\tau} - 1 =
      \RunEndFullPos{j}{\tau}{\Text} - j$,
    \item $\ExpPat{\Pat}{\tau} = \ExpPos{j}{\tau}{\Text}$,
    \item $\TypePat{\Pat}{\tau} = \TypePos{j}{\tau}{\Text}$.
    \end{itemize}
  \item\label{lm:periodic-pos-lce-it-2}
    Let $j' \in \RTwo{\tau}{\Text}$. Then, the following conditions
    are equivalent:
    \begin{itemize}
    \item $\LCE{\Text}{j'}{j} \geq 3\tau - 1$,
    \item $j \in \RTwo{\tau}{\Text}$,
      $\RootPos{j}{\tau}{\Text} = \RootPos{j'}{\tau}{\Text}$, and
      $\HeadPos{j'}{\tau}{\Text} = \HeadPos{j}{\tau}{\Text}$.
    \end{itemize}
    Moreover, if letting $t = \RunEndPos{j}{\tau}{\Text} - j$, it holds
    $\LCE{\Text}{j}{j'} > t$, then:
    \begin{itemize}
    \item $\RunEndPos{j'}{\tau}{\Text} - j' = \RunEndPos{j}{\tau}{\Text} - j$,
    \item $\TailPos{j'}{\tau}{\Text} = \TailPos{j}{\tau}{\Text}$,
    \item $\RunEndFullPos{j'}{\tau}{\Text} - j' =
      \RunEndFullPos{j}{\tau}{\Text} - j$,
    \item $\ExpPos{j'}{\tau}{\Text} = \ExpPos{j}{\tau}{\Text}$,
    \item $\TypePos{j'}{\tau}{\Text} = \TypePos{j}{\tau}{\Text}$.
    \end{itemize}
  \end{enumerate}
\end{lemma}

\begin{lemma}\label{lm:partially-periodic-substring}
  Let $\tau \in [1 \dd \lfloor \tfrac{\Textlen}{2} \rfloor]$. Let $j \in \RTwo{\tau}{\Text}$
  and $\ell > 0$ be such that $\RunEndPos{j}{\tau}{\Text} < j + \ell \leq \Textlen + 1$.
  Then $\Pat := \Text[j \dd j + \ell)$ is a $\tau$-periodic pattern such that
  $\RunEndPat{\Pat}{\tau} \leq |\Pat|$. Moreover:
  \begin{itemize}
  \item $\HeadPat{\Pat}{\tau} = \HeadPos{j}{\tau}{\Text}$,
  \item $\RootPat{\Pat}{\tau} = \RootPos{j}{\tau}{\Text}$,
  \item $\RunEndPat{\Pat}{\tau} - 1 = \RunEndPos{j}{\tau}{\Text} - j$,
  \item $\RunEndFullPat{\Pat}{\tau} - 1 = \RunEndFullPos{j}{\tau}{\Text} - j$,
  \item $\ExpPat{\Pat}{\tau} = \ExpPos{j}{\tau}{\Text}$,
  \item $\TailPat{\Pat}{\tau} = \TailPos{j}{\tau}{\Text}$,
  \item $\TypePat{\Pat}{\tau} = \TypePos{j}{\tau}{\Text}$.
  \end{itemize}
\end{lemma}
\begin{proof}

  Let $\Pat_{\rm suf} = \Text[j \dd \Textlen]$.
  We begin by establishing the properties of $\Pat_{\rm suf}$.
  Note that by \cref{def:sss,def:periodic-pattern}, $\Pat_{\rm suf}$ is $\tau$-periodic
  and, by definition, it holds
  $\RootPat{\Pat_{\rm suf}}{\tau} = \RootPos{j}{\tau}{\Text}$,
  $\HeadPat{\Pat_{\rm suf}}{\tau} = \HeadPos{j}{\tau}{\Text}$,
  $\RunEndPat{\Pat_{\rm suf}}{\tau} = \RunEndPos{j}{\tau}{\Text} - j + 1$,
  $\RunEndFullPat{\Pat_{\rm suf}}{\tau} = \RunEndFullPos{j}{\tau}{\Text} - j + 1$,
  $\TypePat{\Pat_{\rm suf}}{\tau} = \TypePos{j}{\tau}{\Text}$,
  $\ExpPat{\Pat_{\rm suf}}{\tau} = \ExpPos{j}{\tau}{\Text}$, and
  $\TailPat{\Pat_{\rm suf}}{\tau} = \TailPos{j}{\tau}{\Text}$.
  Lastly, observe that by the uniqueness of $\Text[\Textlen]$ in $\Text$, it follows
  that $\RunEndPos{j}{\tau}{\Text} \leq \Textlen$, and hence
  $\RunEndPat{\Pat_{\rm suf}}{\tau} = \RunEndPos{j}{\tau}{\Text} - j + 1 \leq \Textlen - j + 1 = |\Pat_{\rm suf}|$.

  To establish the properties of pattern $\Pat$, first note that
  by \cref{lm:end} and the assumption $\RunEndPos{j}{\tau}{\Text} < j + \ell$, it follows
  that $\ell > \RunEndPos{j}{\tau}{\Text} - j \geq 3\tau - 1$.
  Thus, by $\lcp{\Pat}{\Pat_{\rm suf}} = \lcp{\Text[j \dd j + \ell)}{\Text[j \dd \Textlen]} = \ell$
  and \cref{lm:periodic-pat-lce}, it follows that $\Pat$ is $\tau$-periodic and it holds
  $\RootPat{\Pat}{\tau} = \RootPat{\Pat_{\rm suf}}{\tau} = \RootPos{j}{\tau}{\Text}$
  and $\HeadPat{\Pat}{\tau} = \HeadPat{\Pat_{\rm suf}}{\tau} = \HeadPos{j}{\tau}{\Text}$.
  Moreover, since above observed that
  $\RunEndPat{\Pat_{\rm suf}}{\tau} \leq |\Pat_{\rm suf}|$ and by
  $\RunEndPos{j}{\tau}{\Text} < j + \ell$, it follows that
  $\RunEndPat{\Pat_{\rm suf}}{\tau} = \RunEndPos{j}{\tau}{\Text} - j + 1 \leq \ell = \lcp{\Pat}{\Pat_{\rm suf}}$, we
  obtain from \cref{lm:periodic-pat-lce}, that
  $\RunEndPat{\Pat}{\tau} = \RunEndPat{\Pat_{\rm suf}}{\tau} = \RunEndPos{j}{\tau}{\Text} - j + 1$,
  $\RunEndFullPat{\Pat}{\tau} = \RunEndFullPat{\Pat_{\rm suf}}{\tau} = \RunEndFullPos{j}{\tau}{\Text} - j + 1$,
  $\TypePat{\Pat}{\tau} = \TypePat{\Pat_{\rm suf}}{\tau} = \TypePos{j}{\tau}{\Text}$,
  $\ExpPat{\Pat}{\tau} = \ExpPat{\Pat_{\rm suf}}{\tau} = \ExpPos{j}{\tau}{\Text}$, and
  $\TailPat{\Pat}{\tau} = \TailPat{\Pat_{\rm suf}}{\tau} = \TailPos{j}{\tau}{\Text}$.
  Lastly, note that $\RunEndPos{j}{\tau}{\Text} - j + 1 \leq \ell$ implies $\RunEndPat{\Pat}{\tau} \leq \ell = |\Pat|$.
\end{proof}

\begin{lemma}\label{lm:fully-periodic-substring}
  Let $\tau \in [1 \dd \lfloor \tfrac{\Textlen}{2} \rfloor]$. Let $j \in \RTwo{\tau}{\Text}$
  and $\ell \geq 3\tau - 1$ be such that $j + \ell \leq \RunEndPos{j}{\tau}{\Text}$.
  Then $\Pat := \Text[j \dd j + \ell)$ is a $\tau$-periodic pattern such that $\RunEndPat{\Pat}{\tau} = |\Pat| + 1$.
  Moreover:
  \begin{itemize}
  \item $\HeadPat{\Pat}{\tau} = \HeadPos{j}{\tau}{\Text}$,
  \item $\RootPat{\Pat}{\tau} = \RootPos{j}{\tau}{\Text}$.
  \end{itemize}
\end{lemma}
\begin{proof}

  Let $\Pat_{\rm suf} = \Text[j \dd \Textlen]$.
  By \cref{def:sss,def:periodic-pattern}, $\Pat_{\rm suf}$ is $\tau$-periodic.
  Denote $H = \RootPat{\Pat_{\rm suf}}{\tau}$ and $s = \HeadPat{\Pat_{\rm suf}}{\tau}$.
  By definition of $\RootPos{j}{\tau}{\Text}$ and $\HeadPos{j}{\tau}{\Text}$, it holds
  $\RootPos{j}{\tau}{\Text} = H$ and $\HeadPos{j}{\tau}{\Text} = s$.
  Next, note that by $\lcp{\Pat_{\rm suf}}{\Pat} = \lcp{\Text[j \dd \Textlen]}{\Text[j \dd j + \ell})
  = \ell \geq 3\tau - 1$
  and \cref{lm:periodic-pat-lce}, it follows that $\Pat$ is $\tau$-periodic
  and it holds $\RootPat{\Pat}{\tau} = H = \RootPos{j}{\tau}{\Text}$
  and $\HeadPat{\Pat}{\tau} = s = \HeadPos{j}{\tau}{\Text}$.

  To prove the remaining claim, i.e., that $\RunEndPat{\Pat}{\tau} = |\Pat| + 1$, we proceed as follows.
  By definition, $\RunEndPos{j}{\tau}{\Text} - j = \RunEndPat{\Pat_{\rm suf}}{\tau} - 1 =
  p + \lcp{\Pat_{\rm suf}[1 \dd |\Pat_{\rm suf}|]}{\Pat_{\rm suf}[1 + p \dd |\Pat_{\rm suf}|]} =
  p + \lcp{\Text[j \dd \Textlen]}{\Text[j + p \dd \Textlen]}$.
  By the assumption in the claim, it holds $\RunEndPos{j}{\tau}{\Text} - j \geq \ell$.
  We thus obtain $p + \lcp{\Text[j \dd \Textlen]}{\Text[j + p \dd \Textlen]} \geq \ell$.
  Thus, $p + \lcp{\Text[j \dd j + \ell)}{\Text[j + p \dd j + \ell}) = \ell$.
  Combining $\Pat = \Text[j \dd j + \ell)$ and the definition of $\RunEndPat{\Pat}{\tau}$ with the above
  observations, we thus obtain
  $\RunEndPat{\Pat}{\tau} = 1 + p + \lcp{\Pat[1 \dd \ell]}{\Pat[1 + p \dd \ell]} =
  1 + p + \lcp{\Text[j \dd j + \ell)}{\Text[j + p \dd j + \ell}) = 1 + \ell = 1 + |\Pat|$.
\end{proof}

\begin{lemma}\label{lm:R-lex-block-pat}
  Let $\tau \in [1 \dd \floor{\frac{\Textlen}{2}}]$ and $j \in \RTwo{\tau}{\Text}$.
  Let $\Pat \in \Sigma^{m}$ be a $\tau$-periodic pattern such that
  $\RootPat{\Pat}{\tau} = \RootPos{j}{\tau}{\Text}$ and
  $\HeadPat{\Pat}{\tau} = \HeadPos{j}{\tau}{\Text}$.  Then,
  letting $t_1 = \RunEndPos{j}{\tau}{\Text} - j$ and $t_2 =
  \RunEndPat{\Pat}{\tau} - 1$, it holds $\lcp{\Text[j \dd \Textlen]}{\Pat}
  \geq \min(t_1, t_2)$ and:
  \begin{enumerate}
  \item\label{lm:R-lex-block-pat-it-1}
    If $\TypePos{j}{\tau}{\Text} \neq
    \TypePat{\Pat}{\tau}$ or $t_1 \neq t_2$, then
    $\Text[j {\dd} \Textlen] \neq \Pat$ and
    $\lcp{\Text[j {\dd} \Textlen]}{\Pat} = \min(t_1, t_2)$,
  \item\label{lm:R-lex-block-pat-it-2}
    If $\TypePos{j}{\tau}{\Text} \neq \TypePat{\Pat}{\tau}$,
    then $\Text[j \dd \Textlen] \prec \Pat$ if and only if
    $\TypePos{j}{\tau}{\Text} < \TypePat{\Pat}{\tau}$,
  \item\label{lm:R-lex-block-pat-it-3}
    If $\TypePos{j}{\tau}{\Text} = -1$, then $t_1 < t_2$ implies
    $\Text[j \dd \Textlen] \prec \Pat$,
  \item\label{lm:R-lex-block-pat-it-4}
    If $\TypePos{j}{\tau}{\Text} = +1$, then $t_1 < t_2$ implies
    $\Text[j \dd \Textlen] \succ \Pat$,
  \item\label{lm:R-lex-block-pat-it-5}
    If $\TypePos{j}{\tau}{\Text} = \TypePat{\Pat}{\tau} = -1$, and
    $t_1 \neq t_2$, then $t_1 < t_2$ if and only if
    $\Text[j \dd \Textlen] \prec \Pat$,
  \item\label{lm:R-lex-block-pat-it-6}
    If $\TypePos{j}{\tau}{\Text} = \TypePat{\Pat}{\tau} = +1$, and
    $t_1 \neq t_2$, then $t_1 < t_2$ if and only if
    $\Text[j \dd \Textlen] \succ \Pat$.
  \end{enumerate}
\end{lemma}

\begin{lemma}\label{lm:R-lex-block-pos}
  Let $\tau \in [1 \dd \floor{\frac{\Textlen}{2}}]$ and $j \in \RTwo{\tau}{\Text}$. Let
  $j' \in \RTwo{\tau}{\Text}$ be such that $\RootPos{j'}{\tau}{\Text} =
  \RootPos{j}{\tau}{\Text}$ and $\HeadPos{j'}{\tau}{\Text} =
  \HeadPos{j}{\tau}{\Text}$.  Then, letting $t_1 =
  \RunEndPos{j}{\tau}{\Text} - j$ and $t_2 = \RunEndPos{j'}{\tau}{\Text} - j'$,
  it holds $\LCE{\Text}{j}{j'} \geq \min(t_1, t_2)$ and:
  \begin{enumerate}
  \item\label{lm:R-lex-block-pos-it-1}
    If $\TypePos{j}{\tau}{\Text} \neq \TypePos{j'}{\tau}{\Text}$ or $t_1
    \neq t_2$, then $\LCE{\Text}{j}{j'} = \min(t_1, t_2)$,
  \item\label{lm:R-lex-block-pos-it-2}
    If $\TypePos{j}{\tau}{\Text} \neq \TypePos{j'}{\tau}{\Text}$, then
    $\Text[j \dd \Textlen] \prec \Text[j' \dd \Textlen]$ if and only if
    $\TypePos{j}{\tau}{\Text} <
    \TypePos{j'}{\tau}{\Text}$,
  \item\label{lm:R-lex-block-pos-it-3}
    If $\TypePos{j}{\tau}{\Text} = \TypePos{j'}{\tau}{\Text} = -1$ and
    $t_1 \neq t_2$, then $t_1 < t_2$ if and only if
    $\Text[j \dd \Textlen] \prec \Text[j'
    \dd \Textlen]$,
  \item\label{lm:R-lex-block-pos-it-4}
    If $\TypePos{j}{\tau}{\Text} = \TypePos{j'}{\tau}{\Text} = +1$ and
    $t_1 \neq t_2$, then $t_1 < t_2$ if and only if
    $\Text[j \dd \Textlen] \succ \Text[j'
    \dd \Textlen]$.
  \end{enumerate}
\end{lemma}

\begin{lemma}\label{lm:run-end}
  Let $\tau \in [1 \dd \floor{\frac{\Textlen}{2}}]$.  Let $j, j',
  j'' \in [1 \dd \Textlen]$ be such that $j, j'' \in \RTwo{\tau}{\Text}$, $j'
  \not\in \RTwo{\tau}{\Text}$, and $j < j' < j''$. Then, it holds
  $\RunEndPos{j}{\tau}{\Text} \leq j'' + \tau - 1$.
\end{lemma}

\begin{lemma}\label{lm:runs}
  For every $\tau \in [1 \dd \lfloor \tfrac{\Textlen}{2} \rfloor]$,
  it holds $|\RPrimTwo{\tau}{\Text}| \leq \tfrac{2\Textlen}{\tau}$
  and $\sum_{i \in \RPrimTwo{\tau}{\Text}} \RunEndPos{i}{\tau}{\Text} - i \leq 2\Textlen$.
\end{lemma}

\begin{lemma}\label{lm:efull}
  Let $\tau \in [1 \dd \floor{\frac{\Textlen}{2}}]$.
  For any $j, j' \in
  \RPrimTwo{\tau}{\Text}$, $j \neq j'$ implies $\RunEndFullPos{j}{\tau}{\Text} \neq
  \RunEndFullPos{j'}{\tau}{\Text}$.
\end{lemma}

\begin{lemma}\label{lm:RskH}
  Let $\tau \in [1 \dd \lfloor \tfrac{\Textlen}{2} \rfloor]$.
  Let $H \in \Sigma^{+}$, $p = |H|$, $s \in [0 \dd p)$, and
  $k_{\min} = \lceil \tfrac{3\tau-1-s}{p} \rceil - 1$.
  For every $k \in (k_{\min} \dd \Textlen]$, it holds
  \[
    \RMinusFive{s}{k}{H}{\tau}{\Text} =
      \{\RunEndFullPos{j}{\tau}{\Text} - s - kp :
      j \in \RPrimMinusThree{H}{\tau}{\Text}
      \text{ and }
      s + kp \leq \RunEndFullPos{j}{\tau}{\Text} - j\}.
  \]
\end{lemma}

\begin{definition}\label{def:runs-minus-lex-sorted}
  Let $\tau \in [1 \dd \lfloor \tfrac{\Textlen}{2} \rfloor]$.
  \begin{itemize}
  \item Letting $m = |\RPrimMinusTwo{\tau}{\Text}|$,
    by $\RunsMinusLexSortedTwo{\tau}{\Text}$, we denote
    a sequence $(a_i)_{i \in [1 \dd m]}$
    containing all elements of $\RPrimMinusTwo{\tau}{\Text}$ such
    that for every $i, j \in [1 \dd m]$, $i < j$ implies
    $\RootPos{a_i}{\tau}{\Text} \prec \RootPos{a_j}{\tau}{\Text}$,
    or $\RootPos{a_i}{\tau}{\Text} = \RootPos{a_j}{\tau}{\Text}$ and
    $\Text[\RunEndFullPos{a_i}{\tau}{\Text} \dd \Textlen] \prec
    \Text[\RunEndFullPos{a_j}{\tau}{\Text} \dd \Textlen]$.
  \item For
    every $H \in \Sigma^{+}$, letting
    $m_H = |\RPrimMinusThree{H}{\tau}{\Text}|$, by
    $\RunsMinusLexSortedThree{H}{\tau}{\Text}$, we denote
    a sequence $(a_i)_{i \in [1 \dd m_{H}]}$
    containing all elements of $\RPrimMinusThree{H}{\tau}{\Text}$ such
    that for every $i, j \in [1 \dd m_{H}]$, $i < j$ implies
    $\Text[\RunEndFullPos{a_i}{\tau}{\Text} \dd \Textlen] \prec
    \Text[\RunEndFullPos{a_j}{\tau}{\Text} \dd \Textlen]$.
  \end{itemize}
\end{definition}

\begin{definition}\label{def:runs-minus-text-sorted}
  Let $\tau \in [1 \dd \lfloor \tfrac{\Textlen}{2} \rfloor]$.
  \begin{itemize}
  \item Letting $m = |\RPrimMinusTwo{\tau}{\Text}|$,
    by $\RunsMinusTextSortedTwo{\tau}{\Text}$, we denote
    a sequence $(a_i)_{i \in [1 \dd m]}$
    containing all elements of $\RPrimMinusTwo{\tau}{\Text}$ such
    that for every $i, j \in [1 \dd m]$, $i < j$ implies
    $\RootPos{a_i}{\tau}{\Text} \prec \RootPos{a_j}{\tau}{\Text}$,
    or $\RootPos{a_i}{\tau}{\Text} = \RootPos{a_j}{\tau}{\Text}$ and $a_i < a_j$.
  \item For every $H \in \Sigma^{+}$, letting
    $m_{H} = |\RPrimMinusThree{H}{\tau}{\Text}|$,
    by $\RunsMinusTextSortedThree{H}{\tau}{\Text}$, we denote
    a sequence $(a_i)_{i \in [1 \dd m_{H}]}$
    containing all elements of $\RPrimMinusThree{H}{\tau}{\Text}$ such
    that for every $i, j \in [1 \dd m_{H}]$, $i < j$ implies $a_i < a_j$.
  \end{itemize}
\end{definition}

\begin{definition}\label{def:rmin}
  For every $\tau \in [1 \dd \lfloor \tfrac{\Textlen}{2} \rfloor]$, we
  define
  \[
    \RMinMinusTwo{\tau}{\Text} := \{j \in \RMinusTwo{\tau}{\Text}
      : j = \min \OccTwo{\Text[j \dd \RunEndPos{j}{\tau}{\Text})}{\Text}
                 \cap \RMinusTwo{\tau}{\Text}\}.
  \]
\end{definition}

\begin{definition}\label{def:min-pos-bitvector-minus}
  Let $\tau \in [1 \dd \floor{\tfrac{\Textlen}{2}}]$. By
  $\MinPosBitvectorMinusTwo{\tau}{\Text}$, we denote a bitvector $B[1 \dd \Textlen]$
  defined such that for
  every $i \in [1 \dd \Textlen]$,
  \[
    B[i] =
      \begin{cases}
       1 & \text{if }\SA{\Text}[i] \in \RMinMinusTwo{\tau}{\Text},\\
       0 & \text{otherwise}.\\
   \end{cases}
  \]
  Moreover, for every $H \in \Sigma^{+}$, $s \in [0 \dd |H|)$, and $k \in \Zp$,
  we then denote
  \begin{align*}
    \MinPosBitvectorMinusFour{s}{H}{\tau}{\Text}
      &= B(x \dd y],\\
    \MinPosBitvectorMinusFive{s}{k}{H}{\tau}{\Text}
      &= B(x' \dd y'],\\
  \end{align*}
  where $x, y, x', y' \in [0 \dd \Textlen]$ are such that
  \begin{align*}
    \{\SA{\Text}[i] : i \in (x \dd y]\}
      &= \RMinusFour{s}{H}{\tau}{\Text},\\
    \{\SA{\Text}[i] : i \in (x' \dd y']\}
      &= \RMinusFive{s}{k}{H}{\tau}{\Text}.
  \end{align*}
\end{definition}

\begin{remark}\label{rm:min-pos-bitvector-minus}
  To show that $x$ and $y$ in \cref{def:min-pos-bitvector-minus} are well-defined, note
  that by \cref{lm:periodic-pos-lce}\eqref{lm:periodic-pos-lce-it-2},
  the set $\RFour{s}{H}{\tau}{\Text}$ occupies a contiguous block
  of positions in $\SA{\Text}$. Moreover, by \cref{lm:R-lex-block-pos}\eqref{lm:R-lex-block-pos-it-2}, all
  positions $j \in \RFour{s}{H}{\tau}{\Text}$ satisfying
  $\TypePos{j}{\tau}{\Text} = -1$ precede positions
  $j' \in \RFour{s}{H}{\tau}{\Text}$ satisfying
  $\TypePos{j'}{\tau}{\Text} = +1$. In order words, positions in
  $\RMinusFour{s}{H}{\tau}{\Text}$ occupy a contiguous
  block in $\SA{\Text}$. Thus, $x,y$ are indeed well-defined.

  To show that $x'$ and $y'$ in \cref{def:min-pos-bitvector-minus} are well-defined,
  observe now that letting
  $a,b \in [0 \dd \Textlen]$ be such that $\{\SA{\Text}[i] : i \in (a \dd b]\} = \RMinusFour{s}{H}{\tau}{\Text}$,
  it follows by \cref{lm:R-lex-block-pos}\eqref{lm:R-lex-block-pos-it-3} that for
  every $i \in (a \dd b)$,
  it holds $\RunEndPos{\SA{\Text}[i]}{\tau}{\Text} - \SA{\Text}[i] \leq \RunEndPos{\SA{\Text}[i+1]}{\tau}{\Text} - \SA{\Text}[i+1]$.
  By $\HeadPos{\SA{\Text}[i]}{\tau}{\Text} = \HeadPos{\SA{\Text}[i+1]}{\tau}{\Text} = s$, we thus obtain
  \begin{align*}
    \ExpPos{\SA{\Text}[i]}{\tau}{\Text}
      &= \lfloor \tfrac{\RunEndPos{\SA{\Text}[i]}{\tau}{\Text} - \SA{\Text}[i] - s}{p} \rfloor\\
      &\leq \lfloor \tfrac{\RunEndPos{\SA{\Text}[i+1]}{\tau}{\Text} - \SA{\Text}[i+1] - s}{p} \rfloor\\
      &= \ExpPos{\SA{\Text}[i+1]}{\tau}{\Text}.
  \end{align*}
  Thus, all positions from $\RMinusFive{s}{k}{H}{\tau}{\Text}$ occupy
  a contiguous block in $\SA{\Text}$. The
  values $x',y'$ are thus indeed well-defined.
\end{remark}

\begin{definition}\label{def:emin}
  Let $\tau \in [1 \dd \lfloor \tfrac{\Textlen}{2} \rfloor]$.  For
  every $j \in \RPrimMinusTwo{\tau}{\Text}$, we define
  \[
    \RunMinEndPos{j}{\tau}{\Text} =
      \max\{j' \in [j \dd \Textlen] :
      [j \dd j') \subseteq \RMinMinusTwo{\tau}{\Text}\}.
  \]
\end{definition}

%% file: index/periodic/nav.tex
\subsubsection{Basic Navigation Primitives}\label{sec:cst-periodic-nav}

\begin{proposition}\label{pr:nav-index-periodic}
  Let $\tau = \mu\log_{\AlphabetSize} \Textlen$. Denote
  $(a_i)_{i \in [1 \dd m]} = \RunsMinusLexSortedTwo{\tau}{\Text}$ (\cref{def:runs-minus-lex-sorted})
  and let $(b_i)_{i \in [1 \dd m]}$ be a sequence defined by
  $b_i = \RunEndFullPos{a_i}{\tau}{\Text} - |\PowStr{\tau}{\RootPos{a_i}{\tau}{\Text}}|$.
  Given the packed representation of the text $\Text$, we can in
  $\bigO(\Textlen / \log_{\AlphabetSize} \Textlen)$ time construct a data
  structure, denoted $\NavPeriodic{\Text}$, that supports the following queries:
  \begin{enumerate}
  \item\label{pr:nav-index-periodic-it-1}
    Let $j \in \RTwo{\tau}{\Text}$, $H = \RootPos{j}{\tau}{\Text}$,
    $s = \HeadPos{j}{\tau}{\Text}$, $k = \ExpPos{j}{\tau}{\Text}$, and
    $\deltatext = (\RunEndFullPos{j}{\tau}{\Text} - j) - |\PowStr{\tau}{H}|$.
    \begin{enumerate}
    \item\label{pr:nav-index-periodic-it-1a}
      Given $j$, in $\bigO(1)$ time compute
      \begin{itemize}
      \item $\HeadPos{j}{\tau}{\Text}$,
      \item $\RootPos{j}{\tau}{\Text}$,
      \item $|\RootPos{j}{\tau}{\Text}|$,
      \item $\RunEndPos{j}{\tau}{\Text}$,
      \item $\RunEndFullPos{j}{\tau}{\Text}$,
      \item $\TypePos{j}{\tau}{\Text}$,
      \item a pair $x,y \in [0 \dd \Textlen]$ satisfying
        $\{\SA{\Text}[i] : i \in (x \dd y]\} = \RMinusFive{s}{k}{H}{\tau}{\Text}$,
      \item a pair $x,y \in [0 \dd \Textlen]$ satisfying
        $\{\SA{\Text}[i] : i \in (x \dd y]\} = \RMinusFour{s}{H}{\tau}{\Text}$.
      \end{itemize}
    \item\label{pr:nav-index-periodic-it-1b}
      If $\TypePos{j}{\tau}{\Text} = -1$ and $\ell \geq 3\tau - 1$ is
      such that $\RunEndPos{j}{\tau}{\Text} < j + \ell \leq \Textlen + 1$, then
      given $j$ and $\ell$, in $\bigO(\log \log \Textlen)$ compute
      the pair $(b,e)$ such that
      $b = |\{i \in [1 \dd m] : \Text[b_i \dd \Textlen] \prec \Text[j + \deltatext \dd j + \ell)\}|$ and
      $e - b = |\{i \in [1 \dd m] : \Text[j + \deltatext \dd j + \ell)\text{ is a prefix of }\Text[b_i \dd \Textlen]\}|$.
    \end{enumerate}
  \item\label{pr:nav-index-periodic-it-2}
    Let $\Pat \in \Sigma^{m}$ be a $\tau$-periodic pattern.
    Denote $H = \RootPat{\Pat}{\tau}$,
    $\deltatext = (\RunEndFullPat{\Pat}{\tau} - 1) - |\PowStr{\tau}{H}|$, and
    $\Pat' = \Pat(\deltatext \dd m]$.
    \begin{enumerate}
    \item\label{pr:nav-index-periodic-it-2a}
      Given the packed representation of $\Pat$, in
      $\bigO(1)$ time compute
      \begin{itemize}
      \item $\HeadPat{\Pat}{\tau}$,
      \item $\RootPat{\Pat}{\tau}$,
      \item $|\RootPat{\Pat}{\tau}|$.
      \end{itemize}
    \item\label{pr:nav-index-periodic-it-2b}
      Given the packed representation of $\Pat$, in
      $\bigO(1 + m / \log_{\AlphabetSize} \Textlen)$ time compute
      \begin{itemize}
      \item $\RunEndPat{\Pat}{\tau}$,
      \item $\RunEndFullPat{\Pat}{\tau}$,
      \item $\TypePat{\Pat}{\tau}$.
      \end{itemize}
    \item\label{pr:nav-index-periodic-it-2c}
      If $\TypePat{\Pat}{\tau} = -1$ and $\RunEndPat{\Pat}{\tau} \leq |\Pat|$,
      then given the packed
      representation of $\Pat$, in
      $\bigO(\log \log \Textlen + m / \log_{\AlphabetSize} \Textlen)$
      time compute the pair $(b,e)$ such that
      $b = |\{i \in [1 \dd m] : \Text[b_i \dd \Textlen] \prec \Pat'\}|$ and
      $e - b = |\{i \in [1 \dd m] : \Pat'\text{ is a prefix of }\Text[b_i \dd \Textlen]\}|$.
    \end{enumerate}
  \item\label{pr:nav-index-periodic-it-3}
    Given any $i \in [1 \dd \Textlen]$ such that $\SA{\Text}[i] \in \RTwo{\tau}{\Text}$,
    compute $\SA{\Text}[i]$ in $\bigO(\log \log \Textlen)$ time.
  \item\label{pr:nav-index-periodic-it-4}
    Let $\Pat \in \Sigma^{m}$ be a $\tau$-periodic pattern satisfying $\RunEndPat{\Pat}{\tau} > |\Pat|$.
    Given $\HeadPat{\Pat}{\tau}$, $|\Pat|$, and the packed representation of $\RootPat{\Pat}{\tau}$,
    in $\bigO(\log \log \Textlen)$ time compute the pair
    $(\RangeBegTwo{\Pat}{\Text}, \RangeEndTwo{\Pat}{\Text})$.
  \item\label{pr:nav-index-periodic-it-5}
    For every $H \in \Sigma^{+}$, given the packed representation of $H$, the
    sequence $\RunsMinusLexSortedThree{H}{\tau}{\Text}$ (\cref{def:runs-minus-lex-sorted})
    and any array $A_{\rm pos}[1 \dd q]$ of positions from $\RMinusThree{H}{\tau}{\Text}$,
    in $\bigO(|\RPrimMinusThree{H}{\tau}{\Text}| + |\RMinusThree{H}{\tau}{\Text}|/\log \Textlen + q)$ time
    compute an array $A_{\rm ans}[1 \dd q]$ defined by $A_{\rm ans}[i] = \ISA{\Text}[A_{\rm pos}[i]]$.
  \end{enumerate}
\end{proposition}
\begin{proof}

  $\NavPeriodic{\Text}$ consists of three components:
  \begin{enumerate}
  \item The component of the pattern matching index to handle periodic patterns from~\cite[Section~6.3.2]{breaking}.
    It needs $\bigO(\Textlen / \log_{\AlphabetSize} \Textlen)$ space.
  \item The component of the compressed suffix tree (CST) to handle periodic nodes from~\cite[Section~7.3.1]{breaking}.
    This component also needs $\bigO(\Textlen / \log_{\AlphabetSize} \Textlen)$ space.
  \item The data structure from \cref{th:offline-range-queries-three-sided} for $u = \Textlen$ and $\alpha = 1/2$. If
    needs $\bigO(\Textlen^{1/2}) = \bigO(\Textlen / \log_{\AlphabetSize} \Textlen)$ space.
  \end{enumerate}
  In total, $\NavPeriodic{\Text}$ needs $\bigO(\Textlen / \log_{\AlphabetSize} \Textlen)$ space.

  \paragraph{Implementation of queries}

  Queries~\ref{pr:nav-index-periodic-it-1a}, \ref{pr:nav-index-periodic-it-2a}, and
  \ref{pr:nav-index-periodic-it-2b} are standard navigation queries of the above indexes
  (the computation of $(x,y)$ in Query~\ref{pr:nav-index-periodic-it-1a}
  uses the rank/select queries on a bitvector marking the boundaries between blocks of
  periodic positions with the same exponent; see, e.g.,~\cite[Proposition~5.9]{breaking}).

  Queries~\ref{pr:nav-index-periodic-it-1b} and~\ref{pr:nav-index-periodic-it-2c} are supported similarly
  as in \cref{pr:nav-index-periodic}, i.e., using tries and weighted ancestor queries (which are components of the above
  indexes), except the node in each trie is obtained using navigation primitives for periodic positions and patterns
  (similar to, e.g.,~\cite[Proposition~7.15]{breaking}).

  Queries~\ref{pr:nav-index-periodic-it-3}
  and~\ref{pr:nav-index-periodic-it-4} are described in~\cite[Proposition~5.14 and Proposition~7.17]{breaking}.

  Query~\ref{pr:nav-index-periodic-it-5} is implemented as follows.
  Denote $m_{H} = |\RPrimMinusThree{H}{\tau}{\Text}|$, and assume that the input sequence
  $\RunsMinusLexSortedThree{H}{\tau}{\Text}$ (\cref{def:runs-minus-lex-sorted}) is
  given in an array $A[1 \dd m_{H}]$.
  Let $A_{\rm len}[1 \dd m_{H}]$ be an array defined by
  $A_{\rm len}[i] = \RunEndFullPos{A[i]}{\tau}{\Text} - A[i]$.
  In~\cite[Proposition~5.10 and Proposition~5.11]{breaking}, it is proved that computing $\ISA{\Text}[j]$ for
  $j \in \RMinusThree{H}{\tau}{\Text}$ reduces to three-sided orthogonal range counting queries on $A_{\rm len}$.
  More precisely,
  for every $j \in \RMinusThree{H}{\tau}{\Text}$, there exist indices $B(j), E(j) \in [0 \dd m_{H}]$ and an integer
  $V(j) \geq 0$ that satisfy
  the following properties:
  \begin{itemize}
  \item Given the position $j$ and the above data structures, we can compute $B(j)$, $E(j)$, and $V(j)$ in $\bigO(1)$ time
    (in its original description, the indices $B(j)$ and $E(j)$ are computed with respect to $\RunsMinusLexSortedTwo{\tau}{\Text}$,
    but it is easy to adjust them relative to $\RunsMinusLexSortedThree{H}{\tau}{\Text}$ in $\bigO(1)$ time).
  \item Given the position $j$, the value $\ThreeSidedRangeCount{A_{\rm len}}{B(j)}{E(j)}{V(j)}$, and the above data structures,
    we can compute $\ISA{\Text}[j]$ in $\bigO(1)$ time.
  \end{itemize}
  The query thus proceeds in four steps:
  \begin{enumerate}
  \item Using Query~\ref{pr:nav-index-periodic-it-1a}, in $\bigO(m_{H})$ time we compute the array $A_{\rm len}[1 \dd m_{H}]$.
  \item Using the above indexes, we compute the arrays $Q_{\rm beg}[1 \dd q]$, $Q_{\rm end}[1 \dd q]$, and
    $Q_{\rm val}[1 \dd q]$ defined by $Q_{\rm beg}[i] = B(A_{\rm pos}[i])$, $Q_{\rm end}[i] = E(A_{\rm pos}[i])$, and
    $Q_{\rm val}[i] = V(A_{\rm pos}[i])$. This takes $\bigO(q)$ time.
  \item Denote $s = \sum_{i \in [1 \dd m_{H}]} A_{\rm len}[i]$.
    Using \cref{th:offline-range-queries-three-sided}, in $\bigO(m_{H} + q + s / \log \Textlen)$ time
    we compute an array $A_{\rm rcount}[1 \dd q]$ defined by
    $A_{\rm rcount}[i] = \ThreeSidedRangeCount{A_{\rm len}}{Q_{\rm beg}[i]}{Q_{\rm end}[i]}{Q_{\rm val}[i]}$.
  \item Using the above indexes and the array $A_{\rm rcount}[1 \dd q]$, in $\bigO(q)$ time we compute
    the array $A_{\rm ans}[1 \dd q]$ defined by $A_{\rm ans}[i] = \ISA{\Text}[A_{\rm pos}[i]]$.
  \end{enumerate}
  In total, the computation takes $\bigO(m_{H} + s/\log \Textlen + q)$ time. To bound this time, recall that
  by \cref{lm:R-text-block} and \cref{lm:end}, it holds $\RMinusThree{H}{\tau}{\Text} =
  \bigcup_{j \in \RPrimMinusThree{H}{\tau}{\Text}} [j \dd \RunEndPos{j}{\tau}{\Text} - 3\tau + 2)$ (and all intervals
  in the union are disjoint). Thus,
  \begin{align*}
    s &= \textstyle\sum_{i \in [1 \dd m_{H}]} A_{\rm len}[i]\\
      &= \textstyle\sum_{i \in [1 \dd m_{H}]} \RunEndFullPos{A[i]}{\tau}{\Text} - A[i]\\
      &= \textstyle\sum_{j \in \RPrimMinusThree{H}{\tau}{\Text}} \RunEndFullPos{j}{\tau}{\Text} - j\\
      &\leq \textstyle\sum_{j \in \RPrimMinusThree{H}{\tau}{\Text}} \RunEndPos{j}{\tau}{\Text} - j\\
      &\leq \textstyle\sum_{j \in \RPrimMinusThree{H}{\tau}{\Text}} (\RunEndPos{j}{\tau}{\Text} - 3\tau + 2 - j) + 3\tau\\
      &= |\RMinusThree{H}{\tau}{\Text}| + 3\tau \cdot m_{H}.
  \end{align*}
  Since $\tau = \bigO(\log \Textlen)$, we can thus bound the above runtime
  by $\bigO(|\RPrimMinusThree{H}{\tau}{\Text}| + |\RMinusThree{H}{\tau}{\Text}| / \log \Textlen + q)$.

  \paragraph{Construction algorithm}
  The components of $\NavPeriodic{\Text}$ are constructed as follows:
  \begin{enumerate}
  \item With~\cite[Proposition~6.3 and~6.13]{breaking}, we
    build the first component in
    $\bigO(\Textlen / \log_{\AlphabetSize} \Textlen)$ time.
  \item With~\cite[Proposition~7.7 and~7.23]{breaking}, we
    build the second component in
    $\bigO(\Textlen / \log_{\AlphabetSize} \Textlen)$ time.
  \item The last component is constructed using
    \cref{th:offline-range-queries-three-sided} in
    $\bigO(\Textlen^{1/2}) = \bigO(\Textlen / \log_{\AlphabetSize} \Textlen)$
    time.
  \end{enumerate}
  In total, the construction takes
  $\bigO(\Textlen / \log_{\AlphabetSize} \Textlen)$ time.
\end{proof}

%% file: index/periodic/structure.tex
\subsubsection{The Data Structure}\label{sec:minocc-index-periodic-structure}

\paragraph{Definitions}

For every string $X$ and every $\ell \in \Zp$, we denote $\PowStr{\ell}{X} =
X^{\infty}[1 \dd |X| \lceil \tfrac{\ell}{|X|} \rceil]$.
We also denote $\tau = \mu \log_{\AlphabetSize} \Textlen$, where
$\mu$ is as defined at the beginning of \cref{sec:minocc-index}.

Let $m = |\RPrimMinusTwo{\tau}{\Text}|$ and let $(a_i)_{i \in [1 \dd m]} = \RunsMinusLexSortedTwo{\tau}{\Text}$.
Let $A_{\rm pos}[1 \dd m]$ and $A_{\rm len}[1 \dd m]$ be two arrays defined so that
for every $i \in [1 \dd m]$ it holds
$A_{\rm pos}[i] = \RunEndFullPos{a_i}{\tau}{\Text} - |\PowStr{\tau}{\RootPos{a_i}{\tau}{\Text}}|$ and
$A_{\rm len}[i] = A_{\rm pos}[i] - a_i$.

\paragraph{Components}

The data structure, denoted $\MinOccIndexPeriodic{\Text}$, to handle
$\tau$-periodic positions and patterns consists of
two parts. The first part consists of the following three components:

\begin{enumerate}
\item The data structure $\NavPeriodic{\Text}$ from \cref{pr:nav-index-periodic}.
  It needs $\bigO(\Textlen / \log_{\AlphabetSize} \Textlen)$ space.
\item The plain representations of arrays $A_{\rm pos}[1 \dd m]$ and $A_{\rm len}[1 \dd m]$
  augmented with the data structure from \cref{th:three-sided-rmq}.
  Observe that we can use \cref{th:three-sided-rmq} since, letting
  $m_{\max} = 2\Textlen / \tau$, by \cref{lm:runs} it holds
  $m \in [1 \dd m_{\max}]$,
  $\max_{i=1}^{m} A_{\rm pos}[i] \leq \Textlen = \bigO(m_{\max} \log m_{\max})$,
  and
  $\sum_{i=1}^{m} A_{\rm len}[i] \leq 2\Textlen = \bigO(m_{\max} \log m_{\max})$.
  The arrays need
  $\bigO(m) = \bigO(\tfrac{\Textlen}{\tau}) = \bigO(\Textlen / \log_{\AlphabetSize} \Textlen)$ space, and
  the structure from \cref{th:three-sided-rmq} needs
  $\bigO(m_{\max}) = \bigO(\Textlen / \log_{\AlphabetSize} \Textlen)$ space.
\item The bitvector $\BitvectorMin[1 \dd n] = \MinPosBitvectorMinusTwo{\tau}{\Text}$
  (\cref{def:min-pos-bitvector-minus})
  augmented using \cref{th:bin-rank-select} to support rank and select
  queries in $\bigO(1)$ time. The bitvector needs $\bigO(\Textlen
  / \log \Textlen) = \bigO(n / \log_{\AlphabetSize} \Textlen)$ space,
  and the augmentation of \cref{th:bin-rank-select} does not increase
  the space usage.
\end{enumerate}

The second part of the structure consists of the symmetric counterparts
adapted according to \cref{lm:R-lex-block-pat}.

In total, $\MinOccIndexPeriodic{\Text}$ needs
$\bigO(\Textlen / \log_{\AlphabetSize} \Textlen)$ space.

%% file: index/periodic/queries.tex
\subsubsection{Implementation of Queries}\label{sec:minocc-index-periodic-queries}

\paragraph{Combinatorial Properties}

\begin{lemma}[\cite{breaking}]\label{lm:partially-periodic-pat-occ}
  Let $\tau \in [1 \dd \floor{\tfrac{\Textlen}{2}}]$,
  $(a_i)_{i \in [1 \dd q]} = \RunsMinusLexSortedTwo{\tau}{\Text}$
  (\cref{def:runs-minus-lex-sorted}), and $A_{\rm pos}[1 \dd
  q]$ and $A_{\rm len}[1 \dd q]$ be defined by
  \begin{itemize}
  \item $A_{\rm pos}[i] = \RunEndFullPos{a_i}{\tau}{\Text} -
    |\PowStr{\tau}{\RootPos{a_i}{\tau}{\Text}}|$,
  \item $A_{\rm len}[i] = A_{\rm pos}[i] - a_i$.
  \end{itemize}
  Let $\Pat \in \Sigma^{m}$ be a $\tau$-periodic pattern satisfying
  $\RunEndPat{\Pat}{\tau} \leq m$ and $\TypePat{\Pat}{\tau} =
  -1$. Denote $H = \RootPat{\Pat}{\tau}$, $\deltatext =
  (\RunEndFullPat{\Pat}{\tau} - 1) - |\PowStr{\tau}{H}|$, $\Pat'
  = \Pat(\deltatext \dd m]$, and let $(b, e)$ be such that $b \,{=}\,
  |\{i \,{\in}\, [1 \dd q] : \Text[A_{\rm
  pos}[i] \dd \Textlen] \,{\prec}\, \Pat'\}|$ and $e - b =
  |\{i \,{\in}\, [1 \dd q] : \Pat'\text{ is a prefix of }\Text[A_{\rm
  pos}[i] \dd \Textlen]\}|$. Then,
  \begin{align*}
    \OccTwo{\Pat}{\Text}
      &= \{A_{\rm pos}[i] - \deltatext : i \in (b \dd e]\text{ and }A_{\rm len}[i] \geq \deltatext\}.
  \end{align*}
\end{lemma}

\begin{lemma}\label{lm:partially-periodic-pat-occ-min}
  Let $\tau \in [1 \dd \floor{\tfrac{\Textlen}{2}}]$,
  $(a_i)_{i \in [1 \dd q]} = \RunsMinusLexSortedTwo{\tau}{\Text}$
  (\cref{def:runs-minus-lex-sorted}), and $A_{\rm pos}[1 \dd
  q]$ and $A_{\rm len}[1 \dd q]$ be defined by
  \begin{itemize}
  \item $A_{\rm pos}[i] = \RunEndFullPos{a_i}{\tau}{\Text} -
    |\PowStr{\tau}{\RootPos{a_i}{\tau}{\Text}}|$,
  \item $A_{\rm len}[i] = A_{\rm pos}[i] - a_i$.
  \end{itemize}
  Let $\Pat \in \Sigma^{m}$ be a $\tau$-periodic pattern satisfying
  $\RunEndPat{\Pat}{\tau} \leq m$, $\TypePat{\Pat}{\tau} =
  -1$, and $\OccTwo{\Pat}{\Text} \neq \emptyset$.
  Let $H = \RootPat{\Pat}{\tau}$, $\deltatext =
  (\RunEndFullPat{\Pat}{\tau} - 1) - |\PowStr{\tau}{H}|$, $\Pat'
  = \Pat(\deltatext \dd m]$, and let $(b, e)$ be such that $b =
  |\{i \in [1 \dd q] : \Text[A_{\rm
  pos}[i] \dd \Textlen] \prec \Pat'\}|$ and $e - b = |\{i \in [1 \dd
  q] : \Pat'\text{ is a prefix of }\Text[A_{\rm
  pos}[i] \dd \Textlen]\}|$. Then,
  there exists $i \in (b \dd e]$ such that $A_{\rm len}[i] \geq \deltatext$.
  Moreover:
  \begin{align*}
    \min \OccTwo{\Pat}{\Text}
      &= A_{\rm pos}[\ThreeSidedRMQ{A_{\rm pos}}{A_{\rm len}}{b}{e}{\deltatext}] - \deltatext.
  \end{align*}
\end{lemma}
\begin{proof}
  The first claim follows by $\OccTwo{\Pat}{\Text} \neq \emptyset$ and
  \cref{lm:partially-periodic-pat-occ}. To obtain the second claim, it
  suffices to apply \cref{lm:partially-periodic-pat-occ} and
  \cref{def:three-sided-rmq}, resulting in
  \begin{align*}
    \min \OccTwo{\Pat}{\Text}
      &= \min \{A_{\rm pos}[i] - \deltatext :
         i \in (b \dd e]\text{ and }A_{\rm len}[i] \geq \deltatext\}\\
      &= \min \{A_{\rm pos}[i] :
         i \in (b \dd e]\text{ and }A_{\rm len}[i] \geq \deltatext\} - \deltatext\\
      &= A_{\rm pos}[\ThreeSidedRMQ{A_{\rm pos}}{A_{\rm len}}{b}{e}{\deltatext}] - \deltatext.
      \qedhere
  \end{align*}
\end{proof}

\begin{lemma}[\cite{breaking}]\label{lm:fully-periodic-pat-occ}
  Let $\tau \in [1 \dd \floor{\tfrac{\Textlen}{2}}]$ and
  $\Pat \in \Sigma^{m}$ be a $\tau$-periodic pattern satisfying
  $\RunEndPat{\Pat}{\tau} > m$. Denote $b = \RangeBegTwo{\Pat}{\Text}$,
  $e = \RangeEndTwo{\Pat}{\Text}$, $s = \HeadPat{\Pat}{\tau}$, and
  $H = \RootPat{\Pat}{\tau}$.
  Assume that $\RMinusFour{s}{H}{\tau}{\Text} \neq
  \emptyset$ and
  let $(b', e')$ be such that $\{\SA{\Text}[i] : i \in (b' \dd e']\} =
  \RMinusFour{s}{H}{\tau}{\Text}$. Then,
  \begin{align*}
    \OccTwo{\Pat}{\Text} \cap \RMinusTwo{\tau}{\Text}
      &= \{\SA{\Text}[i] : i \in (b \dd e']\}.
  \end{align*}
\end{lemma}

\begin{lemma}\label{lm:bmin-bit}
  Let $\tau \in [1 \dd \floor{\tfrac{\Textlen}{2}}]$. Let
  $H \in \Sigma^{+}$, $s \in \Zn$, and $b, e \in [0 \dd \Textlen]$ be
  such that $\RMinusFour{s}{H}{\tau}{\Text} = \{\SA{\Text}[i] : i \in (b \dd
  e]\}$.  Let $\BitvectorMin[1 \dd \Textlen]
  = \MinPosBitvectorMinusTwo{\tau}{\Text}$
  (\cref{def:min-pos-bitvector-minus}). If $\BitvectorMin[i] = 1$ for
  some $i \in (b \dd e]$, then for every $i' \in (i \dd e]$, it holds
  $\SA{\Text}[i'] > \SA{\Text}[i]$.
\end{lemma}
\begin{proof}
  Denote $j = \SA{\Text}[i]$. By \cref{def:min-pos-bitvector-minus},
  $\BitvectorMin[i] = 1$ implies that
  $j \in \RMinMinusTwo{\tau}{\Text}$. Letting $\Pat
  = \Text[j \dd \RunEndPos{j}{\tau}{\Text})$, we thus have $j
  = \min \OccTwo{\Pat}{\Text} \cap \RMinusTwo{\tau}{\Text}$
  (see \cref{def:rmin}). Let $i' \in (i \dd e]$ and suppose that
  $\SA{\Text}[i'] < \SA{\Text}[i]$. Denote $j' = \SA{\Text}[i']$ and observe that
  $\RunEndPos{j'}{\tau}{\Text} - j' \geq \RunEndPos{j}{\tau}{\Text} -
  j$, since otherwise
  by \cref{lm:R-lex-block-pos}\eqref{lm:R-lex-block-pos-it-3} we would
  have $\Text[\SA{\Text}[i'] \dd \Textlen] \prec \Text[\SA{\Text}[i] \dd \Textlen]$.
  Denoting $\Pat' = \Text[j' \dd \RunEndPos{j'}{\tau}{\Text})$, we
  thus obtain by $\RootPos{j}{\tau}{\Text}
  = \RootPos{j'}{\tau}{\Text}$ and $\HeadPos{j}{\tau}{\Text}
  = \HeadPos{j'}{\tau}{\Text}$ that $\Pat$ is a prefix of $\Pat'$.
  Thus, $j' \in \OccTwo{\Pat}{\Text}$. Recall that we also have
  $j' \in \RMinusTwo{\tau}{\Text}$, and hence
  $j' \in \OccTwo{\Pat}{\Text} \cap \RMinusTwo{\tau}{\Text}$.  By $j'
  < j$, we thus obtain a contradiction with $j
  = \min \OccTwo{\Pat}{\Text} \cap \RMinusTwo{\tau}{\Text}$.
\end{proof}

\begin{lemma}\label{lm:fully-periodic-pat-min-occ}
  Let $\tau \in [1 \dd \floor{\tfrac{\Textlen}{2}}]$ and
  $\Pat \in \Sigma^{m}$ be a $\tau$-periodic pattern satisfying
  $\RunEndPat{\Pat}{\tau} > m$. Denote $b = \RangeBegTwo{\Pat}{\Text}$
  and $e = \RangeEndTwo{\Pat}{\Text}$.  Let
  $\BitvectorMin[1 \dd \Textlen] = \MinPosBitvectorMinusTwo{\tau}{\Text}$
  (\cref{def:min-pos-bitvector-minus}).  Then,
  $\OccTwo{\Pat}{\Text} \cap \RMinusTwo{\tau}{\Text} \neq \emptyset$
  holds if and only if $\Rank{\BitvectorMin}{1}{e}
  - \Rank{\BitvectorMin}{1}{b} \neq 0$.  Moreover, if
  $\OccTwo{\Pat}{\Text} \cap \RMinusTwo{\tau}{\Text} \neq \emptyset$
  then
  \begin{align*}
    \min \OccTwo{\Pat}{\Text} \cap \RMinusTwo{\tau}{\Text}
      &= \SA{\Text}[k],
  \end{align*}
  where $k = \Select{\BitvectorMin}{1}{\Rank{\BitvectorMin}{1}{b} + 1}$.
\end{lemma}
\begin{proof}

  Denote $s = \HeadPat{\Pat}{\tau}$, $H = \RootPat{\Pat}{\tau}$, and
  $p = |H|$. Recall that
  by \cref{lm:periodic-pos-lce}\eqref{lm:periodic-pos-lce-it-1}, it
  holds $\OccTwo{\Pat[1 \dd 3\tau-1]}{\Text}
  = \RFour{s}{H}{\tau}{\Text}$, and hence $\RFour{s}{H}{\tau}{\Text}$
  occupies a contiguous block of positions in $\SA{\Text}$. Thus, letting $x
  = \RangeBegTwo{\Pat[1 \dd 3\tau-1]}{\Text}$ and $z
  = \RangeEndTwo{\Pat[1 \dd 3\tau-1]}{\Text}$, it holds $\{\SA{\Text}[i] :
  i \in (x \dd z]\} = \RFour{s}{H}{\tau}{\Text}$.  Recall also that
  by \cref{lm:R-lex-block-pos}, all elements of
  $\RMinusFour{s}{H}{\tau}{\Text}$ precede the elements of
  $\RPlusFour{s}{H}{\tau}{\Text}$ in $\SA{\Text}$. Thus, there exists $y \in
  [x \dd z]$ such that $\{\SA{\Text}[i] : i \in (x \dd y]\}
  = \RMinusFour{s}{H}{\tau}{\Text}$, and $\{\SA{\Text}[i] : i \in (y \dd z]\}
  = \RPlusFour{s}{H}{\tau}{\Text}$.  Next, observe that by
  $\OccTwo{\Pat}{\Text} \subseteq \OccTwo{\Pat[1 \dd 3\tau-1]}{\Text}
  = \RFour{s}{H}{\tau}{\Text}$, we have $x \leq b \leq e \leq z$.
  Lastly, note that by \cref{def:min-pos-bitvector-minus}, all bits of
  $\BitvectorMin(y \dd z]$ are $0$.

  Let us first assume that
  $\OccTwo{\Pat}{\Text} \cap \RMinusTwo{\tau}{\Text} \neq \emptyset$.
  Denote $j = \min \OccTwo{\Pat}{\Text} \cap \RMinusTwo{\tau}{\Text}$.
  By $j \in \OccTwo{\Pat}{\Text}$, it holds
  $\lcp{\Pat}{\Text[j \dd \Textlen]} \geq 3\tau - 1$.
  By \cref{lm:periodic-pos-lce}\eqref{lm:periodic-pos-lce-it-1}, we
  then obtain $\HeadPos{j}{\tau}{\Text} = s$ and
  $\RootPos{j}{\tau}{\Text} = H$. Thus,
  $j \in \RMinusFour{s}{H}{\tau}{\Text}$.  Note also that $\Pat$ being
  a prefix of $\Text[j \dd \Textlen]$ implies that
  $\lcp{\Text[j \dd \Textlen]}{\Text[j + p \dd \Textlen]}
  \geq \lcp{\Pat[1 \dd m]}{\Pat[1 + p \dd m]}$. Thus,
  $\RunEndPos{j}{\tau}{\Text} - j
  = \RunEndPat{\Text[j \dd \Textlen]}{\tau} - 1 = p
  + \lcp{\Text[j \dd \Textlen]}{\Text[j + p \dd \Textlen]} \geq p
  + \lcp{\Pat[1 \dd m]}{\Pat[1 + p \dd m]} = \RunEndPat{\Pat}{\tau} -
  1 = m$.  In other words, letting $\Pat'
  = \Text[j \dd \RunEndPos{j}{\tau}{\Text})$, it holds $|\Pat'| \geq
  |\Pat|$. Observe that this implies that $j
  = \min \OccTwo{\Pat'}{\Text} \cap \RMinusTwo{\tau}{\Text}$, since
  the existence of
  $j' \in \OccTwo{\Pat'}{\Text} \cap \RMinusTwo{\tau}{\Text}$
  satisfying $j' < j$ would imply
  $j' \in \OccTwo{\Pat}{\Text} \cap \RMinusTwo{\tau}{\Text}$, contradicting
  the definition of $j$. Thus, by \cref{def:rmin},
  $j \in \RMinMinusTwo{\tau}{\Text}$. Next, note that
  $j \in \RMinusFour{s}{H}{\tau}{\Text}$ implies
  $\RMinusFour{s}{H}{\tau}{\Text} \neq \emptyset$.
  By \cref{lm:fully-periodic-pat-occ}, we thus obtain that
  $\OccTwo{\Pat}{\Text} \cap \RMinusTwo{\tau}{\Text} = \{\SA{\Text}[i] :
  i \in (b \dd y]\}$ (in particular, this implies $b < y \leq e$).
  Consequently,
  $j \in \OccTwo{\Pat}{\Text} \cap \RMinusTwo{\tau}{\Text}$ and
  $j \in \RMinMinusTwo{\tau}{\Text}$ imply that there exists $t \in
  (b \dd y]$ such that $\BitvectorMin[t] = 1$
  (see \cref{def:min-pos-bitvector-minus}). Thus, by $b < y \leq e$,
  it follows that $\Rank{\BitvectorMin}{1}{e}
  - \Rank{\BitvectorMin}{1}{b} \neq 0$.

  Let us now assume $\Rank{\BitvectorMin}{1}{e}
  - \Rank{\BitvectorMin}{1}{b} \neq 0$. By $x \leq b \leq e \leq z$
  and all bits in $\BitvectorMin(y \dd z]$ being $0$, it follows that
  there exists $t \in (x \dd y]$ such that $b < t \leq e$. Since as
  noted above, $\{\SA{\Text}[i] : i \in (x \dd y]\}
  = \RMinusFour{s}{H}{\tau}{\Text} \subseteq \RMinusTwo{\tau}{\Text}$,
  we thus obtain that
  $\SA{\Text}[i] \in \OccTwo{\Pat}{\Text} \cap \RMinusTwo{\tau}{\Text}$.
  Hence,
  $\OccTwo{\Pat}{\Text} \cap \RMinusTwo{\tau}{\Text} \neq \emptyset$.

  We now prove the remaining claim. Assume that
  $\OccTwo{\Pat}{\Text} \cap \RMinusTwo{\tau}{\Text} \neq \emptyset$.
  As noted above, we then have
  $\OccTwo{\Pat}{\Text} \cap \RMinusTwo{\tau}{\Text} = \{\SA{\Text}[i] :
  i \in (b \dd y]\}$.  Let $j
  = \min \OccTwo{\Pat}{\Text} \cap \RMinusTwo{\tau}{\Text}$ and $i \in
  (b \dd y]$ be such that $\SA{\Text}[i] = j$. Above we proved that
  $j \in \RMinMinusTwo{\tau}{\Text}$. Thus, $\BitvectorMin[i] = 1$.
  By definition of $k$ and rank/select queries, we thus obtain $k \in
  (b \dd i]$. Suppose $k < i$.  Note that
  $\SA{\Text}[k] \in \OccTwo{\Pat}{\Text} \cap \RMinusTwo{\tau}{\Text}$.  On
  the other hand, by $x \leq b < k < i \leq y$ we obtain
  $\SA{\Text}[k], \SA{\Text}[i] \in \RMinusFour{s}{H}{\tau}{\Text}$. Thus it follows
  by $\BitvectorMin[k] = 1$ and \cref{lm:bmin-bit} that $\SA{\Text}[i]
  > \SA{\Text}[k]$. Thus contradicts $\SA{\Text}[i]
  = \min \OccTwo{\Pat}{\Text} \cap \RMinusTwo{\tau}{\Text}$.  We thus
  have $i = k$, and hence
  $\min \OccTwo{\Pat}{\Text} \cap \RMinusTwo{\tau}{\Text} = \SA{\Text}[i]
  = \SA{\Text}[k]$.
\end{proof}

\paragraph{Query Algorithms}

\begin{proposition}\label{pr:minocc-index-fully-periodic-pattern}
  Let $\tau = \mu\log_{\AlphabetSize} \Textlen$. Let $\Pat \in \Sigma^{m}$
  be a $\tau$-periodic pattern satisfying $\RunEndPat{\Pat}{\tau} > |\Pat|$
  and $\OccTwo{\Pat}{\Text} \neq \emptyset$. Given
  $\MinOccIndexPeriodic{\Text}$ (\cref{sec:minocc-index-periodic-structure}),
  $\HeadPat{\Pat}{\tau}$,
  $|\Pat|$, and the packed representation of $\RootPat{\Pat}{\tau}$, we
  can in $\bigO(\log \log \Textlen)$ time compute
  $\min \OccTwo{\Pat}{\Text}$.
\end{proposition}
\begin{proof}
  The algorithm proceeds in five steps:
  \begin{enumerate}
  \item Initialize the set $\mathcal{A} := \emptyset$.
  \item Using \cref{pr:nav-index-periodic}\eqref{pr:nav-index-periodic-it-4},
    in $\bigO(\log \log \Textlen)$ time we compute
    $(b, e) := (\RangeBegTwo{\Pat}{\Text}, \RangeEndTwo{\Pat}{\Text})$.
  \item Using \cref{th:bin-rank-select}, in $\bigO(1)$ time compute
    $r_b = \Rank{\BitvectorMin}{1}{b}$ and $r_e
    = \Rank{\BitvectorMin}{1}{e}$. If $r_e - r_b = 0$, then by
    \cref{lm:fully-periodic-pat-min-occ}, it holds $\OccTwo{\Pat}{\Text} \cap \RMinusTwo{\tau}{\Text} = \emptyset$,
    and this step is complete. Otherwise, by \cref{lm:fully-periodic-pat-min-occ},
    it holds $\OccTwo{\Pat}{\Text} \cap \RMinusTwo{\tau}{\Text} \neq \emptyset$,
    and we proceed as follows.
    Using \cref{th:bin-rank-select}, in $\bigO(1)$ time we
    compute $k = \Select{\BitvectorMin}{1}{r_b + 1}$. Using
    \cref{pr:nav-index-periodic}\eqref{pr:nav-index-periodic-it-3},
    in $\bigO(\log \log \Textlen)$ time we compute $j := \SA{\Text}[k]$
    (note that \cref{pr:nav-index-periodic}\eqref{pr:nav-index-periodic-it-3}
    uses $\NavPeriodic{\Text}$, which is a component
    of $\MinOccIndexPeriodic{\Text}$; see \cref{sec:minocc-index-periodic-structure}).
    By \cref{lm:fully-periodic-pat-min-occ}, it then holds
    $j = \min \OccTwo{\Pat}{\Text} \cap \RMinusTwo{\tau}{\Text}$.  We add
    $j$ to the set $\mathcal{A}$.
  \item Using the second part of $\MinOccIndexPeriodic{\Text}$ and
    the symmetric variant of \cref{lm:fully-periodic-pat-min-occ},
    analogously as above, in $\bigO(\log \log \Textlen)$ time we check if
    $\OccTwo{\Pat}{\Text} \cap \RPlusTwo{\tau}{\Text} \neq \emptyset$,
    and if so we compute $j'
    = \min \OccTwo{\Pat}{\Text} \cap \RPlusTwo{\tau}{\Text}$, and add
    to $\mathcal{A}$.
  \item Observe that
    by \cref{lm:periodic-pos-lce}\eqref{lm:periodic-pos-lce-it-1}, it
    holds $\OccTwo{\Pat}{\Text} \subseteq \RTwo{\tau}{\Text}$.  Thus,
    the assumption $\OccTwo{\Pat}{\Text} \neq \emptyset$ implies that
    either
    $\OccTwo{\Pat}{\Text} \cap \RMinusTwo{\tau}{\Text} \neq \emptyset$
    or
    $\OccTwo{\Pat}{\Text} \cap \RPlusTwo{\tau}{\Text} \neq \emptyset$.
    Thus, $\mathcal{A} \neq \emptyset$, and we have
    $\min \OccTwo{\Pat}{\Text} = \min \mathcal{A}$.  In $\bigO(1)$
    time we thus compute $p := \min \mathcal{A}$ and return as the
    answer.
  \end{enumerate}
  In total, we spend $\bigO(\log \log \Textlen)$ time.
\end{proof}

\begin{proposition}\label{pr:minocc-index-periodic-query-partially-periodic-minus-pos}
  Let $\tau = \mu\log_{\AlphabetSize} \Textlen$, $j \in \RMinusTwo{\tau}{\Text}$,
  and $\ell \geq 3\tau - 1$ be such that
  $\RunEndPos{j}{\tau}{\Text} < j + \ell \leq \Textlen + 1$. Given
  $\MinOccIndexPeriodic{\Text}$ (\cref{sec:minocc-index-periodic-structure}) and
  the pair $(j,\ell)$, we can in $\bigO(\log \log \Textlen)$ time compute
  $\min \OccTwo{\Pat}{\Text}$, where $\Pat = \Text[j \dd j + \ell)$.
\end{proposition}
\begin{proof}
  The algorithm proceeds in three steps:
  \begin{enumerate}
  \item Using \cref{pr:nav-index-periodic}\eqref{pr:nav-index-periodic-it-1a}, in $\bigO(1)$ time
    we compute $H := \RootPos{j}{\tau}{\Text}$, $p := |\HeadPos{j}{\tau}{\Text}| = |H|$, and
    $e^{\rm full} := \RunEndFullPos{j}{\tau}{\Text}$.
    Note that by \cref{lm:partially-periodic-substring},
    $\Pat$ is $\tau$-periodic (\cref{def:periodic-pattern}) and it holds
    $\RunEndPat{\Pat}{\tau} \leq |\Pat|$,
    $\TypePat{\Pat}{\tau} = -1$,
    $\RootPat{\Pat}{\tau} = H$,
    $\HeadPat{\Pat}{\tau} = p$, and
    $\RunEndFullPat{\Pat}{\tau} - 1 = \RunEndFullPos{j}{\tau}{\Text} - j = e^{\rm full} - j$.
    In $\bigO(1)$ time we compute $t := |\PowStr{\tau}{H}| = p \cdot \lceil \tfrac{\tau}{p} \rceil$
    and $\deltatext = (\RunEndFullPat{\Pat}{\tau} - 1) - |\PowStr{\tau}{H}| = (e^{\rm full} - j) - t$.
  \item Denote $\Pat' = \Pat(\deltatext \dd \ell] = \Text[j + \deltatext \dd j + \ell)$.
    Using \cref{pr:nav-index-periodic}\eqref{pr:nav-index-periodic-it-1b},
    in $\bigO(\log \log \Textlen)$ time, we compute the pair $(b, e)$ defined by
    $b = |\{i \in [1 \dd m] : \Text[b_i \dd \Textlen] \prec
    \Pat'\}|$ and $e - b = |\{i \in [1 \dd m] :
    \Pat'\text{ is a prefix of }\Text[b_i \dd \Textlen]\}|$, where
    $(a_i)_{i \in [1 \dd m]} = \RunsMinusLexSortedTwo{\tau}{\Text}$ (\cref{def:runs-minus-lex-sorted})
    and $(b_i)_{i \in [1 \dd m]}$ is a sequence defined by
    $b_i = \RunEndFullPos{a_i}{\tau}{\Text} - |\PowStr{\tau}{\RootPos{a_i}{\tau}{\Text}}|$.
  \item Using \cref{th:three-sided-rmq}, in $\bigO(\log \log m_{\max}) = \bigO(\log \log \Textlen)$ time
    (where $m_{\max} = 2\Textlen / \tau$; see \cref{sec:minocc-index-periodic-structure}) we compute
    and return as the answer the value
    $j_{\min} := A_{\rm pos}[\ThreeSidedRMQ{A_{\rm pos}}{A_{\rm len}}{b}{e}{\deltatext}] - \deltatext$,
    where $A_{\rm pos}[1 \dd m]$ and $A_{\rm len}[1 \dd m]$ are as in \cref{sec:minocc-index-periodic-structure}.
    By \cref{lm:partially-periodic-pat-occ-min},
    it holds $j_{\min} = \min \OccTwo{\Pat}{\Text}$.
  \end{enumerate}
  In total, we spend $\bigO(\log \log \Textlen)$ time.
\end{proof}

\begin{proposition}\label{pr:minocc-index-periodic-query-partially-periodic-pos}
  Let $\tau = \mu\log_{\AlphabetSize} \Textlen$, $j \in \RTwo{\tau}{\Text}$,
  and $\ell \geq 3\tau - 1$ be such that
  $\RunEndPos{j}{\tau}{\Text} < j + \ell \leq \Textlen + 1$. Given
  $\MinOccIndexPeriodic{\Text}$ (\cref{sec:minocc-index-periodic-structure}) and
  the pair $(j,\ell)$, we can in $\bigO(\log \log \Textlen)$ time compute
  $\min \OccTwo{\Pat}{\Text}$, where $\Pat = \Text[j \dd j + \ell)$.
\end{proposition}
\begin{proof}
  The algorithm consists of two steps:
  \begin{enumerate}
  \item Using \cref{pr:nav-index-periodic}\eqref{pr:nav-index-periodic-it-1a}, in $\bigO(1)$ time
    we compute $\TypePos{j}{\tau}{\Text}$.
  \item If $\TypePos{j}{\tau}{\Text} = -1$, then we compute $\min \OccTwo{\Pat}{\Text}$ using
    \cref{pr:minocc-index-periodic-query-partially-periodic-minus-pos} in $\bigO(\log \log \Textlen)$ time.
    Otherwise (i.e., if $\TypePos{j}{\tau}{\Text} = +1$),
    we compute $\min \OccTwo{\Pat}{\Text}$ using the
    symmetric version of \cref{pr:minocc-index-periodic-query-partially-periodic-minus-pos} and
    the second part of $\MinOccIndexPeriodic{\Text}$ in $\bigO(\log \log \Textlen)$.
  \end{enumerate}
  In total, we spend $\bigO(\log \log \Textlen)$ time.
\end{proof}

\begin{proposition}\label{pr:minocc-index-periodic-query-fully-periodic-pos}
  Let $\tau = \mu\log_{\AlphabetSize} \Textlen$, $j \in \RTwo{\tau}{\Text}$,
  and $\ell \geq 3\tau - 1$ be such that $j + \ell \leq \RunEndPos{j}{\tau}{\Text}$.
  Given $\MinOccIndexPeriodic{\Text}$ (\cref{sec:minocc-index-periodic-structure}) and
  the pair $(j,\ell)$, we can in $\bigO(\log \log \Textlen)$ time compute
  $\min \OccTwo{\Pat}{\Text}$, where $\Pat = \Text[j \dd j + \ell)$.
\end{proposition}
\begin{proof}
  The algorithm proceeds in two steps:
  \begin{enumerate}
  \item Using \cref{pr:nav-index-periodic}\eqref{pr:nav-index-periodic-it-1a}, in $\bigO(1)$ time
    we compute the packed representation of $H := \RootPos{j}{\tau}{\Text}$, and the value
    $s := \HeadPos{j}{\tau}{\Text}$. Note that by \cref{lm:fully-periodic-substring},
    $\Pat$ is $\tau$-periodic (\cref{def:periodic-pattern}) and it holds
    $\RunEndPat{\Pat}{\tau} = |\Pat| + 1$, $\RootPat{\Pat}{\tau} = H$, and $\HeadPat{\Pat}{\tau} = s$.
  \item Using \cref{pr:minocc-index-fully-periodic-pattern}, in $\bigO(\log \log \Textlen)$ time
    we compute and return $\min \OccTwo{\Pat}{\Text}$.
  \end{enumerate}
  In total, we spend $\bigO(\log \log \Textlen)$ time.
\end{proof}

\begin{proposition}\label{pr:minocc-index-periodic-query-pos}
  Let $\tau = \mu\log_{\AlphabetSize} \Textlen$, $j \in \RTwo{\tau}{\Text}$,
  and $\ell \geq 3\tau - 1$ be such that $j + \ell \leq \Textlen + 1$. Given
  $\MinOccIndexPeriodic{\Text}$ (\cref{sec:minocc-index-periodic-structure}) and
  $(j,\ell)$, we can in $\bigO(\log \log \Textlen)$ time compute
  $\min \OccTwo{\Pat}{\Text}$, where $\Pat = \Text[j \dd j + \ell)$.
\end{proposition}
\begin{proof}
  The algorithm consists of two steps:
  \begin{enumerate}
  \item Using \cref{pr:nav-index-periodic}\eqref{pr:nav-index-periodic-it-1a}, in $\bigO(1)$ time
    we compute $e := \RunEndPos{j}{\tau}{\Text}$.
  \item If $e < j + \ell$, then we compute $\min \OccTwo{\Pat}{\Text}$ using
    \cref{pr:minocc-index-periodic-query-partially-periodic-pos} in $\bigO(\log \log \Textlen)$ time.
    Otherwise, we compute $\min \OccTwo{\Pat}{\Text}$ using \cref{pr:minocc-index-periodic-query-fully-periodic-pos}
    in $\bigO(\log \log \Textlen)$ time.
  \end{enumerate}
  In total, we spend $\bigO(\log \log \Textlen)$ time.
\end{proof}

\begin{proposition}\label{pr:minocc-index-periodic-query-partially-periodic-minus-pat}
  Let $\tau = \mu\log_{\AlphabetSize} \Textlen$ and $\Pat \in \Sigma^{m}$ be a $\tau$-periodic
  pattern satisfying $\TypePat{\Pat}{\tau} = -1$, $\RunEndPat{\Pat}{\tau} \leq |\Pat|$, and
  $\OccTwo{\Pat}{\Text} \neq \emptyset$. Given
  $\MinOccIndexPeriodic{\Text}$ (\cref{sec:minocc-index-periodic-structure}) and
  the packed representation of $\Pat$, we can in
  $\bigO(\log \log \Textlen + m / \log_{\AlphabetSize} \Text)$ time compute
  $\min \OccTwo{\Pat}{\Text}$.
\end{proposition}
\begin{proof}
  The algorithm proceeds in three steps:
  \begin{enumerate}
  \item Using \cref{pr:nav-index-periodic}\eqref{pr:nav-index-periodic-it-2a},
    in $\bigO(1)$ time
    we compute $H := \RootPat{\Pat}{\tau}$, $p := |\RootPat{\Pat}{\tau}| = |H|$.
    Using \cref{pr:nav-index-periodic}\eqref{pr:nav-index-periodic-it-2b},
    we determine $e^{\rm full} := \RunEndFullPat{\Pat}{\tau}$ in
    $\bigO(1 + m / \log_{\AlphabetSize} \Textlen)$ time.
    In $\bigO(1)$ time we compute $t := |\PowStr{\tau}{H}| = p \cdot \lceil \tfrac{\tau}{p} \rceil$
    and $\deltatext = (\RunEndFullPat{\Pat}{\tau} - 1) - |\PowStr{\tau}{H}| = (e^{\rm full} - 1) - t$.
  \item Denote $\Pat' = \Pat(\deltatext \dd \ell]$.
    Using \cref{pr:nav-index-periodic}\eqref{pr:nav-index-periodic-it-2c},
    in $\bigO(\log \log \Textlen + m / \log_{\AlphabetSize} \Textlen)$ time,
    we compute the pair $(b, e)$ defined by
    $b = |\{i \in [1 \dd m] : \Text[b_i \dd \Textlen] \prec
    \Pat'\}|$ and $e - b = |\{i \in [1 \dd m] :
    \Pat'\text{ is a prefix of }\Text[b_i \dd \Textlen]\}|$, where
    $(a_i)_{i \in [1 \dd m]} = \RunsMinusLexSortedTwo{\tau}{\Text}$ (\cref{def:runs-minus-lex-sorted})
    and $(b_i)_{i \in [1 \dd m]}$ is a sequence defined by
    $b_i = \RunEndFullPos{a_i}{\tau}{\Text} - |\PowStr{\tau}{\RootPos{a_i}{\tau}{\Text}}|$.
  \item Using \cref{th:three-sided-rmq}, in $\bigO(\log \log m_{\max}) = \bigO(\log \log \Textlen)$ time
    (where $m_{\max} = 2\Textlen / \tau$; see \cref{sec:minocc-index-periodic-structure}) we compute
    and return as the answer the value
    $j_{\min} := A_{\rm pos}[\ThreeSidedRMQ{A_{\rm pos}}{A_{\rm len}}{b}{e}{\deltatext}] - \deltatext$,
    where $A_{\rm pos}[1 \dd m]$ and $A_{\rm len}[1 \dd m]$ are as in \cref{sec:minocc-index-periodic-structure}.
    By \cref{lm:partially-periodic-pat-occ-min},
    it holds $j_{\min} = \min \OccTwo{\Pat}{\Text}$.
  \end{enumerate}
  In total, we spend $\bigO(\log \log \Textlen + m / \log_{\AlphabetSize} \Textlen)$ time.
\end{proof}

\begin{proposition}\label{pr:minocc-index-periodic-query-partially-periodic-pat}
  Let $\tau = \mu\log_{\AlphabetSize} \Textlen$ and $\Pat \in \Sigma^{m}$ be a $\tau$-periodic
  pattern satisfying $\RunEndPat{\Pat}{\tau} \leq |\Pat|$ and
  $\OccTwo{\Pat}{\Text} \neq \emptyset$. Given
  $\MinOccIndexPeriodic{\Text}$ (\cref{sec:minocc-index-periodic-structure}) and
  the packed representation of $\Pat$, we can in
  $\bigO(\log \log \Textlen + m / \log_{\AlphabetSize} \Text)$ time compute
  $\min \OccTwo{\Pat}{\Text}$.
\end{proposition}
\begin{proof}
  The algorithm consists of two steps:
  \begin{enumerate}
  \item Using \cref{pr:nav-index-periodic}\eqref{pr:nav-index-periodic-it-2b},
    in $\bigO(1 + m / \log_{\AlphabetSize} \Textlen)$ time
    we compute $\TypePat{\Pat}{\tau}$.
  \item If $\TypePat{\Pat}{\tau} = -1$, then we compute $\min \OccTwo{\Pat}{\Text}$ using
    \cref{pr:minocc-index-periodic-query-partially-periodic-minus-pat} in
    $\bigO(\log \log \Textlen + m / \log_{\AlphabetSize} \Textlen)$ time.
    Otherwise (i.e., if $\TypePat{\Pat}{\tau} = +1$),
    we compute $\min \OccTwo{\Pat}{\Text}$ using the
    symmetric version of \cref{pr:minocc-index-periodic-query-partially-periodic-minus-pat} and
    the second part of $\MinOccIndexPeriodic{\Text}$ in
    $\bigO(\log \log \Textlen + m / \log_{\AlphabetSize} \Textlen)$.
  \end{enumerate}
  In total, we spend $\bigO(\log \log \Textlen + m / \log_{\AlphabetSize} \Textlen)$ time.
\end{proof}

\begin{proposition}\label{pr:minocc-index-periodic-query-fully-periodic-pat}
  Let $\tau = \mu\log_{\AlphabetSize} \Textlen$ and $\Pat \in \Sigma^{m}$ be a $\tau$-periodic
  pattern satisfying $\RunEndPat{\Pat}{\tau} > |\Pat|$ and
  $\OccTwo{\Pat}{\Text} \neq \emptyset$. Given
  $\MinOccIndexPeriodic{\Text}$ (\cref{sec:minocc-index-periodic-structure}) and
  the packed representation of $\Pat$, we can in
  $\bigO(\log \log \Textlen)$ time compute
  $\min \OccTwo{\Pat}{\Text}$.
\end{proposition}
\begin{proof}
  The algorithm proceeds in two steps:
  \begin{enumerate}
  \item Using \cref{pr:nav-index-periodic}\eqref{pr:nav-index-periodic-it-2a}, in $\bigO(1)$ time
    we compute the packed representation of $H := \RootPat{\Pat}{\tau}$, and the value
    $s := \HeadPat{\Pat}{\tau}$.
  \item Using \cref{pr:minocc-index-fully-periodic-pattern},
    in $\bigO(\log \log \Textlen)$ time
    we compute and return $\min \OccTwo{\Pat}{\Text}$.
  \end{enumerate}
  In total, we spend $\bigO(\log \log \Textlen)$ time.
\end{proof}

\begin{proposition}\label{pr:minocc-index-periodic-query-pat}
  Let $\tau = \mu\log_{\AlphabetSize} \Textlen$ and $\Pat \in \Sigma^{m}$ be a $\tau$-periodic
  pattern satisfying $\OccTwo{\Pat}{\Text} \neq \emptyset$. Given
  $\MinOccIndexPeriodic{\Text}$ (\cref{sec:minocc-index-periodic-structure}) and
  the packed representation of $\Pat$, we can in
  $\bigO(\log \log \Textlen + m / \log_{\AlphabetSize} \Textlen)$ time compute
  $\min \OccTwo{\Pat}{\Text}$.
\end{proposition}
\begin{proof}
  The algorithm consists of two steps:
  \begin{enumerate}
  \item Using \cref{pr:nav-index-periodic}\eqref{pr:nav-index-periodic-it-2b},
    in $\bigO(1 + m / \log_{\AlphabetSize} \Textlen)$ time
    we compute $e := \RunEndPat{\Pat}{\tau}$.
  \item If $e \leq m$, then we compute $\min \OccTwo{\Pat}{\Text}$ using
    \cref{pr:minocc-index-periodic-query-partially-periodic-pat}
    in $\bigO(\log \log \Textlen + m / \log_{\AlphabetSize} \Textlen)$ time.
    Otherwise, we compute $\min \OccTwo{\Pat}{\Text}$ using
    \cref{pr:minocc-index-periodic-query-fully-periodic-pat}
    in $\bigO(\log \log \Textlen)$ time.
  \end{enumerate}
  In total, we spend $\bigO(\log \log \Textlen + m / \log_{\AlphabetSize} \Textlen)$ time.
\end{proof}

%% file: index/periodic/construction.tex
\subsubsection{Construction Algorithm}\label{sec:minocc-index-periodic-construction}

\paragraph{Combinatorial Properties}

\begin{definition}\label{def:delete}
  For every $S \in \Sigma^{+}$ and $p \in [1 \dd |S|]$, we define
  $\DeletePos{S}{p} = S'$, such that $S' = XY$,
  where $X = S[1 \dd p)$ and $Y = S(p \dd |S|]$. We then extend it
  as follows. Let $Q = \{p_1, \ldots, p_k\}$, where $k \geq 0$ and
  $1 \leq p_1 < p_2 < \dots < p_k \leq |S|$. We define
  \[
    \DeleteSubseq{S}{Q} =
      \begin{cases}
        S & \text{if }Q = \emptyset,\\
        \DeleteSubseq{\DeletePos{S}{p_k}}{Q \setminus \{p_k\}} & \text{otherwise}.
      \end{cases}
  \]
\end{definition}

\begin{definition}\label{def:insert}
   Let $S \in \Sigma^{*}$ and $Q = \{(p_1,c_1), \ldots, (p_k,c_k)\}
   \subseteq \Zp \times \Sigma$, where
   $k \geq 0$ and $p_1 < p_2 < \dots < p_k$.
   We define $\InsertSubseq{S}{Q}$ as a string $S'$ satisfying the following
   conditions:
   \begin{enumerate}
   \item For every $i \in [1 \dd k]$, it holds $p_i \in [1 \dd |S'|]$ and
    $S'[p_i] = c_i$,
   \item $\DeleteSubseq{S'}{\{p_1, \ldots, p_k\}} = S$.
  \end{enumerate}
\end{definition}

\begin{lemma}\label{lm:rmin-text-block}
  Let $\tau \in [1 \dd \lfloor \tfrac{\Textlen}{2} \rfloor]$.  For
  every $j \in \RMinMinusTwo{\tau}{\Text}$ such that $j -
  1 \in \RTwo{\tau}{\Text}$, it holds $j -
  1 \in \RMinMinusTwo{\tau}{\Text}$.
\end{lemma}
\begin{proof}
  Note that
  $j \in \RMinMinusTwo{\tau}{\Text} \subseteq \RTwo{\tau}{\Text}$ and
  $j-1 \in \RTwo{\tau}{\Text}$ imply by \cref{lm:R-text-block} that
  $\RunEndPos{j-1}{\tau}{\Text} = \RunEndPos{j}{\tau}{\Text}$ and
  $\TypePos{j-1}{\tau}{\Text} = \TypePos{j}{\tau}{\Text} = -1$.  Denote
  $\Pat = \Text[j \dd \RunEndPos{j}{\tau}{\Text})$ and $\Pat'
  = \Text[j-1 \dd \RunEndPos{j-1}{\tau}{\Text})
  = \Text[j-1 \dd \RunEndPos{j}{\tau}{\Text})$.  Suppose that
  $j-1 \not\in \RMinMinusTwo{\tau}{\Text}$. Since we established
  that $j-1 \in \RMinusTwo{\tau}{\Text}$, by
  \cref{def:rmin} there must exist $j' < j-1$ such that
  $j' \in \OccTwo{\Pat'}{\Text} \cap \RMinusTwo{\tau}{\Text}$.  Since
  $\Pat$ is a suffix of $\Pat'$ and $|\Pat| + 1 = |\Pat'|$, this
  implies that $j' + 1 \in \OccTwo{\Pat}{\Text}$. By \cref{lm:end},
  $|\Pat| \geq 3\tau - 1$.  Thus, we have $\LCE{\Text}{j'+1}{j} \geq
  3\tau - 1$. Consequently,
  by \cref{lm:periodic-pos-lce}\eqref{lm:periodic-pos-lce-it-2}, we
  obtain $j' + 1 \in \RTwo{\tau}{\Text}$. By
  $j' \in \RMinusTwo{\tau}{\Text}$ and \cref{lm:R-text-block}, we thus
  obtain $j'+1 \in \RMinusTwo{\tau}{\Text}$.  Combining with the
  above, we therefore have
  $j'+1 \in \OccTwo{\Pat}{\Text} \cap \RMinusTwo{\tau}{\Text}$.  This
  contradicts $j
  = \min \OccTwo{\Pat}{\Text} \cap \RMinusTwo{\tau}{\Text}$, since
  $j'+1<j$.  We have thus proved that
  $j-1 \in \RMinMinusTwo{\tau}{\Text}$.
\end{proof}

\begin{lemma}\label{lm:rmin-block-size}
  Let $\tau \in [1 \dd \lfloor \tfrac{\Textlen}{2} \rfloor]$.
  For every $j \in \RPrimMinusTwo{\tau}{\Text}$, it holds
  $\RunMinEndPos{j}{\tau}{\Text} - j \leq \min(p, r)$, where
  $H = \RootPos{j}{\tau}{\Text}$, $p = |H|$ and
  $r = \RunEndPos{j}{\tau}{\Text} - j - 3\tau + 2$.
\end{lemma}
\begin{proof}
  By \cref{lm:end}, it holds
  $[j \dd j + r) \subseteq \RTwo{\tau}{\Text}$,
  $j-1 \not\in \RTwo{\tau}{\Text}$, and $j+r \not\in \RTwo{\tau}{\Text}$.
  Thus, by \cref{def:emin}, we immediately obtain
  $\RunMinEndPos{j}{\tau}{\Text} \leq j + r$, or equivalently,
  $\RunMinEndPos{j}{\tau}{\Text} - j \leq r$.

  Next, we prove that $\RunMinEndPos{j}{\tau}{\Text} - j \leq p$.
  Suppose that $\RunMinEndPos{j}{\tau}{\Text} > j + p$. Then,
  $[j \dd j + p] \subseteq \RMinMinusTwo{\tau}{\Text}$.
  In particular, $j + p \in \RMinusTwo{\tau}{\Text}$.
  Denote $\Pat = \Text[j+p \dd \RunEndPos{j+p}{\tau}{\Text})$.
  By \cref{lm:R-text-block}, $\RunEndPos{j+p}{\tau}{\Text} =
  \RunEndPos{j}{\tau}{\Text}$. Thus, $\Pat =
  \Text[j+p \dd \RunEndPos{j}{\tau}{\Text})$.
  Recall that $j+p \in \RMinMinusTwo{\tau}{\Text}$ implies
  that $j+p = \min \OccTwo{\Pat}{\Text} \cap \RMinusTwo{\tau}{\Text}$.
  By definition of $\RootPos{j}{\tau}{\Text}$ and $\RunEndPos{j}{\tau}{\Text}$, the substring
  $\Text[j \dd \RunEndPos{j}{\tau}{\Text})$
  has period $p$. Thus, $j \in \OccTwo{\Pat}{\Text}$. Since
  $j \in \RMinMinusTwo{\tau}{\Text}$ also implies $j \in \RMinusTwo{\tau}{\Text}$,
  we thus obtain $j \in \OccTwo{\Pat}{\Text} \cap \RMinusTwo{\tau}{\Text}$.
  This contradicts $j+p = \min \OccTwo{\Pat}{\Text} \cap \RMinusTwo{\tau}{\Text}$.
  We thus must have $\RunMinEndPos{j}{\tau}{\Text} - j \leq p$.

  Combining the above, we obtain the claim
  $\RunMinEndPos{j}{\tau}{\Text} - j \leq \min(p, r)$.
\end{proof}

\begin{lemma}\label{lm:rmin-equivalence}
  Let $\tau \in [1 \dd \lfloor \tfrac{\Textlen}{2} \rfloor]$.  Let
  $j \in \RPrimMinusTwo{\tau}{\Text}$, $H = \RootPos{j}{\tau}{\Text}$,
  $p = |H|$, and $r = \RunEndPos{j}{\tau}{\Text} - j - 3\tau + 2$.
  For every $j' \in [j \dd j + \min(p, r))$, the following conditions
  are equivalent:
  \begin{enumerate}
  \item\label{lm:rmin-equivalence-it-1}
    $j' \in \RMinMinusTwo{\tau}{\Text}$.
  \item\label{lm:rmin-equivalence-it-2}
    For every $i \in \RPrimMinusThree{H}{\tau}{\Text} \cap [1 \dd
    j)$, it holds $\RunEndFullPos{i}{\tau}{\Text} - i
    < \RunEndFullPos{j}{\tau}{\Text} - j' + p$, and either
    $\RunEndPos{i}{\tau}{\Text} - \RunEndFullPos{i}{\tau}{\Text}
    < \RunEndPos{j}{\tau}{\Text} - \RunEndFullPos{j}{\tau}{\Text}$ or
    $\RunEndFullPos{i}{\tau}{\Text} - i
    < \RunEndFullPos{j}{\tau}{\Text} - j'$.
  \end{enumerate}
\end{lemma}
\begin{proof}

  By $j' \in [j \dd j + r)$ and \cref{lm:end}, we have $[j \dd
  j'] \subseteq \RTwo{\tau}{\Text}$.  Thus, \cref{lm:R-text-block}
  implies that
  $\RootPos{j'}{\tau}{\Text} = \RootPos{j}{\tau}{\Text}$,
  $\RunEndPos{j'}{\tau}{\Text} = \RunEndPos{j}{\tau}{\Text}$,
  $\RunEndFullPos{j'}{\tau}{\Text} = \RunEndFullPos{j}{\tau}{\Text}$,
  $\TailPos{j'}{\tau}{\Text} = \TailPos{j}{\tau}{\Text}$, and
  $\TypePos{j'}{\tau}{\Text} = \TypePos{j}{\tau}{\Text} = -1$.
  Denote $s
  = \HeadPos{j'}{\tau}{\Text}$, $k = \ExpPos{j'}{\tau}{\Text}$, and $t
  = \TailPos{j'}{\tau}{\Text}$. Let $H'$ (resp.\ $H''$) be a suffix
  (resp.\ prefix) of $H$ of length $s$ (resp.\ $t$).  Denote $\Pat
  = \Text[j' \dd \RunEndPos{j'}{\tau}{\Text})
  = \Text[j' \dd \RunEndPos{j}{\tau}{\Text}) = H' H^{k} H''$.
  Note that by \cref{lm:end}, $\RunEndPos{j'}{\tau}{\Text} - j' \geq 3\tau - 1$.
  Thus, we have $|\Pat| = \RunEndPos{j}{\tau}{\Text} - j' = \RunEndPos{j'}{\tau}{\Text} - j'
  \geq 3\tau - 1$.

  (\ref{lm:rmin-equivalence-it-1} $\Rightarrow$ \ref{lm:rmin-equivalence-it-2})
  Assume that condition \ref{lm:rmin-equivalence-it-1} holds, i.e., $j' \in \RMinMinusTwo{\tau}{\Text}$.
  Then, $j' = \min \OccTwo{\Pat}{\Text} \cap \RMinusTwo{\tau}{\Text}$ (\cref{def:rmin}).
  Suppose that condition \ref{lm:rmin-equivalence-it-2} does not hold,
  i.e., there exists $i \in \RPrimMinusThree{H}{\tau}{\Text} \cap [1 \dd j)$
  such that either $\RunEndFullPos{i}{\tau}{\Text} - i
  \geq \RunEndFullPos{j}{\tau}{\Text} - j' + p$, or
  $\RunEndPos{i}{\tau}{\Text} - \RunEndFullPos{i}{\tau}{\Text}
  \geq \RunEndPos{j}{\tau}{\Text} - \RunEndFullPos{j}{\tau}{\Text}$ and
  $\RunEndFullPos{i}{\tau}{\Text} - i
  \geq \RunEndFullPos{j}{\tau}{\Text} - j'$. Consider two cases:
  \begin{itemize}
  \item Let us first assume
    that $\RunEndFullPos{i}{\tau}{\Text} - i
    \geq \RunEndFullPos{j}{\tau}{\Text} - j' + p$.
    Let
    \[
      i' = \RunEndFullPos{i}{\tau}{\Text} - 
      (\RunEndFullPos{j}{\tau}{\Text} - j' + p).
    \]
    We will prove that
    $i' \in \OccTwo{\Pat}{\Text} \cap \RMinusTwo{\tau}{\Text}$
    and $i' < j'$, which contradicts
    $j' = \min \OccTwo{\Pat}{\Text} \cap \RMinusTwo{\tau}{\Text}$.
    We proceed in three steps:
    \begin{itemize}
    \item First, we prove that $i' \in [i \dd \RunEndPos{i}{\tau}{\Text} - 3\tau + 2)$.
      Recall that we assumed $\RunEndFullPos{i}{\tau}{\Text} - i \geq
      \RunEndFullPos{j}{\tau}{\Text} - j' + p$. Thus, by definition of $i'$,
      we have
      $i' = \RunEndFullPos{i}{\tau}{\Text} - (\RunEndFullPos{j}{\tau}{\Text} - j' + p)
      \geq \RunEndFullPos{i}{\tau}{\Text} - (\RunEndFullPos{i}{\tau}{\Text} - i) = i$.
      To show $i' < \RunEndPos{i}{\tau}{\Text} - 3\tau + 2$, first recall that
      above we observed that $\RunEndPos{j}{\tau}{\Text} - j' \geq 3\tau - 1$.
      Combining this with the
      definition of $i'$, we thus obtain
      \begin{align*}
        \RunEndPos{i}{\tau}{\Text} - i'
          &\geq \RunEndFullPos{i}{\tau}{\Text} - i'\\
          &=\RunEndFullPos{j}{\tau}{\Text} - j' + p\\
          &> (\RunEndFullPos{j}{\tau}{\Text} - j') +
             (\RunEndPos{j}{\tau}{\Text} - \RunEndFullPos{j}{\tau}{\Text})\\
          &=\RunEndPos{j}{\tau}{\Text} - j' \geq 3\tau - 1,
      \end{align*}
      which we can equivalently write as $i' < \RunEndPos{i}{\tau}{\Text} - 3\tau + 2$.
    \item Next, we prove that $i' \in \OccTwo{\Pat}{\Text} \cap \RMinusTwo{\tau}{\Text}$.
      By $i' \in [i \dd \RunEndPos{i}{\tau}{\Text} - 3\tau + 2)$ and \cref{lm:R-text-block},
      we have
      $\RootPos{i'}{\tau}{\Text} = \RootPos{i}{\tau}{\Text}$,
      $\RunEndFullPos{i'}{\tau}{\Text} = \RunEndFullPos{i}{\tau}{\Text}$, and
      $\TypePos{i'}{\tau}{\Text} = \TypePos{i}{\tau}{\Text} = -1$.
      Combining this with the definition of $i'$, we thus obtain that
      $\RunEndFullPos{i'}{\tau}{\Text} - i' = \RunEndFullPos{i}{\tau}{\Text} - i' = \RunEndFullPos{j}{\tau}{\Text} - j' + p$.
      Recall that $\RunEndFullPos{j}{\tau}{\Text} - j' = \RunEndFullPos{j'}{\tau}{\Text} - j' = s + kp$.
      Thus, $\RunEndFullPos{i'}{\tau}{\Text} - i' = \RunEndFullPos{j}{\tau}{\Text} - j' + p = s + (k+1)p$.
      Recalling that $i', j' \in \RThree{H}{\tau}{\Text}$, we thus obtain
      $\Text[i' \dd \RunEndFullPos{i'}{\tau}{\Text}) = H'H^{k+1}$. Since above we observed that $\Pat = H' H^{k} H''$,
      where $H''$ is a prefix of $H$,
      we thus obtain that $i' \in \OccTwo{\Pat}{\Text}$.
      Combining with  $i' \in \RMinusTwo{\tau}{\Text}$,
      we obtain $i' \in \OccTwo{\Pat}{\Text} \cap \RMinusTwo{\tau}{\Text}$.
    \item Lastly, we prove that $i' < j'$. First, note that by
      $j \in \RPrimTwo{\tau}{\Text}$, we obtain that $j-1 \not\in \RTwo{\tau}{\Text}$.
      On the other hand, we assumed that $i \in \RTwo{\tau}{\Text}$ and $i < j$.
      Thus, we obtain from $[i \dd i'] \subseteq \RTwo{\tau}{\Text}$ that
      $i' < j-1 < j \leq j'$.
    \end{itemize}
    We thus obtain
    $i' \in \OccTwo{\Pat}{\Text} \cap \RMinusTwo{\tau}{\Text}$
    and $i' < j'$, which contradicts $j' = \min \OccTwo{\Pat}{\Text} \cap \RMinusTwo{\tau}{\Text}$.
  \item Let us now assume the second alternative, i.e., that
    $\RunEndPos{i}{\tau}{\Text} - \RunEndFullPos{i}{\tau}{\Text}
    \geq \RunEndPos{j}{\tau}{\Text} - \RunEndFullPos{j}{\tau}{\Text}$
    and $\RunEndFullPos{i}{\tau}{\Text} - i \geq
    \RunEndFullPos{j}{\tau}{\Text} - j'$. Let
    \[
      i' = \RunEndFullPos{i}{\tau}{\Text} - 
      (\RunEndFullPos{j}{\tau}{\Text} - j').
    \]
    We will again prove that
    $i' \in \OccTwo{\Pat}{\Text} \cap \RMinusTwo{\tau}{\Text}$
    and $i' < j'$, which contradicts
    $j' = \min \OccTwo{\Pat}{\Text} \cap \RMinusTwo{\tau}{\Text}$. We proceed
    in three steps:
    \begin{itemize}
    \item First, we prove that $i' \in [i \dd \RunEndPos{i}{\tau}{\Text} - 3\tau + 2)$.
      Recall that we assumed $\RunEndFullPos{i}{\tau}{\Text} - i \geq
      \RunEndFullPos{j}{\tau}{\Text} - j'$. Thus, by definition of $i'$,
      we have
      $i' = \RunEndFullPos{i}{\tau}{\Text} - (\RunEndFullPos{j}{\tau}{\Text} - j')
      \geq \RunEndFullPos{i}{\tau}{\Text} - (\RunEndFullPos{i}{\tau}{\Text} - i) = i$.
      To show $i' < \RunEndPos{i}{\tau}{\Text} - 3\tau + 2$, first recall that
      above we observed that $\RunEndPos{j}{\tau}{\Text} - j' \geq 3\tau - 1$.
      Combining this with the assumption
      $\RunEndPos{i}{\tau}{\Text} - \RunEndFullPos{i}{\tau}{\Text} \geq
      \RunEndPos{j}{\tau}{\Text} - \RunEndFullPos{j}{\tau}{\Text}$
      and plugging the definition of $i'$, we thus obtain
      \begin{align*}
        \RunEndPos{i}{\tau}{\Text} - i'
          &= (\RunEndPos{i}{\tau}{\Text} - \RunEndFullPos{i}{\tau}{\Text}) +
             (\RunEndFullPos{i}{\tau}{\Text} - i')\\
          &\geq (\RunEndPos{j}{\tau}{\Text} - \RunEndFullPos{j}{\tau}{\Text}) +
                (\RunEndFullPos{i}{\tau}{\Text} - i')\\
          &= \RunEndPos{j}{\tau}{\Text} - j' \geq 3\tau - 1,
      \end{align*}
      which we can equivalently write as $i' < \RunEndPos{i}{\tau}{\Text} - 3\tau + 2$.
    \item Next, we prove that $i' \in \OccTwo{\Pat}{\Text} \cap \RMinusTwo{\tau}{\Text}$.
      Recall that $i \in \RPrimMinusThree{H}{\tau}{\Text}$.
      By $i' \in [i \dd \RunEndPos{i}{\tau}{\Text} - 3\tau + 2)$ and
      \cref{lm:end}, we have $[i \dd i'] \subseteq \RTwo{\tau}{\Text}$.
      Thus, by \cref{lm:R-text-block}, it holds $i' \in \RMinusThree{H}{\tau}{\Text}$.
      By definition of $i'$, we have $\RunEndFullPos{i}{\tau}{\Text} - i' = \RunEndFullPos{j}{\tau}{\Text} - j'$.
      Therefore, $\Text[i' \dd \RunEndFullPos{i}{\tau}{\Text}) = \Text[j' \dd \RunEndFullPos{j}{\tau}{\Text})
      = H' H^{k}$. On the other hand, $\RunEndPos{i}{\tau}{\Text} - \RunEndFullPos{i}{\tau}{\Text}
      \geq \RunEndPos{j}{\tau}{\Text} - \RunEndFullPos{j}{\tau}{\Text}$ implies that $H''$
      is a prefix of $\Text[\RunEndFullPos{i}{\tau}{\Text} \dd \RunEndPos{i}{\tau}{\Text})$. Putting the two together, we thus
      have $\Text[i' \dd \RunEndPos{i}{\tau}{\Text}) = H' H^{k} H'' = \Pat$, i.e., $i' \in \OccTwo{\Pat}{\Text}$.
      Combining with  $i' \in \RMinusTwo{\tau}{\Text}$, we obtain
      $i' \in \OccTwo{\Pat}{\Text} \cap \RMinusTwo{\tau}{\Text}$.
    \item By the same argument as above (using $i < j$,
      $[i \dd i'] \subseteq \RTwo{\tau}{\Text}$, and $j \in \RPrimTwo{\tau}{\Text}$)
      we have $i' < j'$.
    \end{itemize}
    We thus obtain
    $i' \in \OccTwo{\Pat}{\Text} \cap \RMinusTwo{\tau}{\Text}$
    and $i' < j'$, which contradicts $j' = \min \OccTwo{\Pat}{\Text} \cap \RMinusTwo{\tau}{\Text}$.
  \end{itemize}

  (\ref{lm:rmin-equivalence-it-2} $\Rightarrow$ \ref{lm:rmin-equivalence-it-1})
  Assume that condition \ref{lm:rmin-equivalence-it-2} holds, i.e.,
  for every $i \in \RPrimMinusThree{H}{\tau}{\Text} \cap [1 \dd
  j)$, it holds $\RunEndFullPos{i}{\tau}{\Text} - i
  < \RunEndFullPos{j}{\tau}{\Text} - j' + p$, and either
  $\RunEndPos{i}{\tau}{\Text} - \RunEndFullPos{i}{\tau}{\Text}
  < \RunEndPos{j}{\tau}{\Text} - \RunEndFullPos{j}{\tau}{\Text}$ or
  $\RunEndFullPos{i}{\tau}{\Text} - i
  < \RunEndFullPos{j}{\tau}{\Text} - j'$.
  Suppose that condition \ref{lm:rmin-equivalence-it-1} does not hold, i.e.,
  $j' \not\in \RMinMinusTwo{\tau}{\Text}$.
  Since above we established that $j' \in \RMinusTwo{\tau}{\Text}$, the
  assumption $j' \not\in \RMinMinusTwo{\tau}{\Text}$ (see \cref{def:rmin}) implies
  that there exists a position
  $i' < j'$ such that $i' \in \OccTwo{\Pat}{\Text} \cap \RMinusTwo{\tau}{\Text}$.
  Recall that by $j' \in \RTwo{\tau}{\Text}$, we have $|\Pat| \geq 3\tau - 1$
  and $\per(\Pat) = p \leq \tfrac{1}{3}\tau$.
  By the synchronization property of primitive
  strings~\cite[Lemma~1.11]{AlgorithmsOnStrings}, we therefore cannot have
  $i' \in (j'-p \dd j')$.
  Thus, $i' \leq j'-p$. Note, however, that we assumed $j'-j < p$.
  Thus, $i' \leq j'-p < j$. Next, observe that by
  $\Pat$ being a prefix of $\Text[j' \dd \Textlen]$ and $i' \in \OccTwo{\Pat}{\Text}$,
  it follows that $\LCE{\Text}{i'}{j'} \geq |\Pat| \geq 3\tau - 1$. Hence,
  by $j' \in \RFour{s}{H}{\tau}{\Text}$ and
  \cref{lm:periodic-pos-lce}\eqref{lm:periodic-pos-lce-it-2}, we obtain
  $i' \in \RFour{s}{H}{\tau}{\Text}$.
  Combining with the assumption $i' \in \RMinusTwo{\tau}{\Text}$,
  we thus obtain $i' \in \RMinusFour{s}{H}{\tau}{\Text}$.
  Recall that $\Pat$ has period $p$.
  Thus, $\Pat$ being a prefix of $\Text[i' \dd \Textlen]$ implies
  $\lcp{\Text[i' \dd \Textlen]}{\Text[i'+p \dd \Textlen]} \geq
  \lcp{\Pat[1 \dd |\Pat|])}{\Pat[1+p \dd |\Pat|]} = |\Pat|-p = \RunEndPos{j'}{\tau}{\Text} - j' - p$.
  We thus obtain $\RunEndPos{i'}{\tau}{\Text}-i' = p+\lcp{\Text[i' \dd \Textlen]}{\Text[i'+p \dd \Textlen]} \geq
  \RunEndPos{j'}{\tau}{\Text}-j'$.
  Denote $k' = \ExpPos{i'}{\tau}{\Text}$.
  Since as noted above, we have $i',j'\in \RFour{s}{H}{\tau}{\Text}$,
  we obtain $k' = \ExpPos{i'}{\tau}{\Text} = \lfloor \tfrac{\RunEndPos{i'}{\tau}{\Text}-i'-s}{p} \rfloor \geq
  \lfloor \tfrac{\RunEndPos{j'}{\tau}{\Text}-j'-s}{p} \rfloor = \ExpPos{j'}{\tau}{\Text} = k$.
  Let us now consider two cases:
  \begin{itemize}
  \item First, let us assume $\ExpPos{i'}{\tau}{\Text} > \ExpPos{j'}{\tau}{\Text}$. Recall
    that $\RunEndFullPos{j}{\tau}{\Text} - j' = \RunEndFullPos{j'}{\tau}{\Text} - j' = s + kp$.
    Thus, the assumption $k' \geq (k+1)$ implies
    that $\RunEndFullPos{i'}{\tau}{\Text} - i' = s + k'p \geq s + (k+1)p
    = (s + kp) + p = \RunEndFullPos{j}{\tau}{\Text} - j' + p$.
    Let $i \in \RPrimTwo{\tau}{\Text}$ be such that $[i \dd i'] \subseteq \RTwo{\tau}{\Text}$.
    By \cref{lm:R-text-block}, we then have
    $\RootPos{i}{\tau}{\Text} = \RootPos{i'}{\tau}{\Text}$,
    $\RunEndPos{i}{\tau}{\Text} = \RunEndPos{i'}{\tau}{\Text}$,
    $\RunEndFullPos{i}{\tau}{\Text} = \RunEndFullPos{i'}{\tau}{\Text}$, and
    $\TypePos{i}{\tau}{\Text} = \TypePos{i'}{\tau}{\Text} = -1$.
    Note also that
    $\RunEndFullPos{i}{\tau}{\Text} - i = \RunEndFullPos{i'}{\tau}{\Text} - i \geq 
    \RunEndFullPos{i'}{\tau}{\Text} - i' \geq \RunEndFullPos{j}{\tau}{\Text} - j' + p$.
    We have thus proved the existence of $i \in \RPrimMinusThree{H}{\tau}{\Text} \cap [1 \dd j)$
    satisfying $\RunEndFullPos{i}{\tau}{\Text} - i \geq \RunEndFullPos{j}{\tau}{\Text} - j' + p$.
    This contradicts the assumed condition \ref{lm:rmin-equivalence-it-2}.
  \item Let us now assume $\ExpPos{i'}{\tau}{\Text} = \ExpPos{j'}{\tau}{\Text}$.
    We then have $\RunEndFullPos{i'}{\tau}{\Text} - i' = s + k'p = s + kp = \RunEndFullPos{j'}{\tau}{\Text} - j'
    = \RunEndFullPos{j}{\tau}{\Text} - j'$. Thus, we have
    \begin{align*}
      \RunEndPos{i'}{\tau}{\Text} - \RunEndFullPos{i'}{\tau}{\Text}
        &= (\RunEndPos{i'}{\tau}{\Text} - i') - (\RunEndFullPos{i'}{\tau}{\Text} - i')\\
        &\geq (\RunEndPos{j'}{\tau}{\Text} - j') - (\RunEndFullPos{i'}{\tau}{\Text} - i')\\
        &= (\RunEndPos{j'}{\tau}{\Text} - j') - (\RunEndFullPos{j'}{\tau}{\Text} - j')\\
        &= \RunEndPos{j'}{\tau}{\Text} - \RunEndFullPos{j'}{\tau}{\Text}\\
        &= \RunEndPos{j}{\tau}{\Text} - \RunEndFullPos{j}{\tau}{\Text}.
    \end{align*}
    Let $i \in \RPrimTwo{\tau}{\Text}$ be such that $[i \dd i'] \subseteq \RTwo{\tau}{\Text}$.
    By \cref{lm:R-text-block}, we then have
    $\RootPos{i}{\tau}{\Text} = \RootPos{i'}{\tau}{\Text}$,
    $\RunEndPos{i}{\tau}{\Text} = \RunEndPos{i'}{\tau}{\Text}$,
    $\RunEndFullPos{i}{\tau}{\Text} = \RunEndFullPos{i'}{\tau}{\Text}$, and
    $\TypePos{i}{\tau}{\Text} = \TypePos{i'}{\tau}{\Text} = -1$.
    By the above we thus have
    $\RunEndPos{i}{\tau}{\Text} - \RunEndFullPos{i}{\tau}{\Text}
    = \RunEndPos{i'}{\tau}{\Text} - \RunEndFullPos{i'}{\tau}{\Text} \geq
    \RunEndPos{j}{\tau}{\Text} - \RunEndFullPos{j}{\tau}{\Text}$.
    Note also that we have
    $\RunEndFullPos{i}{\tau}{\Text} - i = \RunEndFullPos{i'}{\tau}{\Text} - i \geq
    \RunEndFullPos{i'}{\tau}{\Text} - i' = \RunEndFullPos{j}{\tau}{\Text} - j'$.
    We have thus proved the existence of $i \in \RPrimMinusThree{H}{\tau}{\Text} \cap [1 \dd j)$
    satisfying
    $\RunEndPos{i}{\tau}{\Text} - \RunEndFullPos{i}{\tau}{\Text} \geq
    \RunEndPos{j}{\tau}{\Text} - \RunEndFullPos{j}{\tau}{\Text}$ and
    $\RunEndFullPos{i}{\tau}{\Text} - i \geq \RunEndFullPos{j}{\tau}{\Text} - j'$.
    This contradicts the assumed condition \ref{lm:rmin-equivalence-it-2}.
  \end{itemize}
  In both cases we obtain a contradiction. Thus, condition \ref{lm:rmin-equivalence-it-1} must hold.
\end{proof}

\begin{lemma}\label{lm:emin}
  Let $\tau \in [1 \dd \lfloor \tfrac{\Textlen}{2} \rfloor]$,
  $H \in \Sigma^{+}$, $p = |H|$,
  and $(p_i)_{i \in [1 \dd m]} = \RunsMinusTextSortedThree{H}{\tau}{\Text}$ (\cref{def:runs-minus-text-sorted}).
  For every $t \in [1 \dd m]$, letting $r = \RunEndPos{p_t}{\tau}{\Text} - p_t - 3\tau + 2$,
  it holds:
  \begin{align*}
    \RunMinEndPos{p_t}{\tau}{\Text} - p_t =
    \begin{cases}
      0 & \text{if }\ell_t^{\rm pos} \leq \ell_t^{\max},\\
      \min(\ell_t^{\rm pos} - \ell_t^{\max}, p, r) & \text{otherwise},
    \end{cases}
  \end{align*}
  where $\mathcal{S}_t = \{(\RunEndPos{p_{t'}}{\tau}{\Text} - \RunEndFullPos{p_{t'}}{\tau}{\Text},
  \RunEndFullPos{p_{t'}}{\tau}{\Text} - p_{t'}) : t' \in [1 \dd t)\}$ and
  \begin{align*}
    \ell_t^{\rm pos}
      &= \RunEndFullPos{p_t}{\tau}{\Text} - p_t,\\
    \ell_t^{\rm trim}
      &= \max \{\RunEndFullPos{p_{t'}}{\tau}{\Text} - p_{t'} - p : t' \in [1 \dd t)\} \cup \{0\},\\
    \ell_t^{\rm whole}
      &= \max \{v : (k,v) \in \mathcal{S}_t\cup \{(p,0)\}\text{ and }
              k \geq \RunEndPos{p_t}{\tau}{\Text} - \RunEndFullPos{p_t}{\tau}{\Text}\},\\
    \ell_t^{\max}
      &= \max(\ell_t^{\rm trim}, \ell_t^{\rm whole}).
  \end{align*}
\end{lemma}
\begin{proof}
  Let us first observe that
  it follows by $\RunEndPos{p_t}{\tau}{\Text} - p_t \geq 3\tau - 1$ and
  $\RunEndPos{p_t}{\tau}{\Text} - \RunEndFullPos{p_t}{\tau}{\Text} < p \leq \tfrac{1}{3}\tau$ that
  $\ell_t^{\rm pos} \geq 2\tau$.
  Consider two cases:
  \begin{itemize}
  \item First, assume $\ell_t^{\rm pos} \leq \ell_t^{\max}$.
    Note that by $\ell_t^{\rm pos} \geq 2\tau > 0$, this implies that $\ell_t^{\max} > 0$.
    By definition of $\ell_t^{\max}$, it follows that
    either
    $\max \{\RunEndFullPos{p_{t'}}{\tau}{\Text} - p_{t'} - p : t' \in [1 \dd t)\} \cup \{0\} = \ell_t^{\max}$ or
    $\max \{v : (k,v) \in \mathcal{S}_t\cup\{(p,0)\}\text{ and }
    k \geq \RunEndPos{p_t}{\tau}{\Text} - \RunEndFullPos{p_t}{\tau}{\Text}\} = \ell_t^{\max}$.
    Let us consider two subcases:
    \begin{itemize}
    \item First, let us assume that
      $\max \{\RunEndFullPos{p_{t'}}{\tau}{\Text} - p_{t'} - p : t' \in [1 \dd t)\} \cup \{0\} = \ell_t^{\max}$.
      By $\ell_t^{\rm pos} > 0$, this implies that
      there exists $t' \in [1 \dd t)$ satisfying $\RunEndFullPos{p_{t'}}{\tau}{\Text} - p_{t'} - p = \ell_t^{\max}$.
      By the assumption $\ell_t^{\rm pos} \leq \ell_t^{\max}$ and the definition
      of $\ell_t^{\rm pos}$, we thus obtain that
      $\RunEndFullPos{p_{t'}}{\tau}{\Text} - p_{t'} - p \geq \RunEndFullPos{p_t}{\tau}{\Text} - p_t$. Equivalently,
      $\RunEndFullPos{p_{t'}}{\tau}{\Text} - p_{t'} \geq \RunEndFullPos{p_t}{\tau}{\Text} - p_t + p$.
      By \cref{lm:rmin-equivalence}, we thus have $p_t \not\in \RMinMinusTwo{\tau}{\Text}$ (note
      that \cref{lm:rmin-equivalence} requires that $p_t \in [p_t \dd p_t + \min(p,r))$, which holds
      here since $\min(p,r) > 0$). By \cref{def:emin}, we thus have
      $\RunMinEndPos{p_t}{\tau}{\Text} - p_t = 0$.
    \item
      Let us now assume that
      $\max \{v : (k,v) \in \mathcal{S}_t\cup\{(p,0)\}\text{ and }
      k \geq \RunEndPos{p_t}{\tau}{\Text} - \RunEndFullPos{p_t}{\tau}{\Text}\} = \ell_t^{\max}$.
      By $\ell_t^{\max} > 0$ and the definition of $\mathcal{S}_t$, it then follows that there exists
      $t' \in [1 \dd t)$ satisfying
      $\RunEndPos{p_{t'}}{\tau}{\Text} - \RunEndFullPos{p_{t'}}{\tau}{\Text} \geq
      \RunEndPos{p_t}{\tau}{\Text} - \RunEndFullPos{p_t}{\tau}{\Text}$ and
      $\RunEndFullPos{p_{t'}}{\tau}{\Text} - p_{t'} = \ell_t^{\max}$.
      On the other hand, by the assumption $\ell_t^{\rm pos} \leq \ell_t^{\max}$ and the definition
      of $\ell_t^{\rm pos}$, we have
      $\RunEndFullPos{p_{t'}}{\tau}{\Text} - p_{t'} \geq \RunEndFullPos{p_t}{\tau}{\Text} - p_t$.
      By \cref{lm:rmin-equivalence}, we thus have $p_t \not\in \RMinMinusTwo{\tau}{\Text}$ (we
      again use that $\min(p,r) > 0$ and hence $p_t \in [p_t \dd p_t + \min(p,r))$).
      By \cref{def:emin}, we thus have $\RunMinEndPos{p_t}{\tau}{\Text} - p_t = 0$.
    \end{itemize}
  \item Let us now assume that $\ell_t^{\rm pos} > \ell_t^{\max}$. Denote
    $\delta = \min(\ell_t^{\rm pos} - \ell_t^{\max}, p, r)$ and $x = p_t + \delta - 1$.
    Note that $x \in [p_t \dd p_t + \min(p,r))$. First, we prove that
    $x \in \RMinMinusTwo{\tau}{\Text}$. Suppose that $x \not\in \RMinMinusTwo{\tau}{\Text}$.
    By \cref{lm:rmin-equivalence}, there exists $i \in \RPrimMinusThree{H}{\tau}{\Text} \cap [1 \dd p_t)$
    such that $\RunEndFullPos{i}{\tau}{\Text} - i
    \geq \RunEndFullPos{p_t}{\tau}{\Text} - x + p$, or
    $\RunEndPos{i}{\tau}{\Text} - \RunEndFullPos{i}{\tau}{\Text}
    \geq \RunEndPos{p_t}{\tau}{\Text} - \RunEndFullPos{p_t}{\tau}{\Text}$ and
    $\RunEndFullPos{i}{\tau}{\Text} - i
    \geq \RunEndFullPos{p_t}{\tau}{\Text} - x$. Equivalently, there exists $t' \in [1 \dd t)$ such that
    $\RunEndFullPos{p_{t'}}{\tau}{\Text} - p_{t'}
    \geq \RunEndFullPos{p_t}{\tau}{\Text} - x + p$, or
    $\RunEndPos{p_{t'}}{\tau}{\Text} - \RunEndFullPos{p_{t'}}{\tau}{\Text}
    \geq \RunEndPos{p_t}{\tau}{\Text} - \RunEndFullPos{p_t}{\tau}{\Text}$ and
    $\RunEndFullPos{p_{t'}}{\tau}{\Text} - p_{t'}
    \geq \RunEndFullPos{p_t}{\tau}{\Text} - x$.
    Consider two cases:
    \begin{itemize}
    \item First, assume that
      there exists $t' \in [1 \dd t)$ such that
      $\RunEndFullPos{p_{t'}}{\tau}{\Text} - p_{t'}
      \geq \RunEndFullPos{p_t}{\tau}{\Text} - x + p$, or equivalently,
      $\RunEndFullPos{p_{t'}}{\tau}{\Text} - p_{t'} - p \geq \RunEndFullPos{p_t}{\tau}{\Text} - x$.
      This, by definition of $\ell_t^{\max}$ and $\ell_t^{\rm trim}$,
      implies that $\ell_t^{\max} \geq \ell_t^{\rm trim}
      \geq \RunEndFullPos{p_{t'}}{\tau}{\Text} - p_{t'} - p \geq \RunEndFullPos{p_t}{\tau}{\Text} - x$.
    \item Let us now assume the alternative, i.e., that
      there exists $t' \in [1 \dd t)$ such that
      $\RunEndPos{p_{t'}}{\tau}{\Text} - \RunEndFullPos{p_{t'}}{\tau}{\Text}
      \geq \RunEndPos{p_t}{\tau}{\Text} - \RunEndFullPos{p_t}{\tau}{\Text}$ and
      $\RunEndFullPos{p_{t'}}{\tau}{\Text} - p_{t'}
      \geq \RunEndFullPos{p_t}{\tau}{\Text} - x$.
      By $t' \in [1 \dd t)$, we have
      $(\RunEndPos{p_{t'}}{\tau}{\Text} - \RunEndFullPos{p_{t'}}{\tau}{\Text},
        \RunEndFullPos{p_{t'}}{\tau}{\Text} - p_{t'}) \in \mathcal{S}_t$.
      Thus, it follows from
      $\RunEndPos{p_{t'}}{\tau}{\Text} - \RunEndFullPos{p_{t'}}{\tau}{\Text} \geq 
      \RunEndPos{p_t}{\tau}{\Text} - \RunEndFullPos{p_t}{\tau}{\Text}$
      and the definition of $\ell_t^{\rm whole}$ and $\ell_t^{\max}$ that
      $\ell_t^{\max} \geq \ell_t^{\rm whole} \geq
      \RunEndFullPos{p_{t'}}{\tau}{\Text} - p_{t'} \geq \RunEndFullPos{p_t}{\tau}{\Text} - x$.
    \end{itemize}
    In both cases, we obtain $\ell_t^{\max} \geq \RunEndFullPos{p_t}{\tau}{\Text} - x$.
    Expanding the definition of $x$, we thus obtain
    that $\ell_t^{\max} \geq \ell_t^{\rm pos} - \delta + 1$. Equivalently,
    $\delta \geq \ell_t^{\rm pos} - \ell_t^{\max} + 1$. This contradicts
    $\delta \leq \ell_t^{\rm pos} - \ell_t^{\max}$ (following
    from the definition of $\delta$). We thus obtain that $x \in \RMinMinusTwo{\tau}{\Text}$.
    Denote $x' = x+1$.
    Next, we prove that $x' \not\in \RMinMinusTwo{\tau}{\Text}$.
    Note that if $\ell_t^{\rm pos} - \ell_t^{\max} \geq \min(p, r)$, then $\delta = \min(p, r)$, and
    hence $x' = p_t + \delta = p_t + \min(p,r) \not\in \RMinMinusTwo{\tau}{\Text}$
    follows immediately by \cref{lm:rmin-block-size}.
    Let us thus assume $\ell_t^{\rm pos} - \ell_t^{\max} < \min(p, r)$. We then
    have $\delta = \ell_t^{\rm pos} - \ell_t^{\max}$ and
    $x' = p_t + \delta = p_t + \ell_t^{\rm pos} - \ell_t^{\max} \in [p_t \dd p_t + \min(p, r))$.
    Recall also that above we observed that $\ell_t^{\rm pos} \geq 2\tau$.
    On the other hand, $\ell_t^{\rm pow} - \ell_t^{\max} < \min(p, r)$ implies
    $\ell_t^{\rm pow} - \ell_t^{\max} < p \leq \tfrac{1}{3}\tau \leq \tau$.
    Putting it together, we obtain $\ell_t^{\max} > \ell_t^{\rm pow} - \tau \geq 2\tau - \tau = \tau > 0$.
    By definition of $\ell_t^{\max}$, we either have $\ell_t^{\max} = \ell_t^{\rm trim}$
    or $\ell_t^{\max} = \ell_t^{\rm whole}$.
    Consider two cases:
    \begin{itemize}
    \item First, assume that $\ell_t^{\max} = \ell_t^{\rm trim}$. By $\ell_t^{\max} > 0$, this implies
      that there exists $t' \in [1 \dd t)$ satisfying $\ell_t^{\max} = \RunEndFullPos{p_{t'}}{\tau}{\Text} - p_{t'} - p$.
      By $x' = p_t + \ell_t^{\rm pos} - \ell_t^{\max}$, we thus have
      $\RunEndFullPos{p_{t'}}{\tau}{\Text} - p_{t'} = \ell_t^{\max} + p = p_t + \ell_t^{\rm pos} - x' + p
      = p_t + (\RunEndFullPos{p_t}{\tau}{\Text} - p_t) - x' + p = \RunEndFullPos{p_t}{\tau}{\Text} - x' + p$.
      By \cref{lm:rmin-equivalence}, we thus obtain $x' \not\in \RMinMinusTwo{\tau}{\Text}$.
    \item Let us now assume that $\ell_t^{\max} = \ell_t^{\rm whole}$. By $\ell_t^{\max} > 0$, this implies
      that there exists $t' \in [1 \dd t)$ satisfying $\RunEndPos{p_{t'}}{\tau}{\Text} - \RunEndFullPos{p_{t'}}{\tau}{\Text}
      \geq \RunEndPos{p_t}{\tau}{\Text} - \RunEndFullPos{p_t}{\tau}{\Text}$ and
      $\ell_t^{\max} = \RunEndFullPos{p_{t'}}{\tau}{\Text} - p_{t'}$. By
      $x' = p_t + \ell_t^{\rm pos} - \ell_t^{\max}$, we thus have
      $\RunEndFullPos{p_{t'}}{\tau}{\Text} - p_{t'} = \ell_t^{\max} = p_t + \ell_t^{\rm pos} - x' =
      (p_t + \RunEndFullPos{p_t}{\tau}{\Text} - p_t) - x' = \RunEndFullPos{p_t}{\tau}{\Text} - x'$.
      Combining this with $\RunEndPos{p_{t'}}{\tau}{\Text} - \RunEndFullPos{p_{t'}}{\tau}{\Text}
      \geq \RunEndPos{p_t}{\tau}{\Text} - \RunEndFullPos{p_t}{\tau}{\Text}$ and applying
      \cref{lm:rmin-equivalence} implies that $x' \not\in \RMinMinusTwo{\tau}{\Text}$.
    \end{itemize}
    In both cases, we have $x' \not\in \RMinMinusTwo{\tau}{\Text}$. To
    conclude, we have thus proved that $x = p_t + \delta - 1 \in \RMinMinusTwo{\tau}{\Text}$
    and $x' = x + 1 = p_t + \delta \not\in \RMinMinusTwo{\tau}{\Text}$.
    By \cref{lm:rmin-text-block}, we thus have
    $\RunMinEndPos{p_t}{\tau}{\Text} - p_t = \delta = \min(\ell_t^{\rm pos} - \ell_t^{\max}, p, r)$.
    \qedhere
  \end{itemize}
\end{proof}

\begin{lemma}\label{lm:sweep-step}
  Let $\tau \in [1 \dd \lfloor \tfrac{\Textlen}{2} \rfloor]$,
  $H \in \Sigma^{+}$, $p = |H|$, and $s \in [0 \dd p-1)$.
  Let $x, y, x', y' \in [0 \dd \Textlen]$
  be such that
  \begin{itemize}
  \item $\{\SA{\Text}[i] : i \in (x \dd y]\} = \RMinusFour{s}{H}{\tau}{\Text}$,
  \item $\{\SA{\Text}[i] : i \in (x' \dd y']\} = \RMinusFour{s+1}{H}{\tau}{\Text}$.
  \end{itemize}
  Denote (see \cref{def:min-pos-bitvector-minus}):
  \begin{itemize}
  \item $\BitvectorMin = \MinPosBitvectorMinusTwo{\tau}{\Text}$,
  \item $B_{\rm in} = \MinPosBitvectorMinusFour{s}{H}{\tau}{\Text} = \BitvectorMin(x \dd y]$,
  \item $B_{\rm out} = \MinPosBitvectorMinusFour{s+1}{H}{\tau}{\Text} = \BitvectorMin(x' \dd y']$.
  \end{itemize}
  Let us also define
  \begin{itemize}
   \item $I_{0} = \{\RunEndPos{j}{\tau}{\Text} - 3\tau + 1 :
    j \in \RPrimMinusThree{H}{\tau}{\Text}\text{ and }
    [\RunMinEndPos{j}{\tau}{\Text} \dd \RunEndPos{j}{\tau}{\Text} - 3\tau + 2) \neq \emptyset\}$,
  \item $D_{0} = \{\RunMinEndPos{j}{\tau}{\Text} :
    j \in \RPrimMinusThree{H}{\tau}{\Text}\text{ and }
    [\RunMinEndPos{j}{\tau}{\Text} \dd \RunEndPos{j}{\tau}{\Text} - 3\tau + 2) \neq \emptyset\}$,
  \item $I_{1} = \{\RunMinEndPos{j}{\tau}{\Text} - 1 :
    j \in \RPrimMinusThree{H}{\tau}{\Text}\text{ and }
    [j \dd \RunMinEndPos{j}{\tau}{\Text}) \neq \emptyset\}$.
  \item $D_{1} = \{j :
    j \in \RPrimMinusThree{H}{\tau}{\Text}\text{ and }
    [j \dd \RunMinEndPos{j}{\tau}{\Text}) \neq \emptyset\}$.
  \end{itemize}
  Then, it holds
  \[
    B_{\rm out} =
      \InsertSubseq{\DeleteSubseq{B_{\rm in}}{\mathcal{D}}}{\mathcal{I}},
  \]
  where
  \begin{align*}
    \mathcal{D} &= \{\ISA{\Text}[j] - x :
      j \in (D_{0} \cup D_{1}) \cap \RFour{s}{H}{\tau}{\Text}
      \},\\
    \mathcal{I} &=
      \{(\ISA{\Text}[j] - x', 0) :
        j \in I_{0} \cap \RFour{s+1}{H}{\tau}{\Text}
      \}\ \cup\\
    &\hspace{3.1ex}
      \{(\ISA{\Text}[j] - x', 1) :
        j \in I_{1} \cap \RFour{s+1}{H}{\tau}{\Text}
       \}.
  \end{align*}
\end{lemma}
\begin{proof}

  First, note that $(x,y)$ and $(x',y')$ are well-defined by \cref{rm:min-pos-bitvector-minus}.

  Denote
  \begin{align*}
    \mathcal{D}_{\rm text} &= (D_{0} \cup D_{1}) \cap \RFour{s}{H}{\tau}{\Text},\\
    \mathcal{I}_{\rm text} &= (I_{0} \cup I_{1}) \cap \RFour{s+1}{H}{\tau}{\Text},\\
    \mathcal{D}_{\rm lex} &=
      \{\ISA{\Text}[j] - x : j \in \mathcal{D}_{\rm text}\},\\
    \mathcal{I}_{\rm lex} &=
      \{\ISA{\Text}[j] - x' : j \in \mathcal{I}_{\rm text}\}.
  \end{align*}

  First, we prove that
  $\mathcal{D}_{\rm text} \subseteq \RMinusFour{s}{H}{\tau}{\Text}$,
  $\mathcal{I}_{\rm text} \subseteq \RMinusFour{s+1}{H}{\tau}{\Text}$,
  $\mathcal{D}_{\rm lex} \subseteq [1 \dd |B_{\rm in}|]$, and
  $\mathcal{I}_{\rm lex} \subseteq [1 \dd |B_{\rm out}|]$:
  \begin{itemize}
  \item Let $j \in \mathcal{D}_{\rm text}$, i.e., $j \in (D_{0} \cup D_{1}) \cap \RFour{s}{H}{\tau}{\Text}$.
    By \cref{lm:end}, for every $t \in \RPrimTwo{\tau}{\Text}$,
    $[t \dd \RunEndPos{t}{\tau}{\Text} - 3\tau + 2)$ is a maximal block of
    positions from the set $\RTwo{\tau}{\Text}$. By \cref{def:emin}, for
    every $t \in \RPrimMinusTwo{\tau}{\Text}$, we thus have
    $t \leq \RunMinEndPos{t}{\tau}{\Text} \leq \RunEndPos{t}{\tau}{\Text} - 3\tau + 2$.
    By \cref{lm:R-text-block}, and the definition of $D_{0}$ and $D_{1}$, we therefore
    have $j \in \RMinusThree{H}{\tau}{\Text}$. Combining with $j \in \RFour{s}{H}{\tau}{\Text}$,
    we therefore obtain that $j \in \RMinusFour{s}{H}{\tau}{\Text}$. We thus have $\mathcal{D}_{\rm text} \subseteq
    \RMinusFour{s}{H}{\tau}{\Text}$.
  \item Let $j \in \mathcal{I}_{\rm text}$, i.e., $j \in (I_{0} \cup I_{1}) \cap \RFour{s+1}{H}{\tau}{\Text}$.
    By the same argument as above, it follows from the definition of $I_{0}$ and $I_{1}$, and
    \cref{lm:R-text-block}, that we then have $j \in \RMinusThree{H}{\tau}{\Text}$. Combining with
    $j \in \RFour{s+1}{H}{\tau}{\Text}$, we therefore obtain that $j \in \RMinusFour{s+1}{H}{\tau}{\Text}$.
    Hence, $\mathcal{I}_{\rm text} \subseteq \RMinusFour{s+1}{H}{\tau}{\Text}$.
  \item Let
    $p \in \mathcal{D}_{\rm lex}$. Then, there exists $j \in \mathcal{D}_{\rm text}$
    such that $p = \ISA{\Text}[j] - x$. By the above, $j \in \RMinusFour{s}{H}{\tau}{\Text}$. Thus, 
    by definition of $x$ and $y$, it holds $\ISA{\Text}[j] \in (x \dd y]$. Therefore,
    $p = \ISA{\Text}[j] - x \in [1 \dd y-x] = [1 \dd |B_{\rm in}|]$. We thus obtain
    $\mathcal{D}_{\rm lex} \subseteq [1 \dd |B_{\rm in}|]$.
  \item Let $p \in \mathcal{I}_{\rm lex}$. Similarly as above, then there exists $j \in \mathcal{I}_{\rm text}$
    such that $p = \ISA{\Text}[j] - x'$. By $j \in \RMinusFour{s+1}{H}{\tau}{\Text}$ and the definition of $x'$ and $y'$,
    we thus have $\ISA{\Text}[j] \in (x' \dd y']$. Therefore,
    $p = \ISA{\Text}[j] - x' \in [1 \dd y'-x'] = [1 \dd |B_{\rm out}|]$. We thus obtain
    $\mathcal{I}_{\rm lex} \subseteq [1 \dd |B_{\rm out}|]$.
  \end{itemize}

  We are now ready to show the main claim, i.e., $B_{\rm out} =
  \InsertSubseq{\DeleteSubseq{B_{\rm in}}{\mathcal{D}}}{\mathcal{I}}$.
  First, observe that $\mathcal{D}_{\rm lex} \subseteq [1 \dd |B_{\rm in}|]$ implies
  that the string $\DeleteSubseq{B_{\rm in}}{\mathcal{D}_{\rm lex}} = \DeleteSubseq{B_{\rm in}}{\mathcal{D}}$
  is well-defined (see \cref{def:delete}). To show that
  $\InsertSubseq{\DeleteSubseq{B_{\rm in}}{\mathcal{D}}}{\mathcal{I}}$ is also well-defined, we need to first
  show that for every
  $(i, c), (i', c') \in \mathcal{I}$, $(i,c) \neq (i',c')$ implies
  $i \neq i'$ (see \cref{def:insert}).
  Recall that above
  we noted that for every $t \in \RPrimMinusTwo{\tau}{\Text}$, it holds
  $t \leq \RunMinEndPos{t}{\tau}{\Text} \leq \RunEndPos{t}{\tau}{\Text} - 3\tau + 2$.
  By definition of $I_{0}$ and $I_{1}$, we thus obtain that $I_{0} \cap I_{1} = \emptyset$.
  Since for every $j_1, j_2 \in [1 \dd \Textlen]$, $j_1 \neq j_2$ implies
  $\ISA{\Text}[j_1] \neq \ISA{\Text}[j_2]$, we thus obtain that there are no two pairs
  in $\mathcal{I}$ coinciding on the first coordinate, i.e., $\InsertSubseq{\DeleteSubseq{B_{\rm in}}{\mathcal{D}}}{\mathcal{I}}$
  is indeed well-defined. To finalize the proof of
  $B_{\rm out} = \InsertSubseq{\DeleteSubseq{B_{\rm in}}{\mathcal{D}}}{\mathcal{I}}$,
  we need to show that the two conditions in \cref{def:insert} hold. We proceed as follows:
  \begin{enumerate}

  \item First, we prove that for every $(i,c) \in \mathcal{I}$, it holds
    $i \in [1 \dd |B_{\rm out}|]$ and
    $B_{\rm out}[i] = c$.
    Let $(i, c) \in \mathcal{I}$. Note that then $i \in \mathcal{I}_{\rm lex}$.
    Thus, by $\mathcal{I}_{\rm lex} \subseteq [1 \dd |B_{\rm out}|]$, we obtain the first
    part of the claim. To show $B_{\rm out}[i] = c$, we consider two cases:
    \begin{itemize}
    \item First, let us assume that $c = 0$. In this case, by definition of
      $\mathcal{I}$, there exists $j \in I_0 \cap \RFour{s+1}{H}{\tau}{\Text}$.
      such that $i = \ISA{\Text}[j] - x'$. This in turn implies that, by definition of $I_0$, there
      exists $j' \in \RPrimMinusThree{H}{\tau}{\Text}$ satisfying
      $[\RunMinEndPos{j'}{\tau}{\Text} \dd \RunEndPos{j'}{\tau}{\Text} - 3\tau + 2) \neq \emptyset$
      and $j = \RunEndPos{j'}{\tau}{\Text} - 3\tau + 1$.
      By $j \geq \RunMinEndPos{j'}{\tau}{\Text}$, \cref{def:emin}, and \cref{lm:rmin-text-block},
      it then holds $j \not\in \RMinMinusTwo{\tau}{\Text}$. Consequently, by \cref{def:min-pos-bitvector-minus},
      we have $\BitvectorMin[\ISA{\Text}[j]] = 0$. By $i = \ISA{\Text}[j] - x'$ and the definition of $B_{\rm out}$, we thus
      obtain $B_{\rm out}[i] = B_{\rm out}[\ISA{\Text}[j] - x'] = \BitvectorMin[\ISA{\Text}[j]] = 0$.
    \item Let us now assume $c = 1$. In this case, by definition of
      $\mathcal{I}$, there exists $j \in I_1 \cap \RFour{s+1}{H}{\tau}{\Text}$ such that
      $i = \ISA{\Text}[j] - x'$. This in turn implies that, by definition of $I_1$, there
      exists $j' \in \RPrimMinusThree{H}{\tau}{\Text}$ satisfying
      $[j' \dd \RunMinEndPos{j'}{\tau}{\Text}) \neq \emptyset$
      and $j = \RunMinEndPos{j'}{\tau}{\Text} - 1$.
      By $j < \RunMinEndPos{j'}{\tau}{\Text}$ and \cref{def:emin},
      it holds $j \in \RMinMinusTwo{\tau}{\Text}$. Consequently, by \cref{def:min-pos-bitvector-minus},
      we have $\BitvectorMin[\ISA{\Text}[j]] = 1$. By $i = \ISA{\Text}[j] - x'$ and the definition of $B_{\rm out}$, we thus
      obtain $B_{\rm out}[i] = B_{\rm out}[\ISA{\Text}[j] - x'] = \BitvectorMin[\ISA{\Text}[j]] = 1$.
    \end{itemize}
    In both cases, we have thus obtained $B_{\rm out}[i] = c$.

  \item Second, we prove that, denoting
    \begin{align*}
      B'_{\rm in}
        &= \DeleteSubseq{B_{\rm in}}{\mathcal{D}_{\rm lex}},\\
      B'_{\rm out}
        &= \DeleteSubseq{B_{\rm out}}{\mathcal{I}_{\rm lex}},
    \end{align*}
    it holds $B'_{\rm in} = B'_{\rm out}$. Let
    \begin{align*}
      \{d_1, \ldots, d_k\}
        &= \RMinusFour{s}{H}{\tau}{\Text} \setminus \mathcal{D}_{\rm text},\\
      \{i_1, \ldots, i_q\}
        &= \RMinusFour{s+1}{H}{\tau}{\Text} \setminus \mathcal{I}_{\rm text}
    \end{align*}
    be such that for every $t \in [1 \dd k)$, it holds
    $\Text[d_t \dd \Textlen] \prec \Text[d_{t+1} \dd \Textlen]$, and for
    every $t \in [1 \dd q)$, it holds
    $\Text[i_{t} \dd \Textlen] \prec \Text[i_{t+1} \dd \Textlen]$.
    The proof consists of five steps:
    \begin{enumerate}

    \item\label{sweep-step-1}
      In the first step, we prove that $|B'_{\rm in}| = k$ and that for
      every $t \in [1 \dd k]$, it holds $B'_{\rm in}[t] = \BitvectorMin[\ISA{\Text}[d_t]]$.
      To show the first claim, recall that $\mathcal{D}_{\rm text} \subseteq \RMinusFour{s}{H}{\tau}{\Text}$.
      Thus,
      \begin{align*}
        |B'_{\rm in}|
          &= |\DeleteSubseq{B_{\rm in}}{\mathcal{D}_{\rm lex}}|\\
          &= |B_{\rm in}| - |\mathcal{D}_{\rm lex}|\\
          &= |\RMinusFour{s}{H}{\tau}{\Text}| - |\mathcal{D}_{\rm text}|\\
          &= |\RMinusFour{s}{H}{\tau}{\Text} \setminus \mathcal{D}_{\rm text}| = k.
      \end{align*}
      Let $(u_1, \ldots, u_k)$ be an increasing sequence satisfying
      $\{u_1, \ldots, u_k\} = [1 \dd |B_{\rm in}|] \setminus \mathcal{D}_{\rm lex}$.
      Since $\{\SA{\Text}[x+i] : i \in [1 \dd |B_{\rm in}|]\} = \RMinusFour{s}{H}{\tau}{\Text}$
      and $\{\SA{\Text}[x+i] : i \in \mathcal{D}_{\rm lex}\} = \mathcal{D}_{\rm text}$, it follows that
      \begin{align*}
        \{\SA{\Text}[x+u_t] : t \in [1 \dd k]\}
          &= \{\SA{\Text}[x+i] : i \in [1 \dd |B_{\rm in}|] \setminus \mathcal{D}_{\rm lex}\}\\
          &= \RMinusFour{s}{H}{\tau}{\Text} \setminus \mathcal{D}_{\rm text}\\
          &= \{d_1, \ldots, d_k\}.
      \end{align*}
      Recall that for every $t \in [1 \dd k)$, it holds
      $\Text[\SA{\Text}[x + u_t] \dd \Textlen] \prec \Text[\SA{\Text}[x + u_{t+1}] \dd \Textlen]$
      and $\Text[d_t \dd \Textlen] \prec \Text[d_{t+1} \dd \Textlen]$.
      By $\{\SA{\Text}[x+u_t] : t \in [1 \dd k]\} = \{d_1, \ldots, d_k\}$, we thus obtain
      that for every $t \in [1 \dd k]$, it holds $\SA{\Text}[x+u_t] = d_t$.
      In particular, $\BitvectorMin[\ISA{\Text}[\SA{\Text}[x+u_t]]] = \BitvectorMin[\ISA{\Text}[d_t]]$, or
      equivalently, $\BitvectorMin[x+u_t] = \BitvectorMin[\ISA{\Text}[d_t]]$.
      It remains to observe that by \cref{def:delete}, for every
      $t \in [1 \dd k]$, it holds $B'_{\rm in}[t] = B_{\rm in}[u_t]$. Putting
      everything together, we thus obtain that for every $i \in [1 \dd k]$,
      it holds
      \begin{align*}
        B'_{\rm in}[t]
          &= B_{\rm in}[u_t]\\
          &= \BitvectorMin[x + u_t]\\
          &= \BitvectorMin[\ISA{\Text}[d_t]].
      \end{align*}

    \item\label{sweep-step-2}
      In the second step, we prove that $|B'_{\rm out}| = q$ and that for
      every $t \in [1 \dd q]$, it holds $B'_{\rm out}[t] = \BitvectorMin[\ISA{\Text}[i_t]]$.
      To show the first claim, recall that $\mathcal{I}_{\rm text} \subseteq \RMinusFour{s+1}{H}{\tau}{\Text}$.
      Thus, it holds
      $
        |B'_{\rm out}|
          = |\DeleteSubseq{B_{\rm out}}{\mathcal{I}_{\rm lex}}|
          = |B_{\rm out}| - |\mathcal{I}_{\rm lex}|
          = |\RMinusFour{s+1}{H}{\tau}{\Text}| - |\mathcal{I}_{\rm text}|
          = |\RMinusFour{s+1}{H}{\tau}{\Text} \setminus \mathcal{I}_{\rm text}| = q
      $.
      Let $(u_1, \ldots, u_q)$ be an increasing sequence satisfying
      $\{u_1, \ldots, u_q\} = [1 \dd |B_{\rm out}|] \setminus \mathcal{I}_{\rm lex}$.
      Since $\{\SA{\Text}[x'+i] : i \in [1 \dd |B_{\rm out}|]\} = \RMinusFour{s+1}{H}{\tau}{\Text}$
      and $\{\SA{\Text}[x'+i] : i \in \mathcal{I}_{\rm lex}\} = \mathcal{I}_{\rm text}$, it follows that
      $ 
        \{\SA{\Text}[x'+u_t] : t \in [1 \dd q]\}
          = \{\SA{\Text}[x'+i] : i \in [1 \dd |B_{\rm out}|] \setminus \mathcal{I}_{\rm lex}\}
          = \RMinusFour{s+1}{H}{\tau}{\Text} \setminus \mathcal{I}_{\rm text}
          = \{i_1, \ldots, i_q\}
      $.
      Recall that for every $t \in [1 \dd q)$, it holds
      $\Text[\SA{\Text}[x' + u_t] \dd \Textlen] \prec \Text[\SA{\Text}[x' + u_{t+1}] \dd \Textlen]$
      and $\Text[i_t \dd \Textlen] \prec \Text[i_{t+1} \dd \Textlen]$.
      By $\{\SA{\Text}[x'+u_t] : t \in [1 \dd q]\} = \{i_1, \ldots, i_q\}$, we thus obtain
      that for every $t \in [1 \dd q]$, it holds $\SA{\Text}[x'+u_t] = i_t$.
      In particular, $\BitvectorMin[\ISA{\Text}[\SA{\Text}[x'+u_t]]] = \BitvectorMin[\ISA{\Text}[i_t]]$, or
      equivalently, $\BitvectorMin[x'+u_t] = \BitvectorMin[\ISA{\Text}[i_t]]$.
      It remains to observe that by \cref{def:delete}, for every
      $t \in [1 \dd q]$, it holds $B'_{\rm out}[t] = B_{\rm out}[u_t]$. Putting
      everything together, we thus obtain that for every $i \in [1 \dd q]$,
      it holds
      $
        B'_{\rm out}[t]
          = B_{\rm out}[u_t]
          = \BitvectorMin[x' + u_t]
          = \BitvectorMin[\ISA{\Text}[i_t]]
      $.

    \item\label{sweep-step-3}
      In the third step, we prove that $k = q$, and that for every
      $t \in [1 \dd k]$, it holds $i_t + 1 = d_t$. To this end, we first
      prove that it holds
      $\{j' + 1 : j' \in \RMinusFour{s+1}{H}{\tau}{\Text} \setminus \mathcal{I}_{\rm text}\}
      = \RMinusFour{s}{H}{\tau}{\Text} \setminus \mathcal{D}_{\rm text}$.
      \begin{itemize}
      \item First, we prove that
        $\{j' + 1 : j' \in \RMinusFour{s+1}{H}{\tau}{\Text} \setminus \mathcal{I}_{\rm text}\}
        \subseteq \RMinusFour{s}{H}{\tau}{\Text} \setminus \mathcal{D}_{\rm text}$.
        Let $j \in \{j' + 1 : j' \in \RMinusFour{s+1}{H}{\tau}{\Text} \setminus \mathcal{I}_{\rm text}\}$.
        By definition, then there exists $j' \in \RMinusFour{s+1}{H}{\tau}{\Text}$
        such that $j' \not\in \mathcal{I}_{\rm text}$ and $j = j' + 1$.
        Let $j'' \in [1 \dd \Textlen]$ be the smallest position
        such that $[j'' \dd j'] \subseteq \RTwo{\tau}{\Text}$.
        By definition of $\RPrimTwo{\tau}{\Text}$, we then have
        $j'' \in \RPrimTwo{\tau}{\Text}$, and by
        \cref{lm:R-text-block}, it holds $j'' \in \RPrimMinusThree{H}{\tau}{\Text}$.
        Moreover, by \cref{lm:end}, we then have
        $j' \in [j'' \dd \RunEndPos{j''}{\tau}{\Text} - 3\tau + 2)$. Recall
        now that $j' \not\in \mathcal{I}_{\rm text}$. By $j' \in \RFour{s+1}{H}{\tau}{\Text}$,
        this implies $j' \not\in I_{0} \cup I_{1}$, i.e., $j' \not\in I_{0}$ and $j' \not\in I_{1}$.
        Recall that $j'' \leq \RunMinEndPos{j''}{\tau}{\Text} \leq \RunEndPos{j''}{\tau}{\Text} - 3\tau + 2$.
        Consider two cases:
        \begin{itemize}
        \item First, assume that $\RunMinEndPos{j''}{\tau}{\Text} < \RunEndPos{j''}{\tau}{\Text} - 3\tau + 2$.
          Then, $[\RunMinEndPos{j''}{\tau}{\Text} \dd \RunEndPos{j''}{\tau}{\Text} - 3\tau + 2) \neq \emptyset$,
          and hence by $j' \not\in I_{0}$ it follows that $j' \neq \RunEndPos{j''}{\tau}{\Text} - 3\tau + 1$.
        \item Let us now assume that $\RunMinEndPos{j''}{\tau}{\Text} = \RunEndPos{j''}{\tau}{\Text} - 3\tau + 2$.
          Observe that this implies that $[j'' \dd \RunMinEndPos{j''}{\tau}{\Text}) \neq \emptyset$.
          Thus, from the assumption $j' \not\in I_1$, we obtain that
          $j' \neq \RunMinEndPos{j''}{\tau}{\Text} - 1 = \RunEndPos{j''}{\tau}{\Text} - 3\tau + 1$.
        \end{itemize}
        In both cases, we thus obtain that $j' \neq \RunEndPos{j''}{\tau}{\Text} - 3\tau + 1$.
        This implies that $j' < \RunEndPos{j''}{\tau}{\Text} - 3\tau + 1$, and hence
        $j = j' + 1 \in [j'' \dd \RunEndPos{j''}{\tau}{\Text} - 3\tau + 2)$. In particular,
        $j \in \RTwo{\tau}{\Text}$, which by $j' + 1 = j$ and \cref{lm:R-text-block}, implies
        $j \in \RMinusThree{H}{\tau}{\Text}$. Since by \cref{lm:R-text-block} we also have
        $\RunEndFullPos{j}{\tau}{\Text} = \RunEndFullPos{j'}{\tau}{\Text}$, we obtain
        that $\HeadPos{j}{\tau}{\Text} = (\RunEndFullPos{j}{\tau}{\Text} - j) \bmod p
        = ((\RunEndFullPos{j'}{\tau}{\Text} - j') \bmod p) - 1 =
        \HeadPos{j'}{\tau}{\Text} - 1 = s$.
        Hence, $j \in \RMinusFour{s}{H}{\tau}{\Text}$.
        Next, we show that $j \not\in \mathcal{D}_{\rm text}$. Since $j \in \RFour{s}{H}{\tau}{\Text}$,
        we thus need to show that $j \not\in D_{0}$ and $j \not\in D_{1}$.
        To show $j \not\in D_{1}$, it suffices to observe that $j > j' \geq j''$.
        To show $j \not\in D_{0}$, we consider two cases:
        \begin{itemize}
        \item First, let us again assume $\RunMinEndPos{j''}{\tau}{\Text} < \RunEndPos{j''}{\tau}{\Text} - 3\tau + 2$.
          If $\RunMinEndPos{j''}{\tau}{\Text} = j''$, then by $j > j' \geq j'' = \RunMinEndPos{j''}{\tau}{\Text}$, we obtain
          $j \neq \RunMinEndPos{j''}{\tau}{\Text}$. Since $j'$ belongs to the maximal
          block $[j'' \dd \RunEndPos{j''}{\tau}{\Text} - 3\tau + 2)$ of positions from $\RTwo{\tau}{\Text}$
          (see \cref{lm:end}), it follows that $j \not\in D_{0}$. Let us thus assume
          that $[j'' \dd \RunMinEndPos{j''}{\tau}{\Text}) \neq \emptyset$. In that case, the
          assumption $j' \not\in I_{1}$
          implies that $j' \neq \RunMinEndPos{j''}{\tau}{\Text} - 1$. Consequently,
          $j \neq \RunMinEndPos{j''}{\tau}{\Text}$, and hence again $j \not\in D_{0}$.
        \item Let us now assume that $\RunMinEndPos{j''}{\tau}{\Text} = \RunEndPos{j''}{\tau}{\Text} - 3\tau + 2$.
          Then, $[j'' \dd \RunMinEndPos{j''}{\tau}{\Text}) \neq \emptyset$.
          Thus, by $j' \not\in I_1$, we obtain that
          $j' \neq \RunMinEndPos{j''}{\tau}{\Text} - 1$, and hence $j \neq \RunMinEndPos{j''}{\tau}{\Text}$.
          Since $j \in [j'' \dd \RunEndPos{j''}{\tau}{\Text} - 3\tau + 2)$, it follows that $j \not\in D_{0}$.
        \end{itemize}
        In both cases, we obtain $j \not\in D_{0}$. Combining with $j \not\in D_{1}$, we therefore obtain
        $j \not\in \mathcal{D}_{\rm text}$.
        Further combining with $j \in \RMinusFour{s}{H}{\tau}{\Text}$, we thus obtain
        $j \in \RMinusFour{s}{H}{\tau}{\Text} \setminus \mathcal{D}_{\rm text}$.
        This concludes the proof of
        $\{j' + 1 : j' \in \RMinusFour{s+1}{H}{\tau}{\Text} \setminus \mathcal{I}_{\rm text}\}
        \subseteq \RMinusFour{s}{H}{\tau}{\Text} \setminus \mathcal{D}_{\rm text}$.
      \item Next, we prove that
        $\RMinusFour{s}{H}{\tau}{\Text} \setminus \mathcal{D}_{\rm text}
        \subseteq \{j' + 1 : j' \in \RMinusFour{s+1}{H}{\tau}{\Text} \setminus \mathcal{I}_{\rm text}\}$.
        Let $j \in \RMinusFour{s}{H}{\tau}{\Text} \setminus \mathcal{D}_{\rm text}$.
        Let $j'' \in [1 \dd \Textlen]$ be the smallest position such that $[j'' \dd j] \subseteq \RTwo{\tau}{\Text}$.
        By definition, it holds $j'' \in \RPrimTwo{\tau}{\Text}$. By \cref{lm:R-text-block}, we have
        $j'' \in \RPrimMinusThree{H}{\tau}{\Text}$. Moreover, by \cref{lm:end}, we also have
        $j \in [j'' \dd \RunEndPos{j''}{\tau}{\Text} - 3\tau + 2)$. Recall now that $j \not\in \mathcal{D}_{\rm text}$.
        By $j \in \RFour{s}{H}{\tau}{\Text}$, this implies that $j \not\in D_{0} \cup D_{1}$, i.e., $j \not\in D_{0}$
        and $j \not\in D_{1}$.
        Recall that $j'' \leq \RunMinEndPos{j''}{\tau}{\Text} \leq \RunEndPos{j''}{\tau}{\Text} - 3\tau + 2$.
        Consider two cases:
        \begin{itemize}
        \item First, assume that it holds $j'' < \RunMinEndPos{j''}{\tau}{\Text}$. This is equivalent to
          $[j'' \dd \RunMinEndPos{j''}{\tau}{\Text}) \neq \emptyset$, and hence
          by $j \not\in D_{1}$, it follows that $j \neq j''$.
        \item Let us now assume that $j'' = \RunMinEndPos{j''}{\tau}{\Text}$.
          Then, $[\RunMinEndPos{j''}{\tau}{\Text} \dd \RunEndPos{j''}{\tau}{\Text} - 3\tau + 2) \neq \emptyset$,
          and hence by $j \not\in D_{0}$ it follows that $j \neq \RunMinEndPos{j''}{\tau}{\Text} = j''$.
        \end{itemize}
        In both cases, we obtain $j \neq j''$. This implies that $j > j''$, and hence letting $j' = j - 1$, it holds
        $j' \in [j'' \dd \RunEndPos{j''}{\tau}{\Text} - 3\tau + 2)$. In particular, $j' \in \RTwo{\tau}{\Text}$.
        By $j' = j - 1$ and \cref{lm:R-text-block}, it follows that $j' \in \RMinusThree{H}{\tau}{\Text}$.
        Since by \cref{lm:R-text-block} we also have
        $\RunEndFullPos{j}{\tau}{\Text} = \RunEndFullPos{j'}{\tau}{\Text}$, we obtain
        that $\HeadPos{j'}{\tau}{\Text} = (\RunEndFullPos{j'}{\tau}{\Text} - j') \bmod p
        = ((\RunEndFullPos{j}{\tau}{\Text} - j) \bmod p) + 1 =
        \HeadPos{j}{\tau}{\Text}+1 = s+1$ (where in the last equality we used that $s \in [0 \dd p-1)$).
        Hence, $j' \in \RMinusFour{s+1}{H}{\tau}{\Text}$.
        Next, we show that $j' \not\in \mathcal{I}_{\rm text}$. Since $j' \in \RFour{s+1}{H}{\tau}{\Text}$,
        this means that we need to prove that $j' \not\in I_{0}$ and $j' \not\in I_{1}$.
        To show $j' \not\in I_{0}$, it suffices to observe that $j' < j \leq \RunEndPos{j''}{\tau}{\Text} - 3\tau + 1$.
        To show $j' \not\in I_{1}$, we consider two cases:
        \begin{itemize}
        \item First, assume that $j'' < \RunMinEndPos{j''}{\tau}{\Text}$, i.e.,
          $[j'' \dd \RunMinEndPos{j''}{\tau}{\Text}) \neq \emptyset$.
          If $\RunMinEndPos{j''}{\tau}{\Text} = \RunEndPos{j''}{\tau}{\Text} - 3\tau + 2$, then
          $j < \RunEndPos{j''}{\tau}{\Text} - 3\tau + 2 = \RunMinEndPos{j''}{\tau}{\Text}$, we
          obtain $j \neq \RunMinEndPos{j''}{\tau}{\Text}$. Thus, $j' \neq \RunMinEndPos{j''}{\tau}{\Text} - 1$, and hence
          $j' \not\in I_{1}$. Let us thus assume $\RunMinEndPos{j''}{\tau}{\Text} < \RunEndPos{j''}{\tau}{\Text} - 3\tau + 2$,
          i.e., $[\RunMinEndPos{j''}{\tau}{\Text} \dd \RunEndPos{j''}{\tau}{\Text} - 3\tau - 2) \neq \emptyset$.
          By $j \not\in D_{0}$, it follows that $j \neq \RunMinEndPos{j''}{\tau}{\Text}$. Thus,
          $j' \neq \RunMinEndPos{j''}{\tau}{\Text} - 1$, and hence again $j' \not\in I_{1}$.
        \item Let us now assume that $j'' = \RunMinEndPos{j''}{\tau}{\Text}$.
          Then, $[\RunMinEndPos{j''}{\tau}{\Text} \dd \RunEndPos{j''}{\tau}{\Text} - 3\tau + 2) \neq \emptyset$, and hence
          by $j \not\in D_{0}$, we then obtain $j \neq \RunMinEndPos{j''}{\tau}{\Text}$. Thus,
          $j' \neq \RunMinEndPos{j''}{\tau}{\Text} - 1$, which implies $j' \not\in I_{1}$.
        \end{itemize}
        In both cases, we obtain $j' \not\in I_{1}$. Combining with $j' \not\in I_{0}$, we therefore obtain
        $j' \not\in \mathcal{I}_{\rm text}$. Further combining with $j' \in \RMinusFour{s+1}{H}{\tau}{\Text}$,
        we thus obtain $j' \in \RMinusFour{s+1}{H}{\tau}{\Text} \setminus \mathcal{I}_{\rm text}$.
        Lastly, noting that $j'+1 = j$, we thus conclude that
        $j \in \{j'+1 : j' \in \RMinusFour{s+1}{H}{\tau}{\Text} \setminus \mathcal{I}_{\rm text}\}$, which
        concludes the proof of the inclusion
        $\RMinusFour{s}{H}{\tau}{\Text} \setminus \mathcal{D}_{\rm text} \subseteq
        \{j' + 1 : j' \in \RMinusFour{s+1}{H}{\tau}{\Text} \setminus \mathcal{I}_{\rm text}\}$.
      \end{itemize}
      This concludes the proof of the equality
      $\{j' + 1 : j' \in \RMinusFour{s+1}{H}{\tau}{\Text} \setminus \mathcal{I}_{\rm text}\}
      = \RMinusFour{s}{H}{\tau}{\Text} \setminus \mathcal{D}_{\rm text}$.
      In other words, we proved that $k = q$ and $\{i_1 + 1, \ldots, i_q + 1\} = \{d_1, \ldots, d_k\}$.
      Recall now that the sequence $(i_1, \ldots, i_q)$ is sorted according to the lexicographical order
      of the corresponding suffixes, i.e., for every $t', t'' \in [1 \dd q]$, $t' < t''$ implies
      $\Text[i_{t'} \dd \Textlen] \prec \Text[i_{t''} \dd \Textlen]$. On the other hand,
      $\{i_1, \ldots, i_q\} \subseteq \RFour{s+1}{H}{\tau}{\Text}$ implies
      by \cref{lm:periodic-pos-lce}\eqref{lm:periodic-pos-lce-it-2},
      that all positions in the set $\{i_1, \ldots, i_q\}$ are followed by the same character in $\Text$.
      This implies that the sequence $(i_1 + 1, \ldots, i_q + 1)$ is also sorted according to the lexicographical
      order of the corresponding suffixes. Since the same property also holds for the sequence
      $(d_1, \ldots, d_k)$, we obtain by
      $\{i_1 + 1, \ldots, i_q + 1\} = \{d_1, \ldots, d_k\}$, that for every $t \in [1 \dd k]$, it
      holds $i_t + 1 = d_t$.

    \item\label{sweep-step-4}
      Next, we prove that for every $j \in \RMinusFour{s+1}{H}{\tau}{\Text} \setminus \mathcal{I}_{\rm text}$,
      $j \in \RMinMinusTwo{\tau}{\Text}$ holds if and only if $j+1 \in \RMinMinusTwo{\tau}{\Text}$.
      Let $j \in \RMinusFour{s+1}{H}{\tau}{\Text} \setminus \mathcal{I}_{\rm text}$. We then have
      $j \in \RMinusFour{s+1}{H}{\tau}{\Text}$, $j \not\in I_{0}$, and $j \not\in I_{1}$.
      Let $j'' \in [1 \dd \Textlen]$ be the smallest position such that $[j'' \dd j] \subseteq \RTwo{\tau}{\Text}$.
      By definition of $\RPrimTwo{\tau}{\Text}$, we have $j'' \in \RPrimTwo{\tau}{\Text}$. By \cref{lm:R-text-block},
      we moreover have $j'' \in \RPrimMinusThree{H}{\tau}{\Text}$. Lastly, by \cref{lm:end}, we then
      have $j \in [j'' \dd \RunEndPos{j''}{\tau}{\Text} - 3\tau + 2)$.
      \begin{itemize}
      \item Let us first assume that $j \in \RMinMinusTwo{\tau}{\Text}$. By $j \in [j'' \dd \RunEndPos{j''}{\tau}{\Text} - 3\tau + 2)$
        and \cref{lm:rmin-text-block}, this implies that
        $j \in [j'' \dd \RunMinEndPos{j''}{\tau}{\Text})$. The assumption $j \not\in I_{1}$ implies
        $j \neq \RunMinEndPos{j''}{\tau}{\Text} - 1$. Consequently, $j+1 \in [j'' \dd \RunMinEndPos{j''}{\tau}{\Text})$, which
        implies $j+1 \in \RMinMinusTwo{\tau}{\Text}$.
      \item We prove the opposite implication by contraposition. Let us thus assume that $j \not\in \RMinMinusTwo{\tau}{\Text}$.
        By $j \in [j'' \dd \RunEndPos{j''}{\tau}{\Text} - 3\tau + 2)$ and \cref{lm:rmin-text-block},
        $j \in [\RunMinEndPos{j''}{\tau}{\Text} \dd \RunEndPos{j''}{\tau}{\Text} - 3\tau + 2)$. The assumption $j \not\in I_{0}$ implies
        $j \neq \RunEndPos{j''}{\tau}{\Text} - 3\tau + 1$. Consequently, $j+1 \in
        [\RunMinEndPos{j''}{\tau}{\Text} \dd \RunEndPos{j''}{\tau}{\Text} - 3\tau + 2)$. By \cref{lm:rmin-text-block}, this implies
        $j+1 \not\in \RMinMinusTwo{\tau}{\Text}$.
      \end{itemize}

    \item\label{sweep-step-5}
      We now put everything together. In Step \ref{sweep-step-1}, we proved that,
      $|B'_{\rm in}| = k$ and that for
      every $t \in [1 \dd k]$, it holds $B'_{\rm in}[t] = \BitvectorMin[\ISA{\Text}[d_t]]$.
      In Step \ref{sweep-step-2}, we proved that
      $|B'_{\rm out}| = q$ and that for
      every $t \in [1 \dd q]$, it holds $B'_{\rm out}[t] = \BitvectorMin[\ISA{\Text}[i_t]]$.
      In Step \ref{sweep-step-3}, we proved that $k = q$, and that for every
      $t \in [1 \dd k]$, it holds $i_t + 1 = d_t$.
      Lastly, in Step \ref{sweep-step-4}, we proved that
      for every $j \in \RMinusFour{s+1}{H}{\tau}{\Text} \setminus \mathcal{I}_{\rm text}$,
      $j \in \RMinMinusTwo{\tau}{\Text}$ holds if and only if $j+1 \in \RMinMinusTwo{\tau}{\Text}$.
      Putting together Steps \ref{sweep-step-3} and \ref{sweep-step-4}, and recalling that
      $\{i_1, \ldots, i_q\} = \RMinusFour{s+1}{H}{\tau}{\Text} \setminus \mathcal{I}_{\rm text}$,
      we obtain by \cref{def:min-pos-bitvector-minus}
      that for every $t \in [1 \dd k]$, it holds $\BitvectorMin[\ISA{\Text}[i_t]] = \BitvectorMin[\ISA{\Text}[d_t]]$.
      Combining with Steps \ref{sweep-step-1} and \ref{sweep-step-2}, we therefore obtain that
      for every $t \in [1 \dd k]$, it holds
      \begin{align*}
        B'_{\rm in}[t]
          &= \BitvectorMin[\ISA{\Text}[d_t]]\\
          &= \BitvectorMin[\ISA{\Text}[i_t]]\\
          &= B'_{\rm out}[t].
      \end{align*}
      It remains to observe that $|B'_{\rm in}| = k = q = |B'_{\rm out}|$.
      We thus obtain that $B'_{\rm in} = B'_{\rm out}$.
    \end{enumerate}
  \end{enumerate}
  We have thus proved both conditions in \cref{def:insert}, i.e., that for
  every $(i,c) \in \mathcal{I}$, it holds $i \in [1 \dd |B_{\rm out}|]$ and
  $B_{\rm out}[i] = c$, and that $B'_{\rm in} = B'_{\rm out}$, i.e.,
  $\DeleteSubseq{B_{\rm in}}{\mathcal{D}_{\rm lex}} =
  \DeleteSubseq{B_{\rm out}}{\mathcal{I}_{\rm lex}}$. Since $\mathcal{D}_{\rm lex} = \mathcal{D}$,
  and $\mathcal{I}_{\rm lex} = \{i : (i,c) \in \mathcal{I}\}$, it thus follows that
  $B_{\rm out} = \InsertSubseq{\DeleteSubseq{B_{\rm in}}{\mathcal{D}}}{\mathcal{I}}$, i.e., the claim.
\end{proof}

\begin{lemma}\label{lm:sweep-init}
  Let $\tau \in [1 \dd \lfloor \tfrac{\Textlen}{2} \rfloor]$,
  $H \in \Sigma^{+}$, $p = |H|$, $s \in [0 \dd p)$, and $k > 0$.
  Let $x, y, x', y' \in [0 \dd \Textlen]$
  be such that
  \begin{itemize}
  \item $\{\SA{\Text}[i] : i \in (x \dd y]\} = \RMinusFive{s}{k}{H}{\tau}{\Text}$,
  \item $\{\SA{\Text}[i] : i \in (x' \dd y']\} = \RMinusFive{s}{k+1}{H}{\tau}{\Text}$.
  \end{itemize}
  Denote (see \cref{def:min-pos-bitvector-minus}):
  \begin{itemize}
  \item $\BitvectorMin = \MinPosBitvectorMinusTwo{\tau}{\Text}$,
  \item $B_{\rm in} = \MinPosBitvectorMinusFive{s}{k}{H}{\tau}{\Text} = \BitvectorMin(x \dd y]$,
  \item $B_{\rm out} = \MinPosBitvectorMinusFive{s}{k+1}{H}{\tau}{\Text} = \BitvectorMin(x' \dd y']$.
  \end{itemize}
  Let us also define
  \begin{itemize}[itemsep=1.5ex]
   \item $I_{0} = \{\max\;[\RunMinEndPos{j}{\tau}{\Text} \dd \RunEndPos{j}{\tau}{\Text} - 3\tau + 2) \cap \RFour{s}{H}{\tau}{\Text} :\\
     \hspace*{1.5cm}j \in \RPrimMinusThree{H}{\tau}{\Text}\text{ and }
     [\RunMinEndPos{j}{\tau}{\Text} \dd \RunEndPos{j}{\tau}{\Text} - 3\tau + 2) \cap \RFour{s}{H}{\tau}{\Text} \neq \emptyset\}$,
  \item $D_{0} = \{\min\;[\RunMinEndPos{j}{\tau}{\Text} \dd \RunEndPos{j}{\tau}{\Text} - 3\tau + 2) \cap \RFour{s}{H}{\tau}{\Text} :\\
    \hspace*{1.5cm}j \in \RPrimMinusThree{H}{\tau}{\Text}\text{ and }
    [\RunMinEndPos{j}{\tau}{\Text} \dd \RunEndPos{j}{\tau}{\Text} - 3\tau + 2) \cap \RFour{s}{H}{\tau}{\Text} \neq \emptyset\}$,
  \item $I_{1} = \{\max\;[j \dd \RunMinEndPos{j}{\tau}{\Text}) \cap \RFour{s}{H}{\tau}{\Text} :\\
    \hspace*{1.5cm}j \in \RPrimMinusThree{H}{\tau}{\Text}\text{ and }
    [j \dd \RunMinEndPos{j}{\tau}{\Text}) \cap \RFour{s}{H}{\tau}{\Text} \neq \emptyset\}$,
  \item $D_{1} = \{\min\;[j \dd \RunMinEndPos{j}{\tau}{\Text}) \cap \RFour{s}{H}{\tau}{\Text} :\\
    \hspace*{1.5cm}j \in \RPrimMinusThree{H}{\tau}{\Text}\text{ and }
    [j \dd \RunMinEndPos{j}{\tau}{\Text}) \cap \RFour{s}{H}{\tau}{\Text} \neq \emptyset\}$.
  \end{itemize}
  Then, it holds
  \[
    B_{\rm out} =
      \InsertSubseq{\DeleteSubseq{B_{\rm in}}{\mathcal{D}}}{\mathcal{I}},
  \]
  where
  \begin{align*}
    \mathcal{D} &= \{\ISA{\Text}[j] - x :
      j \in (D_{0} \cup D_{1}) \cap \RFive{s}{k}{H}{\tau}{\Text}
      \},\\
    \mathcal{I} &=
      \{(\ISA{\Text}[j] - x', 0) :
        j \in I_{0} \cap \RFive{s}{k+1}{H}{\tau}{\Text}
      \}\ \cup\\
    &\hspace{3.1ex}
      \{(\ISA{\Text}[j] - x', 1) :
        j \in I_{1} \cap \RFive{s}{k+1}{H}{\tau}{\Text}
       \}.
  \end{align*}
\end{lemma}
\begin{proof}

  The proof below is similar to the proof of \cref{lm:sweep-step}, and hence we focus on highlighting the
  key differences. In \cref{lm:sweep-step}, we show
  how to obtain $\MinPosBitvectorMinusFour{s+1}{H}{\tau}{\Text}$ from $\MinPosBitvectorMinusFour{s}{H}{\tau}{\Text}$.
  Here we show how to obtain $\MinPosBitvectorMinusFive{s}{k+1}{H}{\tau}{\Text}$ from
  $\MinPosBitvectorMinusFive{s}{k}{H}{\tau}{\Text}$.
  Both results are utilized in the construction of $\MinPosBitvectorMinusTwo{\tau}{\Text}$.

  First, note that the pairs $(x,y)$ and $(x',y')$ are well-defined by \cref{rm:min-pos-bitvector-minus}.

  Denote
  \begin{align*}
    \mathcal{D}_{\rm text} &= (D_{0} \cup D_{1}) \cap \RFive{s}{k}{H}{\tau}{\Text},\\
    \mathcal{I}_{\rm text} &= (I_{0} \cup I_{1}) \cap \RFive{s}{k+1}{H}{\tau}{\Text},\\
    \mathcal{D}_{\rm lex} &=
      \{\ISA{\Text}[j] - x : j \in \mathcal{D}_{\rm text}\},\\
    \mathcal{I}_{\rm lex} &=
      \{\ISA{\Text}[j] - x' : j \in \mathcal{I}_{\rm text}\}.
  \end{align*}

  First, we prove that
  $\mathcal{D}_{\rm text} \subseteq \RMinusFive{s}{k}{H}{\tau}{\Text}$,
  $\mathcal{I}_{\rm text} \subseteq \RMinusFive{s}{k+1}{H}{\tau}{\Text}$,
  $\mathcal{D}_{\rm lex} \subseteq [1 \dd |B_{\rm in}|]$, and
  $\mathcal{I}_{\rm lex} \subseteq [1 \dd |B_{\rm out}|]$:
  \begin{itemize}
  \item Let $j \in \mathcal{D}_{\rm text}$, i.e., $j \in (D_{0} \cup D_{1}) \cap \RFive{s}{k}{H}{\tau}{\Text}$.
    By \cref{lm:end}, for every $t \in \RPrimTwo{\tau}{\Text}$,
    $[t \dd \RunEndPos{t}{\tau}{\Text} - 3\tau + 2)$ is a maximal block of
    positions from the set $\RTwo{\tau}{\Text}$. By \cref{def:emin}, for
    every $t \in \RPrimMinusTwo{\tau}{\Text}$, we thus have
    $t \leq \RunMinEndPos{t}{\tau}{\Text} \leq \RunEndPos{t}{\tau}{\Text} - 3\tau + 2$.
    By \cref{lm:R-text-block}, and the definition of $D_{0}$ and $D_{1}$, we therefore
    have $j \in \RMinusTwo{\tau}{\Text}$. Combining with $j \in \RFive{s}{k}{H}{\tau}{\Text}$,
    we therefore obtain that $j \in \RMinusFive{s}{k}{H}{\tau}{\Text}$. We thus have
    $\mathcal{D}_{\rm text} \subseteq \RMinusFive{s}{k}{H}{\tau}{\Text}$.
  \item Let $j \in \mathcal{I}_{\rm text}$, i.e., $j \in (I_{0} \cup I_{1}) \cap \RFive{s}{k+1}{H}{\tau}{\Text}$.
    By the same argument as above, it follows from the definition of $I_{0}$ and $I_{1}$, and
    \cref{lm:R-text-block}, that we then have $j \in \RMinusTwo{\tau}{\Text}$. Combining with
    $j \in \RFive{s}{k+1}{H}{\tau}{\Text}$, we therefore obtain that $j \in \RMinusFive{s}{k+1}{H}{\tau}{\Text}$.
    Hence, $\mathcal{I}_{\rm text} \subseteq \RMinusFive{s}{k+1}{H}{\tau}{\Text}$.
  \item Let $p \in \mathcal{D}_{\rm lex}$. Then, there exists $j \in \mathcal{D}_{\rm text}$
    such that $p = \ISA{\Text}[j] - x$. By the above, $j \in \RMinusFive{s}{k}{H}{\tau}{\Text}$. Thus, 
    by definition of $x$ and $y$, it holds $\ISA{\Text}[j] \in (x \dd y]$. Therefore,
    $p = \ISA{\Text}[j] - x \in [1 \dd y-x] = [1 \dd |B_{\rm in}|]$. We thus obtain
    $\mathcal{D}_{\rm lex} \subseteq [1 \dd |B_{\rm in}|]$.
  \item Let $p \in \mathcal{I}_{\rm lex}$. Similarly as above, then there exists $j \in \mathcal{I}_{\rm text}$
    such that $p = \ISA{\Text}[j] - x'$. By $j \in \RMinusFive{s}{k+1}{H}{\tau}{\Text}$ and the definition of $x'$ and $y'$,
    we thus have $\ISA{\Text}[j] \in (x' \dd y']$. Therefore,
    $p = \ISA{\Text}[j] - x' \in [1 \dd y'-x'] = [1 \dd |B_{\rm out}|]$. We thus obtain
    $\mathcal{I}_{\rm lex} \subseteq [1 \dd |B_{\rm out}|]$.
  \end{itemize}

  We are now ready to show the main claim, i.e., $B_{\rm out} =
  \InsertSubseq{\DeleteSubseq{B_{\rm in}}{\mathcal{D}}}{\mathcal{I}}$.
  Note that the strings $\DeleteSubseq{B_{\rm in}}{\mathcal{D}_{\rm lex}} = \DeleteSubseq{B_{\rm in}}{\mathcal{D}}$
  and $\InsertSubseq{\DeleteSubseq{B_{\rm in}}{\mathcal{D}}}{\mathcal{I}}$ are
  well-defined by the same arguments as in the proof of \cref{lm:sweep-step}.
  To finalize the proof of $B_{\rm out} = \InsertSubseq{\DeleteSubseq{B_{\rm in}}{\mathcal{D}}}{\mathcal{I}}$,
  we need to show that the two conditions in \cref{def:insert} hold. We proceed as follows:
  \begin{enumerate}

  \item First, we prove that for every $(i,c) \in \mathcal{I}$, it holds
    $i \in [1 \dd |B_{\rm out}|]$ and
    $B_{\rm out}[i] = c$.
    Let $(i, c) \in \mathcal{I}$. Note that then $i \in \mathcal{I}_{\rm lex}$.
    Thus, by $\mathcal{I}_{\rm lex} \subseteq [1 \dd |B_{\rm out}|]$, we obtain the first
    part of the claim. To show $B_{\rm out}[i] = c$, we consider two cases:
    \begin{itemize}
    \item First, let us assume that $c = 0$. In this case, by definition of
      $\mathcal{I}$, there exists $j \in I_0 \cap \RFive{s}{k+1}{H}{\tau}{\Text}$.
      such that $i = \ISA{\Text}[j] - x'$. This in turn implies that, by definition of $I_0$, there
      exists $j' \in \RPrimMinusThree{H}{\tau}{\Text}$ satisfying
      $[\RunMinEndPos{j'}{\tau}{\Text} \dd \RunEndPos{j'}{\tau}{\Text} - 3\tau + 2) \cap \RFour{s}{H}{\tau}{\Text} \neq \emptyset$
      and $j = \max\;[\RunMinEndPos{j'}{\tau}{\Text} \dd \RunEndPos{j'}{\tau}{\Text} - 3\tau + 2) \cap \RFour{s}{H}{\tau}{\Text}$.
      By $j \geq \RunMinEndPos{j'}{\tau}{\Text}$, \cref{def:emin}, and \cref{lm:rmin-text-block},
      it then holds $j \not\in \RMinMinusTwo{\tau}{\Text}$. Consequently, by \cref{def:min-pos-bitvector-minus},
      we have $\BitvectorMin[\ISA{\Text}[j]] = 0$. By $i = \ISA{\Text}[j] - x'$ and the definition of $B_{\rm out}$, we thus
      obtain $B_{\rm out}[i] = B_{\rm out}[\ISA{\Text}[j] - x'] = \BitvectorMin[\ISA{\Text}[j]] = 0$.
    \item Let us now assume $c = 1$. In this case, by definition of
      $\mathcal{I}$, there exists $j \in I_1 \cap \RFive{s}{k+1}{H}{\tau}{\Text}$ such that
      $i = \ISA{\Text}[j] - x'$. This in turn implies that, by definition of $I_1$, there
      exists $j' \in \RPrimMinusThree{H}{\tau}{\Text}$ satisfying
      $[j' \dd \RunMinEndPos{j'}{\tau}{\Text}) \cap \RFour{s}{H}{\tau}{\Text} \neq \emptyset$
      and $j = \max\;[j' \dd \RunMinEndPos{j'}{\tau}{\Text}) \cap \RFour{s}{H}{\tau}{\Text}$.
      By $j < \RunMinEndPos{j'}{\tau}{\Text}$ and \cref{def:emin},
      it holds $j \in \RMinMinusTwo{\tau}{\Text}$. Consequently, by \cref{def:min-pos-bitvector-minus},
      we have $\BitvectorMin[\ISA{\Text}[j]] = 1$. By $i = \ISA{\Text}[j] - x'$ and the definition of $B_{\rm out}$, we thus
      obtain $B_{\rm out}[i] = B_{\rm out}[\ISA{\Text}[j] - x'] = \BitvectorMin[\ISA{\Text}[j]] = 1$.
    \end{itemize}
    In both cases, we have thus obtained $B_{\rm out}[i] = c$.

  \item Second, we prove that, denoting
    \begin{align*}
      B'_{\rm in}
        &= \DeleteSubseq{B_{\rm in}}{\mathcal{D}_{\rm lex}},\\
      B'_{\rm out}
        &= \DeleteSubseq{B_{\rm out}}{\mathcal{I}_{\rm lex}},
    \end{align*}
    it holds $B'_{\rm in} = B'_{\rm out}$. Let
    \begin{align*}
      \{d_1, \ldots, d_k\}
        &= \RMinusFive{s}{k}{H}{\tau}{\Text} \setminus \mathcal{D}_{\rm text},\\
      \{i_1, \ldots, i_q\}
        &= \RMinusFive{s}{k+1}{H}{\tau}{\Text} \setminus \mathcal{I}_{\rm text}
    \end{align*}
    be such that for every $t \in [1 \dd k)$, it holds
    $\Text[d_t \dd \Textlen] \prec \Text[d_{t+1} \dd \Textlen]$, and for
    every $t \in [1 \dd q)$, it holds
    $\Text[i_{t} \dd \Textlen] \prec \Text[i_{t+1} \dd \Textlen]$.
    The proof consists of five steps:
    \begin{enumerate}

    \item\label{sweep-init-1}
      In the first step, we prove that $|B'_{\rm in}| = k$ and that for
      every $t \in [1 \dd k]$, it holds $B'_{\rm in}[t] = \BitvectorMin[\ISA{\Text}[d_t]]$.
      To show the first claim, recall that $\mathcal{D}_{\rm text} \subseteq \RMinusFive{s}{k}{H}{\tau}{\Text}$.
      Thus,
      $
        |B'_{\rm in}|
          = |\DeleteSubseq{B_{\rm in}}{\mathcal{D}_{\rm lex}}|
          = |B_{\rm in}| - |\mathcal{D}_{\rm lex}|
          = |\RMinusFive{s}{k}{H}{\tau}{\Text}| - |\mathcal{D}_{\rm text}|
          = |\RMinusFive{s}{k}{H}{\tau}{\Text} \setminus \mathcal{D}_{\rm text}| = k
      $.
      Let $(u_1, \ldots, u_k)$ be an increasing sequence satisfying
      $\{u_1, \ldots, u_k\} = [1 \dd |B_{\rm in}|] \setminus \mathcal{D}_{\rm lex}$.
      Since $\{\SA{\Text}[x+i] : i \in [1 \dd |B_{\rm in}|]\} = \RMinusFive{s}{k}{H}{\tau}{\Text}$
      and $\{\SA{\Text}[x+i] : i \in \mathcal{D}_{\rm lex}\} = \mathcal{D}_{\rm text}$, it follows that
      $
        \{\SA{\Text}[x+u_t] : t \in [1 \dd k]\}
          = \{\SA{\Text}[x+i] : i \in [1 \dd |B_{\rm in}|] \setminus \mathcal{D}_{\rm lex}\}
          = \RMinusFive{s}{k}{H}{\tau}{\Text} \setminus \mathcal{D}_{\rm text}
          = \{d_1, \ldots, d_k\}
      $.
      Recall that for every $t \in [1 \dd k)$, it holds
      $\Text[\SA{\Text}[x + u_t] \dd \Textlen] \prec \Text[\SA{\Text}[x + u_{t+1}] \dd \Textlen]$
      and $\Text[d_t \dd \Textlen] \prec \Text[d_{t+1} \dd \Textlen]$.
      By $\{\SA{\Text}[x+u_t] : t \in [1 \dd k]\} = \{d_1, \ldots, d_k\}$, we thus obtain
      that for every $t \in [1 \dd k]$, it holds $\SA{\Text}[x+u_t] = d_t$.
      In particular, $\BitvectorMin[\ISA{\Text}[\SA{\Text}[x+u_t]]] = \BitvectorMin[\ISA{\Text}[d_t]]$, or
      equivalently, $\BitvectorMin[x+u_t] = \BitvectorMin[\ISA{\Text}[d_t]]$.
      It remains to observe that by \cref{def:delete}, for every
      $t \in [1 \dd k]$, it holds $B'_{\rm in}[t] = B_{\rm in}[u_t]$. Putting
      everything together, we thus obtain that for every $i \in [1 \dd k]$,
      it holds
      $
        B'_{\rm in}[t]
          = B_{\rm in}[u_t]
          = \BitvectorMin[x + u_t]
          = \BitvectorMin[\ISA{\Text}[d_t]]
      $.

    \item\label{sweep-init-2}
      In the second step, we prove that $|B'_{\rm out}| = q$ and that for
      every $t \in [1 \dd q]$, it holds $B'_{\rm out}[t] = \BitvectorMin[\ISA{\Text}[i_t]]$.
      To show the first claim, recall that $\mathcal{I}_{\rm text} \subseteq \RMinusFive{s}{k+1}{H}{\tau}{\Text}$.
      Thus, it holds
      $
        |B'_{\rm out}|
          = |\DeleteSubseq{B_{\rm out}}{\mathcal{I}_{\rm lex}}|
          = |B_{\rm out}| - |\mathcal{I}_{\rm lex}|
          = |\RMinusFive{s}{k+1}{H}{\tau}{\Text}| - |\mathcal{I}_{\rm text}|
          = |\RMinusFive{s}{k+1}{H}{\tau}{\Text} \setminus \mathcal{I}_{\rm text}| = q
      $.
      Let $(u_1, \ldots, u_q)$ be an increasing sequence satisfying
      $\{u_1, \ldots, u_q\} = [1 \dd |B_{\rm out}|] \setminus \mathcal{I}_{\rm lex}$.
      Since $\{\SA{\Text}[x'+i] : i \in [1 \dd |B_{\rm out}|]\} = \RMinusFive{s}{k+1}{H}{\tau}{\Text}$
      and $\{\SA{\Text}[x'+i] : i \in \mathcal{I}_{\rm lex}\} = \mathcal{I}_{\rm text}$, it follows that
      $
        \{\SA{\Text}[x'+u_t] : t \in [1 \dd q]\}
          = \{\SA{\Text}[x'+i] : i \in [1 \dd |B_{\rm out}|] \setminus \mathcal{I}_{\rm lex}\}
          = \RMinusFive{s}{k+1}{H}{\tau}{\Text} \setminus \mathcal{I}_{\rm text}
          = \{i_1, \ldots, i_q\}
      $.
      Recall that for every $t \in [1 \dd q)$, it holds
      $\Text[\SA{\Text}[x' + u_t] \dd \Textlen] \prec \Text[\SA{\Text}[x' + u_{t+1}] \dd \Textlen]$
      and $\Text[i_t \dd \Textlen] \prec \Text[i_{t+1} \dd \Textlen]$.
      By $\{\SA{\Text}[x'+u_t] : t \in [1 \dd q]\} = \{i_1, \ldots, i_q\}$, we thus obtain
      that for every $t \in [1 \dd q]$, it holds $\SA{\Text}[x'+u_t] = i_t$.
      In particular, $\BitvectorMin[\ISA{\Text}[\SA{\Text}[x'+u_t]]] = \BitvectorMin[\ISA{\Text}[i_t]]$, or
      equivalently, $\BitvectorMin[x'+u_t] = \BitvectorMin[\ISA{\Text}[i_t]]$.
      It remains to observe that by \cref{def:delete}, for every
      $t \in [1 \dd q]$, it holds $B'_{\rm out}[t] = B_{\rm out}[u_t]$. Putting
      everything together, we thus obtain that for every $i \in [1 \dd q]$,
      it holds
      $
        B'_{\rm out}[t]
          = B_{\rm out}[u_t]
          = \BitvectorMin[x' + u_t]
          = \BitvectorMin[\ISA{\Text}[i_t]]
      $.

    \item\label{sweep-init-3}
      In the third step, we prove that $k = q$, and that for every
      $t \in [1 \dd k]$, it holds $i_t + p = d_t$. To this end, we first
      prove that it holds
      $\{j' + p : j' \in \RMinusFive{s}{k+1}{H}{\tau}{\Text} \setminus \mathcal{I}_{\rm text}\}
      = \RMinusFive{s}{k}{H}{\tau}{\Text} \setminus \mathcal{D}_{\rm text}$.
      \begin{itemize}
      \item First, we prove that
        $\{j' + p : j' \in \RMinusFive{s}{k+1}{H}{\tau}{\Text} \setminus \mathcal{I}_{\rm text}\}
        \subseteq \RMinusFive{s}{k}{H}{\tau}{\Text} \setminus \mathcal{D}_{\rm text}$.
        Let $j \in \{j' + p : j' \in \RMinusFive{s}{k+1}{H}{\tau}{\Text} \setminus \mathcal{I}_{\rm text}\}$.
        By definition, then there exists $j' \in \RMinusFive{s}{k+1}{H}{\tau}{\Text}$
        such that $j' \not\in \mathcal{I}_{\rm text}$ and $j = j' + p$.
        Let $j'' \in [1 \dd \Textlen]$ be the smallest position
        such that $[j'' \dd j'] \subseteq \RTwo{\tau}{\Text}$.
        By definition of $\RPrimTwo{\tau}{\Text}$, we then have
        $j'' \in \RPrimTwo{\tau}{\Text}$, and by
        \cref{lm:R-text-block}, it holds $j'' \in \RPrimMinusThree{H}{\tau}{\Text}$.
        Moreover, by \cref{lm:end}, we then have
        $j' \in [j'' \dd \RunEndPos{j''}{\tau}{\Text} - 3\tau + 2)$. Recall
        now that $j' \not\in \mathcal{I}_{\rm text}$. By $j' \in \RFive{s}{k+1}{H}{\tau}{\Text}$,
        this implies $j' \not\in I_{0} \cup I_{1}$, i.e., $j' \not\in I_{0}$ and $j' \not\in I_{1}$.
        Recall that $j'' \leq \RunMinEndPos{j''}{\tau}{\Text} \leq \RunEndPos{j''}{\tau}{\Text} - 3\tau + 2$.
        Consider two cases:
        \begin{itemize}
        \item First, assume that $j' \in [\RunMinEndPos{j''}{\tau}{\Text} \dd \RunEndPos{j''}{\tau}{\Text} - 3\tau + 2)$.
          Since we also have $j' \in \RFour{s}{H}{\tau}{\Text}$, 
          it thus follows by $j' \not\in I_{0}$ that, letting $t =
          \max\;[\RunMinEndPos{j''}{\tau}{\Text} \dd \RunEndPos{j''}{\tau}{\Text} - 3\tau + 2) \cap \RFour{s}{H}{\tau}{\Text}$
          it holds $j' < t$.
        \item Let us now assume that $j' \in [j'' \dd \RunMinEndPos{j''}{\tau}{\Text})$. Since we also have
          $j' \in \RFour{s}{H}{\tau}{\Text}$, it follows by $j' \not\in I_{1}$ that, letting
          $t = \max\;[j'' \dd \RunMinEndPos{j''}{\tau}{\Text}) \cap \RFour{s}{H}{\tau}{\Text}$, it holds
          $j' < t$.
        \end{itemize}
        In both cases we thus obtain that there exists $t \in [j'' \dd \RunEndPos{j''}{\tau}{\Text} - 3\tau + 2)$ satisfying
        $t \in \RFour{s}{H}{\tau}{\Text}$ and $j' < t$. Recall now that by \cref{lm:R-text-block}, for every
        $t', t'' \in [j'' \dd \RunEndPos{j''}{\tau}{\Text} - 3\tau + 2)$, it holds $\RunEndFullPos{t'}{\tau}{\Text} 
        = \RunEndFullPos{t''}{\tau}{\Text}$. On the other hand, for every $t' \in [j'' \dd \RunEndPos{j''}{\tau}{\Text} - 3\tau + 2)$
        we have $\HeadPos{t'}{\tau}{\Text} = (\RunEndFullPos{t'}{\tau}{\Text} - t') \bmod p$. Consequently, for
        every $t' \in [j'' \dd \RunEndPos{j''}{\tau}{\Text} - 3\tau + 2)$, $\HeadPos{t'}{\tau}{\Text} = s$ holds if and only if
        $(\RunEndFullPos{t'}{\tau}{\Text} - t') \bmod p = (\RunEndFullPos{j'}{\tau}{\Text} - j') \bmod p$,
        which in turn is equivalent
        to $p \mid (t' - j')$. By $j' < t$, this implies that $j = j' + p \leq t$, and hence
        $j = j' + p  \in [j'' \dd \RunEndFullPos{j''}{\tau}{\Text} - 3\tau + 2)$.
        Consequently, $j \in \RMinusFour{s}{H}{\tau}{\Text}$.
        If remains to observe that
        \begin{align*}
          \ExpPos{j}{\tau}{\Text}
            &= \lfloor \tfrac{\RunEndFullPos{j}{\tau}{\Text} - j}{p} \rfloor\\
            &= \lfloor \tfrac{\RunEndFullPos{j'}{\tau}{\Text} - j' - p}{p} \rfloor\\
            &= \lfloor \tfrac{\RunEndFullPos{j'}{\tau}{\Text} - j'}{p} \rfloor - 1\\
            &= \ExpPos{j'}{\tau}{\Text} - 1 = k.
        \end{align*}
        Hence, $j \in \RMinusFive{s}{k}{H}{\tau}{\Text}$.
        Next, we show that $j \not\in \mathcal{D}_{\rm text}$. Since $j \in \RFive{s}{k}{H}{\tau}{\Text}$,
        we thus need to show that $j \not\in D_{0}$ and $j \not\in D_{1}$.
        To show $j \not\in D_{1}$, it suffices to observe that we either have
        $[j'' \dd \RunMinEndPos{j''}{\tau}{\Text}) \cap \RFour{s}{H}{\tau}{\Text} = \emptyset$ (in which case,
        $j \not\in D_{1}$ follows immediately), or
        $[j'' \dd \RunMinEndPos{j''}{\tau}{\Text}) \cap \RFour{s}{H}{\tau}{\Text} \neq \emptyset$,
        and then by
        $j' \in \RFour{s}{H}{\tau}{\Text}$ and $j'' \leq j' < j$, it follows that
        $\min\;[j'' \dd \RunMinEndPos{j''}{\tau}{\Text}) \cap \RFour{s}{H}{\tau}{\Text} \leq j' < j$ (which again yields
        $j \not\in D_{1}$). We now show that $j \not\in D_{0}$. First, we prove that it is not possible that
        $j' < \RunMinEndPos{j''}{\tau}{\Text} \leq j$. Suppose that this is the case, and recall from above that for
        every $t', t'' \in [j'' \dd \RunEndPos{j''}{\tau}{\Text}) \cap \RFour{s}{H}{\tau}{\Text}$, it holds
        $p \mid t' - t''$. By $j = j' + p$,
        this implies that $j' = \max\;[j'' \dd \RunMinEndPos{j''}{\tau}{\Text}) \cap \RFour{s}{H}{\tau}{\Text}$,
        which contradicts the assumption $j' \not\in I_{1}$. Thus, we must either have $j < \RunMinEndPos{j''}{\tau}{\Text}$
        or $\RunMinEndPos{j''}{\tau}{\Text} \leq j'$. We consider each of the two cases separately:
        \begin{itemize}
        \item First, assume that $j < \RunMinEndPos{j''}{\tau}{\Text}$. By definition of $D_{0}$, this immediately implies
          $j \not\in D_{0}$.
        \item Let us now assume that $\RunMinEndPos{j''}{\tau}{\Text} \leq j'$.
          By $j' \in [j'' \dd \RunEndPos{j''}{\tau}{\Text} - 3\tau + 2)$ and $j' \in \RFour{s}{H}{\tau}{\Text}$, it then
          follows that $\min\;[\RunMinEndPos{j''}{\tau}{\Text} \dd \RunEndPos{j''}{\tau}{\Text} - 3\tau + 2)
          \cap \RFour{s}{H}{\tau}{\Text} \leq j' < j$. Consequently, $j \not\in D_{0}$.
        \end{itemize}
        In both cases, we obtain $j \not\in D_{0}$. Combining with $j \not\in D_{1}$, we therefore obtain
        $j \not\in \mathcal{D}_{\rm text}$.
        Further combining with $j \in \RMinusFive{s}{k}{H}{\tau}{\Text}$, we thus obtain
        $j \in \RMinusFive{s}{k}{H}{\tau}{\Text} \setminus \mathcal{D}_{\rm text}$.
        This concludes the proof of
        $\{j' + p : j' \in \RMinusFive{s}{k+1}{H}{\tau}{\Text} \setminus \mathcal{I}_{\rm text}\}
        \subseteq \RMinusFive{s}{k}{H}{\tau}{\Text} \setminus \mathcal{D}_{\rm text}$.
      \item Next, we prove that
        $\RMinusFive{s}{k}{H}{\tau}{\Text} \setminus \mathcal{D}_{\rm text}
        \subseteq \{j' + p : j' \in \RMinusFive{s}{k+1}{H}{\tau}{\Text} \setminus \mathcal{I}_{\rm text}\}$.
        Let $j \in \RMinusFive{s}{k}{H}{\tau}{\Text} \setminus \mathcal{D}_{\rm text}$.
        Let $j'' \in [1 \dd \Textlen]$ be the smallest position such that $[j'' \dd j] \subseteq \RTwo{\tau}{\Text}$.
        By definition, it holds $j'' \in \RPrimTwo{\tau}{\Text}$. By \cref{lm:R-text-block}, we have
        $j'' \in \RPrimMinusThree{H}{\tau}{\Text}$. Moreover, by \cref{lm:end}, we also have
        $j \in [j'' \dd \RunEndPos{j''}{\tau}{\Text} - 3\tau + 2)$. Recall now that $j \not\in \mathcal{D}_{\rm text}$.
        By $j \in \RFive{s}{k}{H}{\tau}{\Text}$, this implies that $j \not\in D_{0} \cup D_{1}$, i.e., $j \not\in D_{0}$
        and $j \not\in D_{1}$.
        Recall that $j'' \leq \RunMinEndPos{j''}{\tau}{\Text} \leq \RunEndPos{j''}{\tau}{\Text} - 3\tau + 2$.
        Consider two cases:
        \begin{itemize}
        \item First, assume that $j \in [\RunMinEndPos{j''}{\tau}{\Text} \dd \RunEndPos{j''}{\tau}{\Text} - 3\tau + 2)$.
          Since we also have $j \in \RFour{s}{H}{\tau}{\Text}$, 
          it thus follows by $j \not\in D_{0}$ that, letting $t =
          \min\;[\RunMinEndPos{j''}{\tau}{\Text} \dd \RunEndPos{j''}{\tau}{\Text} - 3\tau + 2) \cap \RFour{s}{H}{\tau}{\Text}$
          it holds $t < j$.
        \item Let us now assume that $j \in [j'' \dd \RunMinEndPos{j''}{\tau}{\Text})$. Since we also have
          $j \in \RFour{s}{H}{\tau}{\Text}$, it follows by $j \not\in D_{1}$ that, letting
          $t = \min\;[j'' \dd \RunMinEndPos{j''}{\tau}{\Text}) \cap \RFour{s}{H}{\tau}{\Text}$, it holds
          $t < j$.
        \end{itemize}
        In both cases we thus obtain that there exists $t \in [j'' \dd \RunEndPos{j''}{\tau}{\Text} - 3\tau + 2)$ satisfying
        $t \in \RFour{s}{H}{\tau}{\Text}$ and $t < j$. Recall now that above we observed that for every $t', t'' \in
        [j'' \dd \RunEndPos{j''}{\tau}{\Text} - 3\tau + 2) \cap \RFour{s}{H}{\tau}{\Text}$, it holds $p \mid t' - t''$.
        This implies that $t \leq j - p$, and hence letting $j' = j - p$, we obtain that
        $j' \in [j'' \dd \RunEndPos{j''}{\tau}{\Text} - 3\tau + 2)$. By \cref{lm:R-text-block}
        we thus obtain $j' \in \RMinusThree{H}{\tau}{\Text}$. Since also by \cref{lm:R-text-block}, it holds
        $\RunEndFullPos{j'}{\tau}{\Text} = \RunEndFullPos{j}{\tau}{\Text}$, we thus obtain
        $\HeadPos{j'}{\tau}{\Text} = (\RunEndFullPos{j'}{\tau}{\Text} - j') \bmod p = (\RunEndFullPos{j}{\tau}{\Text} - j + p) \bmod p
        = (\RunEndFullPos{j}{\tau}{\Text} - j) \bmod p = \HeadPos{j}{\tau}{\Text} = s$.
        Consequently, $j' \in \RMinusFour{s}{H}{\tau}{\Text}$.
        It remains to observe that 
        \begin{align*}
          \ExpPos{j'}{\tau}{\Text}
            &= \lfloor \tfrac{\RunEndFullPos{j'}{\tau}{\Text} - j'}{p} \rfloor\\
            &= \lfloor \tfrac{\RunEndFullPos{j}{\tau}{\Text} - (j - p)}{p} \rfloor\\
            &= \lfloor \tfrac{\RunEndFullPos{j}{\tau}{\Text} - j}{p} \rfloor + 1\\
            &= \ExpPos{j}{\tau}{\Text} + 1 = k + 1.
        \end{align*}
        Hence, $j' \in \RMinusFive{s}{k+1}{H}{\tau}{\Text}$.
        Next, we show that $j' \not\in \mathcal{I}_{\rm text}$. Since $j' \in \RFive{s}{k+1}{H}{\tau}{\Text}$,
        this means that we need to prove that $j' \not\in I_{0}$ and $j' \not\in I_{1}$.
        To show $j' \not\in I_{0}$, it suffices to observe that
        we either have
        $[\RunMinEndPos{j''}{\tau}{\Text} \dd \RunEndPos{j''}{\tau}{\Text} - 3\tau + 2) \cap \RFour{s}{H}{\tau}{\Text} =
        \emptyset$ (in which case,
        $j' \not\in I_{0}$ follows immediately), or
        $[\RunMinEndPos{j''}{\tau}{\Text} \dd \RunEndPos{j''}{\tau}{\Text} - 3\tau + 2) \cap \RFour{s}{H}{\tau}{\Text}
        \neq \emptyset$, and then by $j \in \RFour{s}{H}{\tau}{\Text}$ and $j' < j < \RunEndPos{j''}{\tau}{\Text} - 3\tau + 2$,
        it follows that $j' < j \leq \max\;[\RunMinEndPos{j''}{\tau}{\Text} \dd \RunEndPos{j''}{\tau}{\Text} - 3\tau + 2) \cap
        \RFour{s}{H}{\tau}{\Text}$ (which again yields $j' \not\in I_{0}$).
        We now show that $j \not\in I_{1}$. First, we prove that it is not possible that
        $j' < \RunMinEndPos{j''}{\tau}{\Text} \leq j$. Suppose that this is the case, and recall from above that for
        every $t', t'' \in [j'' \dd \RunEndPos{j''}{\tau}{\Text}) \cap \RFour{s}{H}{\tau}{\Text}$, it holds
        $p \mid t' - t''$. By $j = j' + p$,
        this implies that $j = \min\;[\RunMinEndPos{j''}{\tau}{\Text} \dd \RunEndPos{j''}{\tau}{\Text} - 3\tau + 2)
        \cap \RFour{s}{H}{\tau}{\Text}$,
        which contradicts the assumption $j \not\in D_{0}$. Thus, we must either have $j < \RunMinEndPos{j''}{\tau}{\Text}$
        or $\RunMinEndPos{j''}{\tau}{\Text} \leq j'$. We consider each of the two cases separately:
        \begin{itemize}
        \item First, assume that it holds $j < \RunMinEndPos{j''}{\tau}{\Text}$.
          By $j \in \RFour{s}{H}{\tau}{\Text}$, it then
          follows that $\max\;[j'' \dd \RunMinEndPos{j''}{\tau}{\Text}) \cap \RFour{s}{H}{\tau}{\Text}
          \geq j > j'$. Consequently, $j' \not\in I_{1}$.
        \item Assume now $\RunMinEndPos{j''}{\tau}{\Text} \leq j'$.
          By definition of $I_{1}$, this immediately implies $j' \not\in I_{1}$.
        \end{itemize}
        In both cases, we obtain $j' \not\in I_{1}$. Combining with $j' \not\in I_{0}$, we therefore obtain
        $j' \not\in \mathcal{I}_{\rm text}$.
        Further combining with $j' \in \RMinusFive{s}{k+1}{H}{\tau}{\Text}$, we thus obtain
        $j' \in \RMinusFive{s}{k+1}{H}{\tau}{\Text} \setminus \mathcal{I}_{\rm text}$.
        This concludes the proof of
        $\RMinusFive{s}{k}{H}{\tau}{\Text} \setminus \mathcal{D}_{\rm text} \subseteq
        \{j' + p : j' \in \RMinusFive{s}{k+1}{H}{\tau}{\Text} \setminus \mathcal{I}_{\rm text}\}$.
      \end{itemize}
      This concludes the proof of the equality
      $\{j' + p : j' \in \RMinusFive{s}{k+1}{H}{\tau}{\Text} \setminus \mathcal{I}_{\rm text}\}
      = \RMinusFive{s}{k}{H}{\tau}{\Text} \setminus \mathcal{D}_{\rm text}$.
      In other words, we proved that $k = q$ and $\{i_1 + p, \ldots, i_q + p\} = \{d_1, \ldots, d_k\}$.
      Recall now that the sequence $(i_1, \ldots, i_q)$ is sorted according to the lexicographical order
      of the corresponding suffixes, i.e., for every $t', t'' \in [1 \dd q]$, $t' < t''$ implies
      $\Text[i_{t'} \dd \Textlen] \prec \Text[i_{t''} \dd \Textlen]$. On the other hand,
      $\{i_1, \ldots, i_q\} \subseteq \RFour{s}{H}{\tau}{\Text}$ implies
      by \cref{lm:periodic-pos-lce}\eqref{lm:periodic-pos-lce-it-2},
      that all positions in the set $\{i_1, \ldots, i_q\}$ are followed by the same length-$p$ string in $\Text$.
      This implies that the sequence $(i_1 + p, \ldots, i_q + p)$ is also sorted according to the lexicographical
      order of the corresponding suffixes. Since the same property also holds for the sequence
      $(d_1, \ldots, d_k)$, we obtain by
      $\{i_1 + p, \ldots, i_q + p\} = \{d_1, \ldots, d_k\}$, that for every $t \in [1 \dd k]$, it
      holds $i_t + p = d_t$.

    \item\label{sweep-init-4}
      Next, we prove that for every $j \in \RMinusFive{s}{k+1}{H}{\tau}{\Text} \setminus \mathcal{I}_{\rm text}$,
      $j \in \RMinMinusTwo{\tau}{\Text}$ holds if and only if $j+p \in \RMinMinusTwo{\tau}{\Text}$.
      Let $j \in \RMinusFive{s}{k+1}{H}{\tau}{\Text} \setminus \mathcal{I}_{\rm text}$. We then have
      $j \in \RMinusFive{s}{k+1}{H}{\tau}{\Text}$, $j \not\in I_{0}$, and $j \not\in I_{1}$.
      Let $j'' \in [1 \dd \Textlen]$ be the smallest position such that $[j'' \dd j] \subseteq \RTwo{\tau}{\Text}$.
      By definition of $\RPrimTwo{\tau}{\Text}$, we have $j'' \in \RPrimTwo{\tau}{\Text}$. By \cref{lm:R-text-block},
      we moreover have $j'' \in \RPrimMinusThree{H}{\tau}{\Text}$. Lastly, by \cref{lm:end}, we then
      have $j \in [j'' \dd \RunEndPos{j''}{\tau}{\Text} - 3\tau + 2)$.
      \begin{itemize}
      \item First, assume that $j \in \RMinMinusTwo{\tau}{\Text}$. By $j \in [j'' \dd \RunEndPos{j''}{\tau}{\Text} - 3\tau + 2)$
        and \cref{lm:rmin-text-block}, this implies that
        $j \in [j'' \dd \RunMinEndPos{j''}{\tau}{\Text})$. The assumption $j \not\in I_{1}$ implies
        $j \neq \max\;[j'' \dd \RunMinEndPos{j''}{\tau}{\Text}) \cap \RFour{s}{H}{\tau}{\Text}$. Since by the above,
        for every $t', t'' \in [j'' \dd \RunEndPos{j''}{\tau}{\Text} - 3\tau + 2) \cap \RFour{s}{H}{\tau}{\Text}$, it holds
        $p \mid t' - t''$, it follows that $j + p \leq \max\;[j'' \dd \RunMinEndPos{j''}{\tau}{\Text}) \cap \RFour{s}{H}{\tau}{\Text}$.
        Consequently, $j + p \in [j'' \dd \RunMinEndPos{j''}{\tau}{\Text})$, which implies
        $j + p \in \RMinMinusTwo{\tau}{\Text}$.
      \item We prove the opposite implication by contraposition. Let us thus assume that $j \not\in \RMinMinusTwo{\tau}{\Text}$.
        By $j \in [j'' \dd \RunEndPos{j''}{\tau}{\Text} - 3\tau + 2)$ and \cref{lm:rmin-text-block},
        $j \in [\RunMinEndPos{j''}{\tau}{\Text} \dd \RunEndPos{j''}{\tau}{\Text} - 3\tau + 2)$. The assumption $j \not\in I_{0}$ implies
        $j \neq \max\;[\RunMinEndPos{j''}{\tau}{\Text} \dd \RunEndPos{j''}{\tau}{\Text} - 3\tau + 2) \cap \RFour{s}{H}{\tau}{\Text}$.
        Using again the characterization of $[j'' \dd \RunEndPos{j''}{\tau}{\Text} - 3\tau + 2) \cap \RFour{s}{H}{\tau}{\Text}$, this
        implies that $j + p \leq \max\;[\RunMinEndPos{j''}{\tau}{\Text} \dd \RunEndPos{j''}{\tau}{\Text} - 3\tau + 2) \cap
        \RFour{s}{H}{\tau}{\Text}$. Consequently, $j+p \in [\RunMinEndPos{j''}{\tau}{\Text} \dd \RunEndPos{j''}{\tau}{\Text} - 3\tau + 2)$.
        By \cref{lm:rmin-text-block}, this implies $j+p \not\in \RMinMinusTwo{\tau}{\Text}$.
      \end{itemize}

    \item\label{sweep-init-5}
      We now put everything together. In Step \ref{sweep-init-1}, we proved that,
      $|B'_{\rm in}| = k$ and that for
      every $t \in [1 \dd k]$, it holds $B'_{\rm in}[t] = \BitvectorMin[\ISA{\Text}[d_t]]$.
      In Step \ref{sweep-init-2}, we proved that
      $|B'_{\rm out}| = q$ and that for
      every $t \in [1 \dd q]$, it holds $B'_{\rm out}[t] = \BitvectorMin[\ISA{\Text}[i_t]]$.
      In Step \ref{sweep-init-3}, we proved that $k = q$, and that for every
      $t \in [1 \dd k]$, it holds $i_t + p = d_t$.
      Lastly, in Step \ref{sweep-init-4}, we proved that
      for every $j \in \RMinusFive{s}{k+1}{H}{\tau}{\Text} \setminus \mathcal{I}_{\rm text}$,
      $j \in \RMinMinusTwo{\tau}{\Text}$ holds if and only if $j+p \in \RMinMinusTwo{\tau}{\Text}$.
      Putting together Steps \ref{sweep-init-3} and \ref{sweep-init-4}, and recalling that
      $\{i_1, \ldots, i_q\} = \RMinusFive{s}{k+1}{H}{\tau}{\Text} \setminus \mathcal{I}_{\rm text}$,
      we obtain by \cref{def:min-pos-bitvector-minus}
      that for every $t \in [1 \dd k]$, it holds $\BitvectorMin[\ISA{\Text}[i_t]] = \BitvectorMin[\ISA{\Text}[d_t]]$.
      Combining with Steps \ref{sweep-init-1} and \ref{sweep-init-2}, we therefore obtain that
      for every $t \in [1 \dd k]$, it holds
      \begin{align*}
        B'_{\rm in}[t]
          &= \BitvectorMin[\ISA{\Text}[d_t]]\\
          &= \BitvectorMin[\ISA{\Text}[i_t]]\\
          &= B'_{\rm out}[t].
      \end{align*}
      It remains to observe that $|B'_{\rm in}| = k = q = |B'_{\rm out}|$.
      We thus obtain that $B'_{\rm in} = B'_{\rm out}$.
    \end{enumerate}
  \end{enumerate}
  We have thus proved both conditions in \cref{def:insert}, i.e., that for
  every $(i,c) \in \mathcal{I}$, it holds $i \in [1 \dd |B_{\rm out}|]$ and
  $B_{\rm out}[i] = c$, and that $B'_{\rm in} = B'_{\rm out}$, i.e.,
  $\DeleteSubseq{B_{\rm in}}{\mathcal{D}_{\rm lex}} =
  \DeleteSubseq{B_{\rm out}}{\mathcal{I}_{\rm lex}}$. Since $\mathcal{D}_{\rm lex} = \mathcal{D}$,
  and $\mathcal{I}_{\rm lex} = \{i : (i,c) \in \mathcal{I}\}$, it thus follows that
  $B_{\rm out} = \InsertSubseq{\DeleteSubseq{B_{\rm in}}{\mathcal{D}}}{\mathcal{I}}$, i.e., the claim.
\end{proof}

\paragraph{Algorithms}

\begin{proposition}[\cite{breaking}]\label{pr:runs-minus}
  Let $\tau = \mu\log_{\AlphabetSize} \Textlen$.
  Given the packed representation of $\Text$, we can compute the sequence
  $\RunsMinusLexSortedTwo{\tau}{\Text}$ (\cref{def:runs-minus-lex-sorted})
  in $\bigO(\Textlen / \log_{\AlphabetSize} \Textlen)$ time.
\end{proposition}

\begin{proposition}\label{pr:delete}
  In the word RAM model with word size $w$,
  given the packed representation of a nonempty string $S \in \IntegerAlphabet^m$, and a sequence
  $(p_1, \ldots, p_k)$ such that $k \geq 1$ and $1 \leq p_1 < \ldots < p_k \leq m$,
  we can compute the packed representation of a
  string $\DeleteSubseq{S}{\{p_1, \ldots, p_k\}}$ (\cref{def:insert}) in
  $\bigO(k + (m \log \AlphabetSize) / w)$ time.
\end{proposition}
\begin{proof}
  Let $p_0 = 0$ and $p_{k+1} = m+1$. We scan the sequence $(p_1, \ldots, p_{k+1})$
  left to right, and for every $i \in [1 \dd k+1]$, we append
  the string $S(p_{i-1} \dd p_i)$ to the output. Copying $t$ characters of
  a packed string takes $\bigO(1 + (t \log \AlphabetSize) / w)$ time, hence
  in total it takes
  $\sum_{i=1}^{k+1} \bigO(1 + ((p_{i} - p_{i-1} - 1) \log \AlphabetSize) / w)
  = \bigO(k + \tfrac{\log \AlphabetSize}{w} \sum_{i=1}^{k+1} (p_{i} - p_{i-1} - 1))
  = \bigO(k + (m \log \AlphabetSize) / w)$ time to copy all substrings. Including
  the $\bigO(k)$ time to scan the sequence yields the claim.
\end{proof}

\begin{proposition}\label{pr:insert}
  In the word RAM model with word size $w$, given the packed representation
  of a nonempty string $S \in \IntegerAlphabet^m$, and a
  sequence $((p_1, c_1), \ldots, (p_k, c_k))$ such that $k \geq 1$ and
  $1 \leq p_1 < \ldots < p_k \leq m + k$, we can compute the packed
  representation of a string $\InsertSubseq{S}{\{(p_1,c_1), \ldots, (p_k,c_k)\}}$ (\cref{def:insert})
  in $\bigO(k + (m \log \AlphabetSize) / w)$ time.
\end{proposition}
\begin{proof}
  Let $S'$ denote the output string of length $m + k$.
  Let $p_0 = 0$ and $p_{k+1} = m+k+1$. We scan the sequence $(p_1, \ldots, p_{k+1})$
  left to right, and for every $i \in [1 \dd k+1]$, we copy the next $p_i - p_{i-1} - 1$
  symbols from $S$ to $S'(p_{i-1} \dd p_i)$. Copying $t$ characters of
  a packed string takes $\bigO(1 + (t \log \AlphabetSize) / w)$ time, hence
  in total it takes
  $\sum_{i=1}^{k+1} \bigO(1 + ((p_{i} - p_{i-1} - 1) \log \AlphabetSize) / w)
  = \bigO(k + \tfrac{\log \AlphabetSize}{w} \sum_{i=1}^{k+1} (p_{i} - p_{i-1} - 1))
  = \bigO(k + (m \log \AlphabetSize) / w)$ time to copy all substrings, where
  we used that $\sum_{i=1}^{k+1} (p_{i} - p_{i-1} - 1) = p_{k+1} - p_{0} - (k+1) = m$.
  Next, we scan the sequence $((p_1, c_1), \ldots, (p_k, c_k))$, and in $\bigO(k)$ time
  we set $S'[p_i] = c_i$ for every $i \in [1 \dd k]$. In total, we spend
  $\bigO(k + (m \log \AlphabetSize) / w)$ time.
\end{proof}

\begin{proposition}\label{pr:emin-text-order}
  Let $\tau = \mu\log_{\AlphabetSize} \Textlen$, $H \in \Sigma^{+}$,
  and let $\alpha \in (0,1)$ be a constant.
  Given $\NavPeriodic{\Text}$ (\cref{pr:nav-index-periodic}) and an array $A_{\rm runs}[1 \dd m]$
  containing the sequence
  $\RunsMinusTextSortedThree{H}{\tau}{\Text}$ (\cref{def:runs-minus-text-sorted}),
  we can in $\bigO(\Textlen^{\alpha} + m)$ time compute an array $A_{\rm emin}[1 \dd m]$
  defined by $A_{\rm emin}[i] = \RunMinEndPos{A_{\rm runs}[i]}{\tau}{\Text}$
  (\cref{def:emin}).
\end{proposition}
\begin{proof}
  Denote $(p_i)_{i \in [1 \dd m]} = \RunsMinusTextSortedThree{H}{\tau}{\Text}$.
  First, using \cref{pr:nav-index-periodic}\eqref{pr:nav-index-periodic-it-1a},
  in $\bigO(1)$ time we compute $p := |H| = |\RootPos{p_1}{\tau}{\Text}|$.
  We then in $\bigO(\Textlen^{\alpha})$ time initialize the data structure
  from \cref{pr:narrow-range-max} for $u = \Textlen$ and $h = p$
  (note that $p > 0$ and $p \leq \tfrac{1}{3}\tau = \tfrac{1}{3}\mu\log_{\AlphabetSize} \Textlen
  = \bigO(\log \Textlen)$, and hence $h$ satisfies the assumption in \cref{pr:narrow-range-max}).
  For every $t \in [1 \dd m]$, let $\mathcal{S}_t$ denote the set maintained by the structure after $t-1$ insertions.
  Note that $\mathcal{S}_1 = \emptyset$.
  We also set $\ell_1^{\rm trim} := 0$. We then process the sequence $(p_t)_{t \in [1 \dd m]}$ left-to-right.
  For $t=1,\ldots,m$, we execute the following steps:
  \begin{enumerate}
  \item Using \cref{pr:nav-index-periodic}\eqref{pr:nav-index-periodic-it-1a},
    in $\bigO(1)$ time we compute
    $e_t := \RunEndPos{p_t}{\tau}{\Text}$ and $e^{\rm full}_t := \RunEndFullPos{p_t}{\tau}{\Text}$.
    We then set $r := \RunEndPos{p_t}{\tau}{\Text} - p_t - 3\tau + 2 = e_t - p_t - 3\tau + 2$
    and $\ell_t^{\rm pos} := \RunEndFullPos{p_t}{\tau}{\Text} - p_t = e^{\rm full}_t - p_t$.
  \item Using \cref{pr:narrow-range-max} and $q = e_t - e_t^{\rm full}$, we compute
    $\ell_t^{\rm whole} := \max \{v : (k,v) \in \mathcal{S}_t \cup \{(p,0)\}\text{ and }k \geq q\}$.
    Note that the application of the query of \cref{pr:narrow-range-max} is well-defined since
    $q \in [0 \dd p) = [0 \dd h)$.
  \item In $\bigO(1)$ time, we compute $\ell_t^{\max} = \max(\ell_t^{\rm trim}, \ell_t^{\rm whole})$.
    Note that $\ell_t^{\rm trim}$ was either computed at the beginning of algorithm (if $t = 1$), or
    for the previous element of the sequence $(p_t)_{t \in [1 \dd m]}$ (if $t > 1$).
  \item By \cref{lm:emin}, if $\ell_t^{\rm pos} \leq \ell_t^{\max}$, then
    $\RunMinEndPos{p_t}{\tau}{\Text} - p_t = 0$. Otherwise (i.e., if $\ell_t^{\rm pos} > \ell_t^{\max}$),
    it holds $\RunMinEndPos{p_t}{\tau}{\Text} - p_t = \min(\ell_t^{\rm pos} - \ell_t^{\max}, p, r)$.
    Thus, we can compute $A_{\rm emin}[t] = \RunMinEndPos{p_t}{\tau}{\Text}$ in $\bigO(1)$ time.
  \item In $\bigO(1)$ time, we compute
    $\ell_{t+1}^{\rm trim} := \max(\ell_t^{\rm trim}, e_t^{\rm full} - p_t - p)$ (in preparation for the next iteration).
  \item We insert the pair $(e_t - e_t^{\rm full}, e_t^{\rm full} - p_t)$ into
    the structure from \cref{pr:narrow-range-max} (in preparation for the next iteration).
    Note that then
    $\mathcal{S}_{t+1} = \{(\RunEndPos{p_{t'}}{\tau}{\Text} - \RunEndFullPos{p_{t'}}{\tau}{\Text},
        \RunEndFullPos{p_{t'}}{\tau}{\Text} - p_{t'}) : t' \in [1 \dd t]\}$.
  \end{enumerate}
  All operations above, including $2m$ operations on the structure from \cref{pr:narrow-range-max}, take
  $\bigO(m)$ time in total.
  Including the preprocessing for \cref{pr:narrow-range-max}, in total we spend
  $\bigO(\Textlen^{\alpha} + m)$ time.
\end{proof}

\begin{proposition}\label{pr:emin-any-order}
  Let $\tau = \mu\log_{\AlphabetSize} \Textlen$, $H \in \Sigma^{+}$,
  and let $\alpha \in (0,1)$ be a constant.
  Let $A_{\rm runs}[1 \dd m]$ be such that $m = |\RPrimMinusThree{H}{\tau}{\Text}|$ and
  $\{A_{\rm runs}[i] : i \in [1 \dd m]\} = \RPrimMinusThree{H}{\tau}{\Text}$.
  Given $\NavPeriodic{\Text}$, and the array $A_{\rm runs}[1 \dd m]$ as input,
  we can in $\bigO(\Textlen^{\alpha} + m)$ time compute the array
  $A_{\rm emin}[1 \dd m]$ defined by $A_{\rm emin}[i] =
  \RunMinEndPos{A_{\rm runs}[i]}{\tau}{\Text}$
  (\cref{def:emin}).
\end{proposition}
\begin{proof}
  The algorithm proceeds in four steps:
  \begin{enumerate}
  \item We compute the array $A_{\rm perm}[1 \dd m]$ containing the permutation
    of $\{1, \ldots, m\}$ such that for every $i, j \in [1 \dd m]$,
    letting $x = A_{\rm runs}[A_{\rm perm}[i]]$ and
    $y = A_{\rm runs}[A_{\rm perm}[j]]$, $i < j$ implies that $x < y$.
    To this end, we first in $\bigO(m)$ time initialize an array $A_{\rm sort}[1 \dd m]$
    defined by $A_{\rm sort}[i] = (A_{\rm runs}[i], i)$. We then sort it by the first coordinate.
    Note that by $\RPrimMinusThree{H}{\tau}{\Text} \subseteq [1 \dd \Textlen - 3\tau + 2]$,
    the first coordinate in $A_{\rm sort}$ is always smaller than $\Textlen$.
    Thus, we can sort $A_{\rm sort}[1 \dd m]$ using $(1/\alpha)$-round radix
    sort in $\bigO(\Textlen^{\alpha} + m)$ time. The resulting
    array contains $A_{\rm perm}$ on the second coordinate.
  \item In $\bigO(m)$ time,
    we compute the array $A_{\rm incr}[1 \dd m]$
    defined by $A_{\rm incr}[i] = A_{\rm runs}[A_{\rm perm}[i]]$.
    Observe that the array $A_{\rm incr}[1 \dd m]$ contains the
    sequence $\RunsMinusTextSortedThree{H}{\tau}{\Text}$ (\cref{def:runs-minus-text-sorted}).
  \item We apply \cref{pr:emin-text-order}
    to $A_{\rm incr}[1 \dd m]$. It takes $\bigO(\Textlen^{\alpha} + m)$ time, and results in an
    array $A_{\rm ans}[1 \dd m]$ defined by $A_{\rm ans}[i] = \RunMinEndPos{A_{\rm incr}[i]}{\tau}{\Text}
    = \RunMinEndPos{A_{\rm runs}[A_{\rm perm}[i]]}{\tau}{\Text}$.
  \item For $i \in [1 \dd m]$, we set $A_{\rm emin}[A_{\rm perm}[i]] = A_{\rm ans}[i]$. This takes $\bigO(m)$ time.
  \end{enumerate}
  In total, the computation of $A_{\rm emin}[1 \dd m]$ takes $\bigO(\Textlen^{\alpha} + m)$ time.
\end{proof}

\begin{proposition}\label{pr:min-bv-first}
  Let $\tau = \mu\log_{\AlphabetSize} \Textlen$, $H \in \Sigma^{+}$,
  $p = |H|$, and $s \in [0 \dd p)$. Let $\alpha \in (0,1)$ be a constant.
  Given $\NavPeriodic{\Text}$ (\cref{pr:nav-index-periodic}) and the
  sequence $\RunsMinusLexSortedThree{H}{\tau}{\Text}$ (\cref{def:runs-minus-lex-sorted}),
  we can compute the packed representation of
  $\MinPosBitvectorMinusFour{s}{H}{\tau}{\Text}$ (\cref{def:min-pos-bitvector-minus})
  in $\bigO(\Textlen^{\alpha} + |\RPrimMinusThree{H}{\tau}{\Text}| +
  |\RMinusThree{H}{\tau}{\Text}|/\log \Textlen)$ time.
\end{proposition}
\begin{proof}
  Suppose that the sequence $\RunsMinusLexSortedThree{H}{\tau}{\Text}$ is given
  as an array $A_{\rm runs}[1 \dd m]$, where $m = |\RPrimMinusThree{H}{\tau}{\Text}|$.
  Let $I_0$, $I_1$, $D_0$, and $D_1$ be defined as in \cref{lm:sweep-init}.
  The computation consists of four steps:
  \begin{enumerate}
  \item We compute the arrays $A_{I_0}[1 \dd |I_{0}|]$, $A_{I_1}[1 \dd |I_{1}|]$,
    $A_{D_0}[1 \dd |D_{0}|]$, and $A_{D_1}[1 \dd |D_{1}|]$ containing, respectively,
    all elements of sets $I_{0}$, $I_{1}$, $D_{0}$, and $D_{1}$. The computation
    proceeds in three steps:
    \begin{enumerate}
    \item Using \cref{pr:nav-index-periodic}\eqref{pr:nav-index-periodic-it-1a},
      in $\bigO(m)$ time compute
      $A_{\rm end}[1 \dd m]$ defined by
      $A_{\rm end}[i] = \RunEndPos{A_{\rm runs}[i]}{\tau}{\Text}$.
    \item Using \cref{pr:emin-any-order}, in $\bigO(\Textlen^{\alpha} + m)$ time
      compute $A_{\rm emin}[1 \dd m]$, 
      $A_{\rm emin}[i] = \RunMinEndPos{A_{\rm runs}[i]}{\tau}{\Text}$.
    \item Observe that for every $i, j \in [1 \dd \Textlen]$ such that
      $i \leq j$ and $[i \dd j) \subseteq \RMinusThree{H}{\tau}{\Text}$,
      $[i \dd j) \cap \RFour{s}{H}{\tau}{\Text} \neq \emptyset$ holds if
      and only if $\delta < j-i$, where $\delta = (s'-s) \bmod p$ and
      $s' = \HeadPos{i}{\tau}{\Text}$. Moreover, if $[i \dd j) \cap \RFour{s}{H}{\tau}{\Text} \neq \emptyset$, then
      $\min\; [i \dd j) \cap \RFour{s}{H}{\tau}{\Text} = i + \delta$ and
      $\max\; [i \dd j) \cap \RFour{s}{H}{\tau}{\Text} = (j - 1) - \delta'$, where
      $\delta' = (s-s'') \bmod p$ and $s'' = \HeadPos{j-1}{\tau}{\Text}$. Consequently,
      using \cref{pr:nav-index-periodic}\eqref{pr:nav-index-periodic-it-1a}, for every $t \in [1 \dd m]$, given
      $i = A_{\rm runs}[t]$, $j = \RunMinEndPos{i}{\tau}{\Text} = A_{\rm emin}[t]$,
      and $k = \RunEndPos{i}{\tau}{\Text} - 3\tau + 2 = A_{\rm end}[t] - 3\tau + 2$,
      we can determine in $\bigO(1)$ time
      if the sets $[i \dd j) \cap \RFour{s}{H}{\tau}{\Text}$ and $[j \dd k) \cap \RFour{s}{H}{\tau}{\Text}$
      are nonempty, and if so, compute their smallest and largest elements. Thus, we can
      compute the arrays $A_{I_0}$, $A_{I_1}$, $A_{D_0}$, and $A_{D_1}$ in $\bigO(m)$ time.
    \end{enumerate}
    In total, the computation of $A_{I_0}$, $A_{I_1}$, $A_{D_0}$, and $A_{D_1}$
    takes $\bigO(\Textlen^{\alpha} + m) = \bigO(\Textlen^{\alpha} + |\RPrimMinusThree{H}{\tau}{\Text}|)$
    time.
  \item For every $k \in [1 \dd \Textlen]$, denote
    \begin{align*}
      \mathcal{D}_k
        &= \{\ISA{\Text}[j] - x : j \in (D_0 \cup D_1) \cap \RFive{s}{k}{H}{\tau}{\Text}\},\\
      \mathcal{I}_k
        &= \{(\ISA{\Text}[j] - x, 0) : j \in I_0 \cap \RFive{s}{k}{H}{\tau}{\Text}\}\, \cup\\
        &\hspace{3.1ex} \{(\ISA{\Text}[j] - x, 1) : j \in I_1 \cap \RFive{s}{k}{H}{\tau}{\Text}\},
    \end{align*}
    where $x \in [0 \dd \Textlen]$ is such that for some $y \in [0 \dd \Textlen]$, it holds
    $\{\SA{\Text}[i] : i \in (x \dd y]\} = \RMinusFive{s}{k}{H}{\tau}{\Text}$ (recall that such
    $x$ and $y$ always exist; see \cref{rm:min-pos-bitvector-minus}). We also denote
    \begin{align*}
      \mathcal{D}_{\rm all}
        &= \textstyle\bigcup_{k \in [1 \dd \Textlen]} \{(k, i) : i \in \mathcal{D}_k\},\\
      \mathcal{I}_{\rm all}
        &= \textstyle\bigcup_{k \in [1 \dd \Textlen]} \{(k, i, c) : (i, c) \in \mathcal{I}_k\}.
    \end{align*}
    We compute arrays $A_{\mathcal{D}_{\rm all}}$ and $A_{\mathcal{I}_{\rm all}}$ containing,
    respectively, all elements of $\mathcal{D}_{\rm all}$ and $\mathcal{I}_{\rm all}$ sorted
    lexicographically.
    The computation of $A_{\mathcal{D}_{\rm all}}$ proceeds in four steps:
    \begin{enumerate}
    \item Using \cref{pr:nav-index-periodic}\eqref{pr:nav-index-periodic-it-5},
      we compute the arrays $A_{{\rm isa},0}[1 \dd |D_0|]$
      and $A_{{\rm isa},1}[1 \dd |D_1|]$ defined by $A_{{\rm isa},0}[i] = \ISA{\Text}[A_{D_0}[i]]$
      and $A_{{\rm isa},1}[i] = \ISA{\Text}[A_{D_1}[i]]$. It takes
      $\bigO(|\RPrimMinusThree{H}{\tau}{\Text}| + |\RMinusThree{H}{\tau}{\Text}| / \log \Textlen + m)$ time.
      Note that \cref{pr:nav-index-periodic}\eqref{pr:nav-index-periodic-it-5}
      requires as input the sequence
      $\RunsMinusLexSortedThree{H}{\tau}{\Text}$ (\cref{def:runs-minus-lex-sorted}), which
      is available here.
    \item Using \cref{pr:nav-index-periodic}\eqref{pr:nav-index-periodic-it-1a},
      in $\bigO(m)$ time we compute
      arrays $A_{{\rm exp},0}[1 \dd |D_0|]$ and $A_{{\rm exp},1}[1 \dd |D_1|]$ defined
      by $A_{{\rm exp},0}[i] = \ExpPos{A_{D_0}[i]}{\tau}{\Text}$ and
      $A_{{\rm exp},1}[i] = \ExpPos{A_{D_1}[i]}{\tau}{\Text}$.
    \item Using \cref{pr:nav-index-periodic}\eqref{pr:nav-index-periodic-it-1a},
      in $\bigO(m)$ time we compute arrays
      $A_{{\rm beg},0}[1 \dd |D_0|]$ and $A_{{\rm beg},1}[1 \dd |D_1|]$ defined by
      $A_{{\rm beg},0}[i] = x_0$ and $A_{{\rm beg},1}[i] = x_1$, where
      $x_0 \in [0 \dd \Textlen]$ and $x_1 \in [0 \dd \Textlen]$ are such that for some
      $y_0 \in [0 \dd \Textlen]$ and $y_1 \in [0 \dd \Textlen]$, it holds
      $\{\SA{\Text}[i] : i \in (x_0 \dd y_0]\} = \RMinusFive{s}{k_0}{H}{\tau}{\Text}$
      and $\{\SA{\Text}[i] : i \in (x_1 \dd y_1]\} = \RMinusFive{s}{k_1}{H}{\tau}{\Text}$
      (where $k_0 = \ExpPos{A_{D_0}[i]}{\tau}{\Text} = A_{{\rm exp},0}[i]$ and
      $k_1 = \ExpPos{A_{D_1}[i]}{\tau}{\Text} = A_{{\rm exp},1}[i]$).
    \item In $\bigO(m)$ time we initialize the array $A_{\rm sort}[1 \dd |D_0| + |D_1|]$ to
      contain all pairs from the sets $\{(A_{{\rm exp},0}[i], A_{{\rm isa},0}[i] - A_{{\rm beg},0}[i]) : i \in [1 \dd |D_0|]\}$
      and $\{(A_{{\rm exp},1}[i], A_{{\rm isa},1}[i] - A_{{\rm beg},1}[i]) : i \in [1 \dd |D_1|]\}$.
      We then sort $A_{\rm sort}$ lexicographically.
      Since each of the pairs contains positive integers smaller than $\Textlen$, using $(2/\alpha)$-round radix
      sort, we spend $\bigO(\Textlen^{\alpha})$ time. The resulting array is equal to $A_{\mathcal{D}_{\rm all}}$.
    \end{enumerate}
    The computation of $A_{\mathcal{I}_{\rm all}}$ proceeds analogously,
    except in the last step, in each tuple we additionally
    include a symbol indicating whether the pair corresponds to $I_0$ or $I_1$ (see the definition of $\mathcal{I}_k$).
    In total, the computation of $A_{\mathcal{D}_{\rm all}}$ and $A_{\mathcal{I}_{\rm all}}$ takes
    $\bigO(\Textlen^{\alpha} + |\RPrimMinusThree{H}{\tau}{\Text}| + |\RMinusThree{H}{\tau}{\Text}|/\log \Textlen + m) =
    \bigO(\Textlen^{\alpha} + |\RPrimMinusThree{H}{\tau}{\Text}| + |\RMinusThree{H}{\tau}{\Text}|/\log \Textlen)$ time.
  \item In $\bigO(\Textlen^{\alpha})$ time we construct the structure from \cref{pr:packed-copy}.
  \item We are now ready to compute the packed representation of
    $\MinPosBitvectorMinusFour{s}{H}{\tau}{\Text}$ (\cref{def:min-pos-bitvector-minus}).
    Recall that by \cref{lm:R-lex-block-pos} (see also \cref{rm:min-pos-bitvector-minus}), it holds
    \begin{align*}
      \MinPosBitvectorMinusFour{s}{H}{\tau}{\Text}
        &= \textstyle\bigodot_{i=1}^{\Textlen} \MinPosBitvectorMinusFive{s}{i}{H}{\tau}{\Text}.
    \end{align*}
    Let $n_{\mathcal{D}} = |\mathcal{D}_{\rm all}|$ and $n_{\mathcal{I}} = |\mathcal{I}_{\rm all}|$.
    For every $t \in [1 \dd n_{\mathcal{D}}]$ (resp.\ $t \in [1 \dd n_{\mathcal{I}}]$), denote
    $A_{\mathcal{D}_{\rm all}}[t] = (k^{\rm del}_t, i^{\rm del}_t)$ (resp.\ $A_{\mathcal{I}_{\rm all}}[t]
    = (k^{\rm ins}_t, i^{\rm ins}_t, c^{\rm ins}_t)$).
    We execute the following algorithm, maintaining the following invariant at the beginning of each iteration:
    \begin{itemize}
    \item $k \in [1 \dd \Textlen]$,
    \item $b_d = \sum_{t \in [1 \dd k)} |\mathcal{D}_t|$ and $b_i = \sum_{t \in [1 \dd k]} |\mathcal{I}_t|$,
    \item $B_{\rm cur}$ is a packed representation of bitvector $\MinPosBitvectorMinusFive{s}{k}{H}{\tau}{\Text}$,
    \item $B_{\rm out}$ is a packed representation of bitvector $\bigodot_{i=1}^{k} \MinPosBitvectorMinusFive{s}{i}{H}{\tau}{\Text}$.
    \end{itemize}
    To ensure the invariant holds at the beginning of the first step, note that
    $|\MinPosBitvectorMinusFive{s}{1}{H}{\tau}{\Text}| = |\mathcal{I}_{1}| = 0$, and hence we set $k = 1$, $b_d = 0$, $b_i = 0$,
    $B_{\rm cur} = \emptystring$, and $B_{\rm out} = \emptystring$.
    Then, as long as $b_d < n_{\mathcal{D}}$ or $b_i < n_{\mathcal{I}}$, we apply the following
    procedure consisting of five steps:
    \begin{enumerate}
    \item Compute the smallest $k' > k$ such that $\mathcal{D}_{k'-1} \neq \emptyset$ or
      $\mathcal{I}_{k'} \neq \emptyset$. By definition of the arrays $A_{\mathcal{D}_{\rm all}}$ and $A_{\mathcal{I}_{\rm all}}$,
      we can in $\bigO(1)$ accomplish this as follows:
      \begin{itemize}
      \item If $b_d = n_{\mathcal{D}}$, then $k' = k^{\rm inc}_{b_i + 1}$.
      \item If $b_i = n_{\mathcal{I}}$, then $k' = k^{\rm del}_{b_d + 1} + 1$.
      \item If $b_d < n_{\mathcal{D}}$ and $b_i < n_{\mathcal{I}}$, then
        $k' = \min(k^{\rm inc}_{b_i + 1}, k^{\rm del}_{b_d + 1} + 1)$.
     \end{itemize}
  \item Using \cref{pr:packed-copy}, compute the packed representation of
    bitvector $B_{\rm cur}^{k'-k-1}$, and then append it to $B_{\rm out}$.
    Observe that by definition of $k'$ and \cref{lm:sweep-init}, for every $i \in (k \dd k')$,
    it holds $\MinPosBitvectorMinusFive{s}{i}{H}{\tau}{\Text} = \MinPosBitvectorMinusFive{s}{k}{H}{\tau}{\Text}$.
    Consequently, it holds
    \[
        B_{\rm cur}^{k'-k-1} = \textstyle\bigodot_{i=k+1}^{k'-1} \MinPosBitvectorMinusFive{s}{i}{H}{\tau}{\Text},
    \]
    and hence after the update of $B_{\rm out}$ is complete, we have
    $B_{\rm out} = \bigodot_{i=1}^{k'-1} \MinPosBitvectorMinusFive{s}{i}{H}{\tau}{\Text}$.
    This step takes $\bigO(1 + (\sum_{i \in (k \dd k')}|\MinPosBitvectorMinusFive{s}{i}{H}{\tau}{\Text}|) / \log \Textlen)$ time.
  \item Next, we determine integers $e_d$ and $e_i$ satisfying $e_d - b_d = |\mathcal{D}_{k'-1}|$ and $e_i - b_i = |\mathcal{I}_{k'}|$.
    Observe that, by definition of arrays $A_{\mathcal{D}_{\rm all}}$ and $A_{\mathcal{I}_{\rm all}}$, it holds
    $\mathcal{D}_{k'-1} = \{i^{\rm del}_t : t \in (b_d \dd e_d]\}$ and
    $\mathcal{I}_{k'} = \{(i^{\rm ins}_t, c^{\rm ins}_t) : t \in (b_i \dd e_i]\}$.
    Moreover, for $t \in (b_d \dd e_d]$ (resp.\ $t \in (b_i \dd e_i]$), we have $k^{\rm del}_t = k' - 1$
    (resp.\ $k^{\rm ins}_t = k'$).
    Thus, using $A_{\mathcal{D}_{\rm all}}$ and $A_{\mathcal{I}_{\rm all}}$, both
    $e_d$ and $e_i$ can be computed in total $\bigO(1 + |\mathcal{D}_{k'-1}| + |\mathcal{I}_{k'}|)$ time.
  \item Next, we compute the packed representation of bitvector
      $B_{\rm next} = \MinPosBitvectorMinusFive{s}{k'}{H}{\tau}{\Text}$.
    Observe, that by
    \cref{lm:sweep-init}, and the above observations, it holds
    \begin{align*}
     \MinPosBitvectorMinusFive{s}{k'}{H}{\tau}{\Text}
       &= \InsertSubseq{\DeleteSubseq{\MinPosBitvectorMinusFive{s}{k'-1}{H}{\tau}{\Text}}{\mathcal{D}_{k'-1}}}{\mathcal{I}_{k'}}\\
       &= \InsertSubseq{\DeleteSubseq{\MinPosBitvectorMinusFive{s}{k}{H}{\tau}{\Text}}{\mathcal{D}_{k'-1}}}{\mathcal{I}_{k'}}\\
       &= \InsertSubseq{\DeleteSubseq{B_{\rm cur}}{\mathcal{D}_{k'-1}}}{\mathcal{I}_{k'}}.
    \end{align*}
    We proceed as follows:
    \begin{itemize}
    \item In the first substep, we compute the packed representation of
      $B_{\rm aux} = \DeleteSubseq{B_{\rm cur}}{\mathcal{D}_{k'-1}}$.
      If $e_d - b_d = 0$, then we simply copy the bitvector $B_{\rm cur}$ to $B_{\rm aux}$ in
      $\bigO(1 + |B_{\rm cur}| / \log \Textlen) =
      \bigO(1 + |\MinPosBitvectorMinusFive{s}{k}{H}{\tau}{\Text}| / \log \Textlen)$ time.
      Otherwise, we compute $B_{\rm aux}$ using \cref{pr:delete}.
      Recall that above we observed that $\mathcal{D}_{k'-1} = \{i^{\rm del}_t : t \in (b_d \dd e_d]\}$.
      Moreover, note that for every $t \in (b_d \dd e_d)$,
      it holds $i^{\rm del}_t < i^{\rm del}_{t+1}$. Thus, we can indeed apply \cref{pr:delete}. It takes
      $\bigO(|\mathcal{D}_{k'-1}| + |B_{\rm cur}| / \log \Textlen) = \bigO(|\mathcal{D}_{k'-1}| +
      |\MinPosBitvectorMinusFive{s}{k}{H}{\tau}{\Text}| / \log \Textlen)$ time.
    \item In the second substep, we compute the packed representation of
      $B_{\rm next} = \InsertSubseq{B_{\rm aux}}{\mathcal{I}_{k'}}$.
      If $e_i - b_i = 0$, then we simply copy $B_{\rm aux}$ to $B_{\rm next}$ in
      $\bigO(1 + |B_{\rm aux}| / \log \Textlen) = \bigO(1 + |B_{\rm cur}|/ \log \Textlen) =
      \bigO(1 + |\MinPosBitvectorMinusFive{s}{k}{H}{\tau}{\Text}| / \log \Textlen)$ time.
      Otherwise, we compute $B_{\rm next}$ using \cref{pr:insert}.
      Recall that above we observed that $\mathcal{I}_{k'} = \{(i^{\rm ins}_t, c^{\rm ins}_t) : t \in (b_i \dd e_i]\}$.
      Moreover, note that for every
      $t \in (b_i \dd e_i)$, it holds $i^{\rm ins}_t < i^{\rm ins}_{t+1}$. Thus, we can indeed apply \cref{pr:insert}. It takes
      $\bigO(|\mathcal{I}_{k'}| + |B_{\rm aux}| / \log \Textlen) =
      \bigO(|\mathcal{I}_{k'}| + |B_{\rm cur}| / \log \Textlen)
      = \bigO(|\mathcal{I}_{k'}| + |\MinPosBitvectorMinusFive{s}{k}{H}{\tau}{\Text}| / \log \Textlen)$ time.
    \end{itemize}
    By the above, $B_{\rm next} =
    \InsertSubseq{\DeleteSubseq{B_{\rm cur}}{\mathcal{D}_{k'-1}}}{\mathcal{I}_{k'}}
    = \MinPosBitvectorMinusFive{s}{k'}{H}{\tau}{\Text}$.
    In total, we spend $\bigO(1 + |\mathcal{D}_{k'-1}| + |\mathcal{I}_{k'}| +
    |\MinPosBitvectorMinusFive{s}{k}{H}{\tau}{\Text}| / \log \Textlen)$ time.
  \item In preparation for the next iteration, we now perform the following steps:
    \begin{itemize}
    \item In $\bigO(1)$ time we set $k := k'$, $b_d := e_d$, and $b_i := e_i$.
    \item We swap the pointers to $B_{\rm next}$ and $B_{\rm cur}$, so that
      $B_{\rm cur} = \MinPosBitvectorMinusFive{s}{k'}{H}{\tau}{\Text}$. In
      $\bigO(1 + |\MinPosBitvectorMinusFive{s}{k}{H}{\tau}{\Text}| / \log \Textlen)$ time
      we then release the space used by the bitvector
      $\MinPosBitvectorMinusFive{s}{k}{H}{\tau}{\Text}$.
    \item Finally, we append $B_{\rm cur}$ to $B_{\rm out}$ in
      $\bigO(1 + |\MinPosBitvectorMinusFive{s}{k'}{H}{\tau}{\Text}| / \log \Textlen)$ time.
      Since by \cref{lm:sweep-init}, $|\MinPosBitvectorMinusFive{s}{k'}{H}{\tau}{\Text}|
      \leq |\MinPosBitvectorMinusFive{s}{k}{H}{\tau}{\Text}| + |\mathcal{I}_{k'}|$, we can
      also bound the time as
      $\bigO(1 + |\mathcal{I}_{k'}| + |\MinPosBitvectorMinusFive{s}{k}{H}{\tau}{\Text}| / \log \Textlen)$.
    \end{itemize}
    In total, the above steps take
    $\bigO(1 + |\mathcal{I}_{k'}| + |\MinPosBitvectorMinusFive{s}{k}{H}{\tau}{\Text}| / \log \Textlen)$
    time.
  \end{enumerate}
  Summing the time for all steps, during a single iteration of the above procedure we spend
  $\bigO(1 + |\mathcal{D}_{k'-1}| + |\mathcal{I}_{k'}| +
  (\sum_{i=k}^{k'}|\MinPosBitvectorMinusFive{s}{i}{H}{\tau}{\Text}|) / \log \Textlen)$ time.
  Letting $\mathcal{K} = \{i \in [2 \dd \Textlen] :
  \mathcal{D}_{i-1} \neq \emptyset\text{ or }\mathcal{I}_{i} \neq \emptyset\}$, and noting
  that
  $|\mathcal{K}| \leq n_{\mathcal{D}} + n_{\mathcal{I}} \leq |D_0| + |D_1| + |I_0| + |I_1| \leq 4m$
  and $\sum_{i \in \mathcal{K}} |\mathcal{D}_{i-1}| + |\mathcal{I}_i| \leq n_{\mathcal{D}} + n_{\mathcal{I}} \leq 4m$,
  the total time
  spent over all iterations is thus
  \begin{align*}
    &\bigO(|\mathcal{K}| + (\textstyle\sum_{i \in \mathcal{K}} |\mathcal{D}_{i-1}| + |\mathcal{I}_{i}|) +
        (\textstyle\sum_{i=1}^{\Textlen}|\MinPosBitvectorMinusFive{s}{i}{H}{\tau}{\Text}|) / \log \Textlen)\\
    &\hspace{2cm}= \bigO(m + |\MinPosBitvectorMinusFour{s}{H}{\tau}{\Text}| / \log \Textlen)\\
    &\hspace{2cm}= \bigO(m + |\RMinusFour{s}{H}{\tau}{\Text}| / \log \Textlen)\\
    &\hspace{2cm}= \bigO(m + |\RMinusThree{H}{\tau}{\Text}| / \log \Textlen)\\
    &\hspace{2cm}= \bigO(|\RPrimMinusThree{H}{\tau}{\Text}| + |\RMinusThree{H}{\tau}{\Text}| / \log \Textlen).
  \end{align*}
  \end{enumerate}
  In total, the computation of $\MinPosBitvectorMinusFour{s}{H}{\tau}{\Text}$
  takes $\bigO(\Textlen^{\alpha} + |\RPrimMinusThree{H}{\tau}{\Text}| + |\RMinusThree{H}{\tau}{\Text}| / \log \Textlen)$ time.
\end{proof}

\begin{proposition}\label{pr:min-bv-rest}
  Let $\tau = \mu\log_{\AlphabetSize} \Textlen$, $H \in \Sigma^{+}$, and
  $p = |H|$. Let $\alpha \in (0,1)$ be a constant.
  Given $\NavPeriodic{\Text}$ (\cref{pr:nav-index-periodic}), the
  sequence $\RunsMinusLexSortedThree{H}{\tau}{\Text}$ (\cref{def:runs-minus-lex-sorted}),
  and the packed representation of $\MinPosBitvectorMinusFour{0}{H}{\tau}{\Text}$,
  in $\bigO(\Textlen^{\alpha} + |\RPrimMinusThree{H}{\tau}{\Text}| +
  |\RMinusThree{H}{\tau}{\Text}|/\log \Textlen)$ time
  we can compute the packed representation of all bitvectors in
  $\{\MinPosBitvectorMinusFour{i}{H}{\tau}{\Text}\}_{i \in [1 \dd p)}$
  (\cref{def:min-pos-bitvector-minus}).
\end{proposition}
\begin{proof}
  Suppose that the sequence $\RunsMinusLexSortedThree{H}{\tau}{\Text}$ is given
  as an array $A_{\rm runs}[1 \dd m]$, where $m = |\RPrimMinusThree{H}{\tau}{\Text}|$.
  Let $I_0$, $I_1$, $D_0$, and $D_1$ be defined as in \cref{lm:sweep-step}.
  The computation consists of three steps:
  \begin{enumerate}
  \item We compute the arrays $A_{I_0}[1 \dd |I_{0}|]$, $A_{I_1}[1 \dd |I_{1}|]$,
    $A_{D_0}[1 \dd |D_{0}|]$, and $A_{D_1}[1 \dd |D_{1}|]$ containing, respectively,
    all elements of $I_{0}$, $I_{1}$, $D_{0}$, and $D_{1}$. The computation
    proceeds in three steps:
    \begin{enumerate}
    \item Using \cref{pr:nav-index-periodic}\eqref{pr:nav-index-periodic-it-1a},
      in $\bigO(m)$ time compute
      $A_{\rm end}[1 \dd m]$ defined by
      $A_{\rm end}[i] = \RunEndPos{A_{\rm runs}[i]}{\tau}{\Text}$.
    \item Using \cref{pr:emin-any-order}, in $\bigO(\Textlen^{\alpha} + m)$ time
      compute $A_{\rm emin}[1 \dd m]$, 
      $A_{\rm emin}[i] = \RunMinEndPos{A_{\rm runs}[i]}{\tau}{\Text}$.
    \item For every $t \in [1 \dd m]$, given
      $i = A_{\rm runs}[t]$, $j = \RunMinEndPos{i}{\tau}{\Text} = A_{\rm emin}[t]$,
      and $k = \RunEndPos{i}{\tau}{\Text} - 3\tau + 2 = A_{\rm end}[t] - 3\tau + 2$,
      we can determine in $\bigO(1)$ time if the sets $[i \dd j)$ and $[j \dd k)$ are nonempty, and if so, compute their
      smallest and largest elements.
      Thus, we can compute the arrays $A_{I_0}$, $A_{I_1}$, $A_{D_0}$, and $A_{D_1}$ in $\bigO(m)$ time.
    \end{enumerate}
    In total, the computation of $A_{I_0}$, $A_{I_1}$, $A_{D_0}$, and $A_{D_1}$
    takes $\bigO(\Textlen^{\alpha} + m) = \bigO(\Textlen^{\alpha} + |\RPrimMinusThree{H}{\tau}{\Text}|)$
    time.
  \item For every $s \in [0 \dd p-1)$, denote
    \begin{align*}
      \mathcal{D}_s
        &= \{\ISA{\Text}[j] - x : j \in (D_0 \cup D_1) \cap \RFour{s}{H}{\tau}{\Text}\},\\
      \mathcal{I}_{s+1}
        &= \{(\ISA{\Text}[j] - x', 0) : j \in I_0 \cap \RFour{s+1}{H}{\tau}{\Text}\}\, \cup\\
        &\hspace{3.1ex} \{(\ISA{\Text}[j] - x', 1) : j \in I_1 \cap \RFour{s+1}{H}{\tau}{\Text}\},
    \end{align*}
    where $x \in [0 \dd \Textlen]$ and $x' \in [0 \dd \Textlen]$ are such that for some
    $y \in [0 \dd \Textlen]$ and $y' \in [0 \dd \Textlen]$, it holds
    $\{\SA{\Text}[i] : i \in (x \dd y]\} = \RMinusFour{s}{H}{\tau}{\Text}$ and
    $\{\SA{\Text}[i] : i \in (x' \dd y']\} = \RMinusFour{s+1}{H}{\tau}{\Text}$
    (recall that such $x, y, x', y'$ always exist; see
    \cref{rm:min-pos-bitvector-minus}). We compute the arrays
    $A_{\mathcal{D}_0}[1 \dd |\mathcal{D}_0|], \ldots,
    A_{\mathcal{D}_{p-2}}[1 \dd |\mathcal{D}_{p-2}|]$,
    $A_{\mathcal{I}_{1}}[1 \dd |\mathcal{I}_1|], \ldots,
    A_{\mathcal{I}_{p-1}}[1 \dd |\mathcal{I}_{p-1}|]$ containing, respectively,
    all elements of $\mathcal{D}_0, \ldots, \mathcal{D}_{p-2}, \mathcal{I}_1, \ldots, \mathcal{I}_{p-1}$ sorted
    by the first coordinate.
    The computation of $A_{\mathcal{D}_0}, \ldots, A_{\mathcal{D}_{p-2}}$ proceeds in four steps:
    \begin{enumerate}
    \item Using \cref{pr:nav-index-periodic}\eqref{pr:nav-index-periodic-it-5},
      we compute arrays $A_{{\rm isa},0}[1 \dd |D_0|]$
      and $A_{{\rm isa},1}[1 \dd |D_1|]$ defined by $A_{{\rm isa},0}[i] = \ISA{\Text}[A_{D_0}[i]]$
      and $A_{{\rm isa},1}[i] = \ISA{\Text}[A_{D_1}[i]]$. It takes
      $\bigO(|\RPrimMinusThree{H}{\tau}{\Text}| + |\RMinusThree{H}{\tau}{\Text}| / \log \Textlen + m)$ time.
      Note that \cref{pr:nav-index-periodic}\eqref{pr:nav-index-periodic-it-5}
      requires as input the sequence
      $\RunsMinusLexSortedThree{H}{\tau}{\Text}$ (\cref{def:runs-minus-lex-sorted}), which
      is available here.
    \item Using \cref{pr:nav-index-periodic}\eqref{pr:nav-index-periodic-it-1a},
      in $\bigO(m)$ time we compute
      arrays $A_{{\rm head},0}[1 \dd |D_0|]$ and $A_{{\rm head},1}[1 \dd |D_1|]$ defined
      by $A_{{\rm head},0}[i] = \HeadPos{A_{D_0}[i]}{\tau}{\Text}$ and
      $A_{{\rm head},1}[i] = \HeadPos{A_{D_1}[i]}{\tau}{\Text}$.
    \item Using \cref{pr:nav-index-periodic}\eqref{pr:nav-index-periodic-it-1a},
      in $\bigO(m)$ time we compute arrays
      $A_{{\rm beg},0}[1 \dd |D_0|]$ and $A_{{\rm beg},1}[1 \dd |D_1|]$ defined by
      $A_{{\rm beg},0}[i] = x_0$ and $A_{{\rm beg},1}[i] = x_1$, where
      $x_0 \in [0 \dd \Textlen]$ and $x_1 \in [0 \dd \Textlen]$ are such that for some
      $y_0 \in [0 \dd \Textlen]$ and $y_1 \in [0 \dd \Textlen]$, it holds
      $\{\SA{\Text}[i] : i \in (x_0 \dd y_0]\} = \RMinusFour{s_0}{H}{\tau}{\Text}$
      and $\{\SA{\Text}[i] : i \in (x_1 \dd y_1]\} = \RMinusFour{s_1}{H}{\tau}{\Text}$
      (where $s_0 = \HeadPos{A_{D_0}[i]}{\tau}{\Text} = A_{{\rm head},0}[i]$ and
      $s_1 = \HeadPos{A_{D_1}[i]}{\tau}{\Text} = A_{{\rm head},1}[i]$).
    \item In $\bigO(m)$ time we initialize the array $A_{\rm sort}[1 \dd |D_0| + |D_1|]$ to
      contain all pairs from the sets $\{(A_{{\rm head},0}[i], A_{{\rm isa},0}[i] - A_{{\rm beg},0}[i]) : i \in [1 \dd |D_0|]\}$
      and $\{(A_{{\rm head},1}[i], A_{{\rm isa},1}[i] - A_{{\rm beg},1}[i]) : i \in [1 \dd |D_1|]\}$.
      We then sort $A_{\rm sort}$ lexicographically.
      Since each of the pairs contains positive integers smaller than $\Textlen$, using $(2/\alpha)$-round radix
      sort, we spend $\bigO(\Textlen^{\alpha})$ time. With a single scan of
      $A_{\rm sort}$, we can compute the arrays $A_{\mathcal{D}_0}, \ldots, A_{\mathcal{D}_{p-2}}$
      (during the scan, we ignore elements with the first coordinate equal to $p-1$). Note that initializing
      the arrays takes $\bigO(p) = \bigO(\tau) = \bigO(\log \Textlen)$ time.
    \end{enumerate}
    The computation of $A_{\mathcal{I}_{1}}, \ldots, A_{\mathcal{I}_{p-1}}$ proceeds analogously,
    except in the last step, in each tuple we additionally
    include a symbol indicating whether the pair corresponds to $I_0$ or $I_1$ (see the definition of $\mathcal{I}_{s+1}$).
    In total, the computation of $A_{\mathcal{D}_0}, \ldots, A_{\mathcal{D}_{p-2}}$ and
    $A_{\mathcal{I}_{1}}, \ldots, A_{\mathcal{I}_{p-1}}$ takes
    $\bigO(\log \Textlen + \Textlen^{\alpha} + |\RPrimMinusThree{H}{\tau}{\Text}| + |\RMinusThree{H}{\tau}{\Text}|/\log \Textlen + m) =
    \bigO(\Textlen^{\alpha} + |\RPrimMinusThree{H}{\tau}{\Text}| + |\RMinusThree{H}{\tau}{\Text}|/\log \Textlen)$ time.
  \item We now compute the packed representation of
    bitvectors in the set
    $\{\MinPosBitvectorMinusFour{i}{H}{\tau}{\Text}\}_{i \in [1 \dd p)}$
    (\cref{def:min-pos-bitvector-minus}).
    Let $s \in [0 \dd p-1)$, and assume that we already have the packed
    representation of bitvectors in the set $\{\MinPosBitvectorMinusFour{i}{H}{\tau}{\Text}\}_{i \in [0 \dd s]}$.
    Note that this is satisfied for the first iteration ($s = 0$), since the packed representation
    of $\MinPosBitvectorMinusFour{0}{H}{\tau}{\Text}$ is given as input. Denote
    $B_{\rm cur} = \MinPosBitvectorMinusFour{s}{H}{\tau}{\Text}$. We then
    compute the packed representation of $\MinPosBitvectorMinusFour{s+1}{H}{\tau}{\Text}$ as follows:
    \begin{enumerate}
    \item First, we compute the packed representation of bitvector
      $B_{\rm aux} = \DeleteSubseq{B_{\rm cur}}{\mathcal{D}_{s}}$.
      If $\mathcal{D}_{s} = \emptyset$, we simply copy bitvector $B_{\rm cur}$ to $B_{\rm aux}$
      in $\bigO(1 + |B_{\rm cur}| / \log \Textlen) = \bigO(1 + |\MinPosBitvectorMinusFour{s}{H}{\tau}{\Text}| / \log \Textlen)$ time.
      Otherwise, we compute $B_{\rm aux}$ using \cref{pr:delete}, using the fact that elements of $\mathcal{D}_{s}$ are stored
      in sorted order in $A_{\mathcal{D}_{s}}$. Applying \cref{pr:delete}
      takes $\bigO(|\mathcal{D}_s| + |B_{\rm cur}| / \log \Textlen) =
      \bigO(|\mathcal{D}_s| + |\MinPosBitvectorMinusFour{s}{H}{\tau}{\Text}| / \log \Textlen)$ time.
    \item Next, we compute the packed representation of bitvector
      $B_{\rm next} = \InsertSubseq{B_{\rm aux}}{\mathcal{I}_{s+1}}$.
      If $\mathcal{I}_{s+1} = \emptyset$, we simply copy bitvector $B_{\rm aux}$ to $B_{\rm next}$
      in $\bigO(1 + |B_{\rm aux}| / \log \Textlen) = \bigO(1 + |B_{\rm cur}| / \log \Textlen) =
      \bigO(1 + |\MinPosBitvectorMinusFour{s}{H}{\tau}{\Text}| / \log \Textlen)$ time.
      Otherwise, we compute $B_{\rm next}$ using \cref{pr:insert}, using the fact that
      elements of $\mathcal{I}_{s+1}$ are stored in $A_{\mathcal{I}_{s+1}}$, and moreover, are ordered by the first coordinate.
      Applying \cref{pr:insert} takes
      $\bigO(|\mathcal{I}_{s+1}| + |B_{\rm aux}| / \log \Textlen) =
      \bigO(|\mathcal{I}_{s+1}| + |B_{\rm cur}| / \log \Textlen) =
      \bigO(|\mathcal{I}_{s+1}| + |\MinPosBitvectorMinusFour{s}{H}{\tau}{\Text}| / \log \Textlen)$ time.
    \end{enumerate}
    After the above steps are complete, it holds by \cref{lm:sweep-step} that:
    \begin{align*}
      B_{\rm next}
        &= \InsertSubseq{B_{\rm aux}}{\mathcal{I}_{s+1}}\\
        &= \InsertSubseq{\DeleteSubseq{B_{\rm cur}}{\mathcal{D}_s}}{\mathcal{I}_{s+1}}\\
        &= \InsertSubseq{\DeleteSubseq{\MinPosBitvectorMinusFour{s}{H}{\tau}{\Text}}{\mathcal{D}_{s}}}{\mathcal{I}_{s+1}}\\
        &= \MinPosBitvectorMinusFour{s+1}{H}{\tau}{\Text}.
    \end{align*}
    In total, the computation of the packed representation of $\MinPosBitvectorMinusFour{s+1}{H}{\tau}{\Text}$
    takes $\bigO(1 + |\mathcal{D}_s| + |\mathcal{I}_{s+1}| + |\MinPosBitvectorMinusFour{s}{H}{\tau}{\Text}| / \log \Textlen)$ time.
    Summing over all $s \in [0 \dd p-1)$, and recalling that $\sum_{s \in [0 \dd p-1)} |\mathcal{D}_{s}| \leq |D_0| + |D_1| \leq 2m$
    and $\sum_{s \in [0 \dd p-1)} |\mathcal{I}_{s+1}| \leq |I_0| + |I_1| \leq 2m$,
    the total computation time for the packed representation of bitvectors in
    $\{\MinPosBitvectorMinusFour{i}{H}{\tau}{\Text}\}_{i \in [1 \dd p)}$ is thus:
    \begin{align*}
      &\textstyle\sum_{s \in [0 \dd p-1)}
        \bigO(1 + |\mathcal{D}_s| + |\mathcal{I}_{s+1}| + |\MinPosBitvectorMinusFour{s}{H}{\tau}{\Text}| / \log \Textlen)\\
      &\hspace{1cm}=
        \bigO(p + \textstyle\sum_{s \in [0 \dd p-1)}|\mathcal{D}_s| +
                  \textstyle\sum_{s \in [0 \dd p-1)}|\mathcal{I}_{s+1}| +
                  \tfrac{1}{\log \Textlen}\textstyle\sum_{s \in [0 \dd p-1)} |\MinPosBitvectorMinusFour{s}{H}{\tau}{\Text}|)\\
      &\hspace{1cm}=
        \bigO(\log \Textlen + m + \tfrac{1}{\log \Textlen}\textstyle\sum_{s \in [0 \dd p-1)}|\RMinusFour{s}{H}{\tau}{\Text}|)\\
      &\hspace{1cm}=
        \bigO(\log \Textlen + |\RPrimMinusThree{H}{\tau}{\Text}| + |\RMinusThree{H}{\tau}{\Text}| / \log \Textlen)\\
      &\hspace{1cm}=
        \bigO(\Textlen^{\alpha} + |\RPrimMinusThree{H}{\tau}{\Text}| + |\RMinusThree{H}{\tau}{\Text}| / \log \Textlen).
    \end{align*}
  \end{enumerate}
  In total, the computation takes
  $\bigO(\Textlen^{\alpha} +
  |\RPrimMinusThree{H}{\tau}{\Text}| + |\RMinusThree{H}{\tau}{\Text}| / \log \Textlen)$ time.
\end{proof}

\begin{proposition}\label{pr:min-bv-all}
  Let $\tau = \mu\log_{\AlphabetSize} \Textlen$, $H \in \Sigma^{+}$, and
  $p = |H|$. Let $\alpha \in (0,1)$ be a constant.
  Given $\NavPeriodic{\Text}$ (\cref{pr:nav-index-periodic}) and the
  sequence $\RunsMinusLexSortedThree{H}{\tau}{\Text}$ (\cref{def:runs-minus-lex-sorted}),
  we can compute the packed representation of all bitvectors in
  $\{\MinPosBitvectorMinusFour{i}{H}{\tau}{\Text}\}_{i \in [0 \dd p)}$
  (\cref{def:min-pos-bitvector-minus}) in $\bigO(\Textlen^{\alpha} + |\RPrimMinusThree{H}{\tau}{\Text}| +
  |\RMinusThree{H}{\tau}{\Text}|/\log \Textlen)$ time.
\end{proposition}
\begin{proof}
  The algorithm proceeds in two steps:
  \begin{enumerate}
  \item Using \cref{pr:min-bv-first}, we compute the packed representation of
    $\MinPosBitvectorMinusFour{0}{H}{\tau}{\Text}$.
  \item Using \cref{pr:min-bv-rest} (and the result from the first step
    as input), we compute the packed representation of
    bitvectors in the set $\{\MinPosBitvectorMinusFour{i}{H}{\tau}{\Text}\}_{i \in [1 \dd p)}$.
  \end{enumerate}
  Both steps take $\bigO(\Textlen^{\alpha} + |\RPrimMinusThree{H}{\tau}{\Text}| +
  |\RMinusThree{H}{\tau}{\Text}|/\log \Textlen)$ time and result
  in the collection of bitvectors from the claim. Note that both
  steps need $\RunsMinusLexSortedThree{H}{\tau}{\Text}$ as input,
  which is available here.
\end{proof}

\begin{proposition}\label{pr:min-bv}
  Let $\tau = \mu\log_{\AlphabetSize} \Textlen$.
  Given $\NavPeriodic{\Text}$ (\cref{pr:nav-index-periodic}) and
  the packed representation of $\Text$, we can compute the packed representation
  of $\MinPosBitvectorMinusTwo{\tau}{\Text}$ (\cref{def:min-pos-bitvector-minus})
  in $\bigO(\Textlen / \log_{\AlphabetSize} \Textlen)$ time.
\end{proposition}
\begin{proof}
  Let $\alpha < 1-\mu$ be a constant. The algorithm proceeds in two steps:
  \begin{enumerate}
  \item Using \cref{pr:runs-minus}, we compute the sequence
    $\RunsMinusLexSortedTwo{\tau}{\Text}$ (\cref{def:runs-minus-lex-sorted}) in
    $\bigO(\Textlen / \log_{\AlphabetSize} \Textlen)$ time. Assume that
    the sequence is stored in the array $A_{\rm runs}[1 \dd m]$, where
    $m = |\RPrimMinusTwo{\tau}{\Text}|$.
  \item We are now ready to compute the packed representation of
    $\MinPosBitvectorMinusTwo{\tau}{\Text}$.
    In $\bigO(\Textlen / \log \Textlen)$ time, we initialize
    the packed representation of bitvector $B := {\tt 0}^{\Textlen}$,
    which we will use to store the output.
    We set $b := 0$. Then, as long as $b < m$, we repeat the following
    procedure consisting of four steps:
    \begin{enumerate}
    \item Compute the largest
      $e \in (b \dd m]$ such that for every $i \in (b \dd e)$, it holds
      $\RootPos{A_{\rm runs}[i]}{\tau}{\Text} = \RootPos{A_{\rm runs}[i + 1]}{\tau}{\Text}$.
      Observe that then, letting
      $H = \RootPos{A_{\rm runs}[e]}{\tau}{\Text}$, the subarray
      $A_{\rm runs}(b \dd e]$ contains the sequence $\RunsMinusLexSortedThree{H}{\tau}{\Text}$
      (\cref{def:runs-minus-lex-sorted}). Using \cref{pr:nav-index-periodic}\eqref{pr:nav-index-periodic-it-1a},
      the computation of $e$ takes $\bigO(e - b) = \bigO(|\RPrimMinusThree{H}{\tau}{\Text}|)$ time.
    \item Let $p = |H|$. Using \cref{pr:min-bv-all} (and $A_{\rm runs}(b \dd e]$ as input),
      compute the packed representation of
      bitvectors $\{\MinPosBitvectorMinusFour{i}{H}{\tau}{\Text}\}_{i \in [0 \dd p)}$
      in $\bigO(\Textlen^{\alpha} + |\RPrimMinusThree{H}{\tau}{\Text}| +
      |\RMinusThree{H}{\tau}{\Text}|/\log \Textlen)$ time.
    \item We copy all bitvectors in the set $\{\MinPosBitvectorMinusFour{i}{H}{\tau}{\Text}\}_{i \in [0 \dd p)}$
      into their correct location in $B$. To this end, for every $s \in [0 \dd p)$ such that
      $\MinPosBitvectorMinusFour{i}{H}{\tau}{\Text} \neq \emptyset$:
      \begin{itemize}
      \item Using \cref{pr:nav-index-periodic}\eqref{pr:nav-index-periodic-it-1a}
        (with $A_{\rm runs}[e]$ as an element of $\RMinusFour{s}{H}{\tau}{\Text}$),
        we first in $\bigO(1)$ time compute $b', e' \in [0 \dd \Textlen]$
        such that $\{\SA{\Text}[i] : i \in (b' \dd e']\} = \RMinusFour{s}{H}{\tau}{\Text}$.
      \item In $\bigO(1 + |\MinPosBitvectorMinusFour{s}{H}{\tau}{\Text}| / \log \Textlen) =
        \bigO(1 + |\RMinusFour{s}{H}{\tau}{\Text}| / \log \Textlen)$ time, we then
        copy $\MinPosBitvectorMinusFour{s}{H}{\tau}{\Text}$ to $B(b' \dd e']$.
      \end{itemize}
      In total, we spend
      $\bigO(p + \textstyle\sum_{s \in [0 \dd p)}|\RMinusFour{s}{H}{\tau}{\Text}| / \log \Textlen)
      = \bigO(\log \Textlen + \tfrac{1}{\log \Textlen} \textstyle\sum_{s \in [0 \dd p)}|\RMinusFour{s}{H}{\tau}{\Text}|)
      = \bigO(\Textlen^{\alpha} + |\RMinusThree{H}{\tau}{\Text}| / \log \Textlen)$
      time.
    \item In preparation for the next iteration, we set $b := e$.
    \end{enumerate}
    Denote $\mathcal{H} = \{H \in \Sigma^{+} : \RMinusThree{H}{\tau}{\Text} \neq \emptyset\}$.
    Recall (see \cref{sec:minocc-index-periodic-prelim}),
    that for every $H \in \mathcal{H}$, it holds $H \in \IntegerAlphabet^{\leq \tau}$.
    Thus, by $\tau = \mu\log_{\AlphabetSize} \Textlen$, it holds
    $|\mathcal{H}| = \bigO(\AlphabetSize^{\mu\log_{\AlphabetSize} \Textlen})
    = \bigO(\Textlen^{\mu})$. Note also that by
    \cref{lm:runs}, we have $|\RPrimMinusTwo{\tau}{\Text}| =
    \bigO(\Textlen/\tau) = \bigO(\Textlen / \log_{\AlphabetSize} \Textlen)$.
    Lastly, recall that
    $\sum_{H \in \mathcal{H}} |\RMinusThree{H}{\tau}{\Text}| = |\RMinusTwo{\tau}{\Text}|$ and
    $\sum_{H \in \mathcal{H}} |\RPrimMinusThree{H}{\tau}{\Text}| = |\RPrimMinusTwo{\tau}{\Text}|$.
    Thus, over all iterations, the above procedure takes
    \begin{align*}
      &\textstyle\sum_{H \in \mathcal{H}} \bigO(\Textlen^{\alpha} + |\RPrimMinusThree{H}{\tau}{\Text}| +
        |\RMinusThree{H}{\tau}{\Text}| / \log \Textlen)\\
      &\hspace{1cm}=\bigO(|\mathcal{H}| \cdot \Textlen^{\alpha} + \textstyle\sum_{H \in \mathcal{H}} |\RPrimMinusThree{H}{\tau}{\Text}| +
        \tfrac{1}{\log \Textlen}\textstyle\sum_{H \in \mathcal{H}} |\RMinusThree{H}{\tau}{\Text}|)\\
      &\hspace{1cm}=\bigO(\Textlen^{\mu+\alpha} + |\RPrimMinusTwo{\tau}{\Text}| + |\RMinusTwo{\tau}{\Text}| / \log \Textlen)\\
      &\hspace{1cm}=\bigO(\Textlen / \log_{\AlphabetSize} \Textlen)
    \end{align*}
    time, where in the last equality we exploit that $\alpha < 1 - \mu$ implies
    that $\bigO(\Textlen^{\mu+\alpha}) = \bigO(\Textlen / \log_{\AlphabetSize} \Textlen)$.
    Including the initialization of $B$ in $\bigO(\Textlen / \log \Textlen)$ time, we thus
    spend $\bigO(\Textlen / \log_{\AlphabetSize} \Textlen)$ time in total.
  \end{enumerate}
  In total, the computation takes
  $\bigO(\Textlen / \log_{\AlphabetSize} \Textlen)$ time.
\end{proof}

\begin{proposition}\label{pr:minocc-index-periodic-construction}
  Given the packed representation of $\Text$, we can construct
  $\MinOccIndexPeriodic{\Text}$ (\cref{sec:minocc-index-periodic-structure}) in
  $\bigO(\Textlen / \log_{\AlphabetSize} \Textlen)$ time.
\end{proposition}
\begin{proof}
  Let $\tau = \mu\log_{\AlphabetSize} \Textlen$ be defined as in
  \cref{sec:minocc-index-periodic-structure}.
  We construct the components of the first part of
  $\MinOccIndexPeriodic{\Text}$ (\cref{sec:minocc-index-periodic-structure})
  as follows:
  \begin{enumerate}
  \item Using \cref{pr:nav-index-periodic}, we construct
    $\NavPeriodic{\Text}$ in $\bigO(\Textlen / \log_{\AlphabetSize} \Textlen)$
    time.
  \item Next, we construct the structure from \cref{th:three-sided-rmq}
    for arrays $A_{\rm pos}[1 \dd q]$ and $A_{\rm len}[1 \dd q]$ (defined
    as in \cref{sec:minocc-index-periodic-structure}). The construction proceeds in
    three steps:
    \begin{enumerate}
    \item First, using \cref{pr:runs-minus}, in
      $\bigO(\Textlen / \log_{\AlphabetSize} \Textlen)$ time we compute
      the sequence $(a_i)_{i \in [1 \dd q]} = \RunsMinusLexSortedTwo{\tau}{\Text}$
      (\cref{def:runs-minus-lex-sorted}).
    \item Next, we construct the arrays $A_{\rm pos}[1 \dd q]$ and
      $A_{\rm len}[1 \dd q]$. To this end, for every $i \in [1 \dd q]$,
      using \cref{pr:nav-index-periodic}\eqref{pr:nav-index-periodic-it-1a},
      in $\bigO(1)$ time we
      compute $e := \RunEndFullPos{a_i}{\tau}{\Text}$ and $p
      := \RootPos{a_i}{\tau}{\Text}$. We then set $\ell :=
      |\PowStr{\tau}{\RootPos{a_i}{\tau}{\Text}}| =
      p \cdot \lceil \tfrac{\tau}{p} \rceil$, $A_{\rm pos}[i] := e
      - \ell$, and $A_{\rm len}[i] := (e - \ell) - a_i$. Recall that
      by the above, it holds $q = \bigO(\Textlen
      / \log_{\AlphabetSize} \Textlen)$. Thus, this step takes
      $\bigO(1 + q) = \bigO(\Textlen / \log_{\AlphabetSize} \Textlen)$
      time.
    \item We apply \cref{th:three-sided-rmq} to arrays $A_{\rm
      pos}[1 \dd q]$ and $A_{\rm len}[1 \dd q]$. As noted
      in \cref{sec:minocc-index-periodic-structure}, for $m = 2\Textlen/\tau$
      (where $\tau$ is as in \cref{sec:minocc-index-periodic-structure}) it
      holds $q \in [1 \dd m]$,
      $\max_{i=1}^{q} A_{\rm pos}[i] \leq \Textlen = \bigO(m \log m)$, and
      $\sum_{i=1}^{q} A_{\rm len}[i] \leq 2\Textlen = \bigO(m \log m)$.
      Thus, \cref{th:three-sided-rmq} takes $\bigO(m) = \bigO(\Textlen
      / \log_{\AlphabetSize} \Textlen)$ time.
    \end{enumerate}
    In total, we spend $\bigO(\Textlen / \log_{\AlphabetSize} \Textlen)$
    time.
  \item With \cref{pr:min-bv},
    we construct the packed representation of $\MinPosBitvectorMinusTwo{\tau}{\Text}$
    in $\bigO(\Textlen / \log_{\AlphabetSize} \Textlen)$ time.
    We then augment it using \cref{th:bin-rank-select} in
    $\bigO(\Textlen / \log \Textlen) = \bigO(\Textlen / \log_{\AlphabetSize} \Textlen)$ time.
  \end{enumerate}
  In total, the construction of the first part of
  $\MinOccIndexPeriodic{\Text}$ takes
  $\bigO(\Textlen / \log_{\AlphabetSize} \Textlen)$ time.
  We then construct the second part analogously. In total,
  the construction takes $\bigO(\Textlen / \log_{\AlphabetSize} \Textlen)$
  time.
\end{proof}

%% file: index/final.tex
\subsection{The Final Data Structure}\label{sec:minocc-index-final}

\input{index/final/structure.tex}

\input{index/final/queries.tex}

\input{index/final/construction.tex}

%% file: index/final/structure.tex
\subsubsection{The Data Structure}\label{sec:minocc-index-final-structure}

\paragraph{Components}

The data structure, denoted $\MinOccIndex{\Text}$, consists of three components:
\begin{enumerate}
\item The structure $\MinOccIndexCore{\Text}$ (\cref{sec:minocc-index-core-structure}).
  It needs $\bigO(\Textlen / \log_{\AlphabetSize} \Textlen)$ space.
\item The structure $\MinOccIndexNonperiodic{\Text}$ (\cref{sec:minocc-index-nonperiodic-structure}).
  It needs $\bigO(\Textlen / \log_{\AlphabetSize} \Textlen)$ space.
\item The structure $\MinOccIndexPeriodic{\Text}$ (\cref{sec:minocc-index-periodic-structure}).
  It needs $\bigO(\Textlen / \log_{\AlphabetSize} \Textlen)$ space.
\end{enumerate}

In total, $\MinOccIndex{\Text}$ needs
$\bigO(\Textlen / \log_{\AlphabetSize} \Textlen)$ space.

%% file: index/final/queries.tex
\subsubsection{Implementation of Queries}\label{sec:minocc-index-final-queries}

\begin{proposition}\label{pr:minocc-index-final-query-pos}
  Let $j \in [1 \dd \Textlen]$,
  and $\ell > 0$ be such that $j + \ell \leq \Textlen + 1$. Given
  $\MinOccIndex{\Text}$ (\cref{sec:minocc-index-final-structure}) and
  $(j,\ell)$, we can in $\bigO(\log^{\epsilon} \Textlen)$ time compute
  $\min \OccTwo{\Pat}{\Text}$, where $\Pat = \Text[j \dd j + \ell)$.
\end{proposition}
\begin{proof}
  Let $\tau = \mu\log_{\AlphabetSize} \Textlen$.
  We consider two cases:
  \begin{itemize}
  \item If $\ell < 3\tau - 1$, then we compute $\min \OccTwo{\Pat}{\Text}$
    in $\bigO(1)$ time using \cref{pr:minocc-index-core-query-pos}.
  \item Let us now assume that $\ell \geq 3\tau - 1$. Note that by the
    assumption $j + \ell \leq \Textlen + 1$, we then have
    $j \leq \Textlen - \ell + 1 \leq \Textlen - 3\tau + 2$.
    In $\bigO(1)$ time we check if $j \in \RTwo{\tau}{\Text}$ using
    \cref{pr:nav-index-core}\eqref{pr:nav-index-core-it-1}. We then consider two cases:
    \begin{itemize}
    \item If $j \not\in \RTwo{\tau}{\Text}$, we compute
      $\min \OccTwo{\Pat}{\Text}$ using \cref{pr:minocc-index-nonperiodic-query-pos}
      in $\bigO(\log^{\epsilon} \Textlen)$ time.
    \item Otherwise, we compute
      $\min \OccTwo{\Pat}{\Text}$ using \cref{pr:minocc-index-periodic-query-pos}
      in $\bigO(\log \log \Textlen)$ time.
    \end{itemize}
  \end{itemize}
  In total, we spend $\bigO(\log^{\epsilon} \Textlen)$ time.
\end{proof}

\begin{proposition}\label{pr:minocc-index-final-query-pat}
  Let $\Pat \in \Sigma^{m}$ be a nonempty pattern
  such that $\OccTwo{\Pat}{\Text} \neq \emptyset$. Given
  $\MinOccIndex{\Text}$ (\cref{sec:minocc-index-final-structure}) and
  the packed representation of $\Pat$, we can in
  $\bigO(\log^{\epsilon} \Textlen + m / \log_{\AlphabetSize} \Textlen)$ time compute
  $\min \OccTwo{\Pat}{\Text}$.
\end{proposition}
\begin{proof}
  Let $\tau = \mu\log_{\AlphabetSize} \Textlen$.
  We consider two cases:
  \begin{itemize}
  \item If $m < 3\tau - 1$, then we compute $\min \OccTwo{\Pat}{\Text}$
    in $\bigO(1)$ time using \cref{pr:minocc-index-core-query-pat}.
  \item Let us now assume that $m \geq 3\tau - 1$.
    In $\bigO(1)$ time we check if $\Pat$ is $\tau$-periodic using
    \cref{pr:nav-index-core}\eqref{pr:nav-index-core-it-2}. We then consider two cases:
    \begin{itemize}
    \item If $\Pat$ is not $\tau$-periodic, we compute
      $\min \OccTwo{\Pat}{\Text}$ using \cref{pr:minocc-index-nonperiodic-query-pat}
      in $\bigO(\log^{\epsilon} \Textlen + m / \log_{\AlphabetSize} \Textlen)$ time.
    \item Otherwise, we compute
      $\min \OccTwo{\Pat}{\Text}$ using \cref{pr:minocc-index-periodic-query-pat}
      in $\bigO(\log \log \Textlen + m / \log_{\AlphabetSize} \Textlen)$ time.
    \end{itemize}
  \end{itemize}
  In total, we spend $\bigO(\log^{\epsilon} \Textlen + m / \log_{\AlphabetSize} \Textlen)$ time.
\end{proof}

%% file: index/final/construction.tex
\subsubsection{Construction Algorithm}\label{sec:minocc-index-final-construction}

\begin{proposition}\label{pr:minocc-index-final-construction}
  Given the packed representation of $\Text$, we can construct
  $\MinOccIndex{\Text}$ (\cref{sec:minocc-index-final-structure}) in
  $\bigO(\Textlen \min(1, \log \AlphabetSize / \sqrt{\log \Textlen}))$ time and
  using $\bigO(\Textlen / \log_{\AlphabetSize} \Textlen)$ working space.
\end{proposition}
\begin{proof}
  We construct the components of $\MinOccIndex{\Text}$ as follows:
  \begin{enumerate}
  \item Using \cref{pr:minocc-index-core-construction}, we construct
    $\MinOccIndexCore{\Text}$ in $\bigO(\Textlen / \log_{\AlphabetSize} \Textlen)$ time.
  \item Using \cref{pr:minocc-index-nonperiodic-construction}, we construct
    $\MinOccIndexNonperiodic{\Text}$ in $\bigO(\Textlen \min(1, \log \AlphabetSize / \sqrt{\log \Textlen}))$ time
    and using $\bigO(\Textlen / \log_{\AlphabetSize} \Textlen)$ working space.
  \item Using \cref{pr:minocc-index-periodic-construction}, we construct
    $\MinOccIndexPeriodic{\Text}$ in $\bigO(\Textlen / \log_{\AlphabetSize} \Textlen)$ time.
  \end{enumerate}
  In total, the construction takes
  $\bigO(\Textlen \min(1, \log \AlphabetSize / \sqrt{\log \Textlen}))$ time
  and uses $\bigO(\Textlen / \log_{\AlphabetSize} \Textlen)$ working space.
\end{proof}

%% file: index/summary.tex
\subsection{Summary}

By combining \cref{pr:minocc-index-final-query-pos}, \cref{pr:minocc-index-final-query-pat},
and \cref{,pr:minocc-index-final-construction},
we obtain the following result.

\begin{theorem}\label{th:minocc-index}
  Given any constant $\epsilon \in (0, 1)$ and the packed representation of a text
  $\Text \in \IntegerAlphabet^{\Textlen}$ with $2 \leq \AlphabetSize < \Textlen^{1/7}$,
  we can in $\bigO(\Textlen \min(1, \log \AlphabetSize / \sqrt{\log \Textlen}))$ time
  and using $\bigO(\Textlen / \log_{\AlphabetSize} \Textlen)$ working space construct
  a data structure of size $\bigO(\Textlen / \log_{\AlphabetSize} \Textlen)$ that
  supports the following queries:
  \begin{itemize}
  \item Given any position $j \in [1 \dd \Textlen]$ and any length $\ell > 0$ such that $j + \ell \leq \Textlen + 1$,
    in $\bigO(\log^{\epsilon} \Textlen)$ time compute the position $\min \OccTwo{\Pat}{\Text}$, where $\Pat = \Text[j \dd j + \ell)$.
  \item Given the packed representation of any pattern $\Pat \in \IntegerAlphabet^{m}$ that satisfies $\OccTwo{\Pat}{\Text} \neq \emptyset$,
    in $\bigO(\log^{\epsilon} \Textlen + m / \log_{\AlphabetSize} \Textlen)$ time compute the position $\min \OccTwo{\Pat}{\Text}$.
  \end{itemize}
\end{theorem}

We also immediately obtain the following general reduction.

\begin{theorem}\label{th:minocc-index-general}
  Consider a data structure answering prefix RMQ queries that, for any
  sequence of $k$ length-$\ell$ strings over alphabet $\IntegerAlphabet$,
  achieves the following complexities:
  \begin{enumerate}
  \item Space usage $S(k,\ell,\AlphabetSize)$,
  \item Preprocessing time $P_t(k,\ell,\AlphabetSize)$,
  \item Preprocessing space $P_s(k,\ell,\AlphabetSize)$,
  \item Query time $Q(k,\ell,\AlphabetSize)$.
  \end{enumerate}
  For every $\Text \in \IntegerAlphabet^{n}$ with $2 \leq \AlphabetSize < \Textlen^{1/7}$,
  there exists $k = \bigO(\Textlen / \log_{\AlphabetSize} \Textlen)$ and $\ell = \bigO(\log_{\AlphabetSize} \Textlen)$
  such that, given the packed representation of $\Text$,
  we can in $\bigO(\Textlen / \log_{\AlphabetSize} \Textlen + P_t(k,\ell,\AlphabetSize))$ time
  and $\bigO(\Textlen / \log_{\AlphabetSize} \Textlen + P_s(k,\ell,\AlphabetSize))$ working space build
  a data structure of size $\bigO(\Textlen / \log_{\AlphabetSize} \Textlen + S(k,\ell,\AlphabetSize))$ that
  supports the following queries:
  \begin{itemize}
  \item Given any position $j \in [1 \dd \Textlen]$ and any length $\ell > 0$ such that $j + \ell \leq \Textlen + 1$,
    in $\bigO(\log \log \Textlen + Q(k,\ell,\AlphabetSize))$ time compute the
    position $\min \OccTwo{\Pat}{\Text}$, where $\Pat = \Text[j \dd j + \ell)$.
  \item Given the packed representation of any pattern $\Pat \in \IntegerAlphabet^{m}$ that satisfies
    $\OccTwo{\Pat}{\Text} \neq \emptyset$,
    in $\bigO(\log \log \Textlen + Q(k,\ell,\AlphabetSize) + m / \log_{\AlphabetSize} \Textlen)$
    time compute the position $\min \OccTwo{\Pat}{\Text}$.
  \end{itemize}
\end{theorem}

%% file: index/applications.tex
\subsection{Applications}\label{sec:minocc-index-applications}

The index presented in this section can be used to augmented the
compressed suffix tree (CST) presented in~\cite{breaking} with
a new operation $\MinOcc{v}$, that given a representation
$\Repr{v}$ of any explicit node of the
suffix tree $\SuffixTree{\Text}$, returns the position
$\min \OccTwo{\Pat}{\Text}$, where
$\Pat = \Str{v}$ is the string obtained by concatenating edge labels
on the path from the root to $v$. The representation $\Repr{v}$ is defined
as a pair $\Repr{v} = (\RangeBegTwo{\Pat}{\Text}, \RangeEndTwo{\Pat}{\Text})$.

\begin{theorem}\label{th:cst-minocc}
  Given any constant $\epsilon \in (0, 1)$ and the packed representation of a text $\Text \in \IntegerAlphabet^{\Textlen}$, where
  $2 \leq \AlphabetSize < \Textlen^{1/7}$, we can in $\bigO(\Textlen \min(1, \log \AlphabetSize / \sqrt{\log \Textlen}))$ time
  and $\bigO(\Textlen / \log_{\AlphabetSize} \Textlen)$ working space construct an augmented compressed suffix tree
  occupying $\bigO(\Textlen / \log_{\AlphabetSize} \Textlen)$ space that, in addition to all standard operations
  (see~\cite[Table~1]{breaking}), supports the $\MinOcc{v}$ operation in $\bigO(\log^{\epsilon} \Textlen)$ time.
\end{theorem}
\begin{proof}

  The augmented compressed suffix tree consists of two components:
  \begin{enumerate}
  \item The compressed suffix tree presented in~\cite[Theorem~7.1]{breaking} using
    $\bigO(\Textlen / \log_{\AlphabetSize} \Textlen)$ space.
  \item The index for finding the leftmost occurrences from \cref{th:minocc-index}.
    It needs $\bigO(\Textlen / \log_{\AlphabetSize} \Textlen)$ space.
  \end{enumerate}
  In total, the augmented compressed suffix tree needs $\bigO(\Textlen / \log_{\AlphabetSize} \Textlen)$ space.

  Let $v$ be an explicit node of $\SuffixTree{\Text}$. The query algorithm to compute $\MinOcc{v}$ works as follows.
  Denote $\Pat = \Str{v}$ and $(b, e) = \Repr{v} = (\RangeBegTwo{\Pat}{\Text}, \RangeEndTwo{\Pat}{\Text})$.
  We proceed in three steps:
  \begin{enumerate}
  \item Given $(b,e)$, in $\bigO(\log^{\epsilon} \Textlen)$ time we compute an
    element $j \in \OccTwo{\Pat}{\Text}$ (using the operation $\Index{v}$).
  \item Given $(b,e)$, in $\bigO(\log^{\epsilon} \Textlen)$ time we compute
    $\ell := |\Str{v}|$ (using the operation $\SDepth{v}$).
  \item Using \cref{th:minocc-index}, in
    $\bigO(\log^{\epsilon} \Textlen)$ time we compute and return $j_{\min} := \min \OccTwo{\Pat}{\Text} = \MinOcc{v}$.
  \end{enumerate}
  In total, the computation of $\MinOcc{v}$ takes $\bigO(\log^{\epsilon} \Textlen)$ time.

  The components of the augmented compressed suffix tree are constructed as follows:
  \begin{enumerate}
  \item The compressed suffix tree (with the given parameter $\epsilon$) is
    constructed in $\bigO(\Textlen \min(1, \log \AlphabetSize / \sqrt{\log \Textlen}))$ time
    and using $\bigO(\Textlen / \log_{\AlphabetSize} \Textlen)$ working space, as described in~\cite{breaking}.
  \item The index for leftmost occurrence is constructed in the same time and working space as above using
    \cref{th:minocc-index}.
  \end{enumerate}
  In total, the construction takes
  $\bigO(\Textlen \min(1, \log \AlphabetSize / \sqrt{\log \Textlen}))$ time
  and uses $\bigO(\Textlen / \log_{\AlphabetSize} \Textlen)$ working space.
\end{proof}

%% file: lpf.tex
\section{Indexes for Longest Previous Factors}\label{sec:lpf-indexes}

\subsection{LPF with Self-Overlaps}\label{sec:LPF}

\begin{definition}\label{def:LPF}
  Let $\Text \in \Sigma^{\Textlen}$. We define
  $\LPF{\Text}[1 \dd \Textlen]$ as an array such that $\LPF{\Text}[1] = 0$,
  and for every $j \in [2 \dd \Textlen]$,
  \[
    \LPF{\Text}[j] = \max\{\ell \in [0 \dd \Textlen - j + 1] :
      \min \OccTwo{\Text[j \dd j+\ell)}{\Text} < j\}.
  \]
  We also let $\LPFMinOcc{\Text}[1 \dd \Textlen]$ be such that
  for every $j \in [1 \dd \Textlen]$, letting $\ell_j = \LPF{\Text}[j]$,
  \[
    \LPFMinOcc{\Text}[j] =
    \begin{cases}
      \Text[j] & \text{if }\ell_j = 0,\\
      \min\OccTwo{\Text[j \dd j + \ell_j)}{\Text} & \text{otherwise}.
    \end{cases}
  \]
\end{definition}

\begin{observation}\label{ob:LPF}
  Let $\Text \in \Sigma^{\Textlen}$. For every $j \in [2 \dd \Textlen]$, it holds
  $\LPF{\Text}[j] \geq \LPF{\Text}[j-1] - 1$.
\end{observation}

\begin{lemma}\label{lm:LPF}
  Let $\Text \in \Sigma^{\Textlen}$.
  For every $i, j, k \in [1 \dd \Textlen]$ satisfying $i \leq j \leq k$, it holds
  \[
     \LPF{\Text}[i] - (j-i) \leq \LPF{\Text}[j] \leq \LPF{\Text}[k] + (k-j).
  \]
\end{lemma}
\begin{proof}
  By repeatedly applying \cref{ob:LPF} (which is straightforward to formalize
  using induction), it holds
  $\LPF{\Text}[j] \geq \LPF{\Text}[j-1]-1 \geq \LPF{\Text}[j-2]-2 \geq \cdots \geq \LPF{\Text}[i]-(j-i)$.
  Analogously, it holds $\LPF{\Text}[k] \geq \LPF{\Text}[j]-(k-j)$, or equivalently,
  $\LPF{\Text}[j] \leq \LPF{\Text}[k]+(k-j)$.
\end{proof}

\begin{theorem}\label{th:LPF-index-small-alphabet}
  Given any constant $\epsilon \in (0, 1)$ and the packed
  representation of a text $\Text \in \IntegerAlphabet^{\Textlen}$, where $2 \leq \AlphabetSize < \Textlen^{1/7}$,
  we can in $\bigO(\Textlen \min(1, \log \AlphabetSize / \sqrt{\log \Textlen}))$ time and
  $\bigO(\Textlen / \log_{\AlphabetSize} \Textlen)$ working space construct a data structure of size
  $\bigO(\Textlen / \log_{\AlphabetSize} \Textlen)$ that, given any
  $j \in [1 \dd \Textlen]$, returns the values $\LPF{\Text}[j]$ and $\LPFMinOcc{\Text}[j]$
  (\cref{def:LPF}) in $\bigO(\log^{\epsilon} \Textlen)$ time.
\end{theorem}
\begin{proof}

  We use the following definitions.
  Let $b = \lceil \log^{3} \Textlen \rceil$, $b' = \lceil \log^{6} \Textlen \rceil$,
  and $m = \lfloor \tfrac{\Textlen}{b} \rfloor$.
  Let $A[0 \dd m]$ be an array such that $A[0] = 0$ and for every $i \in [1 \dd m]$,
  it holds $A[i] = \LPF{\Text}[ib]$. Let $B[1 \dd m]$ be a bitvector defined such that for
  every $i \in [1 \dd m]$, $B[i] = 1$ holds if and only if $A[i] - A[i-1] \geq b' - b$.
  Finally, let $n' = \Rank{B}{1}{m}$ and $L[1 \dd bn']$ be an array such that for
  every $i \in [1 \dd n']$ and $\delta \in [1 \dd b]$, it holds
  $L[(i-1)b + \delta] = \LPF{\Text}[(i'-1)b + \delta]$, where $i' = \Select{B}{1}{i}$.
  In other words, assuming we partition the first $mb$ entries of $\LPF{\Text}$ into
  blocks of length $b$, the array $L$ stores the contents of blocks marked in bitvector $B$.

  The value $n'$ can be bounded as follows. For every $j \in [1 \dd \Textlen]$, let us
  denote $f(j) := j + \LPF{\Text}[j]$. Let also $f(0) := 0$.
  Note that by \cref{ob:LPF}, for every $j \in [2 \dd \Textlen]$,
  it holds $f(j-1) = (j-1) + \LPF{\Text}[j-1] \leq j+\LPF{\Text}[j] = f(j)$.
  Moreover, since for every $j \in [1 \dd \Textlen]$, it holds $\LPF{\Text}[j] \leq \Textlen - j + 1$, it follows
  that $f(0) \leq f(1) \leq \cdots \leq f(\Textlen) \leq \Textlen + 1$.
  For $j \in [1 \dd \Textlen]$, denote $d(j) := f(j) - f(j-1) \geq 0$.
  Note that $\sum_{j=1}^{\Textlen} d(j) = f(\Textlen) - f(0) \leq \Textlen + 1$.
  Note also that for every $i \in [1 \dd m]$, $B[i] = 1$ implies that
  \begin{align*}
    \textstyle\sum_{\delta=1}^{b} d((i-1)b + \delta)
      &= \textstyle\sum_{\delta=1}^{b} f((i-1)b + \delta) - f((i-1)b + \delta-1)\\
      &= f(ib) - f((i-1)b)\\
      &= (ib + A[i]) - ((i-1)b + A[i-1])\\
      &= b + (A[i] - A[i-1])\\
      &\geq b + (b' - b) = b'.
  \end{align*}
  Thus, if $k$ bits of $B$ are set to one, then $\sum_{j=1}^{\Textlen} d(j) \geq kb'$.
  Consequently, we have $n'b' \leq \sum_{j=1}^{\Textlen} d(j)$.
  Combining this with the upper bound $\sum_{j=1}^{\Textlen} d(j) \leq \Textlen + 1$, we
  obtain $n'b' \leq \Textlen + 1$, and hence $n' = \bigO(n/b')$.

  \paragraph{Components}

  The structure consists of the following five components:
  \begin{enumerate}
  \item The packed representation of $\Text$ using $\bigO(\Textlen / \log_{\AlphabetSize} \Textlen)$ space.
  \item The index from \cref{th:minocc-index} constructed for the parameter $\epsilon' := \epsilon/2$. It needs
    $\bigO(\Textlen / \log_{\AlphabetSize} \Textlen)$ space.
  \item The array $A[0 \dd m]$ it plain form. It needs $\bigO(m) = \bigO(\Textlen / \log^{3} \Textlen)
    = \bigO(\Textlen / \log_{\AlphabetSize} \Textlen)$ space.
  \item The bitvector $B[1 \dd m]$ augmented using \cref{th:bin-rank-select} to support rank and select queries in $\bigO(1)$ time.
    The bitvector needs $\bigO(m/\log \Textlen) = \bigO(\Textlen / \log^4 \Textlen) =
    \bigO(\Textlen / \log_{\AlphabetSize} \Textlen)$ space,
    and the augmentation of \cref{th:bin-rank-select} does not increase the space usage.
  \item The array $L[1 \dd bn']$ stored in plain form. By definition of $b$, and the upper bound on $n'$, it needs
    $\bigO(bn') = \bigO(b\Textlen / b') = \bigO(\Textlen / \log^{3} \Textlen) =
    \bigO(\Textlen / \log_{\AlphabetSize} \Textlen)$ space.
  \end{enumerate}
  In total, the data structure takes $\bigO(\Textlen / \log_{\AlphabetSize} \Textlen)$ space.

  \paragraph{Implementation of queries}

  Let $j \in [1 \dd \Textlen]$. The computation
  of $\LPF{\Text}[j]$ and $\LPFMinOcc{\Text}[j]$ proceeds as follows:
  \begin{enumerate}
  \item In the first step, we compute $\LPF{\Text}[j]$. If $j = 1$, then we have $\LPF{\Text}[1] = 0$. Let us thus
    assume $j \geq 2$. First, in $\bigO(1)$ time we compute $i := \lceil j/b \rceil$ and $\delta = j - (i-1)b \in [1 \dd b]$.
    In $\bigO(1)$ time we then lookup the value $B[i]$ and consider two cases:
    \begin{itemize}
    \item First, assume $B[i] = 1$. Then, $\LPF{\Text}[j]$ is precomputed and stored in $L$.
      To locate it, in $\bigO(1)$ time we compute $i' = \Rank{B}{1}{i}$, and then
      obtain $\LPF{\Text}[j] = \LPF{\Text}[(i-1)b + \delta] = L[(i'-1)b + \delta]$.
    \item Let us now assume that $B[i] = 0$. This implies that $A[i] - A[i-1] < b' - b$,
      or equivalently, $A[i] < A[i-1] + (b'-b)$. By \cref{lm:LPF}, it holds
      $A[i-1] - b \leq \LPF{\Text}[j] \leq A[i] + b < A[i-1] + b'$.
      We thus proceed as follows. First, in $\bigO(1)$ time we set $\ell_{\min} := \max(0, A[i-1] - b)$ and
      $\ell_{\max} = \min(\Textlen - j + 1, \ell_{\min} + b' + b)$.
      Then, using the index from \cref{th:minocc-index}, we binary search for the value
      $\ell := \max\{\ell \in [\ell_{\min} \dd \ell_{\max}] : \min \OccTwo{\Text[j \dd j + \ell)}{\Text} < j\}$,
      which by the above discussion is equal to $\LPF{\Text}[j]$. To see that this is correct, note that
      if for some $t > 0$, it holds $\min \OccTwo{\Text[j \dd j + t)}{\Text} < j$, then for all
      $t' \in [0 \dd t)$, we also have $\min \OccTwo{\Text[j \dd j + t')}{\Text} < j$.
      The computation of $\ell$ takes
      $\bigO(\log^{\epsilon'} \Textlen \cdot \log (\ell_{\max} - \ell_{\min} + 1))
      = \bigO(\log^{\epsilon'} \Textlen \cdot \log (\log^{6} \Textlen)) = \bigO(\log^{\epsilon} \Textlen)$ time.
    \end{itemize}
  \item In the second step, we compute $\LPFMinOcc{\Text}[j]$. Let $\ell = \LPF{\Text}[j]$.
    If $\ell = 0$, then we obtain $\LPFMinOcc{\Text}[j] = \Text[j]$ in $\bigO(1)$ time.
    Otherwise, we compute $\LPFMinOcc{\Text}[j] = \min \OccTwo{\Text[j \dd j + \ell)}{\Text}$
    in $\bigO(\log^{\epsilon'} \Textlen) = \bigO(\log^{\epsilon} \Textlen)$ time using the index
    from \cref{th:minocc-index}.
  \end{enumerate}
  In total, the query takes $\bigO(\log^{\epsilon} \Textlen)$ time.

  \paragraph{Construction algorithm}

  The components of the data structure are constructed as follows:
  \begin{enumerate}
  \item We save the packed representation of $\Text$ in $\bigO(\Textlen / \log_{\AlphabetSize} \Textlen)$ time.
  \item We construct the index from \cref{th:minocc-index} in
    $\bigO(\Textlen \min(1, \log \AlphabetSize / \sqrt{\log \Textlen}))$ time and using
    $\bigO(\Textlen / \log_{\AlphabetSize} \Textlen)$ working space.
  \item As noted above, any value of $\LPF{\Text}[j]$ can be computed in
    $\bigO(\log^{1+\epsilon'} \Textlen)$ time using
    binary search and the index from \cref{th:minocc-index}.
    Thus, constructing $A$ takes $\bigO(m \log^{1 + \epsilon'} \Textlen) =
    \bigO((\Textlen / b) \log^{1+\epsilon'} \Textlen) =
    \bigO(\Textlen / \log \Textlen) =
    \bigO(\Textlen / \log_{\AlphabetSize} \Textlen)$ time.
  \item Given the array $A$, the bitvector $B$ is easily constructed
    in $\bigO(m) = \bigO(\Textlen / \log_{\AlphabetSize} \Textlen)$ time,
    and the augmentation of \cref{th:bin-rank-select}
    takes $\bigO(m / \log \Textlen) =
    \bigO(\Textlen / \log_{\AlphabetSize} \Textlen)$ time.
  \item Given the bitvector $B$, the construction of $L$ reduces
    to computing $\bigO(bn') = \bigO(b\Textlen / b')$ values of $\LPF{\Text}$.
    Each value takes $\bigO(\log^{1+\epsilon'} \Textlen)$ time to compute, and hence
    in total, we spend $\bigO((b\Textlen / b') \log^{1+\epsilon'} \Textlen)
    = \bigO(\Textlen / \log \Textlen) = \bigO(\Textlen / \log_{\AlphabetSize} \Textlen)$
    time.
    \qedhere
  \end{enumerate}
\end{proof}

\subsection{LPF without Self-Overlaps}\label{sec:LPnF}

\begin{definition}\label{def:LPnF}
  Let $\Text \in \Sigma^{\Textlen}$. We define
  $\LPnF{\Text}[1 \dd \Textlen]$ as an array such that $\LPnF{\Text}[1] = 0$,
  and for every $j \in [2 \dd \Textlen]$,
  \[
    \LPnF{\Text}[j] = \max\{\ell \in [0 \dd \Textlen - j + 1] :
      \min \OccTwo{\Text[j \dd j+\ell)}{\Text} + \ell \leq j\}.
  \]
  We also let $\LPnFMinOcc{\Text}[1 \dd \Textlen]$ be such that
  for every $j \in [1 \dd \Textlen]$, letting $\ell_j = \LPnF{\Text}[j]$,
  \[
    \LPnFMinOcc{\Text}[j] =
    \begin{cases}
      \Text[j] & \text{if }\ell_j = 0,\\
      \min\OccTwo{\Text[j \dd j + \ell_j)}{\Text} & \text{otherwise}.
    \end{cases}
  \]
\end{definition}

\begin{observation}\label{ob:LPnF}
  Let $\Text \in \Sigma^{\Textlen}$. For every $j \in [2 \dd \Textlen]$, it holds
  $\LPnF{\Text}[j] \geq \LPnF{\Text}[j-1] - 1$.
\end{observation}

\begin{lemma}\label{lm:LPnF}
  Let $\Text \in \Sigma^{\Textlen}$.
  For every $i, j, k \in [1 \dd \Textlen]$ satisfying $i \leq j \leq k$, it holds
  \[
     \LPnF{\Text}[i] - (j-i) \leq \LPnF{\Text}[j] \leq \LPnF{\Text}[k] + (k-j).
  \]
\end{lemma}
\begin{proof}
  The proof proceeds as in \cref{lm:LPF}, except instead of
  \cref{ob:LPF}, we use \cref{ob:LPnF}.
\end{proof}

\begin{theorem}\label{th:LPnF-index-small-alphabet}
  Given any constant $\epsilon \in (0, 1)$ and the packed
  representation of a text $\Text \in \IntegerAlphabet^{\Textlen}$, where $2 \leq \AlphabetSize < \Textlen^{1/7}$,
  we can in $\bigO(\Textlen \min(1, \log \AlphabetSize / \sqrt{\log \Textlen}))$ time and
  $\bigO(\Textlen / \log_{\AlphabetSize} \Textlen)$ working space construct a data structure of size
  $\bigO(\Textlen / \log_{\AlphabetSize} \Textlen)$ that, given any
  $j \in [1 \dd \Textlen]$, returns the values $\LPnF{\Text}[j]$ and $\LPnFMinOcc{\Text}[j]$
  (\cref{def:LPF}) in $\bigO(\log^{\epsilon} \Textlen)$ time.
\end{theorem}
\begin{proof}
  The proof proceeds as in \cref{th:LPF-index-small-alphabet}, except instead of
  \cref{ob:LPF} and \cref{lm:LPF}, we use \cref{ob:LPnF} and \cref{lm:LPnF}, respectively.
\end{proof}

%% file: lz.tex
\section{Sublinear LZ77 Factorization}\label{sec:lz}

\subsection{LZ77 with Self-Overlaps}\label{sec:lz-ov}

\begin{theorem}[{\cite[Theorem~2]{LZ76}}]\label{th:lz-size}
  For every text $\Text \in \IntegerAlphabet^{\Textlen}$, it holds
  $\LZSize{\Text} = \bigO(\Textlen / \log_{\AlphabetSize} \Textlen)$.
\end{theorem}

\begin{theorem}\label{th:lz-from-text}
  Given the packed representation
  of a text $\Text \in \IntegerAlphabet^{\Textlen}$, where $2 \leq \AlphabetSize < \Textlen^{1/7}$, we can
  construct the LZ77 factorization of $\Text$ in
  $\bigO((\Textlen \log \AlphabetSize) / \sqrt{\log \Textlen})$ time using
  $\bigO(\Textlen / \log_{\AlphabetSize} \Textlen)$ working space.
\end{theorem}
\begin{proof}
  We proceed as follows:
  \begin{enumerate}
  \item Consider any constant $\epsilon \in (0, \tfrac{1}{2}]$.
    In the first step, we construct the index from \cref{th:LPF-index-small-alphabet} in
    $\bigO(\Textlen \min(1, \log \AlphabetSize / \sqrt{\log \Textlen})) =
    \bigO((\Textlen \log \AlphabetSize) / \sqrt{\log \Textlen})$ time
    and using $\bigO(\Textlen / \log_{\AlphabetSize} \Textlen)$ working space.
  \item By repeatedly using the query of the index constructed in the previous step,
    we then construct the
    LZ77 factorization in $\bigO(\LZSize{\Text} \log^{\epsilon} \Textlen)$ time and using
    $\bigO(\Textlen / \log_{\AlphabetSize} \Textlen)$ working space. By
    \cref{th:lz-size}, this time can be bounded as
    $\bigO(\LZSize{\Text} \log^{\epsilon} \Textlen)
    = \bigO((\Textlen \log \AlphabetSize) / \log^{1-\epsilon} \Textlen)
    = \bigO((\Textlen \log \AlphabetSize) / \sqrt{\log \Textlen})$.
    \qedhere
  \end{enumerate}
\end{proof}

\subsection{LZ77 without Self-Overlaps}\label{sec:lz-nonov}

\begin{theorem}\label{th:lz-nonov-size}
  For every text $\Text \in \IntegerAlphabet^{\Textlen}$, it holds
  $\LZNonOvSize{\Text} = \bigO(\Textlen / \log_{\AlphabetSize} \Textlen)$.
\end{theorem}
\begin{proof}
  Let $g^{*}(\Text)$ denote the size of the smallest grammar encoding
  of $\Text$, and let $g^{*}_{\rm irr}(\Text)$ be the
  size of the smallest irreducible grammar encoding $\Text$~\cite{charikar}.
  Then:
  \begin{itemize}
  \item In~\cite[Theorem~1]{Rytter03}, it is proved that
    $\LZNonOvSize{\Text} \leq g^{*}(\Text)$.
  \item By definition of $g^{*}_{\rm irr}(\Text)$, it holds
    $g^{*}(\Text) \leq g^{*}_{\rm irr}(\Text)$.
  \item In~\cite[Lemma~4]{OchoaN19}, it is proved that
    $g^{*}_{\rm irr}(\Text) = \bigO(\Textlen / \log_{\AlphabetSize} \Textlen)$.
  \end{itemize}
  Combining the above inequalities yields the claim.
\end{proof}

\begin{theorem}\label{th:lz-nonov-from-text}
  Given the packed representation
  of a text $\Text \in \IntegerAlphabet^{\Textlen}$, where $2 \leq \AlphabetSize < \Textlen^{1/7}$, we can
  construct the non-overlapping variant of LZ77 factorization of $\Text$ in
  $\bigO((\Textlen \log \AlphabetSize) / \sqrt{\log \Textlen})$ time using
  $\bigO(\Textlen / \log_{\AlphabetSize} \Textlen)$ working space.
\end{theorem}
\begin{proof}
  The algorithm proceeds analogously as in \cref{th:lz-from-text}, except instead
  of \cref{th:LPF-index-small-alphabet} and \cref{th:lz-size}, we
  use \cref{th:LPnF-index-small-alphabet} and \cref{th:lz-nonov-size}, respectively.
\end{proof}

%% file: paper.bbl
\newcommand{\etalchar}[1]{$^{#1}$}
\begin{thebibliography}{FMG{\etalchar{+}}22}

\bibitem[ABBK17]{AbboudBBK17}
Amir Abboud, Arturs Backurs, Karl Bringmann, and Marvin K{\"{u}}nnemann.
\newblock Fine-grained complexity of analyzing compressed data: Quantifying
  improvements over decompress-and-solve.
\newblock In Chris Umans, editor, {\em 58th {IEEE} Annual Symposium on
  Foundations of Computer Science, {FOCS} 2017}, pages 192--203. {IEEE}
  Computer Society, 2017.
\newblock \href {https://doi.org/10.1109/FOCS.2017.26}
  {\path{doi:10.1109/FOCS.2017.26}}.

\bibitem[ACI{\etalchar{+}}12]{Al-HafeedhCIKSTY12}
Anisa Al{-}Hafeedh, Maxime Crochemore, Lucian Ilie, Evguenia Kopylova,
  William~F. Smyth, German Tischler, and Munina Yusufu.
\newblock A comparison of index-based {L}empel-{Z}iv {LZ77} factorization
  algorithms.
\newblock {\em {ACM} Computing Surveys}, 45(1):5:1--5:17, 2012.
\newblock \href {https://doi.org/10.1145/2379776.2379781}
  {\path{doi:10.1145/2379776.2379781}}.

\bibitem[AFF{\etalchar{+}}18]{brotli}
Jyrki Alakuijala, Andrea Farruggia, Paolo Ferragina, Eugene Kliuchnikov, Robert
  Obryk, Zoltan Szabadka, and Lode Vandevenne.
\newblock {B}rotli: {A} general-purpose data compressor.
\newblock {\em ACM Transactions on Information Systems}, 37(1), 2018.
\newblock \href {https://doi.org/10.1145/3231935} {\path{doi:10.1145/3231935}}.

\bibitem[ALU02]{AmirLU02}
Amihood Amir, Gad~M. Landau, and Esko Ukkonen.
\newblock Online timestamped text indexing.
\newblock {\em Information Processing Letters}, 82(5):253--259, 2002.
\newblock \href {https://doi.org/10.1016/S0020-0190(01)00275-7}
  {\path{doi:10.1016/S0020-0190(01)00275-7}}.

\bibitem[BCFG17]{BilleCFG17}
Philip Bille, Patrick~Hagge Cording, Johannes Fischer, and Inge~Li G{\o}rtz.
\newblock {L}empel-{Z}iv compression in a sliding window.
\newblock In Juha K{\"{a}}rkk{\"{a}}inen, Jakub Radoszewski, and Wojciech
  Rytter, editors, {\em 28th Annual Symposium on Combinatorial Pattern
  Matching, {CPM} 2017}, volume~78 of {\em LIPIcs}, pages 15:1--15:11. Schloss
  Dagstuhl--Leibniz-Zentrum f{\"{u}}r Informatik, 2017.
\newblock \href {https://doi.org/10.4230/LIPICS.CPM.2017.15}
  {\path{doi:10.4230/LIPICS.CPM.2017.15}}.

\bibitem[BCG{\etalchar{+}}21]{blocktree}
Djamal Belazzougui, Manuel C{\'{a}}ceres, Travis Gagie, Paweł Gawrychowski,
  Juha K{\"{a}}rkk{\"{a}}inen, Gonzalo Navarro, Alberto~Ord{\'{o}}{\~{n}}ez
  Pereira, Simon~J. Puglisi, and Yasuo Tabei.
\newblock Block trees.
\newblock {\em Journal of Computer and System Sciences}, 117:1--22, 2021.
\newblock \href {https://doi.org/10.1016/j.jcss.2020.11.002}
  {\path{doi:10.1016/j.jcss.2020.11.002}}.

\bibitem[BEGV18]{BilleEGV18}
Philip Bille, Mikko~Berggren Ettienne, Inge~Li G{\o}rtz, and Hjalte~Wedel
  Vildh{\o}j.
\newblock Time-space trade-offs for {L}empel-{Z}iv compressed indexing.
\newblock {\em Theoretical Computer Science}, 713:66--77, 2018.
\newblock \href {https://doi.org/10.1016/J.TCS.2017.12.021}
  {\path{doi:10.1016/J.TCS.2017.12.021}}.

\bibitem[BGKS15]{WaveletSuffixTree}
Maxim Babenko, Pawe{\l} Gawrychowski, Tomasz Kociumaka, and Tatiana
  Starikovskaya.
\newblock Wavelet trees meet suffix trees.
\newblock In {\em 26th Annual {ACM-SIAM} Symposium on Discrete Algorithms,
  {SODA} 2015}, pages 572--591, 2015.
\newblock \href {https://doi.org/10.1137/1.9781611973730.39}
  {\path{doi:10.1137/1.9781611973730.39}}.

\bibitem[BKW19]{BringmannWK19}
Karl Bringmann, Marvin K{\"{u}}nnemann, and Philip Wellnitz.
\newblock Few matches or almost periodicity: Faster pattern matching with
  mismatches in compressed texts.
\newblock In Timothy~M. Chan, editor, {\em 30th Annual {ACM-SIAM} Symposium on
  Discrete Algorithms, {SODA} 2019}, pages 1126--1145. {SIAM}, 2019.
\newblock \href {https://doi.org/10.1137/1.9781611975482.69}
  {\path{doi:10.1137/1.9781611975482.69}}.

\bibitem[BLR{\etalchar{+}}15]{BLRSRW15}
Philip Bille, Gad~M. Landau, Rajeev Raman, Kunihiko Sadakane, Srinivasa~Rao
  Satti, and Oren Weimann.
\newblock Random access to grammar-compressed strings and trees.
\newblock {\em SIAM Journal on Computing}, 44(3):513--539, 2015.
\newblock \href {https://doi.org/10.1137/130936889}
  {\path{doi:10.1137/130936889}}.

\bibitem[BP16]{BelazzouguiP16}
Djamal Belazzougui and Simon~J. Puglisi.
\newblock Range predecessor and {L}empel-{Z}iv parsing.
\newblock In Robert Krauthgamer, editor, {\em 27th Annual {ACM-SIAM} Symposium
  on Discrete Algorithms, {SODA} 2016}, pages 2053--2071. {SIAM}, 2016.
\newblock \href {https://doi.org/10.1137/1.9781611974331.CH143}
  {\path{doi:10.1137/1.9781611974331.CH143}}.

\bibitem[BW94]{bwt}
Michael Burrows and David~J. Wheeler.
\newblock A block-sorting lossless data compression algorithm.
\newblock Technical Report 124, Digital Equipment Corporation, Palo Alto,
  California, 1994.
\newblock URL:
  \url{https://www.hpl.hp.com/techreports/Compaq-DEC/SRC-RR-124.pdf}.

\bibitem[CEK{\etalchar{+}}21]{ChristiansenEKN21}
Anders~Roy Christiansen, Mikko~Berggren Ettienne, Tomasz Kociumaka, Gonzalo
  Navarro, and Nicola Prezza.
\newblock Optimal-time dictionary-compressed indexes.
\newblock {\em ACM Transactions on Algorithms}, 17(1):8:1--8:39, 2021.
\newblock \href {https://doi.org/10.1145/3426473} {\path{doi:10.1145/3426473}}.

\bibitem[CHL07]{AlgorithmsOnStrings}
Maxime Crochemore, Christophe Hancart, and Thierry Lecroq.
\newblock {\em Algorithms on strings}.
\newblock Cambridge University Press, Cambridge, UK, 2007.
\newblock \href {https://doi.org/10.1017/cbo9780511546853}
  {\path{doi:10.1017/cbo9780511546853}}.

\bibitem[CI08]{CrochemoreI08}
Maxime Crochemore and Lucian Ilie.
\newblock Computing longest previous factor in linear time and applications.
\newblock {\em Information Processing Letters}, 106(2):75--80, 2008.
\newblock \href {https://doi.org/10.1016/J.IPL.2007.10.006}
  {\path{doi:10.1016/J.IPL.2007.10.006}}.

\bibitem[CIR09]{CrochemoreIR09}
Maxime Crochemore, Lucian Ilie, and Wojciech Rytter.
\newblock Repetitions in strings: Algorithms and combinatorics.
\newblock {\em Theoretical Computer Science}, 410(50):5227--5235, 2009.
\newblock \href {https://doi.org/10.1016/J.TCS.2009.08.024}
  {\path{doi:10.1016/J.TCS.2009.08.024}}.

\bibitem[CIS08]{CrochemoreIS08}
Maxime Crochemore, Lucian Ilie, and William~F. Smyth.
\newblock A simple algorithm for computing the {L}empel {Z}iv factorization.
\newblock In {\em 2008 Data Compression Conference, {DCC} 2008}, pages
  482--488. {IEEE} Computer Society, 2008.
\newblock \href {https://doi.org/10.1109/DCC.2008.36}
  {\path{doi:10.1109/DCC.2008.36}}.

\bibitem[CKW20]{CKW20}
Panagiotis Charalampopoulos, Tomasz Kociumaka, and Philip Wellnitz.
\newblock Faster approximate pattern matching: {A} unified approach.
\newblock In Sandy Irani, editor, {\em 61st {IEEE} Annual Symposium on
  Foundations of Computer Science, {FOCS} 2020}, pages 978--989. {IEEE}
  Computer Society, 2020.
\newblock \href {https://doi.org/10.1109/FOCS46700.2020.00095}
  {\path{doi:10.1109/FOCS46700.2020.00095}}.

\bibitem[CKW22]{Charalampopoulos22}
Panagiotis Charalampopoulos, Tomasz Kociumaka, and Philip Wellnitz.
\newblock Faster pattern matching under edit distance : {A} reduction to
  dynamic puzzle matching and the seaweed monoid of permutation matrices.
\newblock In {\em 63rd {IEEE} Annual Symposium on Foundations of Computer
  Science, {FOCS} 2022}, pages 698--707. {IEEE}, 2022.
\newblock \href {https://doi.org/10.1109/FOCS54457.2022.00072}
  {\path{doi:10.1109/FOCS54457.2022.00072}}.

\bibitem[Cla98]{Clark98}
David~R. Clark.
\newblock {\em Compact Pat Trees}.
\newblock PhD thesis, University of Waterloo, 1998.
\newblock URL: \url{http://hdl.handle.net/10012/64}.

\bibitem[CLL{\etalchar{+}}05]{charikar}
Moses Charikar, Eric Lehman, Ding Liu, Rina Panigrahy, Manoj Prabhakaran, Amit
  Sahai, and Abhi Shelat.
\newblock The smallest grammar problem.
\newblock {\em IEEE Transactions on Information Theory}, 51(7):2554--2576,
  2005.
\newblock \href {https://doi.org/10.1109/TIT.2005.850116}
  {\path{doi:10.1109/TIT.2005.850116}}.

\bibitem[CLZ02]{CrochemoreLZ02}
Maxime Crochemore, Gad~M. Landau, and Michal Ziv{-}Ukelson.
\newblock A sub-quadratic sequence alignment algorithm for unrestricted cost
  matrices.
\newblock In David Eppstein, editor, {\em 13th Annual {ACM-SIAM} Symposium on
  Discrete Algorithms, {SODA} 2002}, pages 679--688. {ACM/SIAM}, 2002.
\newblock URL: \url{http://dl.acm.org/citation.cfm?id=545381.545472}.

\bibitem[CPS07]{ChenPS07}
Gang Chen, Simon~J. Puglisi, and William~F. Smyth.
\newblock Fast and practical algorithms for computing all the runs in a string.
\newblock In Bin Ma and Kaizhong Zhang, editors, {\em 18th Annual Symposium on
  Combinatorial Pattern Matching, {CPM} 2007}, volume 4580 of {\em LNCS}, pages
  307--315. Springer, 2007.
\newblock \href {https://doi.org/10.1007/978-3-540-73437-6\_31}
  {\path{doi:10.1007/978-3-540-73437-6\_31}}.

\bibitem[CR91]{CrochemoreR91}
Maxime Crochemore and Wojciech Rytter.
\newblock Efficient parallel algorithms to test square-freeness and factorize
  strings.
\newblock {\em Information Processing Letters}, 38(2):57--60, 1991.
\newblock \href {https://doi.org/10.1016/0020-0190(91)90223-5}
  {\path{doi:10.1016/0020-0190(91)90223-5}}.

\bibitem[Cro86]{Crochemore86}
Maxime Crochemore.
\newblock Transducers and repetitions.
\newblock {\em Theoretical Computer Science}, 45(1):63--86, 1986.
\newblock \href {https://doi.org/10.1016/0304-3975(86)90041-1}
  {\path{doi:10.1016/0304-3975(86)90041-1}}.

\bibitem[CT11]{CrochemoreT11}
Maxime Crochemore and German Tischler.
\newblock Computing longest previous non-overlapping factors.
\newblock {\em Information Processing Letters}, 111(6):291--295, 2011.
\newblock \href {https://doi.org/10.1016/J.IPL.2010.12.005}
  {\path{doi:10.1016/J.IPL.2010.12.005}}.

\bibitem[DKK{\etalchar{+}}04]{DuvalKKLL04}
Jean{-}Pierre Duval, Roman Kolpakov, Gregory Kucherov, Thierry Lecroq, and
  Arnaud Lefebvre.
\newblock Linear-time computation of local periods.
\newblock {\em Theoretical Computer Science}, 326(1-3):229--240, 2004.
\newblock \href {https://doi.org/10.1016/J.TCS.2004.06.024}
  {\path{doi:10.1016/J.TCS.2004.06.024}}.

\bibitem[EFP23]{EllertFP23}
Jonas Ellert, Johannes Fischer, and Max~Rish{\o}j Pedersen.
\newblock New advances in rightmost {L}empel-{Z}iv.
\newblock In Franco~Maria Nardini, Nadia Pisanti, and Rossano Venturini,
  editors, {\em 30th International Symposium on String Processing and
  Information Retrieval, {SPIRE} 2023}, volume 14240 of {\em LNCS}, pages
  188--202. Springer, 2023.
\newblock \href {https://doi.org/10.1007/978-3-031-43980-3\_15}
  {\path{doi:10.1007/978-3-031-43980-3\_15}}.

\bibitem[EGG23]{EllertGG23}
Jonas Ellert, Pawel Gawrychowski, and Garance Gourdel.
\newblock Optimal square detection over general alphabets.
\newblock In Nikhil Bansal and Viswanath Nagarajan, editors, {\em 34th Annual
  {ACM-SIAM} Symposium on Discrete Algorithms, {SODA} 2023}, pages 5220--5242.
  {SIAM}, 2023.
\newblock \href {https://doi.org/10.1137/1.9781611977554.CH189}
  {\path{doi:10.1137/1.9781611977554.CH189}}.

\bibitem[Ell23]{Ellert23}
Jonas Ellert.
\newblock Sublinear time {L}empel-{Z}iv {(LZ77)} factorization.
\newblock In Franco~Maria Nardini, Nadia Pisanti, and Rossano Venturini,
  editors, {\em 30th International Symposium on String Processing and
  Information Retrieval, {SPIRE} 2023}, volume 14240 of {\em LNCS}, pages
  171--187. Springer, 2023.
\newblock \href {https://doi.org/10.1007/978-3-031-43980-3\_14}
  {\path{doi:10.1007/978-3-031-43980-3\_14}}.

\bibitem[FG15]{wexp}
Johannes Fischer and Pawe{\l} Gawrychowski.
\newblock Alphabet-dependent string searching with {W}exponential search trees.
\newblock In {\em 26th Annual Symposium on Combinatorial Pattern Matching,
  {CPM} 2015}, pages 160--171, 2015.
\newblock Full version: \url{https://arxiv.org/abs/1302.3347}.
\newblock \href {https://doi.org/10.1007/978-3-319-19929-0_14}
  {\path{doi:10.1007/978-3-319-19929-0_14}}.

\bibitem[FGGK15]{FischerGGK15}
Johannes Fischer, Travis Gagie, Pawel Gawrychowski, and Tomasz Kociumaka.
\newblock Approximating {LZ77} via small-space multiple-pattern matching.
\newblock In Nikhil Bansal and Irene Finocchi, editors, {\em 23rd Annual
  European Symposium on Algorithms, {ESA} 2015}, volume 9294 of {\em LNCS},
  pages 533--544. Springer, 2015.
\newblock \href {https://doi.org/10.1007/978-3-662-48350-3\_45}
  {\path{doi:10.1007/978-3-662-48350-3\_45}}.

\bibitem[FGHP14]{hybrid}
H{\'e}ctor Ferrada, Travis Gagie, Tommi Hirvola, and Simon~J Puglisi.
\newblock Hybrid indexes for repetitive datasets.
\newblock {\em Philosophical Transactions of the Royal Society A}, 372, 2014.
\newblock \href {https://doi.org/10.1098/rsta.2013.0137}
  {\path{doi:10.1098/rsta.2013.0137}}.

\bibitem[FH11]{FischerH11}
Johannes Fischer and Volker Heun.
\newblock Space-efficient preprocessing schemes for range minimum queries on
  static arrays.
\newblock {\em SIAM Journal on Computing}, 40(2):465--492, 2011.
\newblock \href {https://doi.org/10.1137/090779759}
  {\path{doi:10.1137/090779759}}.

\bibitem[FIK15]{FischerIK15}
Johannes Fischer, Tomohiro I, and Dominik K{\"{o}}ppl.
\newblock {L}empel {Z}iv computation in small space {(LZ-CISS)}.
\newblock In Ferdinando Cicalese, Ely Porat, and Ugo Vaccaro, editors, {\em
  26th Annual Symposium on Combinatorial Pattern Matching, {CPM} 2015}, volume
  9133 of {\em LNCS}, pages 172--184. Springer, 2015.
\newblock \href {https://doi.org/10.1007/978-3-319-19929-0\_15}
  {\path{doi:10.1007/978-3-319-19929-0\_15}}.

\bibitem[FM95]{FarachM95}
Martin Farach and S.~Muthukrishnan.
\newblock Optimal parallel dictionary matching and compression (extended
  abstract).
\newblock In Charles~E. Leiserson, editor, {\em 7th Annual {ACM} Symposium on
  Parallel Algorithms and Architectures, {SPAA} 1995}, pages 244--253. {ACM},
  1995.
\newblock \href {https://doi.org/10.1145/215399.215451}
  {\path{doi:10.1145/215399.215451}}.

\bibitem[FMG{\etalchar{+}}22]{FerraginaMGKNST22}
Paolo Ferragina, Giovanni Manzini, Travis Gagie, Dominik K{\"{o}}ppl, Gonzalo
  Navarro, Manuel Striani, and Francesco Tosoni.
\newblock Improving matrix-vector multiplication via lossless
  grammar-compressed matrices.
\newblock {\em Proceedings of the {VLDB} Endowment}, 15(10):2175--2187, 2022.
\newblock URL: \url{https://www.vldb.org/pvldb/vol15/p2175-tosoni.pdf}.

\bibitem[FNV08]{bitlz}
Paolo Ferragina, Igor Nitto, and Rossano Venturini.
\newblock Bit-optimal {L}empel-{Z}iv compression, 2008.
\newblock URL: \url{http://arxiv.org/abs/0802.0835}, \href
  {http://arxiv.org/abs/0802.0835} {\path{arXiv:0802.0835}}.

\bibitem[Gaw11]{Gaw11}
Paweł Gawrychowski.
\newblock Pattern matching in {L}empel-{Z}iv compressed strings: Fast, simple,
  and deterministic.
\newblock In Camil Demetrescu and Magn{\'{u}}s~M. Halld{\'{o}}rsson, editors,
  {\em 19th Annual European Symposium on Algorithms, {ESA} 2011}, volume 6942
  of {\em LNCS}, pages 421--432. Springer, 2011.
\newblock \href {https://doi.org/10.1007/978-3-642-23719-5_36}
  {\path{doi:10.1007/978-3-642-23719-5_36}}.

\bibitem[Gaw12]{Gawrychowski12}
Pawel Gawrychowski.
\newblock Faster algorithm for computing the edit distance between
  {SLP}-compressed strings.
\newblock In Liliana Calder{\'{o}}n{-}Benavides, Cristina~N.
  Gonz{\'{a}}lez{-}Caro, Edgar Ch{\'{a}}vez, and Nivio Ziviani, editors, {\em
  19th International Symposium on String Processing and Information Retrieval,
  {SPIRE} 2012}, volume 7608 of {\em LNCS}, pages 229--236. Springer, 2012.
\newblock \href {https://doi.org/10.1007/978-3-642-34109-0\_24}
  {\path{doi:10.1007/978-3-642-34109-0\_24}}.

\bibitem[GB13]{GotoB13}
Keisuke Goto and Hideo Bannai.
\newblock Simpler and faster {L}empel {Z}iv factorization.
\newblock In Ali Bilgin, Michael~W. Marcellin, Joan Serra{-}Sagrist{\`{a}}, and
  James~A. Storer, editors, {\em 2013 Data Compression Conference, {DCC} 2013},
  pages 133--142. {IEEE}, 2013.
\newblock \href {https://doi.org/10.1109/DCC.2013.21}
  {\path{doi:10.1109/DCC.2013.21}}.

\bibitem[GB14]{GotoB14}
Keisuke Goto and Hideo Bannai.
\newblock Space efficient linear time {L}empel-{Z}iv factorization for small
  alphabets.
\newblock In Ali Bilgin, Michael~W. Marcellin, Joan Serra{-}Sagrist{\`{a}}, and
  James~A. Storer, editors, {\em 2024 Data Compression Conference, {DCC} 2014},
  pages 163--172. {IEEE}, 2014.
\newblock \href {https://doi.org/10.1109/DCC.2014.62}
  {\path{doi:10.1109/DCC.2014.62}}.

\bibitem[GBT84]{GBT84}
Harold~N. Gabow, Jon~Louis Bentley, and Robert~Endre Tarjan.
\newblock Scaling and related techniques for geometry problems.
\newblock In Richard~A. DeMillo, editor, {\em 16th Annual {ACM} Symposium on
  Theory of Computing, {STOC} 1984}, pages 135--143. {ACM}, 1984.
\newblock \href {https://doi.org/10.1145/800057.808675}
  {\path{doi:10.1145/800057.808675}}.

\bibitem[GG22]{GanardiG22}
Moses Ganardi and Paweł Gawrychowski.
\newblock Pattern matching on grammar-compressed strings in linear time.
\newblock In Joseph~(Seffi) Naor and Niv Buchbinder, editors, {\em 33rd Annual
  {ACM-SIAM} Symposium on Discrete Algorithms, {SODA} 2022}, pages 2833--2846.
  {SIAM}, 2022.
\newblock \href {https://doi.org/10.1137/1.9781611977073.110}
  {\path{doi:10.1137/1.9781611977073.110}}.

\bibitem[GGK{\etalchar{+}}12]{GagieGKNP12}
Travis Gagie, Paweł Gawrychowski, Juha K{\"{a}}rkk{\"{a}}inen, Yakov Nekrich,
  and Simon~J. Puglisi.
\newblock A faster grammar-based self-index.
\newblock In Adrian{-}Horia Dediu and Carlos Mart{\'{\i}}n{-}Vide, editors,
  {\em 6th International Conference on Language and Automata Theory and
  Applications, {LATA} 2012}, volume 7183 of {\em LNCS}, pages 240--251.
  Springer, 2012.
\newblock \href {https://doi.org/10.1007/978-3-642-28332-1_21}
  {\path{doi:10.1007/978-3-642-28332-1_21}}.

\bibitem[GGK{\etalchar{+}}14]{GagieGKNP14}
Travis Gagie, Paweł Gawrychowski, Juha K{\"{a}}rkk{\"{a}}inen, Yakov Nekrich,
  and Simon~J. Puglisi.
\newblock {LZ77}-based self-indexing with faster pattern matching.
\newblock In Alberto Pardo and Alfredo Viola, editors, {\em 11th Latin American
  Symposium on Theoretical Informatics, {LATIN} 2014}, volume 8392 of {\em
  LNCS}, pages 731--742. Springer, 2014.
\newblock \href {https://doi.org/10.1007/978-3-642-54423-1_63}
  {\path{doi:10.1007/978-3-642-54423-1_63}}.

\bibitem[GGP15]{GagieGP15}
Travis Gagie, Pawel Gawrychowski, and Simon~J. Puglisi.
\newblock Approximate pattern matching in {LZ77}-compressed texts.
\newblock {\em Journal of Discrete Algorithms}, 32:64--68, 2015.
\newblock \href {https://doi.org/10.1016/J.JDA.2014.10.003}
  {\path{doi:10.1016/J.JDA.2014.10.003}}.

\bibitem[GHN20]{Gao0N20}
Younan Gao, Meng He, and Yakov Nekrich.
\newblock Fast preprocessing for optimal orthogonal range reporting and range
  successor with applications to text indexing.
\newblock In {\em 28th Annual European Symposium on Algorithms, {ESA} 2020},
  volume 173 of {\em LIPIcs}, pages 54:1--54:18. Schloss
  Dagstuhl--Leibniz-Zentrum f{\"{u}}r Informatik, 2020.
\newblock \href {https://doi.org/10.4230/LIPICS.ESA.2020.54}
  {\path{doi:10.4230/LIPICS.ESA.2020.54}}.

\bibitem[GJKT24]{quantumlz}
Daniel Gibney, Ce~Jin, Tomasz Kociumaka, and Sharma~V. Thankachan.
\newblock Near-optimal quantum algorithms for bounded edit distance and
  {L}empel-{Z}iv factorization.
\newblock In {\em 35th Annual {ACM-SIAM} Symposium on Discrete Algorithms,
  {SODA} 2024}, pages 3302--3332, 2024.
\newblock \href {https://doi.org/10.1137/1.9781611977912.118}
  {\path{doi:10.1137/1.9781611977912.118}}.

\bibitem[GJL21]{balancing}
Moses Ganardi, Artur Jeż, and Markus Lohrey.
\newblock Balancing straight-line programs.
\newblock {\em Journal of the {ACM}}, 68(4):27:1--27:40, 2021.
\newblock \href {https://doi.org/10.1145/3457389} {\path{doi:10.1145/3457389}}.

\bibitem[GKK{\etalchar{+}}18]{dynstr}
Paweł Gawrychowski, Adam Karczmarz, Tomasz Kociumaka, Jakub Łącki, and Piotr
  Sankowski.
\newblock Optimal dynamic strings.
\newblock In Artur Czumaj, editor, {\em 29th Annual {ACM-SIAM} Symposium on
  Discrete Algorithms, {SODA} 2018}, pages 1509--1528. {SIAM}, 2018.
\newblock \href {https://doi.org/10.1137/1.9781611975031.99}
  {\path{doi:10.1137/1.9781611975031.99}}.

\bibitem[GKLS22]{GaneshKLS22}
Arun Ganesh, Tomasz Kociumaka, Andrea Lincoln, and Barna Saha.
\newblock How compression and approximation affect efficiency in string
  distance measures.
\newblock In Joseph~(Seffi) Naor and Niv Buchbinder, editors, {\em 33rd Annual
  {ACM-SIAM} Symposium on Discrete Algorithms, {SODA} 2022}, pages 2867--2919.
  {SIAM}, 2022.
\newblock \href {https://doi.org/10.1137/1.9781611977073.112}
  {\path{doi:10.1137/1.9781611977073.112}}.

\bibitem[GKM23]{GawrychowskiKM23}
Pawel Gawrychowski, Maria Kosche, and Florin Manea.
\newblock On the number of factors in the {LZ}-{E}nd factorization.
\newblock In Franco~Maria Nardini, Nadia Pisanti, and Rossano Venturini,
  editors, {\em 30th International Symposium on String Processing and
  Information Retrieval, {SPIRE} 2023}, volume 14240 of {\em LNCS}, pages
  253--259. Springer, 2023.
\newblock \href {https://doi.org/10.1007/978-3-031-43980-3\_20}
  {\path{doi:10.1007/978-3-031-43980-3\_20}}.

\bibitem[GNP18]{GNPlatin18}
Travis Gagie, Gonzalo Navarro, and Nicola Prezza.
\newblock On the approximation ratio of {L}empel-{Z}iv parsing.
\newblock In Michael~A. Bender, Martin Farach{-}Colton, and Miguel~A. Mosteiro,
  editors, {\em 13th Latin American Symposium on Theoretical Informatics,
  {LATIN} 2018}, volume 10807 of {\em LNCS}, pages 490--503. Springer, 2018.
\newblock \href {https://doi.org/10.1007/978-3-319-77404-6_36}
  {\path{doi:10.1007/978-3-319-77404-6_36}}.

\bibitem[GS04]{GusfieldS04}
Dan Gusfield and Jens Stoye.
\newblock Linear time algorithms for finding and representing all the tandem
  repeats in a string.
\newblock {\em Journal of Computer and System Sciences}, 69(4):525--546, 2004.
\newblock \href {https://doi.org/10.1016/J.JCSS.2004.03.004}
  {\path{doi:10.1016/J.JCSS.2004.03.004}}.

\bibitem[Gus97]{gusfield}
Dan Gusfield.
\newblock {\em Algorithms on Strings, Trees, and Sequences: {C}omputer Science
  and Computational Biology}.
\newblock Cambridge University Press, Cambridge, UK, 1997.
\newblock \href {https://doi.org/10.1017/cbo9780511574931}
  {\path{doi:10.1017/cbo9780511574931}}.

\bibitem[Hag98]{Hagerup98}
Torben Hagerup.
\newblock Sorting and searching on the word {RAM}.
\newblock In Michel Morvan, Christoph Meinel, and Daniel Krob, editors, {\em
  15th Annual Symposium on Theoretical Aspects of Computer Science, {STACS}
  1998}, volume 1373 of {\em LNCS}, pages 366--398. Springer, 1998.
\newblock \href {https://doi.org/10.1007/BFb0028575}
  {\path{doi:10.1007/BFb0028575}}.

\bibitem[HLLW13]{HermelinLLW13}
Danny Hermelin, Gad~M. Landau, Shir Landau, and Oren Weimann.
\newblock Unified compression-based acceleration of edit-distance computation.
\newblock {\em Algorithmica}, 65(2):339--353, 2013.
\newblock \href {https://doi.org/10.1007/s00453-011-9590-6}
  {\path{doi:10.1007/s00453-011-9590-6}}.

\bibitem[HLN22]{HanLN22}
Ling~Bo Han, Bin Lao, and Ge~Nong.
\newblock Succinct parallel {L}empel-{Z}iv factorization on a multicore
  computer.
\newblock {\em Journal of Supercomputing}, 78(5):7278--7303, 2022.
\newblock \href {https://doi.org/10.1007/S11227-021-04165-W}
  {\path{doi:10.1007/S11227-021-04165-W}}.

\bibitem[HRB23]{Hong0B23}
Aaron Hong, Massimiliano Rossi, and Christina Boucher.
\newblock {LZ77} via prefix-free parsing.
\newblock In Gonzalo Navarro and Julian Shun, editors, {\em 25th Symposium on
  Algorithm Engineering and Experiments, {ALENEX} 2023}, pages 123--134.
  {SIAM}, 2023.
\newblock \href {https://doi.org/10.1137/1.9781611977561.CH11}
  {\path{doi:10.1137/1.9781611977561.CH11}}.

\bibitem[HT84]{HT84}
Dov Harel and Robert~Endre Tarjan.
\newblock Fast algorithms for finding nearest common ancestors.
\newblock {\em SIAM Journal on Computing}, 13(2):338--355, 1984.
\newblock \href {https://doi.org/10.1137/0213024} {\path{doi:10.1137/0213024}}.

\bibitem[I17]{tomohiro-lce}
Tomohiro I.
\newblock Longest common extensions with recompression.
\newblock In Juha K{\"{a}}rkk{\"{a}}inen, Jakub Radoszewski, and Wojciech
  Rytter, editors, {\em 28th Annual Symposium on Combinatorial Pattern
  Matching, {CPM} 2017}, volume~78 of {\em LIPIcs}, pages 18:1--18:15. Schloss
  Dagstuhl--Leibniz-Zentrum f{\"{u}}r Informatik, 2017.
\newblock \href {https://doi.org/10.4230/LIPIcs.CPM.2017.18}
  {\path{doi:10.4230/LIPIcs.CPM.2017.18}}.

\bibitem[IEE04]{milestone}
IEEE.
\newblock Milestones: {Lempel-Ziv Data Compression Algorithm}, 1977, 2004.
\newblock URL:
  \url{https://ethw.org/Milestones:Lempel-Ziv_Data_Compression_Algorithm,_1977}.

\bibitem[IEE21]{medal}
IEEE.
\newblock {R}ecipients of {IEEE} {M}edal of {H}onor, 2021.
\newblock URL:
  \url{https://corporate-awards.ieee.org/recipients/ieee-medal-of-honor-recipients/}.

\bibitem[Jac89]{Jac89}
Guy Jacobson.
\newblock Space-efficient static trees and graphs.
\newblock In {\em 30th IEEE Annual Symposium on Foundations of Computer
  Science, {FOCS} 1989}, pages 549--554, 1989.
\newblock \href {https://doi.org/10.1109/SFCS.1989.63533}
  {\path{doi:10.1109/SFCS.1989.63533}}.

\bibitem[Je{\.{z}}15]{Jez2015}
Artur Je{\.{z}}.
\newblock Faster fully compressed pattern matching by recompression.
\newblock {\em ACM Transactions on Algorithms}, 11(3):20:1--20:43, 2015.
\newblock \href {https://doi.org/10.1145/2631920} {\path{doi:10.1145/2631920}}.

\bibitem[Je{\.{z}}16]{Jez16}
Artur Je{\.{z}}.
\newblock A really simple approximation of smallest grammar.
\newblock {\em Theoretical Computer Science}, 616:141--150, 2016.
\newblock \href {https://doi.org/10.1016/J.TCS.2015.12.032}
  {\path{doi:10.1016/J.TCS.2015.12.032}}.

\bibitem[Kem19]{Kempa19}
Dominik Kempa.
\newblock Optimal construction of compressed indexes for highly repetitive
  texts.
\newblock In Timothy~M. Chan, editor, {\em 30th Annual {ACM-SIAM} Symposium on
  Discrete Algorithms, {SODA} 2019}, pages 1344--1357. {SIAM}, 2019.
\newblock \href {https://doi.org/10.1137/1.9781611975482.82}
  {\path{doi:10.1137/1.9781611975482.82}}.

\bibitem[KK99]{KolpakovK99}
Roman~M. Kolpakov and Gregory Kucherov.
\newblock Finding maximal repetitions in a word in linear time.
\newblock In {\em 40th IEEE Annual Symposium on Foundations of Computer
  Science, {FOCS} 1999}, pages 596--604. {IEEE} Computer Society, 1999.
\newblock \href {https://doi.org/10.1109/SFFCS.1999.814634}
  {\path{doi:10.1109/SFFCS.1999.814634}}.

\bibitem[KK00]{KolpakovK00}
Roman~M. Kolpakov and Gregory Kucherov.
\newblock Finding repeats with fixed gap.
\newblock In Pablo de~la Fuente, editor, {\em 7th International Symposium on
  String Processing and Information Retrieval, {SPIRE} 2000}, pages 162--168.
  {IEEE} Computer Society, 2000.
\newblock \href {https://doi.org/10.1109/SPIRE.2000.878192}
  {\path{doi:10.1109/SPIRE.2000.878192}}.

\bibitem[KK03]{KolpakovK03}
Roman~M. Kolpakov and Gregory Kucherov.
\newblock Finding approximate repetitions under {H}amming distance.
\newblock {\em Theoretical Computer Science}, 303(1):135--156, 2003.
\newblock \href {https://doi.org/10.1016/S0304-3975(02)00448-6}
  {\path{doi:10.1016/S0304-3975(02)00448-6}}.

\bibitem[KK17a]{KempaK17b}
Dominik Kempa and Dmitry Kosolobov.
\newblock {LZ}-{E}nd parsing in compressed space.
\newblock In Ali Bilgin, Michael~W. Marcellin, Joan Serra{-}Sagrist{\`{a}}, and
  James~A. Storer, editors, {\em 2017 Data Compression Conference, {DCC} 2017},
  pages 350--359. {IEEE}, 2017.
\newblock \href {https://doi.org/10.1109/DCC.2017.73}
  {\path{doi:10.1109/DCC.2017.73}}.

\bibitem[KK17b]{KempaK17}
Dominik Kempa and Dmitry Kosolobov.
\newblock {LZ}-{E}nd parsing in linear time.
\newblock In Kirk Pruhs and Christian Sohler, editors, {\em 25th Annual
  European Symposium on Algorithms, {ESA} 2017}, volume~87 of {\em LIPIcs},
  pages 53:1--53:14. Schloss Dagstuhl--Leibniz-Zentrum f{\"{u}}r Informatik,
  2017.
\newblock \href {https://doi.org/10.4230/LIPICS.ESA.2017.53}
  {\path{doi:10.4230/LIPICS.ESA.2017.53}}.

\bibitem[KK19]{sss}
Dominik Kempa and Tomasz Kociumaka.
\newblock String synchronizing sets: Sublinear-time {BWT} construction and
  optimal {LCE} data structure.
\newblock In Moses Charikar and Edith Cohen, editors, {\em 51st Annual {ACM}
  {SIGACT} Symposium on Theory of Computing, {STOC} 2019}, pages 756--767.
  {ACM}, 2019.
\newblock \href {https://doi.org/10.1145/3313276.3316368}
  {\path{doi:10.1145/3313276.3316368}}.

\bibitem[KK20]{resolution}
Dominik Kempa and Tomasz Kociumaka.
\newblock Resolution of the {B}urrows-{W}heeler {T}ransform conjecture.
\newblock In Sandy Irani, editor, {\em 61st {IEEE} Annual Symposium on
  Foundations of Computer Science, {FOCS} 2020}, pages 1002--1013. {IEEE}
  Computer Society, 2020.
\newblock \href {https://doi.org/10.1109/FOCS46700.2020.00097}
  {\path{doi:10.1109/FOCS46700.2020.00097}}.

\bibitem[KK23a]{breaking}
Dominik Kempa and Tomasz Kociumaka.
\newblock Breaking the ${O(n)}$-barrier in the construction of compressed
  suffix arrays and suffix trees.
\newblock In Nikhil Bansal and Viswanath Nagarajan, editors, {\em 34th Annual
  {ACM-SIAM} Symposium on Discrete Algorithms, SODA 2023}, pages 5122--5202.
  {SIAM}, 2023.
\newblock \href {https://doi.org/10.1137/1.9781611977554.ch187}
  {\path{doi:10.1137/1.9781611977554.ch187}}.

\bibitem[KK23b]{collapsing}
Dominik Kempa and Tomasz Kociumaka.
\newblock Collapsing the hierarchy of compressed data structures: Suffix arrays
  in optimal compressed space.
\newblock In {\em 64th {IEEE} Annual Symposium on Foundations of Computer
  Science, {FOCS} 2023}, pages 1877--1886. {IEEE}, 2023.
\newblock \href {https://doi.org/10.1109/FOCS57990.2023.00114}
  {\path{doi:10.1109/FOCS57990.2023.00114}}.

\bibitem[KKP13a]{KarkkainenKP13}
Juha K{\"{a}}rkk{\"{a}}inen, Dominik Kempa, and Simon~J. Puglisi.
\newblock Lightweight {L}empel-{Z}iv parsing.
\newblock In Vincenzo Bonifaci, Camil Demetrescu, and Alberto
  Marchetti{-}Spaccamela, editors, {\em 12th International Symposium on
  Experimental Algorithms, SEA 2013}, volume 7933 of {\em LNCS}, pages
  139--150. Springer, 2013.
\newblock \href {https://doi.org/10.1007/978-3-642-38527-8\_14}
  {\path{doi:10.1007/978-3-642-38527-8\_14}}.

\bibitem[KKP13b]{KarkkainenKP13b}
Juha K{\"{a}}rkk{\"{a}}inen, Dominik Kempa, and Simon~J. Puglisi.
\newblock Linear time {L}empel-{Z}iv factorization: Simple, fast, small.
\newblock In Johannes Fischer and Peter Sanders, editors, {\em 24th Annual
  Symposium on Combinatorial Pattern Matching, {CPM} 2013}, volume 7922 of {\em
  LNCS}, pages 189--200. Springer, 2013.
\newblock \href {https://doi.org/10.1007/978-3-642-38905-4\_19}
  {\path{doi:10.1007/978-3-642-38905-4\_19}}.

\bibitem[KKP14]{KarkkainenKP14}
Juha K{\"{a}}rkk{\"{a}}inen, Dominik Kempa, and Simon~J. Puglisi.
\newblock {L}empel-{Z}iv parsing in external memory.
\newblock In Ali Bilgin, Michael~W. Marcellin, Joan Serra{-}Sagrist{\`{a}}, and
  James~A. Storer, editors, {\em 2014 Data Compression Conference, {DCC} 2014},
  pages 153--162. {IEEE}, 2014.
\newblock \href {https://doi.org/10.1109/DCC.2014.78}
  {\path{doi:10.1109/DCC.2014.78}}.

\bibitem[KKR{\etalchar{+}}20]{KociumakaKRRW20}
Tomasz Kociumaka, Marcin Kubica, Jakub Radoszewski, Wojciech Rytter, and Tomasz
  Waleń.
\newblock A linear-time algorithm for seeds computation.
\newblock {\em ACM Transactions on Algorithms}, 16(2):27:1--27:23, 2020.
\newblock \href {https://doi.org/10.1145/3386369} {\path{doi:10.1145/3386369}}.

\bibitem[KMS{\etalchar{+}}03]{collage}
Takuya Kida, Tetsuya Matsumoto, Yusuke Shibata, Masayuki Takeda, Ayumi
  Shinohara, and Setsuo Arikawa.
\newblock Collage system: A unifying framework for compressed pattern matching.
\newblock {\em Theoretical Computer Science}, 298(1):253--272, 2003.
\newblock \href {https://doi.org/10.1016/S0304-3975(02)00426-7}
  {\path{doi:10.1016/S0304-3975(02)00426-7}}.

\bibitem[KN10]{kreft2010navarro}
Sebastian Kreft and Gonzalo Navarro.
\newblock {LZ}77-like compression with fast random access.
\newblock In {\em 2010 Data Compression Conference, DCC 2010}, pages 239--248.
  {IEEE} Computer Society, 2010.
\newblock \href {https://doi.org/10.1109/DCC.2010.29}
  {\path{doi:10.1109/DCC.2010.29}}.

\bibitem[KN13]{KreftN13}
Sebastian Kreft and Gonzalo Navarro.
\newblock On compressing and indexing repetitive sequences.
\newblock {\em Theoretical Computer Science}, 483:115--133, 2013.
\newblock \href {https://doi.org/10.1016/J.TCS.2012.02.006}
  {\path{doi:10.1016/J.TCS.2012.02.006}}.

\bibitem[KNO22]{KociumakaNO22}
Tomasz Kociumaka, Gonzalo Navarro, and Francisco Olivares.
\newblock Near-optimal search time in {\(\delta\)}-optimal space, and vice
  versa.
\newblock {\em Algorithmica}, 13568(4):1031--1056, 2022.
\newblock \href {https://doi.org/10.1007/S00453-023-01186-0}
  {\path{doi:10.1007/S00453-023-01186-0}}.

\bibitem[KNP23]{delta}
Tomasz Kociumaka, Gonzalo Navarro, and Nicola Prezza.
\newblock Towards a definitive compressibility measure for repetitive
  sequences.
\newblock {\em IEEE Transactions on Information Theory}, 69(4):2074--2092,
  2023.
\newblock \href {https://doi.org/10.1109/TIT.2022.3224382}
  {\path{doi:10.1109/TIT.2022.3224382}}.

\bibitem[Kos15a]{Kosolobov15}
Dmitry Kosolobov.
\newblock Faster lightweight {L}empel-{Z}iv parsing.
\newblock In Giuseppe~F. Italiano, Giovanni Pighizzini, and Donald Sannella,
  editors, {\em 40th International Symposium on Mathematical Foundations of
  Computer Science, {MFCS} 2015}, volume 9235 of {\em LNCS}, pages 432--444.
  Springer, 2015.
\newblock \href {https://doi.org/10.1007/978-3-662-48054-0\_36}
  {\path{doi:10.1007/978-3-662-48054-0\_36}}.

\bibitem[Kos15b]{Kosolobov15b}
Dmitry Kosolobov.
\newblock {L}empel-{Z}iv factorization may be harder than computing all runs.
\newblock In Ernst~W. Mayr and Nicolas Ollinger, editors, {\em 32nd
  International Symposium on Theoretical Aspects of Computer Science, {STACS}
  2015}, volume~30 of {\em LIPIcs}, pages 582--593. Schloss
  Dagstuhl--Leibniz-Zentrum f{\"{u}}r Informatik, 2015.
\newblock \href {https://doi.org/10.4230/LIPICS.STACS.2015.582}
  {\path{doi:10.4230/LIPICS.STACS.2015.582}}.

\bibitem[KP13]{KempaP13}
Dominik Kempa and Simon~J. Puglisi.
\newblock {L}empel-{Z}iv factorization: Simple, fast, practical.
\newblock In Peter Sanders and Norbert Zeh, editors, {\em 15th Meeting on
  Algorithm Engineering and Experiments, {ALENEX} 2013}, pages 103--112.
  {SIAM}, 2013.
\newblock \href {https://doi.org/10.1137/1.9781611972931.9}
  {\path{doi:10.1137/1.9781611972931.9}}.

\bibitem[KP18]{attractors}
Dominik Kempa and Nicola Prezza.
\newblock At the roots of dictionary compression: String attractors.
\newblock In Ilias Diakonikolas, David Kempe, and Monika Henzinger, editors,
  {\em 50th Annual {ACM} {SIGACT} Symposium on Theory of Computing, {STOC}
  2018}, pages 827--840. {ACM}, 2018.
\newblock \href {https://doi.org/10.1145/3188745.3188814}
  {\path{doi:10.1145/3188745.3188814}}.

\bibitem[KS16]{KopplS16}
Dominik K{\"{o}}ppl and Kunihiko Sadakane.
\newblock {L}empel-{Z}iv computation in compressed space {(LZ-CICS)}.
\newblock In Ali Bilgin, Michael~W. Marcellin, Joan Serra{-}Sagrist{\`{a}}, and
  James~A. Storer, editors, {\em 2016 Data Compression Conference, {DCC} 2016},
  pages 3--12. {IEEE}, 2016.
\newblock \href {https://doi.org/10.1109/DCC.2016.38}
  {\path{doi:10.1109/DCC.2016.38}}.

\bibitem[KS22]{KempaS22}
Dominik Kempa and Barna Saha.
\newblock An upper bound and linear-space queries on the {LZ}-end parsing.
\newblock In Joseph~(Seffi) Naor and Niv Buchbinder, editors, {\em 33rd Annual
  {ACM-SIAM} Symposium on Discrete Algorithms, {SODA} 2022}, pages 2847--2866.
  {SIAM}, 2022.
\newblock \href {https://doi.org/10.1137/1.9781611977073.111}
  {\path{doi:10.1137/1.9781611977073.111}}.

\bibitem[KVNP20]{KosolobovVNP20}
Dmitry Kosolobov, Daniel Valenzuela, Gonzalo Navarro, and Simon~J. Puglisi.
\newblock {L}empel-{Z}iv-like parsing in small space.
\newblock {\em Algorithmica}, 82(11):3195--3215, 2020.
\newblock \href {https://doi.org/10.1007/s00453-020-00722-6}
  {\path{doi:10.1007/s00453-020-00722-6}}.

\bibitem[KW05]{KleinW05}
Shmuel~Tomi Klein and Yair Wiseman.
\newblock Parallel {L}empel {Z}iv coding.
\newblock {\em Discrete Applied Mathematics}, 146(2):180--191, 2005.
\newblock \href {https://doi.org/10.1016/J.DAM.2004.04.013}
  {\path{doi:10.1016/J.DAM.2004.04.013}}.

\bibitem[Lar14]{Larsson14}
N.~Jesper Larsson.
\newblock Most recent match queries in on-line suffix trees.
\newblock In Alexander~S. Kulikov, Sergei~O. Kuznetsov, and Pavel~A. Pevzner,
  editors, {\em 25th Annual Symposium on Combinatorial Pattern Matching, {CPM}
  2014}, volume 8486 of {\em LNCS}, pages 252--261. Springer, 2014.
\newblock \href {https://doi.org/10.1007/978-3-319-07566-2\_26}
  {\path{doi:10.1007/978-3-319-07566-2\_26}}.

\bibitem[LNCW16]{LiuNCW16}
Weijun Liu, Ge~Nong, Wai~Hong Chan, and Yi~Wu.
\newblock Improving a lightweight {LZ77} computation algorithm for running
  faster.
\newblock {\em Software: Practice and Experience}, 46(9):1201--1217, 2016.
\newblock \href {https://doi.org/10.1002/SPE.2377}
  {\path{doi:10.1002/SPE.2377}}.

\bibitem[LZ76]{LZ76}
Abraham Lempel and Jacob Ziv.
\newblock On the complexity of finite sequences.
\newblock {\em IEEE Transactions on Information Theory}, 22(1):75--81, 1976.
\newblock \href {https://doi.org/10.1109/TIT.1976.1055501}
  {\path{doi:10.1109/TIT.1976.1055501}}.

\bibitem[Mah]{ltcb}
Matt Mahoney.
\newblock {L}arge {T}ext {C}ompression {B}enchmark.
\newblock Accessed: 2024-03-20.
\newblock URL: \url{http://mattmahoney.net/dc/text.html}.

\bibitem[Mai89]{main1989detecting}
Michael~G Main.
\newblock Detecting leftmost maximal periodicities.
\newblock {\em Discrete Applied Mathematics}, 25(1-2):145--153, 1989.
\newblock \href {https://doi.org/10.1016/0166-218X(89)90051-6}
  {\path{doi:10.1016/0166-218X(89)90051-6}}.

\bibitem[MNN17]{MunroNN17}
J.~Ian Munro, Gonzalo Navarro, and Yakov Nekrich.
\newblock Space-efficient construction of compressed indexes in deterministic
  linear time.
\newblock In Philip~N. Klein, editor, {\em 28th Annual {ACM-SIAM} Symposium on
  Discrete Algorithms, {SODA} 2017}, pages 408--424. {SIAM}, 2017.
\newblock \href {https://doi.org/10.1137/1.9781611974782.26}
  {\path{doi:10.1137/1.9781611974782.26}}.

\bibitem[MNV16]{MunroNV16}
J.~Ian Munro, Yakov Nekrich, and Jeffrey~Scott Vitter.
\newblock Fast construction of wavelet trees.
\newblock {\em Theoretical Computer Science}, 638:91--97, 2016.
\newblock \href {https://doi.org/10.1016/j.tcs.2015.11.011}
  {\path{doi:10.1016/j.tcs.2015.11.011}}.

\bibitem[Nao91]{Naor91}
Moni Naor.
\newblock String matching with preprocessing of text and pattern.
\newblock In Javier~Leach Albert, Burkhard Monien, and Mario
  Rodr{\'{\i}}guez{-}Artalejo, editors, {\em 18th International Colloquium on
  Automata, Languages and Programming, ICALP 1991}, volume 510 of {\em LNCS},
  pages 739--750. Springer, 1991.
\newblock \href {https://doi.org/10.1007/3-540-54233-7\_179}
  {\path{doi:10.1007/3-540-54233-7\_179}}.

\bibitem[Nav21a]{NavarroMeasures}
Gonzalo Navarro.
\newblock Indexing highly repetitive string collections, part {I}:
  Repetitiveness measures.
\newblock {\em ACM Computing Surveys}, 54(2):29:1--29:31, 2021.
\newblock \href {https://doi.org/10.1145/3434399} {\path{doi:10.1145/3434399}}.

\bibitem[Nav21b]{NavarroIndexes}
Gonzalo Navarro.
\newblock Indexing highly repetitive string collections, part {II}: Compressed
  indexes.
\newblock {\em ACM Computing Surveys}, 54(2):26:1--26:32, 2021.
\newblock \href {https://doi.org/10.1145/3432999} {\path{doi:10.1145/3432999}}.

\bibitem[Nek21]{Nekrich21}
Yakov Nekrich.
\newblock New data structures for orthogonal range reporting and range minima
  queries.
\newblock In D{\'{a}}niel Marx, editor, {\em 32nd Annual {ACM-SIAM} Symposium
  on Discrete Algorithms, {SODA} 2021}, pages 1191--1205. {SIAM}, 2021.
\newblock \href {https://doi.org/10.1137/1.9781611976465.73}
  {\path{doi:10.1137/1.9781611976465.73}}.

\bibitem[NII{\etalchar{+}}16]{NishimotoMFCS}
Takaaki Nishimoto, Tomohiro I, Shunsuke Inenaga, Hideo Bannai, and Masayuki
  Takeda.
\newblock Fully dynamic data structure for {LCE} queries in compressed space.
\newblock In Piotr Faliszewski, Anca Muscholl, and Rolf Niedermeier, editors,
  {\em 41st International Symposium on Mathematical Foundations of Computer
  Science, {MFCS} 2016}, volume~58 of {\em LIPIcs}, pages 72:1--72:15. Schloss
  Dagstuhl--Leibniz-Zentrum f{\"{u}}r Informatik, 2016.
\newblock \href {https://doi.org/10.4230/LIPIcs.MFCS.2016.72}
  {\path{doi:10.4230/LIPIcs.MFCS.2016.72}}.

\bibitem[NII{\etalchar{+}}20]{NishimotoIIBT20}
Takaaki Nishimoto, Tomohiro I, Shunsuke Inenaga, Hideo Bannai, and Masayuki
  Takeda.
\newblock Dynamic index and {LZ} factorization in compressed space.
\newblock {\em Discrete Applied Mathematics}, 274:116--129, 2020.
\newblock \href {https://doi.org/10.1016/J.DAM.2019.01.014}
  {\path{doi:10.1016/J.DAM.2019.01.014}}.

\bibitem[OG11]{OhlebuschG11}
Enno Ohlebusch and Simon Gog.
\newblock {L}empel-{Z}iv factorization revisited.
\newblock In Raffaele Giancarlo and Giovanni Manzini, editors, {\em 22nd Annual
  Symposium on Combinatorial Pattern Matching, {CPM} 2011}, volume 6661 of {\em
  LNCS}, pages 15--26. Springer, 2011.
\newblock \href {https://doi.org/10.1007/978-3-642-21458-5\_4}
  {\path{doi:10.1007/978-3-642-21458-5\_4}}.

\bibitem[ON19]{OchoaN19}
Carlos Ochoa and Gonzalo Navarro.
\newblock {RePair} and all irreducible grammars are upper bounded by high-order
  empirical entropy.
\newblock {\em {IEEE} Transactions on Information Theory}, 65(5):3160--3164,
  2019.
\newblock \href {https://doi.org/10.1109/TIT.2018.2871452}
  {\path{doi:10.1109/TIT.2018.2871452}}.

\bibitem[OS08]{OkanoharaS08}
Daisuke Okanohara and Kunihiko Sadakane.
\newblock An online algorithm for finding the longest previous factors.
\newblock In Dan Halperin and Kurt Mehlhorn, editors, {\em 16th Annual European
  Symposium on Algorithms, {ESA} 2008}, volume 5193 of {\em LNCS}, pages
  696--707. Springer, 2008.
\newblock \href {https://doi.org/10.1007/978-3-540-87744-8\_58}
  {\path{doi:10.1007/978-3-540-87744-8\_58}}.

\bibitem[OS11]{OzsoyS11}
Adnan Ozsoy and D.~Martin Swany.
\newblock {CULZSS:} {LZSS} lossless data compression on {CUDA}.
\newblock In {\em 2011 {IEEE} International Conference on Cluster Computing,
  CLUSTER 2011}, pages 403--411. {IEEE} Computer Society, 2011.
\newblock \href {https://doi.org/10.1109/CLUSTER.2011.52}
  {\path{doi:10.1109/CLUSTER.2011.52}}.

\bibitem[OSC14]{OzsoySC14}
Adnan Ozsoy, D.~Martin Swany, and Arun Chauhan.
\newblock Optimizing {LZSS} compression on {GPGPU}s.
\newblock {\em Future Generation Computer System}, 30:170--178, 2014.
\newblock \href {https://doi.org/10.1016/J.FUTURE.2013.06.022}
  {\path{doi:10.1016/J.FUTURE.2013.06.022}}.

\bibitem[PNB17]{PereiraNB17}
Alberto~Ord{\'{o}}{\~{n}}ez Pereira, Gonzalo Navarro, and Nieves~R. Brisaboa.
\newblock Grammar compressed sequences with rank/select support.
\newblock {\em Journal of Discrete Algorithms}, 43:54--71, 2017.
\newblock \href {https://doi.org/10.1016/j.jda.2016.10.001}
  {\path{doi:10.1016/j.jda.2016.10.001}}.

\bibitem[PP15]{PolicritiP15}
Alberto Policriti and Nicola Prezza.
\newblock Fast online {L}empel-{Z}iv factorization in compressed space.
\newblock In Costas~S. Iliopoulos, Simon~J. Puglisi, and Emine Yilmaz, editors,
  {\em 22nd International Symposium on String Processing and Information
  Retrieval, {SPIRE} 2015}, volume 9309 of {\em LNCS}, pages 13--20. Springer,
  2015.
\newblock \href {https://doi.org/10.1007/978-3-319-23826-5\_2}
  {\path{doi:10.1007/978-3-319-23826-5\_2}}.

\bibitem[Pre19]{Prezza19}
Nicola Prezza.
\newblock Optimal rank and select queries on dictionary-compressed text.
\newblock In Nadia Pisanti and Solon~P. Pissis, editors, {\em 30th Annual
  Symposium on Combinatorial Pattern Matching, {CPM} 2019}, volume 128 of {\em
  LIPIcs}, pages 4:1--4:12. Schloss Dagstuhl--Leibniz-Zentrum f{\"{u}}r
  Informatik, 2019.
\newblock \href {https://doi.org/10.4230/LIPIcs.CPM.2019.4}
  {\path{doi:10.4230/LIPIcs.CPM.2019.4}}.

\bibitem[RPE81]{RodehPE81}
Michael Rodeh, Vaughan~R. Pratt, and Shimon Even.
\newblock Linear algorithm for data compression via string matching.
\newblock {\em Journal of the ACM}, 28(1):16--24, 1981.
\newblock \href {https://doi.org/10.1145/322234.322237}
  {\path{doi:10.1145/322234.322237}}.

\bibitem[Ryt03]{Rytter03}
Wojciech Rytter.
\newblock Application of {L}empel--{Z}iv factorization to the approximation of
  grammar-based compression.
\newblock {\em Theoretical Computer Science}, 302(1--3):211--222, 2003.
\newblock \href {https://doi.org/10.1016/S0304-3975(02)00777-6}
  {\path{doi:10.1016/S0304-3975(02)00777-6}}.

\bibitem[SS82]{macro}
James~A. Storer and Thomas~G. Szymanski.
\newblock Data compression via textual substitution.
\newblock {\em Journal of the ACM}, 29(4):928--951, 1982.
\newblock \href {https://doi.org/10.1145/322344.322346}
  {\path{doi:10.1145/322344.322346}}.

\bibitem[Sta12]{Starikovskaya12}
Tatiana Starikovskaya.
\newblock Computing {L}empel-{Z}iv factorization online.
\newblock In Branislav Rovan, Vladimiro Sassone, and Peter Widmayer, editors,
  {\em 37th International Symposium on Mathematical Foundations of Computer
  Science, {MFCS} 2012}, volume 7464 of {\em LNCS}, pages 789--799. Springer,
  2012.
\newblock \href {https://doi.org/10.1007/978-3-642-32589-2\_68}
  {\path{doi:10.1007/978-3-642-32589-2\_68}}.

\bibitem[SZ13]{ShunZ13}
Julian Shun and Fuyao Zhao.
\newblock Practical parallel {L}empel-{Z}iv factorization.
\newblock In Ali Bilgin, Michael~W. Marcellin, Joan Serra{-}Sagrist{\`{a}}, and
  James~A. Storer, editors, {\em 2013 Data Compression Conference, {DCC} 2013},
  pages 123--132. {IEEE}, 2013.
\newblock \href {https://doi.org/10.1109/DCC.2013.20}
  {\path{doi:10.1109/DCC.2013.20}}.

\bibitem[Tis15]{Tiskin15}
Alexander Tiskin.
\newblock Fast distance multiplication of unit-{M}onge matrices.
\newblock {\em Algorithmica}, 71(4):859--888, 2015.
\newblock \href {https://doi.org/10.1007/s00453-013-9830-z}
  {\path{doi:10.1007/s00453-013-9830-z}}.

\bibitem[Val16]{Valenzuela16}
Daniel Valenzuela.
\newblock {CHICO:} {A} compressed hybrid index for repetitive collections.
\newblock In Andrew~V. Goldberg and Alexander~S. Kulikov, editors, {\em 15th
  International Symposium on Experimental Algorithms, SEA 2016}, volume 9685 of
  {\em LNCS}, pages 326--338. Springer, 2016.
\newblock \href {https://doi.org/10.1007/978-3-319-38851-9\_22}
  {\path{doi:10.1007/978-3-319-38851-9\_22}}.

\bibitem[Wei73]{Weiner73}
Peter Weiner.
\newblock Linear pattern matching algorithms.
\newblock In {\em 14th Annual Symposium on Switching and Automata Theory, {SWAT
  (FOCS)} 1973}, pages 1--11. {IEEE} Computer Society, 1973.
\newblock \href {https://doi.org/10.1109/SWAT.1973.13}
  {\path{doi:10.1109/SWAT.1973.13}}.

\bibitem[YIB{\etalchar{+}}14]{YamamotoIBIT14}
Jun{-}ichi Yamamoto, Tomohiro I, Hideo Bannai, Shunsuke Inenaga, and Masayuki
  Takeda.
\newblock Faster compact on-line {L}empel-{Z}iv factorization.
\newblock In Ernst~W. Mayr and Natacha Portier, editors, {\em 31st
  International Symposium on Theoretical Aspects of Computer Science, {STACS}
  2014}, volume~25 of {\em LIPIcs}, pages 675--686. Schloss
  Dagstuhl--Leibniz-Zentrum f{\"{u}}r Informatik, 2014.
\newblock \href {https://doi.org/10.4230/LIPICS.STACS.2014.675}
  {\path{doi:10.4230/LIPICS.STACS.2014.675}}.

\bibitem[ZH14]{ZuH14}
Yuan Zu and Bei Hua.
\newblock {GLZSS:} {LZSS} lossless data compression can be faster.
\newblock In John Cavazos, Xiang Gong, and David~R. Kaeli, editors, {\em 7th
  Workshop on General Purpose Processing Using GPUs, GPGPU 2014}, page~46.
  {ACM}, 2014.
\newblock URL: \url{https://dl.acm.org/citation.cfm?id=2576785}.

\bibitem[ZL77]{LZ77}
Jacob Ziv and Abraham Lempel.
\newblock A universal algorithm for sequential data compression.
\newblock {\em IEEE Transactions on Information Theory}, 23(3):337--343, 1977.
\newblock \href {https://doi.org/10.1109/TIT.1977.1055714}
  {\path{doi:10.1109/TIT.1977.1055714}}.

\bibitem[ZL78]{LZ78}
Jacob Ziv and Abraham Lempel.
\newblock Compression of individual sequences via variable-rate coding.
\newblock {\em IEEE Transactions on Information Theory}, 24(5):530--536, 1978.
\newblock \href {https://doi.org/10.1109/TIT.1978.1055934}
  {\path{doi:10.1109/TIT.1978.1055934}}.

\end{thebibliography}
